\documentclass[epsf,psfig]{article}

\usepackage{makeidx}   % allows index generation
\usepackage{graphicx}  % standard LaTeX graphics tool
\usepackage{subeqnar}  % subnumbers individual equations
\usepackage{multicol}  % used for the two-column index
\usepackage{cl2emonoNP}  % main Springer style for monographs
\usepackage{physmobb}  % centered layout of captions etc.
\makeindex             % used for the subject index
\usepackage{epsf}
\usepackage{psfig}

\def\mxth{\mathsurround=0pt }
\def\xversim#1#2{\lower2.pt\vbox{\baselineskip0pt \lineskip-.5pt
  \ialign{$\mxth#1\hfil##\hfil$\crcr#2\crcr\sim\crcr}}}             
\def\gtrsim{\mathrel{\mathpalette\xversim >}}                                    
\def\lesssim{\mathrel{\mathpalette\xversim <}}    
\newcommand{\postscript}[2]
{\setlength{\epsfxsize}{#2\hsize}
\centerline{\epsfbox{#1}}}
\def\be{\begin{equation}}
\def\ee{\end{equation}}
\def\bea{\begin{eqnarray}}
\def\eea{\end{eqnarray}}
\def\ie{{\it i.e.}, }
\def\mweak{M_{\rm Weak}}
\def\mplanck{M_{\rm Planck}}
\def\hc{\it {h.c.}}
\newcommand{\mrU}{{U}^{\prime}}
\newcommand{\mrD}{{D}^{\prime}}
\newcommand{\mrQ}{{Q}^{\prime}}
\newcommand{\mrL}{{L}^{\prime}}
\newcommand{\mrE}{{E}^{\prime}}

\newcommand{\sU}{{\widetilde{U}}}
\newcommand{\sD}{{\widetilde{D}}}
\newcommand{\sQ}{{\widetilde{Q}}}
\newcommand{\sL}{{\widetilde{L}}}
\newcommand{\sE}{{\widetilde{E}}}

\newcommand{\mrsU}{{\widetilde{U^{\prime}}}}
\newcommand{\mrsD}{{\widetilde{D^{\prime}}}}
\newcommand{\mrsQ}{{\widetilde{Q^{\prime}}}}
\newcommand{\mrsL}{{\widetilde{L^{\prime}}}}
\newcommand{\mrsE}{{\widetilde{E^{\prime}}}}

\newcommand{\hu}{{H}_{2}}
\newcommand{\hd}{{H}_{1}}
\newcommand{\mrhu}{{H}^{\prime}_{2}}
\newcommand{\mrhd}{{H}^{\prime}_{1}}
\newcommand{\hinou}{{\widetilde{H}}_{2}}
\newcommand{\hinod}{{\widetilde{H}}_{1}}
\newcommand{\mrhinou}{{\widetilde{H}}^{\prime}_{2}}
\newcommand{\mrhinod}{{\widetilde{H}}^{\prime}_{1}}
\newcommand{\bino}{\widetilde{B}}
\newcommand{\wino}{\widetilde{W}}
\newcommand{\gluino}{\widetilde{g}}
\newcommand{\adjoint}{\Phi_{V}}
\newcommand{\sadjoint}{\phi_{V}}
\newcommand{\fadjoint}{\psi_{V}}

\newcommand{\gadjoint}{\Phi_{g}}
\newcommand{\gsadjoint}{\phi_{g}}
\newcommand{\gfadjoint}{\psi_{g}}

\newcommand{\wadjoint}{\Phi_{W}}
\newcommand{\wsadjoint}{\phi_{W}}
\newcommand{\wfadjoint}{\psi_{W}}

\newcommand{\badjoint}{\Phi_{B}}
\newcommand{\bsadjoint}{\phi_{B}}
\newcommand{\bfadjoint}{\psi_{B}}

\begin{document}

%%Frontmatter%%%%%%%%%%%%%%%%%%%%%%%%%%%%%%%%%%%%%%%%%%%%%%%%%%%%

\author{Nir Polonsky }
\title{
Supersymmetry: \\ Structure and  Phenomena\\
%{\small SPIN Springer's internal project number, if known}
}
%\subtitle{Physics -- Monograph\\
%(Editorial W. Beiglb\"ock)}
\subtitle{Published by Springer-Verlag Heidelberg \\
(Editorial W. Beiglb\"ock): \\  
Lecture Notes in Physics Monographs \\ Vol. m 68: N. Polonsky \\
"Supersymmetry: Structure and Phenomena -  \\
Extensions of the Standard Model", \\
XV, 169 pages, 2001.}

%\date{May, 2001}
\date{MIT-CTP-3164}

\maketitle

%ROMAN
\pagenumbering{Roman}
%SET COUNTER
\setcounter{page}{5}

\thispagestyle{empty}
%\vspace*{3.5cm}
\clearemptydoublepage

\chapter*{Preface}
%% * means preface does not appear in Contents
\markboth{Preface}{Preface}

\begin{flushleft}
{\it Moor: point\\
my horse\\
where birds sing.\\}
\vspace*{0.5cm}
Basho (Translation by Lucien Stryk)
\end{flushleft}
\vspace*{1.5cm}

\noindent
%01
These notes provide an introductory
yet comprehensive discussion of the phenomenology associated
with supersymmetric extensions of the Standard Model of electroweak
and strong interactions.
The choice of topics and of their presentation meant to draw
readers with various backgrounds and research interests, 
and to provide a pedagogical tour of  
supersymmetry,  structure and phenomena. 
The intensive and ongoing efforts to understand and discover such
phenomena, or to alternatively falsify the paradigm predicting it, 
and the central role of supersymmetry in particle physics studies, 
render such a journey a well motivated endeavor --  if not a necessity.
This is particularly true as we look forward to embarking into a new era
of discoveries with the forthcoming commissioning of CERN's Large Hadron Collider
(LHC), which is scheduled to operate by the year 2005.
%

%02
Even though our focus is on 
(weak-scale) supersymmetry, it would still be impossible to cover 
within these notes all that it entails.
In the following, we choose to stress the more elementary details over the
technicalities, the physical picture
over its formal derivation. While some issues are covered in
depth, many others are only sketched or referred to, leaving the
details to either optional exercises or further reading.
Our mission is to gain acquaintance with the fundamental picture
(regardless of the reader's background),
and we will attempt to do so in an intuitive  fashion while providing
a broad and honest review of the related issues.
Many reviews, lecture notes, and text books 
(most notably, Weinberg's book \cite{WEINBERG})
are available and references will be given to some of those, as well as to some
early and recent research papers on the subjects discussed.
Some areas are particularly well covered by reviews, e.g., 
the algebra on the one hand and experimental searches on the other hand, 
and hence will not be explored extensively here. 
%

%03
Before discussing supersymmetry and the ``TeV World'',
it is useful to recall and establish our starting point: The Standard Model
of electroweak and strong interactions as characterized and tested
at currently available energies $\sqrt{s} \lesssim$ 200-300 GeV
(per parton).  We therefore begin with a brief summary 
of the main ingredients of the SM.
Hints for an additional structure within experimental reach will be emphasized,
and the known theoretical possibilities
for such a structure, particularly supersymmetry, will be introduced.
This will be done in the first part of the notes.
The second part provides a phenomenological
``bottom-up'' construction of supersymmetry, followed by a brief
discussion of the formalism -- providing intuitive understanding as
well as the necessary terminology. We then supersymmetrize the
Standard Model, paying attention to some of the details 
while only touching upon other.
%

%04
In the third part of the notes we turn
to an exploration of a small number of issues
which relate ultraviolet and infrared structures. These include
renormalization and radiative symmetry breaking, unification,
and the lightness of the Higgs boson, 
issues which underlie the realization of the weak scale.
This particular choice of topics does not only lay the foundation
to supersymmetry phenomenology but also represents some of the successful
predictions of weak-scale supersymmetry.
Subsequently, we discuss in the fourth part
some of the difficulties in extending the standard model, 
the most important of which is the flavor problem.
Its possible resolutions are used as an organizing
principle for the model space, which is then characterized and described.
%

%05 
Throughout these notes we attempt to leave the reader
with a clear view of current research directions and avenues
which stem from the above topics. Some more technical and specialized
issues are explored further in the fifth and last part of the notes.
These include the possible origins of neutrino mass
and mixing and the implications of the different scenarios; 
vacuum stability and the respective constraints;
classifications of supersymmetry breaking operators; 
and an introductory discussion of extended supersymmetry.
These serve as a (subjective) sample of contemporary and potential
avenues of research. The reader is encouraged to explore 
the literature given throughout
the notes for technical discussions of other topics of interest.
%

%06
As physicists, our interest is in
confronting theory with experiment.
Here, we will argue for and motivate supersymmetry
as a reasonable and likely
perturbative extension of the standard model.
Whether supersymmetry is realized in nature, and if so, 
in what form, can and must be determined by experiment.
Experimental efforts in this direction are ongoing
and will hopefully result in an answer or a clear indication within the next decade.
Discovery will only signal the start of a new era 
dedicated to deciphering the ultraviolet theory from the data:
Supersymmetry, by its nature (and as we will demonstrate), relates
the infrared and the ultraviolet regimes.
By doing so it provides us with extraordinary opportunities for a glimpse
at scales we cannot reach and
to  answer fundamental questions regarding
unification, gravity, and much more.
The more one is  familiar with the ultraviolet possibilities
and their infrared implications, the more one can direct
experiment towards a possible discovery.

%07
Clearly, a lot of work is yet to be
done and there is always room for new  ideas to be put forward.
It is my hope that these notes
convince and enable one to follow (if not participate in) the story of 
weak-scale supersymmetry, and new physics in general, as it unveils.
It is not our aim to equip the reader with calculation tools or to
review in these notes each and every aspect or possibility. Rather, we will attempt to
enable  the interested reader to form an educated opinion and identify areas of
interest for further study and exploration. Many of the issues
discussed will be presented from an angle which differs from 
other presentations in the literature,
for the potential benefit of both the novice and expert readers.
Our bibliography is comprehensive, however, it is far from exhaustive or complete.
Most research articles on the subject (which were written in the last decade)
can be found in the Los Alamos National Laboratory electronic archive
{http://xxx.lanl.gov/archive/hep-ph}, 
and Stanford Linear Accelerator Center's Spires electronic database
{http://www.slac.stanford.edu/spires/hep} 
lists relevant articles and reviews. 
%

%08
The core of the manuscript is based on the lecture series ``Essential Supersymmetry''
which appeared in the proceedings of the TASI-98 school \cite{V1}. 
However, the material was revised, updated and significantly extended.
I thank
Francesca Borzumati for her help with the figures, and Howard Haber and
Chris Kolda for providing me with Fig.~\ref{fig:higgs1}
and Fig.~\ref{fig:rsb}, respectively.
I also thank Francesca Borzumati, Lee Hyun Min, Yasanury Nomura, and Myck Schwetz
for their careful  reading of earlier versions of 
the manuscript and for their comments.  
I benefited from works done
in collaboration with Jonathan Bagger, Francesca Borzumati, 
Hsin-Chia Cheng, Jens Erler, Glennys Farrar,
Jonathan Feng, Hong-Jian He, Jean-Loic Kneur, Chris Kolda,
Paul Langacker, Hans-Peter Nilles, Stefan Pokorski, 
Alex Pomarol, Shufang Su, Scott Thomas, Jing Wang, and Ren-Jie Zhang.
This manuscript was completed during my tenure as a Research Scientist
at the Massachusetts  Institute of Technology, and I thank Bob Jaffe for
his encouragement.
Finally, I thank Peter Nilles for his support.

\vspace{1cm}
\begin{flushright}\noindent
%place(s),\hfill {\it Nir Polonsky}\\
%month year\hfill {\it Firstname  Surname}\\
\begin{tabular}{l}
Cambridge,\\
May 2001
\end{tabular}
\hfill {\it Nir Polonsky}\\
%May 2001%\hfill {\it Firstname  Surname}\\
\end{flushright}

\tableofcontents

\clearemptydoublepage

%%ARABIC
\pagenumbering{arabic}
%SET COUNTER
\setcounter{page}{1}

%P1
\part{Stepping Beyond the Standard Model}
\label{p1}

%\include{physch1e}

%\newpage

%C1
\chapter{Basic Ingredients}
\label{c1}

%101
The Standard Model of strong and electroweak interactions (SM)
was extensively outlined and discussed  by many, for example
in Altarelli's TASI lectures \cite{ALT}
where the reader can find an in-depth discussion and references.
The SM has provided the cornerstone of elementary particle physics
for nearly three decades, particularly so after it was soundly established
in recent years by the various high-energy collider experiments.
Even though the reader's familiarity with the SM is assumed throughout these notes,
let us recall some of its main ingredients,
particularly those that force one to look for its extensions.
%

%102
Indeed, the most direct evidence for an extended structure
was provided most recently by experimentally establishing neutrino
flavor oscillations \cite{superK}. 
(See Langacker for details and
references \cite{LAN} as well as recent reviews by Barger \cite{VB}.) 
However, let us refrain from discussing
neutrino mass and mixing until the last part of these notes where neutrino
mass generation within supersymmetric frameworks will be explored.
With the exception of Chapter~\ref{c11},
we will adopt the SM postulation of massless neutrinos.
More intrinsic and fundamental indications for additional structure 
(that will ultimately have to also explain the neutrino spectrum) stem from 
the SM (so-far untested) postulation of a scalar field, the Higgs field,
whose vacuum expectation value ({\it vev})  spontaneously breaks the SM gauge symmetry.
%

%103
We begin in this chapter with a brief review of 
the boson and fermion structures in the SM.
This summary sets  the foundation
for the discussion  of indications for additional structure,
which follows in the next chapter.

\section{Bosons and Fermions}

%111
The SM is a theory of fermions and of gauge bosons
mediating the  ${SU(3)}_{c}\times{ SU(2)}_{L}\times{U(1)}_{{Y}}$ 
color and electroweak gauge interactions of the fermions.
However, the ${SU(2)}_{L}\times{ U(1)}_{{Y}}$ weak isospin $\times$  hypercharge
electroweak symmetry is spontaneously broken and is not respected by the vacuum,
as is manifested in its massive $W$ and $Z$ gauge bosons.  
The spontaneous symmetry breaking is parameterized by
a complex scalar field, the Higgs field, 
which provides the Goldstone bosons of electroweak symmetry breaking.
Its inclusion in the weak-scale spectrum restores
unitarity and perturbative consistency of  $WW$ scattering, for example.
Its gauge-invariant Yukawa interactions
allow one to simultaneously explain, at least technically so,  
the ${SU(2)}_{L}\times {U(1)}_{Y}$ breaking and the chiral-symmetry
violating fermion mass spectrum. 
The SM employs the by-now standard quantum-field theory
tools, and it commutes with, rather than incorporates, gravity. 
%

%112
Table~\ref{table:mattersm}
lists the matter, Higgs and gauge fields which constitute the SM
particle (field) content. (The graviton is listed for
completeness.) The table also serves to establish the relevant notation.
The fermion flavor index $a = 1,\,2,\,3$ indicates the three
identical generations (or families) of fermions  which are distinguished
only by their mass spectrum. It is important to note that the $U(1)$
hypercharge is {\it de facto} assigned such that each family
constitutes an anomaly-free set of fermions.   
All fields (or their linear combinations corresponding to the physical
eigenstates : $B,\,W \rightarrow
\gamma,\, Z,\, W^{\pm}$; see below) aside from the Higgs boson have
been essentially observed. (Indirect evidence  for the existence of
the Higgs boson from precision measurements of electroweak observables,
however, has been  recently acquiring statistical significance.)
It is therefore not surprising that  it is the symmetry breaking sector which is
the least understood. It will occupy most of our attention in the
remaining of Part~\ref{p1}.
%

%113
Given the charge assignments, it is straightforward to write
down the gauge invariant scalar potential and Yukawa interactions.
We begin with the scalar potential and reserve the discussion of 
Yukawa interactions and flavor to the next section.
%

%114
The scalar potential is given by
\begin{equation}
V(H) =  -m^{2}HH^{\dagger} + \lambda (HH^{\dagger})^{2},
\label{VH}
\end{equation}
where $\lambda > 0$ is an arbitrary quartic coupling. If $m^{2} < 0$
then the global minimum of the potential is at the origin
and all symmetries are preserved. However,  for $m^{2} > 0$,
as we shall assume, the Higgs doublet acquires a non-vanishing
{\it vev}  $\langle H \rangle =
(\nu,\,0)/\sqrt{2}$ where $\nu \equiv \sqrt{(m^{2}/\lambda)}$.
%

%115
The electroweak  $SU(2)_{L}\times U(1)_{Y}$ symmetry is now spontaneously broken down to
${U(1)}_{Q}$ of QED with $Q = T_{3} + Y/2$.
(A $SU(2)$ rotation was used to fix the Higgs {\it vev} 
in its conventionally chosen direction). 
The charged and neutral CP-odd Higgs components are 
Goldstone bosons in this case and are absorbed by the electroweak
gauge bosons which, as given by experiment,  are massive with \cite{PDG} 
$M_{W^{\pm}}^{2} = g^{2}\nu^{2}/4 = (80.419 \pm 0.056\,\, {\rm GeV})^{2}$ and 
$M_{Z}^{2} = (g^{2} + g^{\prime 2})\nu^{2}/4 = (91.1882 \pm 0.0022 \,\, {\rm GeV})^{2}$. 
Here, $g \equiv g_{2}$ and $g^{\prime}$
are the conventional notation for the $SU(2)$ and hypercharge gauge
couplings, respectively. (Note that unlike
the symbols $g$ and $g_{2}$ which will be used
interchangeably, the symbols $g^{\prime}$
used here and $g_{1}$ used in Chapter~\ref{c6} differ in  their normalization.
The symbol $g$ will be also used to denote a gauge coupling in general.)
One defines the weak angle $\tan\theta_{W} \equiv g^{\prime}/g$ so that
$Z = \cos\theta_{W}W_{3} - \sin\theta_{W}B$, and the massless photon
of the unbroken QED is given by the orthogonal combination.
(The charged mass eigenstates are $W^{\pm} = (W_{1} \mp iW_{2})/\sqrt{2}$.)
The QED coupling is $e = g\sin\theta_{W}$.
From the Fermi constant, $G_{F} = g^{2}/4\sqrt{2}M_{W}^{2}$,
measured in muon decay, one can extract $\nu = (\sqrt{2}G_{F})^{-1/2} = 246$ GeV.
(Our normalization is the conventional one for the one-Higgs
doublet SM and will be modified when discussing the two-Higgs doublet
supersymmetric extensions.) 
%
 
%116
Note the (tree-level) mass relation $M_{W}^{2} = M_{Z}^{2}\cos^{2}\theta_{W}$.
We will mention  below the quantum corrections to this relation,
measured by  the $\rho$-parameter, $M_{W}^{2} \equiv  \rho M_{Z}^{2}\cos^{2}\theta_{W}$. 
In fact $\rho = 1$ at tree level in the case of Higgs doublets
but not, for example, in the case of Higgs triplets.
Its fitted value (subtracting SM
quantum effects $\sim{\cal{O}}(m_{t}^{2})$), 
$\rho \simeq 1$ \cite{PDG}, therefore severely constrains the possibility of an
electroweak breaking $vev$ of a $SU(2)_{L}$-triplet Higgs field.
In particular, the SM, as well as any
of its extensions, must assume that the Higgs field(s) of electroweak symmetry
breaking is a (are) $SU(2)_{L}$-doublet(s).
%

%T1
\begin{center}
\begin{table}
\caption{The SM field content. Our notation $(Q_{c},Q_{L})_{Q_{Y/2}}$
lists color, weak isospin and hypercharge assignments
of a given field, respectively, and  $Q_{c}=1$, $Q_{L}= 1$ 
or $Q_{Y/2}= 0$ indicate a
singlet under the respective group transformations.
The $T_{3}$ isospin operator is $+1/2$ ($-1/2$) when acting
on the upper (lower) component of an isospin doublet (and zero otherwise).}
\label{table:mattersm} 
\hspace*{0.75cm}
\begin{tabular}{|l|c|l|}\hline
Sector& Spin& Field\\ \hline
&&\\
$SU(3)$ gauge bosons (gluons) & 1& $g \equiv (8,1)_{0}$ \\
&&\\
$SU(2)$ gauge bosons  & & $W \equiv (1,3)_{0}$ \\
&&\\
$U(1)$ gauge boson  & & $B \equiv (1,1)_{0}$ \\
&&\\ \hline 
&&\\
Chiral matter&$\frac{1}{2}$&$Q_{a}\equiv\left(
\begin{tabular}{c}$u$\\$d$\end{tabular}\right)_{L_{a}}\equiv
(3,2)_{\frac{1}{6}}$\\
(Three families: $a = 1,\,2,\,3$)&&\\
&&$U_{a}\equiv u^{c}_{L_{a}}
\equiv(\bar{3},1)_{-\frac{2}{3}}$\\ 
&&\\
&&$D_{a}\equiv
d^{c}_{L_{a}}\equiv(\bar{3},1)_{\frac{1}{3}}$\\ 
&&\\
&&$L_{a}\equiv\left(
\begin{tabular}{c}$\nu$\\$e^{-}$\end{tabular}\right)_{L_{a}}\equiv
(1,2)_{-\frac{1}{2}}$\\
&&\\
&&$E_{a}\equiv(e^{-})^{c}_{L_{a}}\equiv(1,1)_{1}$\\ 
&& \\ \hline
&&\\
Symmetry breaking & 0 & $H
\equiv\left(\begin{tabular}{c}$H^{0}$\\$H^{-}$\end{tabular}\right)
\equiv(1,2)_{-\frac{1}{2}}$  \\
(the Higgs boson) &&\\
&& \\ \hline 
&&\\
Gravity (the graviton) & 2& $G\equiv(1,1)_{0}$ \\
&& \\ \hline
\end{tabular}
\end{table}
\end{center}

\section{Flavor}

%121
The gauge invariant Yukawa interactions are of the from
${\psi_{L}}H\psi_{R}$  or of the form
${\psi_{L}}i\sigma_{2}H^{*}\psi_{R}$,
where we labeled the (left-handed) chiral matter transforming under
$SU(2)_{L}$ and the (right-handed) chiral matter which is a weak-isopspin
singlet with the $L$ and $R$ subscripts, respectively. One has
\begin{equation}
{\cal{L}}_{\rm Yukawa}= y_{l_{ab}}{L}_{a}HE_{b} +
y_{d_{ab}}{Q}_{a}HD_{b} + y_{u_{ab}}{Q}_{a}(i\sigma_{2}H^{*})U_{b} + h.c.,
\label{LYukawaSM}
\end{equation}
where $SU(2)$ indices are implicit and are contracted with 
the antisymmetric tensor
$\epsilon_{ij}$: $\epsilon_{12} = +1 = -\epsilon_{21}$.
Note that the choice of the hypercharge sign for $H$ is somewhat  arbitrary 
since one can always define $\bar{H} = i\sigma_{2}H^{*}$
which carries the opposite hypercharge ($\sigma_{2}$ is the Pauli matrix
and $\bar{H} = (H^{+},\,-H^{0*})^{T}$ given our choice).
This will not be the case in the supersymmetric extension.
One can rewrite the Lagrangian  (\ref{LYukawaSM}) in terms of the physical CP-even
component $\eta$  (with mass $\sqrt{2}m$)
and the  {\it vev} of $H(x) \rightarrow  ((\nu  +
\eta(x),\, 0)/\sqrt{2})^{T}$ in order to find the physical Higgs Yukawa interactions
and the fermion (Dirac) mass terms $m_{f_{ab}} = y_{f_{ab}}\nu/\sqrt{2}$.
%

%122
It is interesting to note that had the spectrum contained  
a SM singlet fermion $N \equiv (1,1)_{0}$ 
(with a lepton number $-1$) then a (lepton-number preserving)
neutrino  Yukawa/mass term $y_{\nu_{ab}}{L}_{a}(i\sigma_{2}H^{*})N_{b} + h.c.$ 
could also be written. $N$ is the right-handed neutrino.
Being a SM singlet it could also have
a gauge-invariant Majorana mass term $\sim N_{a}N_{b}$
which violates lepton number by two units.
However,  the SM contains no right-handed neutrinos,
and the (left-handed) neutrinos are assumed  massless. Furthermore,
lepton $L$ ($L = +1$ for $L_{a}$ and  $L = -1$ for $E_{a},\, N_{a}$)
and baryon $B$ ($B = +1/3$ for $Q_{a}$ and  $B = -1/3$ for $U_{a},\, D_{a}$)
numbers are  automatically conserved 
(in perturbation theory) 
by its  interactions, e.g. see eqs.~(\ref{LYukawaSM}) and (\ref{VH}).
$L$ and $B$ are accidental but exact symmetries of the SM, 
and their conservation holds to all orders in perturbation theory. 
In particular, the lightest baryon, the proton, is predicted to be stable. 
We also note in passing that the introduction of a right-handed neutrino
$N_{a}$ (per family) allows for a new  anomaly-free gauged $U(1)$ symmetry
$U(1)_{Q^{\prime}}$  with $Q^{\prime} = B - L$. It would also allow one
to arrange the SM right-handed fields in isospin doublets of a $SU(2)_{R}$
gauge symmetry \cite{LR} such that $Y = T_{3_{R}} + [(B - L)/2]$ 
(and as before, $Q = T_{3_{L}}+ Y/2$ for QED).  Hence, extended gauge symmetries
and right-handed neutrinos are naturally linked. 
(This is particularly relevant in the case of grand-unified theories
which are discussed in Chapter~\ref{c6}.) We return to the discussion
of neutrinos and lepton number in chapter~\ref{c11}.
%

%123 
Returning to the SM,
both Yukawa couplings and fermion masses are written above as 
$3\times3$ matrices in the corresponding (up, down, lepton) flavor
space. (The dimension of the flavor space is given by the number
of families, $N_{f} = 3$.)
In order to obtain the physical mass eigenstates one has 
to diagonalize the different Yukawa matrices and subsequently
to perform independent unitary transformations on all vectors in
flavor space.
For example, $u_{a_{L}}\equiv (u,\,c,\,t)_{L}$  and 
$d_{a_{L}}\equiv (d,\,s,\,b)_{L}$ (employing
standard flavor symbols which we will use interchangeably with the symbols
defined in Table~\ref{table:mattersm}.) 
are now rotated by different transformations
$U$ and $V$, respectively, which correspond to the diagonalization
of the respective Yukawa matrix.
Hence, the weak (interaction) and mass (physical) 
eigenstates are not identical:
$d_{L_{a}}^{W}= V^{ab}d_{{L}_b}^{m}$ {\it etc.} 
where the superscripts $W$ and $m$
denote weak and mass eigenstates, respectively. 
%

%124
Charged currents $J_{CC} \sim \bar{u}^{W}\Gamma Td^{W} + h.c.$
(interacting with the charged gauge bosons $W^{\pm}J^{\pm}_{CC}$) 
are then proportional, when rotated to the physical mass basis, 
to $U^{\dagger}V \equiv V_{CKM}$.  
Here, $\Gamma$ and $T$ denote Dirac and $SU(2)_{L}$ matrices, respectively,
and subscripts were omitted.
(Note that only the product $V_{CKM}$ appears in physical observables.
One could choose,
for convenience, a basis such  that $U = I$ and $V \equiv V_{CKM}$.) 
The $3\times 3$ Cabibbo-Kobayashi-Maskawa $V_{CKM}$ (unitary) 
matrix contains three independent angles and one phase parameter
and connects the different generations:
Charged currents contain
flavor off-diagonal interaction vertices which lead to flavor changing 
currents. For example,  $V_{CKM}^{us}\bar{c}^{m}\Gamma Td^{m} + h.c.
\subset J_{CC}$
leads to flavor changing $W$ decays, 
$W^{+} \rightarrow c\bar{d}$. 
%

%125
On the other hand, unitarity guarantees that no flavor changing
neutral currents (FCNC) arise (at tree level): $J_{NC} \sim \bar{u}^{W}\Gamma Tu^{W}
+ \bar{d}^{W}\Gamma T d^{W} + h.c.$ (interacting with the $Z$ and with the photon)
is proportional to $U^{\dagger}U,\, V^{\dagger}V \equiv I$ and is basis invariant.
(Loop corrections change the situation due to the flavor
non-conserving nature of the charged currents and lead, for example,
to loop-induced meson mixing and quark decays such as $b \rightarrow s\gamma$.)
In addition, the absence of right-handed electroweak-singlet neutrinos and of the
corresponding Yukawa terms implies
that a flavor rotation of the  neutrinos  is not physical 
(since obviously no neutrino Yukawa/mass term is rotated/diagonalized)
and therefore can always be chosen so that the left-handed 
neutrinos align with the charged leptons. 
Hence no leptonic flavor (\ie $e,\,\mu,\,\tau$) violating (LFV) currents, charged or neutral,
arise in the SM at any order in perturbation theory.
Both predictions provide important arenas for testing extensions of
the SM, which often predict new sources of FCNC 
(which appear only at loop level in the SM) and/or LFV
(which are absent in the SM).
\section{Status of the Standard Model}
%

%131
Even though the SM is extremely successful in explaining all observed
phenomena to date, it does not provide any guidelines in choosing
its various {\it a priori} 18 free parameters
(in addition to its rank and representations): 
the gauge couplings, the number of chiral families,
the Higgs parameters and Yukawa couplings, 
and ultimately, the weak scale $\sim g\nu$.
%

%132
The fermion mass spectrum is indeed successfully  related
to the spontaneous gauge symmetry breaking: The same Higgs $vev$ which
breaks the gauge symmetry also spontaneously breaks 
the chiral symmetries of the SM
(which forbid, if preserved, massive fermions).
Nevertheless, 
the size of the breaking of the chiral symmetries
is put in by hand in (\ref{LYukawaSM}).
Subsequently, all flavor parameters - 
fermion masses and CKM angles and phase -- have to be fixed by
hand: The SM does not guide us as to the origins of flavor (nor as to
the family duplication). Charge quantization and the quark-lepton
distinction also remain mysterious (even though they can be argued to
provide a unique anomaly-free set of fermions, given the gauge symmetries).
%

%133
One may argue that these (and the exclusion
of gravity) are sufficient reasons to view the SM as only a ``low-energy''
limit of an extended ``more fundamental'' theory, 
which may provide some of the answers.
However, an even  stronger motivation to adopt the effective theory point of view 
emerges from the discussion of quantum corrections to the Higgs potential (\ref{VH}),
and in fact, it will restrict the possible frameworks in which one can 
consistently address all other fundamental questions. 
We turn to this subject in the next chapter.
%

%C2
\chapter{The Hierarchy Problem and Beyond}
\label{c2}

\section{The Hierarchy Problem}

%211
Using standard field theory tools, one renormalizes the SM
Lagrangian in order to account for  quantum ({\it i.e.}, loop) corrections. 
Indeed, this procedure is successfully carried out in the case of
the extrapolation of ``the effective'' ${SU(3)}_{c}\times{ U(1)}_{Q}$ theory to
the weak scale and its embedding in the 
${SU(3)}_{c}\times{SU(2)}_{L}\times{ U(1)}_{{Y}}$
SM, as was described by Altarelli \cite{ALT}.
(See also Fig.~\ref{fig:unification} below.)
While most corrections are at most logarithmically divergent
and hence correspond 
%to finite shifts of parameters or to their
a well-understood logarithmic renormalization (and finite shifts\footnote{
In fact, from the effective low-energy theory point of view,
finite corrections which are proportional to the square of the top mass
$m_{t}^{2}$ are quadratically divergent corrections which are cut off by $m_{t}$.
Such corrections appear at the quantum level, for example, in the 
neutral to charged current ratio, the $\rho$-parameter 
$\sim M_{W}^{2}/M_{Z}^{2}\cos^{2}\theta_{W} \sim 1 +{\cal{O}}(m_{t}^{2})$, 
and also in the $Z \rightarrow b\bar{b}$ branching ratio
(which is relevant for the effective theory at $M_{Z}$).}) 
of parameters, there is an important exception.
The model now contains the Higgs
field whose lightness is not protected by either gauge or chiral symmetries.
The leading corrections to the Higgs two-point function (and hence, to its
mass parameter in (\ref{VH})) depend  quadratically
on the ultraviolet cut-off scale.
Adopting  the view that the SM itself is only an effective low-energy theory,
the renormalization  procedure is again understood as extrapolating the model towards
the more fundamental ultraviolet scale. The quadratic dependence on
this  scale, however,  is forcing  the theory to
reside at the ultraviolet scale itself\footnote{Again, an analogy with the top 
quark is in place: The electroweak theory indeed resides near
the cut-off on its quadratic divergences $\sim m_{t}$. 
(Note that quantum corrections to electroweak observables
depend only logarithmically on the Higgs mass, 
which is assumed to be of the same scale.)},
hence leading to an apparent paradox
with crucial implications.  It amounts to a failure to consistently accommodate
fundamental scalar fields within the framework of the SM,
undermining the notion of an ultraviolet independent low-energy theory.
%

\begin{figure}[ht]
\begin{center}
\leavevmode
\epsfxsize= 5.2 truecm
\epsfbox[200 585 422 690]{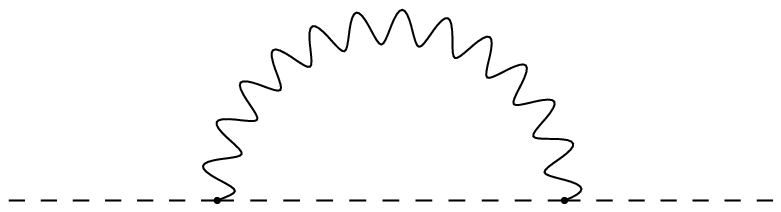}
%\hspace*{1.5truecm}
\hspace*{1.1 truecm}
\epsfxsize= 5.2 truecm
\epsfbox[200 585 422 690]{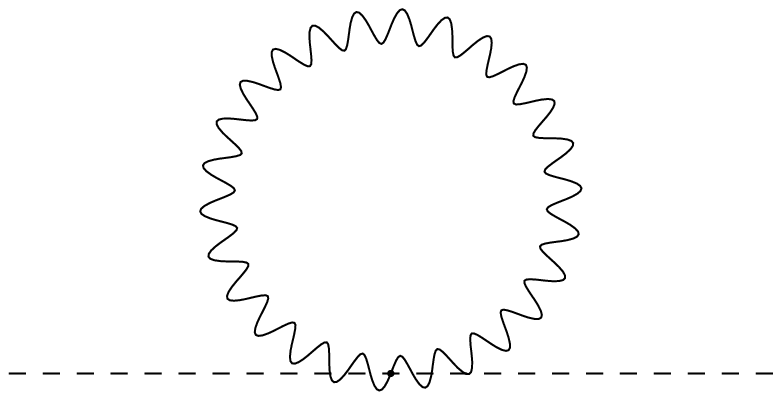}
\end{center}
\caption{The gauge interaction contributions to the quadratic divergence.}
\label{fig:Hdiv1}
\end{figure}

\begin{figure}[ht]
\begin{center}
\hspace*{0.75truecm}
\epsfxsize= 5.2 truecm
\epsfbox[200 545 422 630]{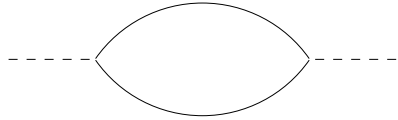}
\end{center}
\caption{The Yukawa interaction contribution to the quadratic divergence.}
\label{fig:Hdiv3}
\end{figure}

\begin{figure}[ht]
\begin{center}
\hspace*{0.75truecm}
\epsfxsize= 5.2 truecm
\epsfbox[200 585 422 690]{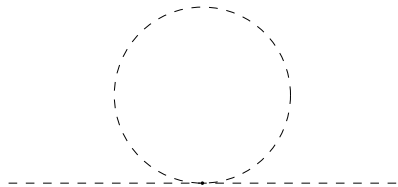}
\end{center}
\caption{The scalar self-interaction contribution to the quadratic divergence.}
\label{fig:Hdiv2}
\end{figure}

%212
More specifically, one needs to consider, at one-loop order, 
the three classes of quadratically
divergent one-loop contributions shown in Figs.~\ref{fig:Hdiv1}--\ref{fig:Hdiv2}. 
The first divergence results from the Higgs $SU(2)\times U(1)$ gauge interactions. 
(Since we are interested in the ultraviolet dependence we will
only consider the full $SU(2) \times U(1)$ theory above the weak scale.) 
The gauge bosons and the Higgs fields themselves 
are propagating in the loops. A naive power
counting gives for the corrections to the Higgs mass
\begin{equation}
\delta m^{2}_{\rm gauge} \sim C_{W} \frac{g^{2}}{16\pi^{2}}\Lambda_{\rm UV}^{2},
\label{qd1}
\end{equation}
where $C_{I}$ are numerical coefficients whose value is immaterial here.
We included only the leading contributions in the ultraviolet cut-off
$\Lambda_{\rm UV}$ (neglecting logarithmic dependences), and similarly
for the hypercharge contribution which we omit.
A second contribution  results from the Higgs Yukawa interactions. 
In this case fermions are propagating in the loops and  one has 
\begin{equation}
\delta m^{2}_{\rm Yukawa}\sim -\sum_{f} C_{f} 
\frac{y^{2}_{f}}{16\pi^{2}}\Lambda_{\rm UV}^{2},
\label{qd2}
\end{equation}
where the summation is over all fermion flavors. 
(Small Yukawa couplings are not negligible once 
$\Lambda_{\rm UV} \rightarrow \infty$!)
Note the negative sign due to the fermion loop. Lastly, the Higgs self
interaction leads to another contribution which is proportional to its
quartic coupling,
\begin{equation}
\delta m^{2}_{\rm Higgs} \sim C_{H} \frac{\lambda}{16\pi^{2}}\Lambda_{\rm UV}^{2}.
\label{qd3}
\end{equation}
Summing all independent contributions one finds 
\begin{equation}
-m^{2}_{\rm tree} + \frac{1}{16\pi^{2}}\left\{C_{W}g^{2} +
C_{H}\lambda - \sum_{f}C_{f}y_{f}^{2}
\right\}\Lambda_{\rm UV}^{2} =
-\lambda(246{\mbox{ GeV}})^{2}.  
\label{qd4}
\end{equation}
The ``one-loop improved'' relation (\ref{qd4}) essentially depends on
all of SM free parameters. (The $SU(3)$ coupling $g_{s}$ enters at
higher loop orders which are not shown here.)
Hence, even though cancellations are technically possible (by adjusting
the tree-level mass
$m_{\rm tree}^{2}$ or $\lambda$) and despite the fact that in the
presence of a tree-level mass one can simply subtract the 
infinity by introducing an appropriate  counter term, such procedures
do not seem reasonable unless $\Lambda_{\rm UV} \sim {\cal{O}}(\nu)$.
For example, for a cut-off of the order of magnitude of the Planck mass
$\Lambda_{\rm UV} \sim M_{\rm Planck} \sim 10^{19}$ GeV
one needs to fine-tune the value of $\nu = 246$ GeV by $10^{17}$ orders of
magnitude (and $\nu^{2}$, which appears in eq.~(\ref{qd4}), 
by $10^{34}$ orders of magnitude)! 
%

%213
This naturalness problem is often referred to as the
hierarchy problem: Naively, one expects that electroweak
symmetry breaking would have occurred near the Planck scale 
$M_{\rm Planck} \sim 10^{19}$ GeV and not at
the electroweak scale $M_{\rm Weak} \sim g\nu \sim 10^{2}$ GeV
(which is measured  with astonishing precision, for example, see 
$M_{Z}$ and $M_{W}$ measurements given in the previous chapter).
Since the SM does not contain the means to address this
issue, it follows that the resolution of the hierarchy problem and any 
understanding of the specific choice of the electroweak scale 
must lie outside the SM. To reiterate,
the instability of the infrared SM Higgs
potential implies that the model cannot be fundamental while it
severely constrains the possible extrapolations and embeddings of the SM.
% 

%F1A

%F2
\vspace*{0.3 truecm}
%\vspace*{0.3 truecm}
%F3
\vspace*{0.3 truecm}

\section{Possible Avenues}

%221
Once it has been argued that the SM cannot be a fundamental theory,
the obvious question arises:
What are the possible avenues that could be pursued 
in order to consistently extend it.
Our discussion of the hierarchy problem clearly indicates two
possibilities: $(i)$ The existence
of a fundamental scale near the electroweak scale corresponding to
new gauge (or gravitational) dynamcis which effectively serves
as a cut-off on any extrapolation of the SM and its parameters.
Such dynamics presumably
render one or more elements (e.g. a fundamental Higgs field, two hierarchically
distinctive scales, perturbative validity) 
of the hierarchy problem obsolete
and the corrections eqs.~(\ref{qd1})--(\ref{qd3}) are cut off by
its low scale and are not dangerous.
$(ii)$ A theory containing a fundamental Higgs field
and which extends and extrapolates the SM within perturbation theory (and in four
dimensions). Such a theory contains all the elements
of the hierarchy problem yet
corrections of the form of eqs.~(\ref{qd1})--(\ref{qd3}) ``miraculously'' vanish.
The essence of the theory in this case is in naturally explaining the ``miracle''.
Both options imply the prediction of a new field content at or near
the electroaweak scale and which is associated
with either the nearby more
fundamental ``cut-off'' scale (let it be resonances associated with 
nearby strong interactions or states associated with the compactification of
an extended space-time geometry near the electroweak scale), 
or alternatively,  with the cancellation of (a larger more complete set of) 
quadratically divergent  corrections in a perturbative theory. 
This realization is a main driving force, side by side with the search
for the Higgs boson itself, in current studies of particle 
physics at electroweak energies and
beyond, and provides the basis for this manuscript.
%
 
%222
Various proposals exist that either
postulate $\Lambda_{\rm UV} \sim 4\pi \nu$
or otherwise provide for a natural cancellation of
divergences. We will sketch the various  ideas which
underlie those frameworks here,
and will further elaborate on the supersymmetry framework
in the next section.

\subsection{Strong Interactions}

%223
Certain frameworks postulate that the Higgs field is a
meson, a composite of fermions, rather than a fundamental field.
In this case
the notion of a cut-off scale and of a confinement scale are
entangled and, in fact, the solution to the hierarchy
problem is the elimination of scalar bosons from the fundamental theory.
It duplicates (and at much higher energies) in its philosophy 
the QCD picture  in which the low-energy
degrees of freedom are mesons and baryons while at high energy these
are the quarks and gluons. Under this category fall the different versions of
technicolor models, top-color models, and their various offsprings.
(For a recent review, see Refs.~\cite{CHIV,LANE,QUIG}.)
%

%224
The models predict not only additional matter, but also strong and rich
dynamics ``nearby'' (not far from current experimental energies).
Strong dynamics, in general, does not tend to completely decouple from
its nearby low-energy limit (the weak scale, in this case),
leading, in most cases, to many  predictions at observable energies.
For example, the models typically predict new FCNC 
as well as (oblique) quantum corrections to the $W$ and $Z$ masses and
interactions. (As mentioned in the previous chapter, 
the latter lead to calculable shifts from SM tree-level relations
such as $M_{W} = M_{Z}\cos\theta_{W}$, which is measured
by the $\rho$-parameter.) 
While a detailed critique  can be found in Altarelli's lectures \cite{ALT},
no corresponding deviations from SM predictions (when including its
own quantum corrections to such observables) were  found, 
seriously undermining the case for such an embedding of the SM.
%

%225
Attempts to solve this contradiction
are often based on distinguishing, for example,
the fermion-mass generation sector from the sector responsible
to electroweak symmetry breaking,
leading to complex multi-layer constructions.
It is also interesting to note that strong dynamics typically 
(counterexamples, however, exist \cite{Hill}) predict
for the Higgs boson mass $m_{\eta} \gg \nu$ (consider the pions in QCD), 
contrary to current implications based on fits to 
electroweak data \cite{PGLER} $m_{\eta} \lesssim \nu$.
(Such fits are based, however, on the assumption that the new physics
decouples from many of the SM observables. A heavier Higgs boson
is allowed otherwise \cite{Pes,N23}.) 
One should bare in mind, however, that by its nature
non-perturbative strong dynamics do not always allow for reliable calculations
of SM quantities (consider the Postmodern Technicolor \cite{PMT} model, for example), 
and it is therefore difficult to rule it out.

\subsection{Extra Spatial Dimensions}
\label{sec:extra}

%226
A different and more recent proposal 
is of  a framework with a low-energy cut-off and
(possibly) a fundamental Higgs field in the background of an extended
space-time geometry (and more then three spatial dimensions).
Theories with extra low-energy (very large)
compacitified dimensions where first argued
to be consistent with all data (for two or more extra dimensions)
by Arkani-Hamed, Dimopoulos and Dvali \cite{extra}.
In short, it is postulated that our four-dimensional universe resides
on a four-dimensional submanifold of an extended space-time.
Such theories must contain towers of Kaluza-Klein states which parameterize   
the excitations of those states propagating in the extra dimensions
and which contain $M \sim {\cal{O}}$(TeV) or heavier states. 
Such towers may include the graviton (and other fields
describing quantum gravity such as various moduli) but
could also include for not ``so low-energy dimensions''
(\ie for extra dimensions smaller than originally proposed
in Ref.~\cite{extra}) towers of the usual and extended gauge fields
and/or the usual and extended matter and Higgs fields,
leading to a large variety of possibilities and constraints
on the size and number compactified dimensions $R \gg (\mplanck^{4d})^{-1}$.
%

%227
The relation between the four-dimensional to 
$4 + n$-dimensional Planck constants reads
\begin{equation}
M_{\rm Planck}^{4d} = (M_{\rm Planck}^{(4+n)d})^{\frac{n + 2}{2}}R^{\frac{n}{2}},
\label{extrad}
\end{equation}
relating the size and number of compactified dimensions. Relation (\ref{extrad})
explains the weakness of gravity as we know it in terms of extra large 
(but with particular size)
dimensions rather than in terms of 
a large Planck mass.
$R^{n} \equiv V$ is the volume of the compactified $n$-dimensional space
(and $R^{-1}$ is the compactification scale).
%

%228
In this framework, the Higgs boson is usually assumed to be 
a fundamental particle, but this need not be the case. Rather,
it depends on the assumed details of the ``bulk'' physics  (where 
the ``bulk'' refers to the the extra-dimension volume $V$).
%

%229
The theoretical limit  of a small volume $V$ ($R^{-1} \sim M_{\rm Planck}^{4d}$)
is also very interesting.
It occurs in what is often called $M$-theory \cite{HPN97,DIEN}
(which is related to the discussion of unification in Chapter~\ref{c6}). 
It also appears in an alternative realization of these ideas which is based on the notion
that space-time geometry is modified such that the full space-time metric
is not factorizable (that is, it cannot be written as  a direct sum
of the metric on the four-dimensional submanifold and of the metric in the extra
dimensions, as is implicitly assumed in deriving eq.~(\ref{extrad})). 
Rather, the four-dimensional metric is assumed to be multiplied by an exponential
function of the additional dimension(s), the warp factor. It was shown 
by Randall and Sundrum \cite{RS2}
that in the case of two four-dimensional manifolds, or ``walls'' 
(one of which corresponds to our world, while gravity is localized on the other
``wall''), in an anti-de Sitter background, 
and which  are separated by one extra dimension, 
it is sufficient to stabilize the size of the compactified 
dimension $R$ 
at a small (yet particular) value in order to realize the hierarchy
as seen in our four-dimensional world: The hierarchy 
is given by $\exp(-2MR\pi)$, 
where $M$ here is some fundamental scale in the full theory
(and is assumed to be of the order of $M_{\rm Planck}$). 
A rescaling of the coordinates by the square root of the
inverse of the warp factor $\exp(MR\pi)$ leads to
a similar picture but now the two scales play the reverse roles
with the fundamental Planck scale being of the order of the electroweak scale,
in the spirit of the original proposal of Ref.~\cite{extra}.
%

%2210
The hierarchy problem is not eliminated in the modified geometry
(extra-dimensional) frameworks but is only rephrased in a
different language. 
The large extra dimensions must be stabilized so that
the volume $V$ is appropriately fixed. 
The stabilization of the extra dimensions
at particular values is the
hierarchy problem of models with extra large (or equivalently,
low-energy) dimensions.
Intuitively, one expects its solution within a framework of quantum gravity. 
On the other hand, the only candidate to describe quantum gravity, 
namely string theory, does not provide the answer, given current knowledge.
(It was shown \cite{RS3}, however, that in the Randall-Sundrum \cite{RS2} setup 
one could consistently take the limit of a non-compactified
infinitely-large extra dimension $R \rightarrow \infty$ while having gravity
localized on the remaining four-dimensional ``wall''.) 
Furthermore, if the $d+n$ Planck mass (or the ``warped'' quantities, in
the alternative picture) are not sufficiently close to the weak-scale 
then the usual hierarchy problem re-appears.
%

%2211
Many issues have been recently
discussed in the context of extra dimensions, including the unification 
of couplings (first discussed in Ref.~\cite{DDG}), 
proton decay (see, for example, Ref.~\cite{extra2}), {\it etc.}
The discussions  are far from conclusive due to the lack of a concrete
framework on the one hand, and strong dependence on the
details of the full theory, \ie  boundary conditions, on the other hand.
(Note in particular  that the issue of
possible new contributions to FCNC
is difficult to address in the absence of concrete 
(and complete) realizations of these ideas.)
Even though it is intriguing that low-energy extra dimensions
may still be  consistent with all data, significant constraints
were derived in the literature in the recent couple of years
and no indications for such a dramatic scheme exist in current data. 
These ideas may be further tested in collider experiments.
A generic  prediction in many of the extra-dimension schemes 
is of large effective modifications of the
SM gauge interactions 
(due, for example,  to graviton emission
into the bulk of the extended volume) 
which can be tested in the next generation of hadron and lepton colliders,
if the $(4+n)d$ Planck mass is not too large. Other effects of low-energy gravity
may include a spin 2 graviton resonance \cite{JLH} and pure gravitational effects
(one may even speculate on balck-hole\footnote{Black hole were also 
argued \cite{Black}
to induce dangerous low-energy operators in these theories.}
production at colliders),
if gravity is indeed sufficiently strong at sufficiently low energies.

\subsection{Conformal Theories}

%2212
On a different note,
it was also argued (but not demonstrated) that one could construct theories which are
conformally (scale) invariant above the weak scale and hence contain 
no other scales, no dimensionful parameters,
and trivially, no hierarchy problem \cite{FV}. 
This proposal combines the notion of a fundamental weak-scale cut-off
(above which nature is truly scale invariant)
with that of a cancellation of all divergences (due to the conformal symmetry).
Again, this is only a rephrasing of the problem
since conformal invariance must be broken and by the right amount. Furthermore,
this approach faces severe and fundamental difficulties \cite{noFV} in constructing
valid examples.

\subsection{Supersymmetry}

%2213
Unlike in all of the above schemes,
one could maintain perturbativity to an arbitrary
high-energy scale if dangerous quantum corrections are canceled
to all orders in perturbation theory 
(a cancellation which does not require conformal 
invariance discussed in the previous section). 
Pursuing this alternative, one arrives at the notion of weak-scale
supersymmetry. Supersymmetry ensures the desired cancellations
by relating bosonic and fermionic {\bf degrees of freedom} (e.g. for any
chiral fermion the spectrum contains a complex scalar field, the s-fermion) 
and {\bf couplings}  (e.g. quartic couplings depend on gauge and Yukawa couplings).
This is done by simply organizing fermions and bosons in ``spin multiplets''
(so that chirality is attached to 
the bosons by their association with the fermions).
Henceforth, supersymmetry is able to ``capitalize'' on the sign difference
between bosonic and fermionic loops, which is insignificant as it stands
in the SM result (\ref{qd4}). 
This is the avenue that will be pursued in these notes, and
the elimination of the dangerous corrections will be shown below explicitly
in a simple example. 
A detailed discussion will follow in the succeeding chapters. 
We conclude this chapter by elaborating
on the status of (weak-scale) supersymmetry, which further motivates
our choice. 

\subsection{Mix and Match}

%2214
We note in passing that it is possible to link together various
avenues. Theories with extra dimensions may very well contain weak-scale
supersymmetry, depending on the relation between the (string)
compactification scale and the supersymmetry-breaking scale \cite{POM}.
In fact, it is difficult to believe that the stabilization of the
large extra dimensions does not involve supersymmetry in some form or another.
It was also suggested that supersymmetric QCD-like theories may play
some role in defining the low-energy theory (for example, that some
fermions and their corresponding sfermions are composites \cite{NS,GK}.)
Of course, strong dynamics may reside in extra dimensions (for a recent proposal,
see Ref.~\cite{TCextra})
and all three avenues may be realized simultaneously.
These possibilities lie beyond the scope of this manuscript. 

\section{More on Supersymmetry}
\label{sec:moreonsusy}

%231
Aside from its elegant and nearly effortless solution
to the hierarchy problem (which will occupy us in the next part),
supersymmetric models  suggest that electroweak symmetry is broken in only
a slightly ``stronger'' fashion ({\it i.e.}, two fundamental Higgs doublets acquire a
{\it vev}) than predicted in the SM (where only one Higgs doublet is
present). Furthermore, in many cases the more complicated Higgs sector is reduced to
and imitates the SM situation with the additional theoretical constraint
$m_{\eta} \lesssim 130 - 180$ GeV.
This is a crucial element in testing the supersymmetry framework
and we will pay special attention to this issue in Part~\ref{p3}.
In fact, one typically finds that the  theory is effectively described by a direct
sum of the SM sector (including the SM-like Higgs boson)
and of a superparticle (sparticle) sector,
where the latter roughly decouples 
(for most practical purposes) from electroweak symmetry breaking
(\ie its spectrum is to a very good approximation
$SU(2)_{L}\times U(1)_{Y}$ conserving
and is nearly independent of electroweak symmetry breaking {\it vev's}).
The sparticle mass scale provides the desired cut-off scale
above which the presence of the sparticles 
and the restoration of supersymmetric relations among couplings
ensure cancellation of quadratically divergent quantum corrections.
Hence, the super-particle mass scale is predicted to be
$M_{\rm SUSY} \sim {\cal{O}}(100 \, {\rm GeV} - 1 \, {\rm TeV})$ with details
varying among models.
The mass scale $M_{\rm SUSY}$ is also a measure, in a sense, of supersymmetry
breaking in nature (e.g. the obvious observation that  a fermion and
the corresponding sfermion boson are not degenerate in mass). Its size  implies that
supersymmetry survives (and describes the effective theory)
down to the weak scale. More correctly, it is $M_{\rm SUSY}$ that
defines in this picture the weak scale $\nu \sim  M_{\rm SUSY}$
(up to a possible loop factor and dimensionless couplings).
These issues will be discussed in some detail in the following chapters.
%

%232
The electroweak symmetry conserving structure of the sparticle spectrum
sharply distinguishes the models from, e.g. technicolor models.
It implies that the heavy sparticles do not contribute significantly
even at the quantum level to electroweak 
(or more correctly, to the related custodial $SU(2)$ symmetry) breaking  effects 
such as the $\rho$-parameter.
(This is generically true if all relevant superparticle masses
are above the 200-300 GeV mark.) 
This is consistent with the  data, which is in nearly perfect agreement with SM
predictions. Returning to the Higgs mass, it is particularly 
interesting to note that the same analysis of various electroweak observables
({\it i.e.}, various cross-sections, partial widths, left-right
and forward-backward asymmetries, etc., including their SM quantum
corrections)  which determines the value of the $\rho$-parameter, suggests
that the SM Higgs boson (or the SM-like Higgs boson $h^{0} \simeq \eta$ in supersymmetry)
is lighter than 200-300 GeV \cite{PGLER}, which is consistent  with 
the predicted range in supersymmetry, and more generally, with the
notion of a fundamental Higgs particle (rather that a composite Higgs with
a mass $m_{\eta} \sim \Lambda_{\rm UV}$ which is a factor of $2-4$ heavier).
Both the data and supersymmetry seem to suggest
\begin{equation}
g_{\rm Weak}\nu \lesssim m_\eta \lesssim \nu,
\label{Hrange}
\end{equation}
where $g_{\rm Weak}\nu \equiv M_{Z}$.
The lower experimental bound of ${\cal{O}}(100)$ GeV is from direct searches.
(The origin of the theoretical bounds will be explored at a later point).
%

%233
It should be stressed that one is able to reach such strong conclusions 
regarding the Higgs mass primarily because  the $t$-quark (top) mass \cite{PDG}
$m_{t} = 174\pm 5$ GeV is measured and well known, hence, reducing
the number of unknown parameters that appear in the expressions
of the quantum mechanically corrected SM predictions for
the different observables. Indeed, it was $m_{t}$ which appears
quadratically in such expressions (rather than only logarithmically
as the Higgs mass) that was initially determined by a similar analysis
long before the $t$-quark  was discovered 
and its mass measured \cite{top} in 1995.
(For example, Langacker and Luo found \cite{PGLLUO} 
$m_{t} = 124 \pm 37$ GeV in 1991.) 
The most important lesson learned from the early analysis was that the top is heavy,
$m_{t} \sim \nu$ (in analogy to the current lesson that the Higgs
boson is light, with light and heavy assuming a different meaning in each case).
Once we discuss in Part~\ref{p3} the renormalization of supersymmetric models
and the dynamical realization of electroweak symmetry breaking
that follows, we will find that a crucial element in this
realization is a sufficiently large $t$-Yukawa coupling,
{\it i.e.}, a sufficiently large $m_{t} \simeq y_{t}\nu$. Indeed, this observation 
corresponds to the  successful prediction of a heavy top
in supersymmetry
(though there are small number of cases in which this requirement can be evaded).
%

%234
A somewhat similar but more speculative issue has to do with the 
extrapolation to high
energy of the SM gauge couplings. Once loops due to virtual
$(i)$ sparticles 
and $(ii)$ the second Higgs doublet are all
included in the extrapolation at their predicted mass scale of
${\cal{O}}(100 \, {\rm GeV} - 1 \, {\rm TeV})$, 
then one finds that the couplings unify two orders of magnitude below
the Planck scale and with a perturbative value (provided that the spectrum contains
exactly two Higgs doublets and their super-partners).
This enables the consistent discussion of grand-unified and string theories, which
claim such a unification, in this framework. 
(Note that supersymmetry is a natural consequence of (super)string theory,
though its survival to low energies is an independent conjecture.)
This result was known qualitatively for a long time, but it was shown
more recently to hold once the increasing precision in the measurements
of the gauge coupling at the weak scale is taken into account,
and it re-focused the attention of many on low-energy supersymmetry.
(As we discuss in Chapter~\ref{c11}, the large unification
scale may play an important  role in the smallness of neutrino masses.)
%

%235
On the other hand, no known extension of the SM
can imitate the simplicity by which FCNC and LFV are suppressed.
Supersymmetry is no exception and 
like in the case of any other low-energy extension, 
new potentially large contributions to
FCNC (and in some cases to LFV) could arise. Their suppression to acceptable
levels is definitely possible, but strongly restricts the 
``model space''. The absence of large low-energy FCNC is the most
important information extracted from the data so far with
regard to supersymmetry, and as we shall see, one's choice
of how to satisfy the FCNC constraints (the ``flavor problem'')
defines to great extent the model.  
%

%236
In conclusion, once large contributions to FCNC are absent,
the whole framework is consistent with all data, and furthermore, it is successful
with regard to certain indirect probes.
Its ultimate test, however,
lies in direct searches in existing and future collider experiments.
(See Zeppenfeld \cite{ZEP} for a discussion of collider searches and
Halzen  \cite{HAZ} for the discussion of possible non-collider searches.)
In the course of these notes all of the above statements and
claims regarding supersymmetry will be discussed in some detail and presented using
the tools and notions that we are about to define and develop.
%
%
%

%P2
\part{Supersymmetry Bottom-Up: Basics}
\label{p2}

%C3
\chapter{Bottom-Up Construction}
\label{c3}

%301
Though supersymmetry corresponds to a self-consistent
field-theoretical framework, one can arrive at its crucial elements
by simply requiring a theory which is at most logarithmically
divergent: Logarithmic divergences correspond to only a weak
dependence on the ultraviolate cut-off $\Lambda_{\rm UV}$ and can be 
consistently absorbed in tree-level quantities, and thus
understood as simply scaling (or renormalization) of the theory.
Unlike quadratic divergences -- which imply strong dependence
on the ultraviolate cut-off and therefore need to be fine-tuned away, 
logarithmic divergences do not destabilize the infrared
theory but only point towards its cut-off scale.
%

%302
As we shall see, the mere
elimination of the quadratic divergences from the
theory renders it ``supersymmetric'', and the importance of supersymmetry 
to the discussion of any consistent high-energy extension of the SM 
naturally follows.
This bottom-up construction of ``supersymmetry'' is an instructive exercise. 
The simple example of  a (conserved) $U(1)$ gauge theory 
will suffice here and will be  considered below, followed 
by the consideration of a Yukawa (singlet) theory. 
This exercise could be
repeated, however, for  the case of a (conserved) non-Abelian gauge theory and
in the case of the SM (with a spontaneously broken non-Abelian gauge
symmetry and non-singlet Yukawa couplings) 
which are technically  more evolved.  
Suggested readings for this section include lecture notes Refs.~\cite{DINE,TATA1,DREES}.
\section{Cancellation of Quadratic Divergences 
\protect\newline in a $U(1)$ Gauge Theory}
\label{sec:s21}
%

%311
We set to build a theory which contains at least one
complex scalar (Higgs) field $\phi^{+}$ (with a $U(1)$ charge of $+1$) and which is at
most logarithmically divergent. In particular, we would like to extend
the field content and fix the couplings such that all quadratic
divergences are eliminated.
For simplicity, assume a massless theory and zero external momenta
$p^{2} = 0$. All terms which are consistent
with the symmetry are allowed.
(The Feynamn gauge, with the gauge boson propagator given by $D_{\mu\nu} =
-ig_{\mu\nu}/q^{2}$, is employed below.) 
%

%312
As argued above, the most obvious ``trouble spot''  is  the scalar
two-point function  (Figs.~\ref{fig:Hdiv1}--\ref{fig:Hdiv2}). 
A $U(1)$ theory with only scalar (Higgs) field(s)
receives at one-loop positive contributions $\propto \Lambda_{\rm UV}^{2}$
from the propagation of the gauge 
and Higgs bosons in the loops (Figs.~\ref{fig:Hdiv1} and~\ref{fig:Hdiv2}). 
They are readily evaluated in the $U(1)$ case to read
\begin{equation}
\delta m^{2}_{\phi}|_{\rm boson} = +\left\{
(4 - 1)g^{2}
+ \lambda\frac{N}{2}
\right\} 
\int{\frac{d^{4}q}{(2\pi)^{4}}\frac{1}{q^{2}}},
\label{s2e1}
\end{equation}
where $g$ and $\lambda$ are the gauge and quartic couplings,
respectively, and $N = 8$ is a combinatorial factor ($N = 2$ for
non-identical external and internal fields). Keeping only leading
terms, the integral is replaced by $\Lambda_{\rm UV}^{2}/16\pi^{2}$.
%

%313
Obviously, there must be a negative contribution to 
$\delta m^{2}_{\phi}$, if cancellation is to be achieved.
A fermion loop equivalent to Fig.~\ref{fig:Hdiv3},
which could cancel the contribution from (\ref{s2e1}),
implies a new  Yukawa interaction. 
%which then leads to a negative contribution equivalent to Fig.~\ref{fig:Hdiv3}.
In order to enable such an interaction and the subsequent
cancellation, we are forced to introduce new fields
to the theory:
$(i)$ A chiral fermion $\psi_{V}$ with  gauge quantum numbers 
identical to those of the gauge boson (\ie neutral).
(It follows that $\psi_{V}$ is a real Majorana fermion.) 
$(ii)$ A chiral fermion $\psi_{\phi}^{+}$
with gauge quantum numbers identical to those
of the Higgs  boson (i.e., with $U(1)$ charge of $+1$).
Given this fermion spectrum, the Lagrangian can now be extended to include
the Yukawa interaction
$y\phi^{+}\bar{\psi_{\phi}}\bar{\psi_{V}}$
where $y$ is a new (Yukawa) coupling. 
Note the equal number (two) of bosonic and fermionic 
degrees of freedom ($d.o.f.$) with the same quantum numbers 
in the resulting spectrum.
In particular, the requirement of identical
gauge quantum number assignments guarantees the gauge invariance
of the Yukawa interaction.
We will adopt the equality of fermion and boson number of $d.o.f.$ as a guiding principle.
%

%314
Let us redefine $\lambda \equiv bg^{2}$ and $y \equiv cg$.
One arrives at
\begin{equation}
\delta m^{2}_{\phi}|_{\rm fermion} = -2
c^{2}g^{2}
\int{\frac{d^{4}q}{(2\pi)^{4}}\frac{1}{q^{2}}}.
\label{s2e2}
\end{equation}
The coefficients $b$ and $c$ must be  predetermined by some principle
or otherwise a cancellation is trivial and would simply correspond to
fine-tuning (of the coefficients, this time).
The principle is scale invariance.
Even though the determination of the proportionality coefficient 
cannot be done unambiguously at the level of our discussion so far, 
it follows from the consideration of the logarithmic divergences and the
requirement that any proportionality relation is scale invariant
(i.e., that it is preserved under renormalization group evolution:
$db/d\ln{\Lambda} = dc/d\ln{\Lambda} = 0$).
However, before doing so we note an 
inconsistency in our construction above which must be addressed
in order to facilitate the discussion of scale invariance.
%

%315
Once fermions are introduced into the $U(1)$ gauge theory, 
one has to ensure that the (massless)
fermion content is such that the theory is anomaly free, that is
Tr$Q_{\psi_{i}} = 0$ (where the trace is taken over the $U(1)$ charges
of the fermions).  The gauge fermion $\psi_{V}$ carries no charge
(it carries charge in non-Abelian theories, but being in the traceless adjoint
representation it still does not contribute) so only the introduction
of one additional chiral  fermion
$\psi_{\phi}^{-}$ with a $U(1)$ charge of $-1$ 
is required in order to render the
theory anomaly free. The additional fermion $\psi_{\phi}^{-}$ 
itself does not lead
to new one-loop quadratically divergent contribution to the $\phi^{+}$
two-point function. However, 
our newly adopted principle of equal number of bosonic
and fermionic $d.o.f.$ suggests that it should be accompanied by a
complex scalar Higgs boson $\phi^{-}$, which also carries a $U(1)$ charge
of $-1$, and it leads to new divergences.
The most general gauge invariant
quartic potential is now that of a two Higgs model,
\begin{eqnarray}
V_{\phi^{4}} &=& + \lambda_{1}|\phi^{+}|^{2}|\phi^{+}|^{2}
+ \lambda_{2}|\phi^{-}|^{2}|\phi^{-}|^{2}
+ \lambda_{3}|\phi^{+}|^{2}|\phi^{-}|^{2}
+ \lambda_{4}|\phi^{+}\phi^{-}|^{2}
\nonumber \\
&& +\left\{
\lambda_{5}(\phi^{+}\phi^{-})^{2} 
+\lambda_{6}|\phi^{+}|^{2}\phi^{+}\phi^{-}
+ \lambda_{7}|\phi^{-}|^{2}\phi^{+}\phi^{-}
+ h.c.
\right\},
\label{s2V4}
\end{eqnarray}
where we adopted standard notation for the couplings,
replacing $\lambda \rightarrow \lambda_{1}$.
In the following we will define 
$\widetilde\lambda_{3} = \lambda_{3} + \lambda_{4} \equiv \widetilde{b}g^{2}$.
%

%316
With the introduction of $\psi^{-}$ the theory is
now consistent, and with the additional introduction of $\phi^{-}$ it is possible to
require scale invariant relations between the different coefficients.
Doing so one finds a unique solution
$b = 1/2$, $\widetilde{b} = -1$, and $c = \sqrt{2}$
(where $\lambda_{1}$ and $\widetilde\lambda_{3}$ one-loop mixing is accounted for).
It is straightforward to show that the
same arguments lead to $\lambda_{2} = \lambda_{1}$.
Similarly, $\lambda_{6,\,7}$ ($\lambda_{5}$)
may be fixed by consideration of one-loop (two-loop)
wave-function mixing between $\phi^{+}$ and $\phi^{-}$, and must vanish.
%

%317
Thus, we finally arrive at the desired result: A vanishing total contribution
\begin{equation}
\delta m^{2}_{\phi}|_{\rm total} = 
g^{2}\left\{ 3 - 2c^{2} + 4b + \widetilde{b}\right\}
\int{\frac{d^{4}q}{(2\pi)^{4}}\frac{1}{q^{2}}} = 0,
\label{s2e3}
\end{equation}
where the last term is from $\phi^{-}$ circulating in the loop (with
$N = 2$). One concludes that in a consistent theory 
with $(i)$ equal number of bosonic and fermionic
$d.o.f.$, and $(ii)$ scale-invariant relations among the respective couplings,
the scalar two-point function is at most
logarithmically divergent! The combination of these two conditions
defines ``supersymmetry''.
%

%318
This is a remarkable result that suggests more generally the
interaction terms, rewritten in  a more compact form, 
\begin{equation}
{\cal{L}}_{\rm int} =
\sqrt{2}g\sum_{i}Q_{i}\left(\phi_{i}\bar{\psi}_{i}\bar{\psi}_{V}
+ h.c. \right) -\frac{g^{2}}{2}\left\{
\sum_{i}Q_{i}|\phi_{i}|^{2} \right\}^{2},
\label{s2Lint}
\end{equation}
where we denoted explicitly the dependence on the $U(1)$ charge $Q_{i}$.
(This dependence is apparent above in $b = Q_{\phi^{+}}^{2}/2$ and
$\widetilde{b} = 2Q_{\phi^{+}}Q_{\phi^{-}}/2$.)
The most important lesson is that the theory contains
only one free coupling: The gauge coupling. This could have been
anticipated from our introductory discussion of divergences, and
further suggests that  the Lagrangian (\ref{s2Lint}) is derived
in a supersymmtric theory from the gauge Lagrangian.
(This will be illustrated in the following chapter.)
%

%319
Our discussion so far dealt with the elimination of the
${\cal{O}}(\Lambda_{\rm UV}^{2})$ terms at one-loop order.
They re-appear at higher loop orders, for example, two-loop 
${\cal{O}}((g^{2}/16\pi^{2})^{2}\Lambda_{\rm UV}^{2})$ contributions
to the two-point function, etc. It is straightforward,
though technically evolved, to show 
that the conditions found above are not only
sufficient for the elimination of the undesired divergences at
one-loop order, but rather suffice for their elimination order by order.  
Thus, ``supersymmetry'', which postulates equal number of bosonic and
fermionic $d.o.f.$ and specific scale-invariant relations among the
various couplings, is indeed free of quadratic divergences at all
orders in perturbation theory. This is a consequence of the
non-renormalization theorems \cite{WEINBERG} of supersymmetry.
\section{Cancellation of Quadratic Divergences 
\protect\newline in a Yukawa Theory}
%

%321
The above prescription extends also to non-gauge interactions.
For example, consider a theory with a gauge-singlet complex scalar and a gauge-singlet
chiral fermion fields, $\phi_{s}$ and $\psi_{s}$, respectively (or in a more
compact terminology, a superfield $S$ with a scalar and a chiral fermion
components $\phi_{s}$ and $\psi_{s}$, respectively):  The theory
contains an equal number of fermionic and bosonic $d.o.f.$ The 
interaction Lagrangian
\begin{equation}
{\cal{L}}_{\rm int} =
y\phi_{s}{\psi}_{s}\bar{\psi}_{s}  + A\phi_{s}^{3} + C\phi_{s}^{*}\phi_{s}^{2}
+ h.c. - \lambda (|\phi_{s}|^{2})^{2}
\label{s2LS}
\end{equation}
leads to quadratically divergent contributions to the scalar two-point
function (from the equivalents of diagrams~\ref{fig:Hdiv2} and~\ref{fig:Hdiv3}), 
\begin{equation}
\delta m^{2}_{\phi_{s}}|_{\rm total} = 
4(\lambda - y^{2})
\int{\frac{d^{4}q}{(2\pi)^{4}}\frac{1}{q^{2}}}.
\label{s2e4}
\end{equation}
(Note the additional factor of 2 in the fermion loop from contraction
of Weyl indices.)
The cancellation of (\ref{s2e4}) leads to the relation  $\lambda =
y^{2}$ among the dimensionless couplings.
Again we find the same lesson of constrained quartic couplings! 
%

%F4
\begin{figure}[t]
\begin{center}
\leavevmode
\epsfxsize= 7 truecm
\epsfbox[160 540 440 700]{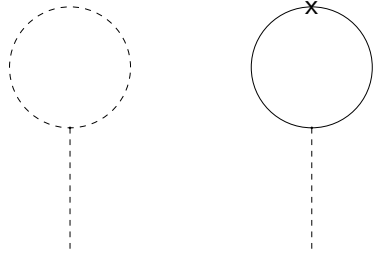}
\end{center}
\caption{The divergent contributions to the one-point function of the singlet.}
\label{fig:tadpole}
\end{figure}
%

%322
The singlet theory contains another potential quadratic divergence,
in the scalar one-point function (the tadpole).
The contribution now depends on the dimensionful trilinear coupling 
$C$. (The symbol $A^{\prime}$ is often used instead of $C$).
Its dimensionality prevents it from
contributing to (\ref{s2e4}), but if non-vanishing it will contribute to the 
quadratically divergent tadpole. Its contribution may be offset by
a fermion loop, 
but only if the fermion is a massive (Dirac) fermion with (a Dirac) mass
$\mu \neq 0$ so that its propagator contains a term  $\propto \mu/q^{2}$
(which, by dimensionality, is the relevant term here).
Both divergent loop contributions to the singlet one-loop
function are illustrated in Fig.~\ref{fig:tadpole}.
Evaluation of the loop integral gives for the leading contributions to
the tadpole 
\begin{equation}
\delta T_{\phi_{s}} = \left\{  y{\rm Tr}I\mu - 4C\right\}
\int{\frac{d^{4}q}{(2\pi)^{4}}\frac{1}{q^{2}}}.
\label{s2e5}
\end{equation}
The trace over the identity is four, leading to the condition
$C = y\mu$ so that the tadpole is canceled (at leading order).
%

%323
As in the gauge theory case, the relevant relations, 
$\lambda = y^{2}$ and $C = y\mu$, 
can be shown to be scale-invariant, and to lead to cancellation of
quadratic divergences order by order in perturbation theory.
These relations extend to non-singlet Yukawa theories such as the SM.
%

%324
Note that the trilinear interaction $A\phi_{s}^{3}$ as well as 
a mass term $m^{2}|\phi_{s}|^{2}$, do not lead to any quadratic
divergences nor upset their cancellations as described above.
Yet, one defines the supersymmetric limit as $m^{2} = \mu^{2}$ and $A = 0$,
corresponding to a fermion-boson mass degeneracy.
%

%325
The singlet Yukawa theory described above is essentially the Wess-Zumino
model \cite{WZ}, which is widely  regarded as the discovery of supersymmetry,
and from which stems the field-theoretical formulation of
supersymmetry.
%
%
%
%
%

%E3
\section*{Exercises}
\addcontentsline{toc}{section}{Exercises}
\markright{Exercises}

\subsubsection*{3.1}
Using the Feynamn rules of a $U(1)$ (QED) theory, confirm the numerical
factors in eqs. (\ref{s2e1}) and (\ref{s2e2}).

\subsubsection*{3.2}
Repeat our exercise for $\phi^{-}$ and show $\lambda_{2} = \lambda_{1}$.

\subsubsection*{3.3}
Repeat the $U(1)$ example in the Landau gauge, $D_{\mu\nu} =
-\frac{i}{q^{2}}\left\{ g_{\mu\nu} - \frac{q_{\mu}q_{\nu}}{q^{2}} \right\}$.

\subsubsection*{3.4}
Show that $\lambda_{6,\,7} = 0$ by considering the
loop corrections to a $m_{+-}^{2}\phi^{+}\phi^{-}$ mass term.
Repeat the exercise for $\lambda_{5}$ which is constrained by
two-loop corrections.

\subsubsection*{3.5}
Show that a term $A\phi_{s}^{3}$ does not lead to any new quadratic divergences.

\subsubsection*{3.6}
Repeat the relevant loop calculations with 
a scalar mass $m^{2} \neq \mu^{2}$ (boson fermion mass non-degeneracy)
and show (for example, by expanding the scalar propagator) 
that it does not lead to any new quadratic divergences (for a finite $m^{2}$)
but only to finite corrections $\propto m^{2} - \mu^{2}$ .

\subsubsection*{3.7}
Above, we implicitly assumed a  unique cut-off scale on both fermionic and bosonic
loops. Assume that $\Lambda_{\rm UV}^{boson} = r\Lambda_{\rm UV}^{fermion}$
for some $r$. Derive the relations between the relevant 
couplings in the last example as a function of $r$.
For what values of $r$ are the relations scale (\ie renormalization-group) invariant? 
%

%
%
%

%C4
\chapter{Supersymmetry}
\label{c4}

%401
In the previous chapter it was shown that in trying to eliminate
quadratic divergences in two- and one-point  functions of a scalar field
one is led to postulate an equal number of bosonic and fermionic 
$d.o.f.$, as well as specific relations among the couplings in
the theory.  (Uniqueness was achieved once the relations were required
to be scale invariant.) From the consideration of the
scalar-vector interaction one is led to postulate new gauge-Yukawa
and gauge-quartic interactions. From the consideration of a
(non-gauge) Yukawa interaction one is able to fix the scalar
trilinear and quartic couplings. These relations suggest a small
number of fundamental couplings, and hence a small number of
fundamental (super)fields: Fermions and bosons may be arranged in spin
multiplets, the superfields, which interact with  each other. 
In this (supersymmetric)
language the Lagrangian will contain a small number of terms.
Nevertheless, when rewriting the interactions and kinetic terms
in terms of the superfield components -- the individual boson and fermion 
fields -- the Lagrangian will
contain a large number of terms which exhibit what may naively seem
to be mysterious correlations.
%

%402
The field theoretical formalism that realizes this elegant idea
is supersymmetry, with the spin multiplets as its fundamental 
building blocks which are subject to the supersymmetric
transformations: Notions which bring to mind low-energy flavor
symmetries and their fundamental building blocks, the isospin
multiplets. Isospin transformations act on the proton-neutron doublet.
supersymmetric ``spin'' transformations act on the boson-fermion 
multiplet, the superfield.
In order to define supersymmetry as a consistent (at this point, global)
symmetry one needs to extend the usual space-time 
to  superspace by the introduction of (not surprisingly
equal number of)  ``fermionic'' (or spinorial) coordinates
\vspace*{0.4cm} 
\begin{center}
$\{x_{\mu}\}_{\mu =1,...,4} \rightarrow
\left\{x_{\mu},  \left(\begin{array}{c}\theta_{\alpha}\\
\bar{\theta}_{\dot\beta}\end{array}\right)
\right\}_{\begin{array}{c}\mu =1,...,4 \\ \alpha,\dot{\beta} = 1,2.
\end{array}}$ 
\end{center}
\vspace*{0.4cm}
The new coordinate $\theta$ is a spin  $1/2$ object, \ie a  
Grassmann variable which carries two $d.o.f.$
(and which, as shown below, carries mass dimensions). 
% 

%403
The formalism of supersymmetric field theories
can be found in the literature: Refs.~\cite{HPN,HPN90,HPN93,JONES,LYK,TATA1,MART}
provide a useful sample of such reviews, while Wess and Bagger \cite{WB} 
provide a complete, though somewhat condensed, discussion.
It is not our purpose to introduce supersymmetry
in an elaborate way, but rather to outline it for completeness
and to illustrate some of its more intuitive
aspects (which are also the more relevant ones for the discussion in the rest of
these notes).
We begin with a brief introduction to Grassmann algebra. Next, we will
define a chiral superfield and use it to illustrate the supersymmetry algebra.
We then will fill in some of the details such as kinetic and interaction terms. We
conclude with brief discussions of a gauge theory and, finally, of gauging
supersymmetry itself. 
\section{Grassmann Algebra}
\label{sec:grassmann}
%

%411
The Grassmann variables are anticommuting variables,
\begin{equation}
\{\theta_{i{\alpha}},\,\theta_{j{\beta}}\} = 0
\label{s4e1}
\end{equation}
with the antisymmetric tensor 
\vspace*{0.4cm} 
\begin{center}
$\epsilon_{\alpha\beta} = \left(\begin{array}{rccc}0&& 1&\\ -1&& 0&
\end{array}\right)$
\end{center}
\vspace*{0.4cm}
as metric.
One has, for example, $\theta_{i \alpha} =
\epsilon_{\alpha\beta}\theta^{\beta}_{i}$,
where Greek indices correspond to spinorial indices while Latin indices
(which are sometimes omitted) label the variable.
It is customary to reserve a special notation to the hermitian
conjugate (as above)
$(\theta^{\alpha}_{i})^{\dagger} = \bar{\theta}_{i}^{\dot{\alpha}}$.
The Pauli matrices, which act on 
Grassmann variables, are then written as $\sigma^{\mu}_{\dot{\alpha}\alpha}$.
%

%412
One defines derivatives and integration with respect
to the Grassmann variables as
\begin{equation}
\frac{d\theta_{i \alpha}}{d\theta_{j \beta}} = \delta_{i}^{j}\delta_{\alpha}^{\beta},
\label{s4e2}
\end{equation}
\begin{equation}
\int d\theta_{i} = 0;\,\,\,
\int d\theta_{i \alpha} \theta_{j\beta}= \delta_{ij}\delta_{\alpha\beta}.
\label{s4e3}
\end{equation}
In eq.~(\ref{s4e3}) the first relation follows from the invariance 
requirement $\int d\theta_{i} f(\theta_{i})= \int 
d\theta_{i} f(\theta_{i} + \theta_{i_{0}})$ where $f$ is an arbitrary function, e.g.
$f(\theta_{i}) = \theta_{i}$.
The second relation is simply a choice of normalization.
The derivative operator is often denoted with the short-hand
notation $\partial_{\theta_{j}}$. Note that derivation
(\ref{s4e2}) and integration (\ref{s4e3})  
of Grassmann variables are equivalent operations.
%

%413
The function $f(\theta_{i})$ can be expanded as a polynomial in $\theta_{i}$.
However, the anticommuting nature of $\theta_{i}$,
\vspace*{0.4cm} 
\begin{center}
$\theta^{\alpha}_{i}\theta_{i\alpha}\theta_{i\alpha} =
\frac{1}{2}\theta^{\alpha}_{i}\left\{\theta_{i\alpha},\, 
\theta_{i\alpha}\right\} = 0$,
\end{center}
\vspace*{0.4cm} 
implies that $f(\theta_{i})$ can at most be a quadratic polynomial in $\theta_{i}$.
In particular, any expansion is finite!
Using its polynomial form and the above definitions, 
the integration-derivation equivalence follows for any function of Gramssmann variables,
\begin{equation}
\int d\theta_{i} f(\theta_{i})= 
\frac{df(\theta_{i})}{d\theta_{i}}.
\label{s4e4}
\end{equation}
Another  useful relation that will be implicitly used in some of the expansions below is
\begin{equation}
\theta^{\alpha}_{i}\theta^{\beta}_{i} = - \frac{1}{2}\epsilon^{\alpha\beta}
\theta^{\alpha}_{i}\theta_{i \alpha}.
\label{id1}
\end{equation}
%

%414
In the following, we will omit Latin indices when discussing 
an expansion in only one Grassmann variable.
\section{The Chiral Superfield: The L-Representation}
\label{sec:chiral}
%

%421
Equipped with the algebraic tools, our obvious task is to
embed the usual bosons and fermions into spin multiplets, the
superfields. Begin by considering a Higgs field, \ie a (complex) scalar
field $\phi(x_{\mu})$. Requiring the corresponding superfield
$\Phi(x_{\mu},\,\theta_{\alpha},\,
\bar{\theta}^{\dot\beta})$ -- which is a function defined in superspace
rather than only in space-time -- to carry the same
quantum numbers and mass dimension as $\phi$, and expanding it, for
example,  in powers of $\theta_{\alpha}$
(with ${x}_{\mu}$-dependent coefficients), 
one arrives at the following embedding:
\begin{equation}
\Phi(x_{\mu},\,\theta_{\alpha}) = \phi(x_{\mu})
+ \theta^{\alpha}\psi_{\alpha}(x_{\mu}) +
\theta^{\alpha}\theta_{\alpha}F(x_{\mu}). 
\label{Phi}
\end{equation}
(We will return to the issue of an expansion in $\bar{\theta}^{\dot\beta}$ below.) 
From the spinorial nature of $\theta_{\alpha}$ one has that
$\psi_{\alpha}$ is also a spin $1/2$ object (so that their product
is a scalar), which can be identified with the usual chiral fermion.
Given that, we can deduce the mass dimensionality of $\theta$: $[\theta]
= [\Phi] - [\psi] = [\phi] - [\psi] = -1/2$. Lastly, the coefficient
of the $\theta^{2}$ term must also be a (complex) scalar.
(Note that if it was a vector field, which could also allow
for the total spin of this term to be zero,  it would have carried
an additional space-time index, unlike $\Phi$.)
Its mass dimension is also uniquely determined
$[F] = [\Phi] - 2[\theta] = 2$.
%

%422
The coefficient $F(x)$ cannot be identified with any known physical  $d.o.f.$
(There are no fundamental scalars whose mass dimension equals two!)
Furthermore, its presence implies four bosonic $vs.$ only two fermionic
$d.o.f.$, seemingly violating the supersymmetric principle found in the
previous chapter.
In order to understand the  role of $F$ and the resolution of this
issue, let us first find the (supersymmetry) transformation law.
%

%423
Consider the transformation $S(\Theta,\,\bar{\Theta})$ acting
on the superfield $\Phi(x,\,\theta)$ of eq.~(\ref{Phi})
(where $\Theta$ is also a Grassmann variable). 
Listing all objects
at our disposal with dimensionality greater than $[\phi] - [\Theta] = 3/2$
but less than  $[F] - [\Theta] = 5/2$  we have  
a spin $1/2$ mass-dimension $3/2$ object $\psi$;
spin zero mass-dimension 2 objects $F$ and $\partial_{\mu}\phi$;
and a spin $1/2$ mass-dimension $5/2$ object $\partial_{\mu}\psi$.
Using the Pauli matrices to contract space-time and spinorial indices,
and matching spin and dimensions of the different objects $O_{I}$:
$[\delta O_{I}] = [\Theta] + [O_{J}]$,
one immediately arrives at the transformation law (up to the bracketed coefficients)
\begin{equation}
\delta\phi = \Theta^{\alpha}\psi_{\alpha},
\label{trans1}
\end{equation}
\begin{equation}
\delta\psi_{\alpha} = (2)\Theta_{\alpha} F +
(2i)\sigma^{\mu}_{\dot\alpha\alpha}
\bar{\Theta}^{\dot\alpha}\partial_{\mu}\phi,
\label{trans2}
\end{equation}
\begin{equation}
\delta F = 
(-i)(\partial_{\mu}\psi^{\alpha})\sigma^{\mu}_{\dot\alpha\alpha}
\bar{\Theta}^{\dot\alpha}.
\label{trans3}
\end{equation}
The ``component fields'' transform into each other, which is possible
since the superspace coordinate carries mass dimensions and spin.
Hence, the superfield $\Phi$ transforms onto itself.
%

%424
The (linearized) transformation law can be written in a more compact
fashion:
\begin{equation}
S(\Theta,\,\bar{\Theta})\Phi(x,\,\theta) = 
\left\{ \Theta\frac{\partial}{\partial\theta}
+ \bar{\Theta}\frac{\partial}{\partial\bar{\theta}}
+ 2i\theta\sigma^{\mu}\bar{\Theta}\partial_{\mu}
\right\}\Phi(x,\,\theta),
\label{trans}
\end{equation}
where the second term on the right-hand side is zero and 
the identity (\ref{id1}) was used to derive (\ref{trans3}) (from
the last term on the right-hand side). Rewriting the differential 
transformation in a standard form
$S(\Theta,\,\bar{\Theta})\Phi(x,\,\theta) = 
\left\{\Theta Q + \bar{Q}\bar{\Theta}\right\}\Phi(x,\,\theta)$,
where $Q$ is the supercharge, \ie the generator of the supersymmetric
algebra, on can identify the generators
\begin{equation}
Q_{\alpha} = \frac{\partial}{\partial\theta_{\alpha}}; \,\,\,
\bar{Q}_{\dot\alpha} = -\frac{\partial}{\partial\bar{\theta}_{\dot\alpha}}
+2i\theta^{\alpha}\sigma^{\mu}_{\dot\alpha\alpha}\partial_{\mu},
\label{Q}
\end{equation}
and $P_{\mu} =  i\partial/\partial x^{\mu} = i\partial_{\mu}$.
%

%425
The generators  are fermions rather than bosons!
Their anticommutation relations can be found by performing successive
transformations on $\Phi$. After carefully re-arranging the result, it reads
\begin{equation}
\left\{Q_{\alpha},\,\bar{Q}_{\dot\beta}\right\} =
2\sigma_{\alpha\dot\beta}^{\mu}P_{\mu},
\label{acr1}
\end{equation}
\begin{equation}
\left\{Q_{\alpha}\, , Q_{\beta}\right\} =
\left\{\bar{Q}_{\dot\alpha}\, , \bar{Q}_{\dot\beta}\right\} = 0.
\label{acr2}
\end{equation}
Relation (\ref{acr1}) implies that in order to close the algebra,  one
needs to include space-time. (This hints towards a relation
to gravity as is indeed one finds in the local case.)
Thus, the algebra must also include the usual
commutation relation
\begin{equation}
\left[P_{\mu},\,P_{\nu}\right] = 0
\end{equation}
as well as the relations
\begin{equation}
\left[Q_{\alpha},\,P_{\nu}\right] = \left[\bar{Q}_{\dot\alpha},\,P_{\nu}\right] = 0.
\label{QP}
\end{equation}
The last relation (\ref{QP})
is only the statement that the supersymmetry is a global symmetry.
(More generally, supersymmetry provides a unique consistent extension
of the Poincare group.)
Such an algebra which includes both commutation and anti-commutation
relations among its generators is referred to as a graded Lie algebra. 
%

%426
The algebra described here has one supersymmetric charge, $Q$. 
Hence, this is a $N = 1$ (global) supersymmetry. In
general, theories with more supersymmetries can be constructed.
However, already $N = 2$ theories do not contain chiral matter,
and $N =4$ theories contain only gauge fields. 
Hence, extended supersymmetric theories,
with certain exceptions in the case of $N=2$, 
cannot describe nature at electroweak energies and 
$N=1$ is unique in this sense.
Discussion of the $N = 2$ case is reserved for the last part of these notes,
and $N > 2$ theories will not be considered.
%

%427
Though it is not our intention to discuss the algebra in great detail,
some general lessons can be drawn by observation. The commutation
between the charge and the Hamiltonian $[Q,\,H] = [Q,\,P_{0}] = 0$
implies that supersymmetric transformations which transform bosons to
fermions and vice versa, do not change the ``energy levels'', \ie
there is a boson-fermion mass degeneracy in these theories. 
The Hamiltonian can be written as 
$H = P_{0} = (1/2)\{Q,\bar{Q}\} = QQ^{\dagger} \geq 0$
and hence is semi-positive definite. (That is, the potential $V$ must vanish,
$\langle V \rangle = 0 $.) If (global) supersymmetry 
is conserved then the vacuum is invariant under the transformation
$Q|0\rangle = 0$ and hence the vacuum energy $\langle0|H|0\rangle = 0$
is zero, regardless whether, for example, the gauge symmetry is
conserved or not. If (global) supersymmetry is broken spontaneously then the vacuum
energy $\sim \| Q|0\rangle\|^{2} > 0$ is positive.
The semi-positive definite 
vacuum energy is an order parameter of (global) supersymmetry breaking.
The Goldstone particle of supersymmetry is given by the operation of
its (super)current $ \sim Q_{\alpha}$ on the vacuum, 
and hence it is a fermion $\psi_{\alpha}$, the Goldstino.
%

%428
The superfield $\Phi(x_{\mu},\,\theta_{\alpha})$ is called a chiral
superfield since it contains a chiral fermion. It constitutes a
representation of the algebra as is evident from the closed relations
that we were able to deduce, mostly by simple arguments.
(The arbitrary coefficients in (\ref{trans1})--(\ref{trans3}) were fixed
by requiring consistency of the algebra.) 
It is called the left- (or L-)representation. The more general chiral
representation, which contains both chiral (left-handed) and
anti-chiral (right-handed) fermions, 
is related to it by
$\Phi(y_{\mu},\,\theta_{\alpha},\,\bar{\theta}_{\dot\alpha})
= \Phi_{L}(y_{\mu} + i\theta_{\alpha}\sigma_{\mu}^{\alpha\dot\alpha}
\bar{\theta}_{\dot\alpha},\,\theta_{\alpha})\, 
[\equiv \Phi_{L}(x_{\mu},\theta_{\alpha})]$, where the shift in
$y_{\mu}$ is to be treated as a translation operation on the superfield, 
and the object that was previously 
denoted by $\Phi$ is now denoted by $\Phi_{L}$.
There is also a right (or R-)representation given by
$\Phi(y_{\mu},\,\theta_{\alpha},\,\bar{\theta}_{\dot\alpha})
= \Phi_{R}(y_{\mu} - i\theta_{\alpha}\sigma_{\mu}^{\alpha\dot\alpha}
\bar{\theta}_{\dot\alpha},\,\bar{\theta}_{\dot\alpha})\,
[\equiv \Phi_{R}(x_{\mu}^{*},\,\bar{\theta}_{\dot\alpha})]$.
In these notes we will mostly use the $L$-representation.
%

%429
It is instructive to verify that the L-representation
is closed. For example, we saw that 
$S(\Theta,\,\bar\Theta)\Phi_{L}(x_{\mu},\,\theta)
= \Phi_{L}^{\prime}(x_{\mu},\,\theta)$. Similarly, it is
straightforward to show that $\Phi^{2} = \phi^{2} +2\theta\phi\psi +
\theta^{2}(2\phi F - (1/2) \psi^{2})$ is also a left-handed field,
where $\theta^{2} = \theta^{\alpha}\theta_{\alpha}$
($\bar\theta^{2} = \bar\theta_{\dot\alpha}\bar\theta^{\dot\alpha}$).
Also, multiplying  $\Phi^{2}\Phi$ one  has  $\Phi^{3} = \cdots + \theta^{2}(\phi^{2}F
- (1/2)\phi\psi^{2})$, a result which will be useful below.
%

%4210
Before returning to
discuss the dimension two scalar field that seems to have appeared in
the spectrum, the vector representation (which also contains such a field) 
will be given. 
\section{The Vector Superfield}
\label{vector}
%

%431
A vector field $V_{\mu}(x_{\mu}) $ is to be embedded in a real object,
a vector superfield
$V(x_{\mu},\,\theta_{\alpha},\,\bar{\theta}_{\dot\alpha}) =
V^{\dagger}(x_{\mu},\,\theta_{\alpha},\,\bar{\theta}_{\dot\alpha})$.
Hence, aside from vector bosons, it can contain only a real scalar configuration 
and Majorana fermions. If it to contain spin zero and/or spin $1/2$
fields, it cannot carry itself a space-time index. Also, it must
depend on real combinations of $\theta$ and $\bar{\theta}$.
In particular, its expansion contains terms
up to a $\theta^{2}\bar{\theta}^{2}$ term.
One then arrives at the following form (coefficients are again fixed 
by the consistency of the algebra)
\begin{eqnarray}
V(x_{\mu},\,\theta_{\alpha},\,\bar{\theta}_{\dot\alpha}) &=&
-\theta_{\alpha}\sigma_{\mu}^{\alpha\dot\alpha}\bar{\theta}_{\dot\alpha}V^{\mu}(x_{\mu}) 
\nonumber\\&&
+ i\theta^{2}\bar{\theta}^{\dot\alpha}\bar{\lambda}_{\dot\alpha}(x_{\mu})
-i\bar{\theta}^{2}{\theta}^{\alpha}{\lambda}_{\alpha}(x_{\mu}) 
+\frac{1}{2}\theta^{2}\bar{\theta}^{2}D(x_{\mu}).
\label{V}
\end{eqnarray}
The vector superfield $V$ is a dimensionless spin zero object!  
Its component fields include, aside from the vector boson, also a
Majorana fermion $\lambda$  (the gaugino) and a real scalar field $D$, again with
mass dimension $[D] = 2$. It again contains four bosonic $vs.$ only
two fermionic $d.o.f.$, a problem we already encountered in the case
of a chiral superfiled, and again we postpone its resolution.
%

%432
In fact, the expansion (\ref{V}) is not the most general possible expansion.
The following terms: $C$, $i\theta\chi -i\bar{\theta}\bar{\chi}$,
and $i\theta^{2}(A +iB) -i\bar{\theta}^{2}(A -iB)$, where $A,\,B,\,C$
are real scalar fields and $\chi$ is again a Majorana fermion,
are dimensionless, spin zero, real terms that could appear in $V$
(in addition to other derivative terms).
We implicitly eliminated the additional $d.o.f.$ by fixing the gauge.
While in ordinary gauge theory the gauge parameter is a real scalar field, in
supersymmetry it has to be a chiral superfiled, hence it contains
four bosonic and two fermionic $d.o.f.$ Subtracting one bosonic $d.o.f.$
to be identified with the usual gauge parameter, three bosonic and two
fermionic $d.o.f.$ are available for fixing the ``supersymmetric
gauge''. These can then be used to eliminate $A,\,B,\,C$ and $\chi$ from
the spectrum. This choice corresponds to the convenient
Wess-Zumino gauge, which is used in these notes. 
%

%433
Similar considerations to the ones used above in order to write the chiral field
transformation laws (\ref{trans1})--(\ref{trans2}) can be used to
write the transformation law $S(\Theta,\,\bar\Theta)V(x_{\mu},\, \theta,\,\bar\theta)$
of the component field of the vector superfiled,
\begin{equation}
\delta V_{\mu} = (i)\left[ 
\Theta\sigma_{\mu}\bar{\lambda} + \bar\Theta\sigma_{\mu}\lambda\right],
\label{trans4}
\end{equation}
\begin{equation}
\delta \lambda = \bar\Theta D +
\Theta\sigma^{\mu\nu}\left[\partial_{\mu}V_{\nu} -
\partial_{\nu}V_{\mu}\right],
\label{trans5}
\end{equation}
\begin{equation}
\delta D = \sigma^{\mu}\partial_{\mu}\left[-\Theta\bar{\lambda} +
\bar\Theta\lambda
\right],
\label{trans6}
\end{equation}
where spinorial indices are suppressed and $\sigma^{\mu\nu} =
(i/2)[\sigma^{\mu},\sigma^{\nu}]$, as usual.
%
 
%434
An important object that transforms in the vectorial representation is
$\Phi_{L}\Phi_{L}^{\dagger}$. In order to convince oneself, 
simply note that it is a real field. In more detail, note that
$\Phi_{L}^{\dagger}$ must depend on $\bar\theta$ only, and is
therefore in the right-handed representation
[$\left\{\Phi_{L}(y_{\mu},\,\theta,\,\bar{\theta})\right\}^{\dagger}
=\left\{\Phi_{L}(x_{\mu},\,{\theta})\right\}^{\dagger} 
=\Phi_{L}^{\dagger}(x_{\mu}^{*},\,\bar{\theta})]$. 
Hence, $\Phi_{L}\Phi_{L}^{\dagger}$ expansion
contains terms up to $\theta^{2}\bar\theta^{2}$.
Of particular interest to our discussion below is the coefficient
of this last term in the $\Phi_{L}\Phi_{L}^{\dagger}$ expansion, 
the $D$-term, $\Phi_{L}\Phi_{L}^{\dagger} = \cdots +
\theta^{2}\bar\theta^{2}D$ where
\begin{equation}
\int d^{2}\theta d^{2}\bar{\theta}\Phi_{L}\Phi_{L}^{\dagger} =
D = FF^{*} - \phi\partial^{\mu}\partial_{\mu}\phi^{*} + 
(i/2)\psi\sigma_{\mu}\partial^{\mu}\bar\psi.
\label{Dkinetic}
\end{equation}
%

%435
In turn, it is also possible to construct a useful chiral object  $W_{\alpha}$
from vectorial fields (which carries spin $1/2$).
For a $U(1)$ theory one has $W_{\alpha} =
(\partial_{\bar\theta})^{2}(\partial_{\theta} + 
2i\sigma_{\mu}\bar\theta\partial^{\mu}) V \sim \lambda_{\alpha}
+ \theta_{\alpha}(F_{\mu\nu} + D) +  \theta^{2}\partial_{\mu}\lambda_{\alpha}$,
and $F_{\mu\nu}$ is as usual the stress tensor. 
It is straightforward to show that $W_{\alpha}$ is a chiral field:
From the anti-commutation one has
$\partial_{\bar\theta}W_{\alpha} = 0$. It follows that $W_{\alpha}$
is independent of $\bar\theta$
and hence is a left-handed chiral superfield.
\section{From the Auxiliary $F$ and $D$ Fields to the Lagrangian}
\label{sec:aux}
%

%441
After discussing chiral and vector superfields, understanding their
expansion in terms of component fields and learning some simple
manipulations of these super-objects, we are in position to do the
final count of the physical $d.o.f.$ and resolve the 
tantalizing access of bosonic $d.o.f.$ which are manifested
in the form of dimension-two scalar
fields. This requires us to derive the usual 
interaction and kinetic terms for the
component fields. We begin with a non-gauge theory.
%

%442
It is not surprising that the two issues are closely related. Observe that
the $F$-term of the bilinear $\Phi^{2}$, given by $\int d^{2}\theta \Phi^{2}$,
includes terms which resemble mass terms, while 
$\int d^{2}\theta \Phi^{3}$ includes Yukawa-like terms. Also,
$\delta F$ given in (\ref{trans3}) is a total derivative, for any
chiral superfield.
That is, the mysterious $F$-field contains interaction and mass terms
on the one hand, and, on the other hand, $\delta\int d^{4}x F = \int d^{4}x ({\rm total-
derivative}) = 0$ so that $\int d^{4}x F$ is 
invariant under supersymmetric transformations and
hence, is a good candidate to describe the potential. Indeed, if it is
a non-propagating (auxiliary) $d.o.f.$ which encodes the potential
then there are no physical spin-two bosonic $d.o.f.$
%

%443
Clearly, one needs  to identify  the kinetic
terms in order to verify this conjecture.
A reasonable guess would be that they are given by the 
self-conjugate combination 
$\Phi\Phi^{\dagger}$. Indeed, the corresponding $D$-term eq.~(\ref{Dkinetic})
contains kinetic terms for $\phi$ and $\psi$, but not for $F$.
The $D$-field itself also transforms 
(for any vector superfield)
as a total derivative, so again it is not a coincidence that
it is a term which could be consistently used to write 
the appropriate  kinetic terms in the Lagrangian. 
The $F$-fields are therefore auxiliary
non-propagating fields that could be eliminated from the
theory. That is, they do not correspond to any physical $d.o.f.$
(But what about the $D$-fields? Are they also auxiliary fields?
The kinetic terms of the gauge-fields are given by the $F$-term
of $W^{2}$, which can be shown to not contain a kinetic term for the $D$-field:
The $D$-term is again an auxiliary field which does not represent 
a physical $d.o.f.$)
%

%444
More specifically, consider a theory described by a Lagrangian
\begin{equation}
{\cal L} = 
\int d^{2}\theta  \left[\mu \Phi^{2} + y\Phi^{3} \right] + h.c. 
+ \int d^{2}\theta d^{2}\bar\theta 
\Phi\Phi^{\dagger},
\label{toylagrangian}
\end{equation}
We can then describe the component (ordinary) field theory
by performing the integrations $\int d^{2}\theta$ and 
$\int d^{2}\theta d^{2}\bar\theta$
over the super-space coordinates and
extracting the corresponding $F$ and $D$ terms, respectively:
\begin{eqnarray}
F_{\Phi^{2}} + F_{\Phi^{2}}^{*} & :& 2\mu\phi F - \frac{1}{2}\mu\psi\psi + h.c.;  \label{Fphi2}\\
F_{\Phi^{3}} + F_{\Phi^{3}}^{*} & :& 
3y\phi^{2} F - \frac{3}{2}y\phi\psi\psi + h.c.; \label{Fphi3}\\
D & : & |\partial_{\mu}\phi|^{2} -
\frac{i}{2}\bar{\psi}{\rlap{$\not$}{\partial}}\psi -FF^{*};
\label{Dterm}
\end{eqnarray}
The Lagrangian ${\cal{L}} = T - V$ is given by their sum. 
%   

%445
Indeed, no kinetic terms appear for $F$. Hence, one can
eliminate the auxiliary $d.o.f.$ by solving $\partial V/\partial F = 0$
and substituting the solution,
\begin{equation}
\frac{\partial V}{\partial F} = 0 \Rightarrow F^{*} = -(2\mu\phi + 3y\phi^{2}),
\label{dVdF}
\end{equation}
back into the potential V. One finds for the scalar potential 
(after reorganizing the conjugate terms)
\begin{equation}
V({\phi}) = FF^{*}  = 4|\mu|^{2}\phi\phi^{*} +
6(y\mu^{*}\phi^{*}\phi^{2} + h.c.) +
9|y|^{2}|\phi\phi^{*}|^{2}. 
\label{Vphi}
\end{equation}
The fermion mass, kinetic and (Yukawa) interaction terms are already given
explicitly in (\ref{Fphi2})--(\ref{Dterm}).
%

%446
It is now instructive to compare our result (\ref{Vphi}) to our
conjectured ``finite'' theory from the previous chapter. In order to
do so it is useful to use standard normalizations, \ie 
$\mu \rightarrow \mu/2$; $y \rightarrow y/3$; and most importantly
$\psi \rightarrow \psi/\sqrt{2}$.  Not surprisingly, our Lagrangian includes
all of (and only) the $d.o.f.$ previously conjectured and it exhibits the
anticipated structure of interactions. In particular trilinear and
quartic couplings are not arbitrary and are given by the masses and
Yukawa couplings, as required. Quartic couplings are (semi-)positive definite.

\section{The Superpotential} 

%451
Let us now define  a function 
\begin{equation}
W = \mu\Phi^{2} + y\Phi^{3},
\end{equation}
then $F^{*} = -\partial W/\partial\Phi$ where 
the substitution $\Phi \rightarrow \phi$ is understood at the last
step. The scalar potential is then simply 
\vspace*{0.4cm} 
\begin{center}
$V = |\partial W/\partial \Phi|^{2}$. 
\end{center}
\vspace*{0.4cm} 
Also note that the potential terms involving the fermions are simply given by 
\vspace*{0.4cm} 
\begin{center}
$-(1/2)\sum_{IJ}(\partial^{2}
W/\partial\Phi_{I}\partial\Phi_{J})\psi_{I}\psi_{J} + h.c.$
\end{center}
\vspace*{0.4cm} 
(These general ``recipes'' will not be justified here.)
The function $W$ is the superpotential which describes
our theory. In general, the superpotential could be any analytic
(holomorphic) function. In particular, it can depend
on any number of chiral superfields $\Phi_{I}$, but not on their
complex conjugates $\Phi_{I}^{\dagger}$ (the holomorphicity property).
%

%452
For example, the most general renormalizable superpotential describing a theory
with one singlet field is $W = c +l\Phi + \mu\Phi^{2} + y\Phi^{3}$.
That is, $W$ has mass dimension $[W] = 3$ and any terms which involve
higher powers of $\Phi$ are necessarily non-renormalizable terms.
Indeed, $W = \kappa\Phi^{4} \Rightarrow V = |\partial
W/\partial\Phi|^{2} = 8\kappa^{2}\phi^{6}$, \ie $[\kappa] = -1$
is a non-renormailzable coupling. 
The holomorphicity property of $W$ is at the core on the
non-renormalization
theorems that protect its parameters from divergences
(and of which some aspects were shown in the previous lecture
but from a different point of view). 

\section{The K\"ahler Potential}

%461
The non-gauge theory is characterized by its superpotential and 
also by its kinetic terms. Above, it was shown that the kinetic terms
are given in the one singlet model simply by $\Phi\Phi^{\dagger}$.
More generally, however, one could write a functional form
\begin{eqnarray}
\Phi\Phi^{\dagger} &\rightarrow& K(\Phi,\,\Phi^{\dagger}) = K_{J}^{I}\Phi_{I}\Phi^{J}
+ \left(H^{IJ}\Phi_{I}\Phi_{J} + h.c.\right) \nonumber \\ && + 
{\mbox{Non-renormalizable terms}}
\end{eqnarray}
(with $\Phi^{J} = \Phi_{J}^{\dagger}$)
which is a dimension $[K] = 2$ real function: The K\"ahler potential. The canonical kinetic
terms $K_{J}^{I}\partial_{\mu}\phi_{I}\partial^{\mu}\phi^{J} = 
\sum_{I}|\partial_{\mu}\phi_{I}|^{2}$ are given 
by the ``minimal'' K\"ahler potential, which is
a renormalizable function with  
$K^{I}_{J} = \delta^{I}_{J}$, as we implicitly assumed above. 
Note  that holomorphic terms 
(\ie terms which do not depend on $\Phi^{\dagger}$'s)
such as $H^{IJ}\Phi_{I}\Phi_{J}$ do
not alter the normalization of the kinetic terms since 
$\int d^{2}\theta d^{2}\bar\theta\left\{H^{IJ}\Phi_{I}\Phi_{J} +
h.c.\right\} = 0$. (These terms, however, could play an interesting
role in the case of local supersymmetry, \ie supergravity.)
%

%462
The K\"ahler potential which
determines the normalization of the kinetic terms cannot share the
holomorphicity of the superpotential. Hence,
$W$ and $K$ are very different functions which together define the
(non-gauge) theory. In particular, the wave function renormalization
of the fields (and therefore of the superpotential couplings)
is determined by the K\"ahler potential which undergoes renormalization
(or scaling).

\section{The Case of a Gauge Theory}
\label{sec:gauge}

%471
The more relevant case of a gauge theory is obviously more
technically evolved. Here, we will only outline
the derivation of the component Lagrangian of
an interacting gauge theory, stressing those elements which
will play a role in the following chapters.
The pure gauge theory is described by
\begin{equation}
\int d^{2}\theta  
\frac{1}{2g^{2}}W^{\alpha}W_{\alpha},
\label{toylagrangian2}
\end{equation}
leading to the usual gauge kinetic terms, as well as kinetic terms for
the gaugino $\lambda$, given by
(in the normalization of eq.~(\ref{toylagrangian2}))
\begin{eqnarray}
F_{W^{2}} & : &  -\frac{1}{4}|F^{\mu\nu}|^{2} + \frac{1}{2}D^{2}
-\frac{i}{2}\bar{\lambda}{\rlap{$\not$}{D}}\lambda, \label{FW2}
\end{eqnarray}
with ${\rlap{$\not$}{D}}$ as usual 
the derivative operator contracted with the Pauli matrices.
There are kinetic term for the physical gauge bosons and
fermions, but not for the $D$ fields. The $D$ fields are indeed
auxiliary fields. 
%

%472
More generally, $(1/2g^{2})W^{\alpha}W_{\alpha} \rightarrow
f_{\alpha\beta}W^{\alpha}W^{\beta}$ and $f_{\alpha\beta}$ is a
holomorphic (analytic) function which determines the normalization
of the gauge kinetic terms: $f_{\alpha\beta}$ is the gauge-kinetic
function. It could be a constant pre-factor or a function
of a dynamical (super)field $f_{\alpha\beta}(S)$. 
$S$ is then the dilaton superfield.
In either case 
(once put in its canonical form) it is essentially 
equivalent to the gauge coupling, $f_{\alpha\beta} = \delta_{\alpha\beta}/2g^{2}$, 
only that in the latter case the gauge couplings is determined by the $vev$ of the 
dilaton $S$.
%

%473
As before, the usual kinetic terms are described by the $D$-terms,
however, they are now written as
$\int d^{2}\theta d^{2}\bar\theta 
\Phi e^{2gQV}\Phi^{\dagger}$
where for simplicity the gauge coupling will be set $g = 1$ (see exercise); 
$Q = 1$ is the charge, and 
$\Phi e^{2V}\Phi^{\dagger} = \Phi\Phi^{\dagger} +
\Phi 2V\Phi^{\dagger} + \cdots$ contains the gauge-covariant kinetic
terms ($\partial_{\mu} \rightarrow D_{\mu} \equiv \partial_{\mu} +
iV_{\mu}$). (More generally, $\Phi e^{2gQV}\Phi^{\dagger} \rightarrow
K(\Phi_{I}, e^{2gQ_{J}V}\Phi^{J})$.) That is, 
\begin{eqnarray}
D & : & |D_{\mu}\phi|^{2} - \frac{i}{2}\bar{\psi}{\rlap{$\not$}{D}}\psi
+ \phi^{*}D\phi + i\phi^{*}\lambda\psi -i\bar{\lambda}\bar{\psi}\phi -FF^{*}.
\label{Dterm2}
\end{eqnarray}
Note that when recovering the explicit dependence on the gauge
coupling, the strength of the gauge Yukawa interaction 
$\phi^{*}\lambda\psi -\bar{\lambda}\bar{\psi}\phi$ is $g$.
(In the standard normalization $\psi \rightarrow \psi/\sqrt{2}$
it is $\sqrt{2}g$.) Note that the Lagrangian of the toy theory
described in Sec.~\ref{sec:s21} is derived here from the gauge Lagrangian
eq.~(\ref{toylagrangian2}), as was already suggested following eq.~(\ref{s2Lint}). 
%

%474
Consider now a theory 
\begin{equation}
\int d^{2}\theta  \left[(\mu \Phi^{+}\Phi^{-} + h.c.)
+ \frac{1}{2}W^{\alpha}W_{\alpha} \right]
+ \int d^{2}\theta d^{2}\bar\theta 
\left\{\Phi^{+} e^{2V}\Phi^{+\,\dagger} + \Phi^{-} e^{-2V}\Phi^{-\, \dagger}\right\},
\label{toylagrangian3}
\end{equation}
where the matter fields carry $+1$ and $-1$ charges.
One has 
\begin{equation}
\frac{\partial V}{\partial F^{+}} = 0 \Rightarrow F^{+\,*} = -\mu\phi^{-} 
\label{dVdFplus}
\end{equation}
\begin{equation}
\frac{\partial V}{\partial F^{-}} = 0 \Rightarrow F^{-\,*} = -\mu\phi^{+} 
\label{dVdFminus}
\end{equation}
\begin{equation}
\frac{\partial V}{\partial D} = 0 \Rightarrow -D = \phi^{+ *}\phi^{+} - \phi^{- *}\phi^{-},
\label{dVdD}
\end{equation}
where in deriving (\ref{dVdD}) we used eqs.~(\ref{FW2}) and
(\ref{Dterm2}). The scalar potential is now readily derived,
\begin{equation}
V({\phi}) = FF^{*} + \frac{1}{2}D^{2} = |\mu|^{2}\phi^{+}\phi^{+\,*} +
|\mu|^{2}\phi^{-}\phi^{-\,*} +
\frac{1}{2}|\phi^{+}\phi^{+\,*} -\phi^{-}\phi^{-\,*}|^{2}.
\label{Vphi2}
\end{equation}
The quartic potential is now dictated by the gauge coupling 
($-D = g(\phi^{+ *}\phi^{+} - \phi^{- *}\phi^{-})$ once reinstating the gauge coupling)
and again is semi-positive definite.
Note that the quartic $D$-potential (the last term in (\ref{Vphi2}))
has a flat direction $\langle \phi^{+} \rangle = 
\langle \phi^{-} \rangle$, possibly destabilizing
the theory: This is a common phenomenon is supersymmetric field theories.
%

%475
The derivation of the interactions in the case of a gauge theory
essentially completes our sketch of the formal derivation of the ingredients
that were required in the previous section in order to guarantee that the
theory is at most logarithmically divergent. Specifically, quartic and gaugino
couplings were found to be proportional to the gauge coupling (or its square)
and with the appropriate coefficients. One is led to conclude that
supersymmetry is indeed a natural habitat of weakly coupled theories.
(Supersymmetry also provides powerful tools for the study
of strongly coupled theories \cite{STRONG}, 
an issue that will not be addressed in these notes.)
The next step is then the supersymmetrization of the Standard Model.
Beforehand, however, we comment on gauging supersymmetry itself
(though for the most part these notes will be concerned 
with global supersymmetry).

\section{Gauging Supersymmetry}
\label{sec:sugra}

%481
The same building blocks are used in the case of local supersymmetry,
which is referred to as supergravity, only that the gravity multiplet
(which includes the graviton and the spin $3/2$ gravitino) is now a
gauge multiplet. The gravitino absorbs the Goldstino once
supersymmetry is spontaneously broken and
acquires a mass 
\begin{equation}
m_{3/2} = \frac{\langle W \rangle}{M_{P}^{2}}e^{\frac{\langle
K\rangle}{2M_{P}^{2}}}.
\label{m32}
\end{equation}
The reduced Planck mass 
$M_{P} = \mplanck/\sqrt{8\pi} \simeq 2.4 \times 10^{18}$ GeV
is the inverse of the supergravity expansion parameter (rather than
the Planck mass itself) and it provides the fundamental mass scale 
in the theory.
Hence, local supersymmetry corresponds to gauging gravity
(and hence, supergravity). 
The appearance of momentum in the anticommutation
relations of global supersymmetry already hinted in this direction.
Gravity is now included automatically, opening the door for
gauge-gravity unification. However, supergravity is a
non-renormalizable theory and is still expected to provide only
a ``low-energy'' limit of the theory of quantum gravity
(for example, of a (super)string theory).
%

%482
The gravitino mass could be given by an expectation value of any  
superpotential term, particularly a term that does not 
describe interactions of known light fields. 
Because supergravity is a theory of gravity, 
all fields in all sectors of the theory ``feel'' the massive
gravitino (since (super)gravity is modified).
For example, one often envisions a scenario
in which supersymmetry is broken spontaneously in a hidden sector --
hidden in the sense that it interacts only gravitationally with the SM
sector -- and the SM sector appears as globally supersymmetric
but with explicit supersymmetry breaking terms that are functions
of the gravitino mass. (See Sec.~\ref{sec:supergravity}.)
%

%483
The $F$-terms are still good order parameters for supersymmetry
breaking
though they take a more complicated form (since they are now given by
a covariant derivative of the superpotential):
\begin{equation}
-F_{\Phi}^{*} = \frac{\partial W}{\partial \Phi} 
+ \frac{\partial K}{\partial \Phi}\frac{W}{M_{P}^{2}},
\end{equation}
where in the  canonical normalization 
one has ${\partial K}/{\partial \Phi} = \phi^{*}$.
Indeed, $|F| > 0$ if supersymmetry is broken.
The scalar potential (or vacuum energy) now reads
\begin{equation}
V = e^{K/M_{P}^{2}}\left\{F_{I}K_{J}^{I}F^{J} - 
\frac{3}{M_{P}^{2}}|W|^{2}\right\}, 
\label{Vsugra}
\end{equation}
and is not a good order parameter as it could take either sign, or
preferably vanish, even if $|F| > 0$. 
This is in fact a blessing in disguise as it allows one to cancel the
cosmological constant (that was fixed in the global case to $\langle V
\rangle = \langle |F|^{2}\rangle $)
by tuning
\begin{equation}
\langle W \rangle  = \frac{1}{\sqrt{3}}\langle F \rangle M_{P}.
\end{equation}
One can define the scale of spontaneous supersymmetry breaking
\begin{equation}
M_{SUSY}^{2} = 
\langle F \rangle {\rm exp}[{\langle K \rangle /2M_{P}^{2}}]. 
\label{Msusy2}
\end{equation}
The cancellation of the cosmological constant then gives
\begin{equation} 
M_{SUSY}^{2} = \sqrt{3} m_{3/2}M_{P}
\end{equation}
as a geometric mean of the
gravitino and supergravity scales. Note that regardless of the size of
$M_{SUSY}$, supersymmetry breaking is communicated to
the whole theory (to the full superpotential)
at the supergravity scale $M_{P}$, as is evident from the form
of the potential (\ref{Vsugra}). 
(The symbol $M_{SUSY}$ itself is also used to denote the mass scale
of the superpartners of the SM particles, as we discuss in the 
following chapter. In that case $M_{SUSY}$ is the scale of
explicit breaking in the SM sector. The two meaning should not be
confused.) 
%

%484
There could be other scales $M < M_{P}$ at which
other mechanisms mediate supersymmetry breaking from some special
(but by definition not truly hidden) sector of the theory to the (supersymmetrized) 
SM fields. In this case there is multiple mediation and one has to examine
whether supergravity or a different ``low-energy'' mechanism
dominates. (We will return to this point when discussing models
of supersymmetry breaking in chapters~\ref{c10} and \ref{c13}.)
%

%485
Of course, the supergravity scalar potential has also the usual $D$-term
contribution $V_{D} = (1/4f_{\alpha\beta})D_{\alpha}D_{\beta}$, 
and the $D$-terms are also order parameters of supersymmetry breaking.
(However, it seems more natural to fine-tune the supepotential to
cancel its first derivative (the $F$-term) in order to eliminate the
cosmological constant, rather than fine-tune it to cancel the square of
the $D$-terms.)
%

%486
We will briefly return to supergravity when discussing 
in Chapter~\ref{c10} the origins
of the explicit supersymmetry breaking in the SM sector,
but otherwise the basic tools of global supersymmetry will suffice 
for our purposes.
%

%E4
\section*{Exercises}
\addcontentsline{toc}{section}{Exercises}
\markright{Exercises}

\subsubsection*{4.1}
Since $\Phi_{L}$ by itself constitutes a representation of the
algebra, its component fields cannot transform for example to
a vectorial object $V_{\mu}$. Show this explicitly from the
consideration of  dimensions, space-time 
and spinorial indices.

\subsubsection*{4.2}
Use the definition of $\Phi_{R}$  to show that it is
independent of $\theta_{\alpha}$.

\subsubsection*{4.3}
Confirm the expansion of $\Phi_{L}^{2}$ and $\Phi_{L}^{3}$.

\subsubsection*{4.4}
Examine the expansion of the vector superfield (\ref{V}),
show that  $V = V^{\dagger}$, and confirm all the assertions regarding
spin and mass dimension of the component fields.

\subsubsection*{4.5}
Derive the transformation law  (\ref{trans4})--(\ref{trans6})
(aside from coefficients) from dimensional and spin arguments.

\subsubsection*{4.6}
Use the relations given in the text to
derive the expression for 
$\Phi(y_{\mu},\,\theta,\, \bar\theta) = \Phi_{L}(y_{\mu} +
i\theta\sigma_{\mu}\bar\theta,\,\theta)
= [1 -\delta y_{\mu} P^{\mu}  + (1/2)(\delta y_{\mu}P^{\mu})^{2}]
\Phi_{L}(y_{\mu},\,\theta)$
with $\delta y_{\mu} = i\theta\sigma_{\mu}\bar\theta$. 
Derive the expression for $\Phi_{L}^{\dagger}$.
Integrate $\int d^{2}\theta d^{2}\bar \theta
\Phi_{L}\Phi_{L}^{\dagger}$ to find the $\Phi_{L}\Phi_{L}^{\dagger}$
$D$-term.

\subsubsection*{4.7}Derive the complete expression for the vector
field in its spinorial-chiral representation $W_{\alpha}$ and confirm
explicitly that it is a spin $1/2$ chiral superfield.
 
\subsubsection*{4.8}
Verify that our derivation of the non-gauge potential agrees with the 
theory conjectured in the previous chapter.

\subsubsection*{4.9}
Find the correctly normalized Yukawa couplings $y$, $W = (y/3)\Phi^{3}_{I}$,
if $K^{I}_{J}= k_{I}\delta^{I}_{J}$. (The fields have to be rescaled so 
that the kinetic terms have their
canonical form.) What about
$W = y_{IJK}\Phi_{I}\Phi_{J}\Phi_{K}$? 
Verify your choice by examining the resulting form of the
quartic potential.

\subsubsection*{4.10}
Show that 
$\int d^{2}\theta d^{2}\bar\theta\left\{H^{IJ}\Phi_{I}\Phi_{J} + h.c.\right\} = 0$. 

\subsubsection*{4.11}
Recover the explicit dependence on the gauge coupling 
and on the charge in the expressions in Sec.~\ref{sec:gauge}. 
Show that the quartic coupling is proportional to $g^{2}/2$
and that self couplings are positive definite.

\subsubsection*{4.12}
Introduce a Yukawa term $y\Phi^{0}\Phi^{+}\Phi^{-}$
to the model described by  eq.~(\ref{toylagrangian3}) and work out the scalar
potential
and the component Yukawa interactions. What are the quartic couplings?

\subsubsection*{4.13}
Show that the Lagrangians (\ref{toylagrangian}), (\ref{toylagrangian2})  and
(\ref{toylagrangian3}) are total derivatives.

\subsubsection*{4.14}
Show that all the supergravity expressions reduce to the global
supersymmetry case in the limit $M_{P} \rightarrow \infty$
in which supergravity effects decouple.
In this limit, for example,  
supersymmetry breaking is not communicated from the
hidden to the SM sector. It may be communicated in this case
from some other sector via means other than gravity.
%
%
%

%C5
\chapter{Supersymmetrizing the Standard Model}
\label{c5}

%501
The discussion in the previous sections suggests that the SM could
still be a sensible theory at the ultraviolet and at the same time
insensitive to the ultraviolet cut-off scale if supersymmetry
is realized in the infrared regime (specifically, near the Fermi scale).
Let us then consider this possibility and construct a (minimal)
supersymmetric extension of the SM.

\section{Preliminaries}
%511
Nearly all of the necessary ingredients are already
present within the SM: scalar bosons, (chiral) fermions and gauge bosons.
Each of the SM gauge-boson multiplets requires the presence of 
a real (Majorana) fermion with the same quantum numbers, its
{\it gaugino} superpartner; each SM fermion requires the presence
of a complex scalar boson with the same quantum numbers, its 
{\it sfermion} superpartner (which also inherits its chirality label);
the Higgs boson requires the introduction of the {\it Higgsino},
its fermion superpartner. It is straightforward to convince oneself that
the theory will not be anomaly free unless two Higgsino doublets (and hence,
by supersymmetry, two Higgs doublets) with opposite hypercharge are introduced
so that the trace over the Higgsino hypercharge vanishes, 
in analogy to the vanishing hypercharge trace of each SM generation.
(Recall our construction in Chapter~\ref{c3}.)
The minimal supersymmetric extension of the Standard Model (MSSM)
as was outlined above
is the extension with minimal new matter content, 
and it must contain a two Higgs doublet model (2HDM) (see eq.~(\ref{s2V4}))
with $H_{1} = ((H_1^{0} +iA_{1}^{0})/\sqrt{2},\, H_{1}^{-})^{T}$  and
$H_{2} = (H_{2}^{+},\, (H_2^{0} +iA_{2}^{0})/\sqrt{2})^{T}$.
Note that the MSSM contains also Majorana fermions.
Discovery of supersymmetry will not only 
establish the existence of fundamental scalar fields but 
also of Majorana fermions! 
%

%512
We list, following our previous notation,
the MSSM field content in Table~\ref{table:mattermssm}. For
completeness we also include the gravity multiplet with the spin 2
graviton $G$ and its spin $3/2$ gravitino superpartner $\widetilde{G}$,
as well as  a generic
supersymmetry breaking scalar superfield $X$ (with an auxiliary component
$\langle F_{X}\rangle \neq 0$)
whose fermion partner $\widetilde{X}$
is the Goldstone particle of supersymmetry breaking, the Goldstino.
The Goldstino is absorbed by the gravitino.
($X$ parameterizes, for example, the hidden sector mentioned in Sec.~\ref{sec:sugra}.)
It is customary to name a sfermion with an ``$s$'' suffix attached to the
name of the corresponding fermion, e.g. {\it top}-quark $\rightarrow$
{\it stop}-squark and $\tau$-lepton $\rightarrow$ {\it stau}-slepton. The
gaugino
partner of a gauge boson $V$ is typically named $V$-ino, for example,
$W$-boson $\rightarrow$ wino. The naming scheme for the mass eigenstates
will be discussed after the diagonalization of the mass matrices.
All superpartners of the SM particles are denoted as s-particles, or
simply {\it sparticles}.
%

%513
Given the above matter content and the constrained relations between
the couplings that follow from supersymmetry, the MSSM is guaranteed
not to contain quadratic divergences. In order to achieve that and 
have a sensible formalism to treat a fundamental scalar
we were forced to introduce a complete scalar
replica of the SM matter which, however, supersymmetry enables us
to understand in terms of a boson-fermion symmetry. 
By doing so, one gives up in a sense spin as a good quantum number.
For example, the Higgs doublet $H_{1}$ and the lepton doublets $L_{a}$
are indistinguishable at this level: In the SM the former is a
scalar field while the latter are fermions. The anomaly
cancellation considerations explained above 
do not allow us to identify $H_{1}$ with $L_{a}$, \ie
the Higgs boson with a slepton (or equivalently, a lepton with a Higgsino).
The possibility of lepton-Higgs mixing arises naturally if no
discrete symmetries are imposed to preserve lepton number, 
a subject that we will return to below. 
Spin and the SM spectrum correspond in this framework  to the low-energy
limit. That is, supersymmetry must be broken at a scale below the
typical cut-off scale that regulates the quadratic divergences $\Lambda \lesssim
4\pi M_{W}$. If the stop-squark, for example, is heavier, then
divergences must be again fine-tuned away --  undermining
our original motivation. (As we shall see, consistency with experiment does not
allow the stop to be much lighter.) 
%
  
%514
As explained in the previous section, the boson-fermion symmetry
and the  duplication of the spectrum
follow naturally once the building blocks used to describe nature
(at electroweak energies and above) are the (chiral and vector) superfields.
However, from the low-energy point of view it is the
{\it component field} interactions which are relevant.   
Let us then elaborate on the structure of the 
{\it component field} interactions, 
which is dictated by the supersymmetry invariant
interactions of the parent superfields. This will be done in the next sections. 
%

%515
Before doing so, it is useful to fix the normalization of the Higgs
fields, as their expectation values appear in essentially all
mass matrices of the MSSM fields.
It is customary to define in 2HDM in general, and in the MSSM in particular,
\vspace*{0.4cm} 
\begin{center}
$\sqrt{\langle{H_{1}^{0}}\rangle^{2}  +\langle{H_{2}^{0}}\rangle^{2}} =
\frac{\sqrt{2}}{g}M_{W}
\simeq 174\,\,{\rm GeV} \equiv \nu$.
\end{center}
\vspace*{0.4cm}
Note that a $1/\sqrt{2}$ factor is now absorbed in the definition of
$H_{i}$ (and $\nu$) in comparison to the Higgs doublet definition in the SM.
The {\it vev} $\nu$ is not a free parameter but is fixed by the
Fermi scale (or equivalently by the $W$ or $Z$ mass).
However, the ratio\footnote{The definition given here corresponds to a physical parameter
only at tree level. The ``physical'' definition
of the angle $\beta$ (see also Chapter~\ref{c8}) is corrected at one-loop 
in a way which depends on the observable used to extract it.} 
of the two {\it vev's} 
\begin{equation}
\tan\beta =
\frac{\langle{H_{2}^{0}}\rangle}{\langle{H_{1}^{0}}\rangle} \equiv
\frac{{\nu_{2}}}{{\nu_{1}}}\gtrsim 1
\label{tanbeta}
\end{equation}
is a free parameter. Its positive sign corresponds a conventional phase choice
that we adopt, and its lower bound stems from pertubativity of Yukawa
couplings $\propto 1/\sin\beta$. (See Chapter~\ref{c7}.)
Correspondingly, one has 
$\langle{H_{1}^{0}}\rangle \equiv \nu_{1} = \nu\cos\beta$  
and $\langle{H_{2}^{0}}\rangle \equiv \nu_{2} = \nu\sin\beta$.
%

%516
The supersymmatrization of the standard model is also discussed
in many of the reviews and notes listed in the previous chapters
(particularly, Refs.~\cite{HPN,TATA1,MART}), as well as in
Refs.~\cite{HK}, \cite{HAB}, and \cite{KAZ}. Recent summaries by the relevant
Tevatron \cite{RUN2} and by  other \cite{FRENCH} working groups 
are also useful, but address particular models. Historical perspective 
on the supersymmetric SM was given recently by Fayet \cite{HISTORY}.
%

%T2
%%%%%%%%%%%%%%%%%%%%%%%TABLE
\begin{center}
\begin{table}
\caption{The Minimal Supersymmetric SM (MSSM) field content. Our notation is
explained in Table~\ref{table:mattersm}.
Each multiplet $(Q_{c},Q_{L})_{Q_{Y/2}}$
is listed according to its  color, weak isospin and hypercharge
assignments. Note that chirality is now associated, by supersymmetry,  
also with the
scalar bosons. In particular,
the superpartner of a fermion $f_{L}^{c}$ is a sfermion
$\tilde{f}^{*}_{R}$ and that of a fermion $f_{L}$ is a sfermion
$\tilde{f}_{L}$.
The model contains two Higgs doublets.
In the case of the matter (Higgs) fields, the same symbol will be used
for the superfield as for its fermion (scalar) component.}
\label{table:mattermssm}
\vspace*{0.3cm}

\begin{center}
\begin{tabular}{|c|c|c|}\hline

Multiplet& Boson & Fermion\\ 
\hline 
\multicolumn{3}{|c|}{Gauge fields} \\ 
\hline
&&\\
$(8,\,1)_{0}$ & $g$ & $\widetilde{g}$ \\
&&\\
$(1,\,3)_{0}$ & $W$ & $\widetilde{W}$ \\
&&\\
$(1,\,1)_{0}$ & $B$ & $\widetilde{B}$ \\ 
&&\\
\hline 
\multicolumn{3}{|c|}{Matter fields} \\ 
\hline 
&&\\
$(3,2)_{\frac{1}{6}}$ & $\widetilde{Q}_{a}$ & $Q_{a}$ \\
&&\\
$(\bar{3},1)_{-\frac{2}{3}}$ & $\widetilde{U}_{a}$ & $U_{a}$ \\
&&\\
$(\bar{3},1)_{\frac{1}{3}}$ & $\widetilde{D}_{a}$ & $D_{a}$ \\
&&\\
$(1,2)_{-\frac{1}{2}}$ & $\widetilde{L}_{a}$ & $L_{a}$ \\
&&\\
$(1,1)_{1}$ & $\widetilde{E}_{a}$ & $E_{a}$ \\
&&\\
\hline 
\multicolumn{3}{|c|}{Symmetry breaking} \\ 
\hline
&&\\ 
$(1,2)_{-\frac{1}{2}}$ & $H_{1}$ & $\widetilde{H}_{1}$ \\
&&\\
$(1,2)_{\frac{1}{2}}$ & $H_{2}$ & $\widetilde{H}_{2}$ \\
&&\\
\hline 
\multicolumn{3}{|c|}{Gravity
and supersymmetry breaking} \\ 
\hline 
&&\\
$(1,\,1)_{0}$ & $G$ & $\widetilde{G}$ \\
&&\\ 
$(1,\,1)_{0}$ & $X$ & $\widetilde{X}$ \\ 
&&\\
\hline
\end{tabular}
\end{center}
\end{table}
\end{center}
\section{Yukawa, Gauge-Yukawa, and Quartic Interactions}
\label{sec:s51}
%

%521
Identifying the gauge kinetic function $f_{\alpha\beta} =
\delta_{\alpha\beta}/2g_{a}^{2}$ for a gauge group $a$, our 
exercise in Sec.~\ref{sec:gauge} 
showed that the fermion $f$ coupling to a vector boson
$V_{a}$ with a coupling $g_{a}$ implies, by supersymmetry, the
following interaction terms:
\begin{center}
 $g_{a}V_{a}\bar{f}T^{i}_{a}f \rightarrow \left\{
\begin{tabular}{c}
\\
$\sqrt{2}g_{a}\lambda_{a}\tilde{f}^{*}T^{i}_{a}f +  h.c.$ \\
\\
$\frac{g_{a}^{2}}{2}\sum_{I}\left|\tilde{f}^{*}_{I}T_{a}^{i}
\tilde{f}_{I}\right|^{2}$
\\
\\
\end{tabular}
\right.$,
\end{center}
in addition to usual gauge interactions of the sfermion $\tilde{f}$. 
Here, we explicitly denote the gauge group generators 
$T_{a}^{i}$ which are taken in the appropriate representation 
(of the fermion $f$)
and which are reduced to the fermion charge in the case of a $U(1)$
considered above. 
The gaugino is denoted by $\lambda_{a}$ and, converting to standard
conventions, it is defined to absorb a factor of $i$ which appeared
previously in the gaugino-fermion-sfermion vertex. Also, the
fermion component of the chiral superfield absorbs a factor of
$1/\sqrt{2}$ as explained in the previous section. 
The gaugino matter interaction is often referred to as 
gauge-Yukawa interaction.
The quartic interaction is given by the square
of the auxiliary $D$ field and hence includes a summation over
all the scalar fields which transform under the gauge group, weighted by their
respective charge. This leads to ``mixed'' quartic interactions, for example,
between two squarks and two Higgs bosons or two squarks and two sleptons.
%

%522
The matter (Yukawa) interactions, and as we learned -- the corresponding
trilinear and quartic terms in the scalar potential, are described by
the superpotential. Recall that the superpotential $W$ is a
holomorphic function so it cannot contain complex conjugate
fields. Particularly, no terms involving $H_{1}^{*}$ can be written.
Indeed, we were already forced to introduce $H_{2} \sim H_{1}^{*}$, 
for the purpose of anomaly cancellation.
The holomorphicity property offers, however, an independent reasoning
for introducing two Higgs doublets with opposite hypercharge:
A Yukawa/mass term for the {\it up} quarks, which $\sim H_{1}^{*}$ in the SM,
can now be written using  $H_{2}$. 
Keeping in mind the generalization $V \sim y\phi_{i}\psi_{j}\psi_{k}
\rightarrow W \sim y\Phi_{I}\Phi_{J}\Phi_{K}$, the MSSM Yukawa
superpotential is readily written
\begin{equation}
W_{\rm Yukawa} = y_{l_{ab}}\epsilon_{ij}H_{1}^{i}L_{a}^{j}E_{b} +
y_{d_{ab}}\epsilon_{ij}H_{1}^{i}Q_{a}^{j}D_{b} 
- y_{u_{ab}}\epsilon_{ij}H_{2}^{i}Q_{a}^{j}U_{b}, 
\label{Wyuk}
\end{equation}
were $SU(2)$ indices are explicitly displayed (while $SU(3)$ indices are
suppressed) and our phase convention is such that all ``mass'' terms
and Yukawa couplings are positive.
%

%523
It is instructive to explicitly write 
all the component interactions encoded in the
superpotential (\ref{Wyuk}). Decoding the $SU(2)$ structure 
and omitting flavor indices, one has
\begin{eqnarray}
W_{\rm Yukawa} &=& y_{l}(H_{1}^{0}E_{L}^{-}E - H_{1}^{-}N_{L}E) +
y_{d}(H_{1}^{0}D_{L}D - H_{1}^{-}U_{L}D) \nonumber \\ 
&&+ y_{u}(H_{2}^{0}U_{L}U - H_{2}^{+}D_{L}U),
\label{Wyuk2}
\end{eqnarray}
with self explanatory definitions of the various superfield symbols.
We note in passing that
explicit decoding of the $SU(2)$ indices need to be carried out cautiously in
certain cases, 
for example, when studying the vacuum structure
(which is generally not invariant under $SU(2)$ rotations).  
However, it is acceptable here.
Considering next, for example, the first term $y_{l}H_{1}^{0}E_{L}^{-}E$,
one can derive the
component interaction $\sim (\partial^{2} W
/\partial\Phi_{I}\Phi_{J})\psi_{I}\psi_{J}+ {h.c.} + |\partial W
/\partial\Phi_{I}|^{2}$ (where the substitution  
$\Phi = \phi + \theta\psi \rightarrow \phi$
in $\partial^{n}W$ is understood),
\begin{eqnarray}
{\cal{L}}_{\rm Yukawa} &=& y_{l}\left\{
H_{1}^{0}E_{L}^{-}E +
\widetilde{H}_{1}^{0}\widetilde{E}_{L}^{-}E +
\widetilde{H}_{1}^{0}E_{L}^{-}\widetilde{E}\right\} + {h.c.},\\
V_{\rm scalar(quartic)} &=&
y_{l}^{2}\left\{ \right|\widetilde{E}_{L}^{-}\widetilde{E}\left|^{2}
+ \right|H_{1}^{0}\widetilde{E}_{L}^{-}\left|^{2} +
\right|H_{1}^{0}\widetilde{E}\left|^{2}
\right\}.
\end{eqnarray}
A single Yukawa interaction in the SM $H_{1}^{0}E_{L}^{-}E$
leads, by supersymmetry, to two additional Yukawa terms and three
quartic terms in the MSSM (or any other supersymmetric extension).
%

%524
It is important to note that the holomorphicity of the superpotential
which forbids terms $\sim H_{1}^{*}QU,\,H_{2}^{*}QD,\,H_{2}^{*}LE$,
automatically implies that the MSSM contains a 2HDM of type II, that
is a model in which each fermion flavor couples only to one of
the Higgs doublets. This is a crucial requirement for suppressing
dangerous tree-level contributions to FCNC from operators such as
$QQUD$ which result from virtual Higgs exchange in a general 2HDM,
but which do not appear in a type II model.
\section{The Higgs Mixing Parameter}
\label{sec:s52}
%

%531
While no ``supersymmetric'' (\ie  a holomorphic superpotential)
mass involving the (SM) matter fields can be written, a Higgs mass
term $\mu\epsilon_{ij}H_{1}^{i}H_{2}^{j}$ is gauge invariant 
and is allowed. The dimension $[\mu] = 1$ parameter mixes the two
doublets and acts as a Dirac mass for the Higgsinos.
The Dirac fermion and the Higgs bosons are then degenerate, as implied
by supersymmetry, with mass $|\mu|$. The $\mu$-parameter can carry an
arbitrary phase. Hence, choosing $\mu > 0$, one has for the MSSM superpotential
\begin{eqnarray}
W_{\rm MSSM} 
&=& 
W_{\rm Yukawa} \pm
\mu\epsilon_{ij}H_{1}^{i}H_{2}^{j} \nonumber \\
& = & y_{l_{ab}}\epsilon_{ij}H_{1}^{i}L_{a}^{j}E_{b} +
y_{d_{ab}}\epsilon_{ij}H_{1}^{i}Q_{a}^{j}D_{b} \nonumber \\
&-& y_{u_{ab}}\epsilon_{ij}H_{2}^{i}Q_{a}^{j}U_{b} \pm 
\mu\epsilon_{ij}H_{1}^{i}H_{2}^{j}.
\label{WMSSM}
\end{eqnarray}
%

%532
Following the standard procedure it is straightforward to derive
the new interaction and mass terms, which complement the usual Yukawa
and quartic potential terms,
\begin{eqnarray}
{\cal{L}}_{\rm Dirac} &=& \mp\mu\widetilde{H}_{1}^{-}\widetilde{H}_{2}^{+}
\pm\mu\widetilde{H}_{1}^{0}\widetilde{H}_{2}^{0} + {h.c.},\\
V_{\rm scalar(mass)} &=& \mu^{2}\left(H_{1}H_{1}^{\dagger}
+ H_{2}H_{2}^{\dagger}\right),\\
V_{\rm scalar(trilinear)} &=& 
\pm\mu y_{l}H_{2}\widetilde{L}^{\dagger}\widetilde{E}^{\dagger}
\pm\mu y_{d}H_{2}\widetilde{Q}^{\dagger}\widetilde{D}^{\dagger}
\mp\mu y_{u}H_{1}\widetilde{Q}^{\dagger}\widetilde{U}^{\dagger}
+ {h.c.},
\end{eqnarray}
where $SU(2)$ indices were also suppressed and the fermion mass terms
were written in $SU(2)$ components, as it will be useful later
when constructing the fermion mass matrices.
Note that the trilinear terms arise from the cross terms in
$|\partial W/\partial H_{2}|^{2} = |\pm\mu H_{1} - y_{u}QU|^{2}$ 
and $|\partial W/\partial H_{1}|^{2} = |\pm\mu H_{2} + y_{l}LE + y_{d}QD|^{2}$.
%

%533
Could it be that $\mu \simeq 0$? Naively, one may expect that in this
case the Higgsinos are massless and therefore $Z$-boson decays
to a Higgsino pair $Z \rightarrow \widetilde{H}\widetilde{H}$
should have been observed at the $Z$ resonance (from total or
invisible width measurements, for example). In fact, once electroweak symmetry breaking effects
are taken into account (see below) the two charged Higgsinos
as well as one neutral Higgsino are 
degenerate in mass with the respective Goldstone bosons and are therefore
massive with $M_{W}$ and $M_{Z}$ masses, respectively. 
Their mass follows from the gauge-Yukawa interaction terms
$\sim g\langle{H_{i}^{0}}\rangle\widetilde{W}\widetilde{H}$
and $g\langle{H_{i}^{0}}\rangle\widetilde{Z}\widetilde{H}$.
While the $Z$ decays are kinematicly forbidden in this case,
charged Higgsinos should have been produced in pairs at the $WW$ 
production threshold at the Large Electron Positron (LEP)
collider at CERN. The absence of anomalous events at the $WW$
threshold allows one to exclude this possibility and to bound $|\mu|$
from below. (This observation will become clearer after defining the fermion
mass eigenstates. For a discussion see Ref.~\cite{FPT}.)
%

%534
Obviously, $|\mu|$ cannot be too large or the Higgs doublets
will decouple, re-introducing a conceptual version of the hierarchy problem.
As we proceed we will see that $\mu$ must encode information on the ultraviolet
in order for it to be a small parameter of the order of the Fermi scale. 
Hence, it may contain one of the keys to unveiling the high-energy theory.
\section{Electroweak vs. Supersymmetry Breaking}
\label{sec:s53}
%

%541
Having established the MSSM matter-matter and matter-gaugino interactions
(in the supersymmetric limit in which only $D$ and $F$ terms are considered),
the new particle spectrum can be written down.
It is instructive to first consider the already available
pattern of symmetry breaking in nature, electroweak symmetry breaking,
and whether it is sufficient to generate a consistent superpartner spectrum. 
This is the  limit 
of the MSSM with no explicit supersymmetry breaking.
If electroweak $SU(2)\times U(1)$ symmetry is conserved,
all particles -- fermions and bosons, new and old -- are massless
in this limit.
Electroweak symmetry breaking (EWSB) is responsible for all mass
generation and, in particular,
is required break the boson - fermion degeneracy.
Indeed, supersymmetry is not conserved but is spontaneously broken 
once electroweak symmetry is broken. 
As will be shown below, the non-vanishing
Higgs {\it vev's} generate non-vanishing $F$ and $D$ {\it vev's},
the order parameters of supersymmetry breaking. 
Let us then postulate for the sake of argument
that the Higgs fields acquire {\it vev's}
which spontaneously break $SU(2)\times U(1)$, and consider the impact
on supersymmetry breaking on the sparticle spectrum.
(As we shall see later, as a matter of fact such {\it vev's} cannot
be generated in this limit.)
%

%542
Consider, for example,  the sfermion spectrum. The spectrum can be
organized by flavor (breaking $SU(2)$ doublets to their components),
each flavor sector constitute a left-handed
$\tilde{f}_{L}$ and right-handed $\tilde{f}_{R}$ sfermions (and in principle
all sfermions with the same QED and QCD quantum numbers could further
mix in a $6\times 6$ subspaces). One has in each sector three
possible mass terms $m^{2}_{LL}\tilde{f}_{L}\tilde{f}_{L}^{*},\, 
m^{2}_{RR}\tilde{f}_{R}^{*}\tilde{f}_{R},\, 
m^{2}_{LR}\tilde{f}_{L}\tilde{f}_{R}^{*} + {h.c.}$, with $m_{LL,\,RR}^{2}
\sim 
\langle D \rangle \sim \nu^{2}$
generated by substituting  the Higgs {\it vev} in the quartic potential
and $m_{LR}^{2} \sim \langle F \rangle \sim \mu\nu$
generated by substituting  the Higgs {\it vev} in the trilinear
potential. The mass matrix for the sfermions 
$(\tilde{f}_{L},\,\tilde{f}_{R}^{*})$ 
\vspace*{0.4cm}
\begin{eqnarray}
M^{2}_{\tilde{f}}
& = &
\left(\begin{tabular}{cc}
$m^{2}_{LL}$&$m^{2}_{LR}$\\
$m^{2\,\,*}_{LR}$& $m_{RR}^{2}$
\end{tabular}\right) 
\nonumber 
\label{sfermion}
\end{eqnarray}
\vspace*{0.4cm}
then reads in this limit
\begin{eqnarray}
\left(\begin{tabular}{cc}
$m^{2}_{f} + M_{Z}^{2}\cos 2\beta\left[T_{3}^{f} - Q_{f_{L}}\sin^{2}\theta_{W}\right]$&
$m_{f}\mu^{*}\tan\beta\,\,\,({\rm or}\,\, 1/\tan\beta)$\\
$m_{f}\mu\tan\beta\,\,\,({\rm or}\,\, 1/\tan\beta)$&
$m^{2}_{f} - (M_{Z}^{2}\cos 2\beta \times Q_{f_{R}}\sin^{2}\theta_{W})$
\end{tabular}\right). &&
\nonumber
\end{eqnarray}
The sneutrino is an exception since there are no singlet
(right-handed) neutrinos and sneutrinos, and its mass is simply
$m^{2}_{LL}\widetilde{\nu}_{L}\widetilde{\nu}_{L}^{*}$.
Hereafter we will simply write 
$m^{2}_{LR}\tilde{f}_{L}\tilde{f}_{R} + {h.c.}$ with $\tilde{f}_{R}$
understood to denote the $SU(2)_{L}$ singlet sfermions
(\ie the conjugation will not be written explicitly).
%

%543
It is instructive to examine all contributions in some detail (see
also exercises). The diagonal fermion mass term is the
$F$-contribution $|y_{f}\langle H^{0}_{i}\rangle\tilde{f}_{L,\,R}|^{2}$.
Note that this particular $F$-term itself has no {\it vev}, 
$\langle F \rangle \sim y_{f} \langle H_{i} \tilde{f}\rangle = 0$,
and hence it does not break supersymmetry but only lead to a degenerate mass
contribution to the fermion and to the respective sfermions. The second
diagonal contribution is that of the $D$-term which has a non-vanishing
{\it vev} 
\begin{equation}
\langle D \rangle^{2} = \frac{g^{2} + g^{\prime\,2}}{8}\left(
\langle{H_{2}}\rangle^{2} - \langle{H_{1}}\rangle^{2}\right)^{2}
= \frac{1}{4}M_{Z}^{2}\nu^{2}|\cos 2\beta|^{2} = \langle V_{\rm MSSM} \rangle^{2},
\label{Dvev}
\end{equation}
where $\cos 2 \beta < 0$ appears only as absolute value
(so that $\langle V_{\rm MSSM}\rangle \geq 0$
as required by (global) supersymmetry); $T_{3} = \pm 1/2$ for the Higgs weak
isospin was used. Relation (\ref{Dvev}) holds in general
(and not only in the supersymmetric limit).
The $D$-term {\it vev} breaks supersymmetry. However,
the limit $\tan\beta \rightarrow 1$ corresponds to
$\langle D \rangle^{2} \rightarrow 0$ and supersymmetry is recovered.
Indeed in this
limit the diagonal sfermion masses squared are given by the respective
fermion mass squared. This limit also corresponds to a flat direction 
in field space (as alluded to in the previous chapter)
along which the (tree-level) potential vanishes.
A flat direction must correspond to a massless (real) scalar field
in the spectrum (its zero mode)
so there is a boson whose (tree-level) mass is proportional
to the $D$-term and vanishes as $\cos 2 \beta \rightarrow 0$.
This is the (model-independent) light Higgs boson of supersymmetry.
This is a crucial point in the
phenomenology of the models which we will return to when discussing
the Higgs sector in Chapter~\ref{c8}. It is then straightforward, after rewriting
the hypercharge in term of electric charge $Q_{f}$, to arrive to the
contribution to the sfermion mass squared. 
One can readily calculate
the numerical coefficients $T_{3} - Q_{f}\sin^{2}\theta_{W}$ using 
$\sin^{2}\theta_{W}\simeq 0.23$ for the weak angle
to find 
$T_{3} - Q_{f}\sin^{2}\theta_{W} \simeq 0.34,\,0.14,\,-0.42,\,
-0.08,\,-0.27,\,-0.23,\,0.5$
for $f = u_{L},\,u_{L}^{*},\,d_{L},\,d_{L}^{*},\,e_{L},\,e_{L}^{*},\,\nu$
(and their generational replicas).
%

%544
Lastly, the off-diagonal terms (often referred to as left-right mixing)
are supersymmetry breaking terms which split the sfermion spectrum from
the fermion spectrum even in the absence of $D$-term contributions.
These terms arise from the cross-terms in $F_{H_{i}}$, as do the 
supersymmetry breaking {\it vev's}
$\langle F_{H_{1,\,2}} \rangle^{2}  = y_{f}^{2}\mu^{2}\langle H_{2,\,1}\rangle^{2}$. 
The $\tan\beta$ dependence in mass squared  matrix assumes  $f = d,\,l$
($f = u$), \ie  a superpotential coupling to $H_{1}$ ($H_{2}$).
The presence of the left-right terms implies
that the sfermion mass eigenstates have no well-defined
chirality association.
%

%545
Since supersymmetry is spontaneously broken by the Higgs {\it vev's},
the Higgsinos provide the Goldstino.
This is the Achilles heal of this scheme:
Let us now define the supertrace function
\vspace*{0.4cm} 
\begin{center}
STR$_{I}{\cal{F}}(O_{I})\equiv \sum_{I}N_{c_{I}}
(-1)^{2S_{I}}(2S_{I} + 1){\cal{F}}(O_{I})$
\end{center}
\vspace*{0.4cm}
which sums over any function ${\cal{F}}$ of an object $O_{I}$
which carries spin $S_{I}$. (The summation is also over color and
isospin factors which here we pre-summed in the color factor $N_{c}$.)
In the $\langle D \rangle \rightarrow 0$
limit, the supertrace summation over the mass eigenvalues of the fermions
and sfermions in each flavor sector is zero! 
After spontaneous supersymmetry breaking (in this limit) one has
$m_{\tilde{f}_{1,\,2}}^{2} = m_{f}^{2} \pm \Delta$ where $\Delta = 
\langle F \rangle$ is given by the $F$-{\it vev} that spontaneously
break supersymmetry, in our case the left-right mixing term, so that ${\rm STR}M = 0$.
Hence, unless $|\langle F \rangle| < m_{e} \sim 0$, which it cannot
be given the lower bound on $|\mu|$ discussed previously,  a sfermion
must acquire a negative mass squared and consequently 
the SM does not correspond to a minimum of the potential 
(and furthermore, QED and QCD could be broken in the vacuum). 
This situation is, of course, intolerable.
It arises because the matter fields couple directly to the Goldstino
supermultiplet.
(It is a general (supertrace) theorem for any spontaneous braking of 
global supersymmetry.) Another general implication is that
the fermions do not feel the spontaneous supersymmetry breaking (at
tree level). 
Turning on the $D$-contributions does not improve the situation. 
Eventhough their sum
over a single sector does not vanish, gauge invariance guarantees that
their sum over each family vanishes, \ie  negative
contributions to some eigenvalues of the mass squared mass, as
indeed we saw above. 
%

%546
This comes as no surprise: The Higgs squared mass matrix 
in this limit simply reads
\begin{eqnarray}
M^{2}_{{H}}
& = &
\left(\begin{tabular}{cc}
$\mu^{2}$&$0$\\
$0$& $\mu^{2}$
\end{tabular}\right) 
\nonumber 
\end{eqnarray}
and has no negative eigenvalues. Thus, it cannot produce the SM Higgs
potential and the Higgs fields do no acquire a {\it vev}, so our
{\it ad hoc} assumption of electroweak symmetry breaking cannot be justified. 
Note that a negative eigenvalue could arise, however, from an
appropriate off-diagonal term.
Had we extended the model by 
including a singlet superfield
$S$, replacing $W = \mu H_{1}H_{2}$ with $W = y_{S}SH_{1}H_{2}  -
\mu_{s}^{2}S$, then electroweak symmetry could be broken while
supersymmetry is preserved.
One  has for the Higgs potential $V = |F_{S}|^{2} = |y_{S}H_{1}H_{2} -
\mu_{s}^{2}|^{2}$. Its minimization gives 
$\langle H_{1}^{0} \rangle =\langle H_{2}^{0} \rangle =
\mu_{S}/\sqrt{y_{S}}$ and $\langle S \rangle = 0$.
Electroweak symmetry is broken along the flat direction 
$V = \langle F_{S}\rangle^{2} = 0 $ so that
supersymmetry is conserved with $m_{\tilde{f}}^{2} = m_{f}^{2}$
(and with no left-right mixing!).
This is a counter example to our observation that in general
electroweak breaking implies supersymmetry breaking.
Models in which $\mu$ is replaced by a dynamical singlet field 
are often called the  next to minimal extension (NMSSM).
The specific NMSSM model described here is, 
however, still far from leading to a realistic model. 
%

%547
Of course, one needs not go through the exercise
of electroweak symmetry breaking in order to convince oneself  
that the models as they are
manifested in this limit are inconsistent. The gluino partner of the
gluon, for example,  cannot receive its mass from the colorless Higgs bosons
and remains massless at tree level, even if electroweak symmetry was successfully 
broken. It would receive a ${\cal{O}}$(GeV) or smaller 
mass from quantum corrections if the winos and bino are massive.
Such a light gluino would alter SM QCD predictions at the level of current
experimental sensitivity \cite{gluino1} while there is no indication
for its existence. (It currently cannot be ruled out 
in certain mass ranges and if it is the
lightest supersymmetric particle \cite{gluino2}, 
in which case it must hadronize.)
%

%548
We must conclude that a realistic model must contain some
other source of supersymmetry breaking. The most straightforward
approach is then to parameterize this source in terms of explicit
breaking in the low-energy effective Lagrangian.

\section{Soft Supersymmetry Breaking}
\label{sec:s54}

%551
Our previous exercises in electroweak and supersymmetry breaking lead
to a very concrete ``shopping list'' of what a realistic model should contain:
\begin{itemize}
\item Positive  squared masses for the sfermions
\item Gaugino masses, particularly for the gluino
\item Possibly an off-diagonal Higgs mass term $m^{2}H_{1}H_{2}$,
as it could lead to the desired negative eigenvalue
\end{itemize}
We can contrast our ``shopping list'' with what we have learned we
cannot have (in order to preserve the cancellation of quadratic divergences):
\begin{itemize}
\item Arbitrary quartic couplings
\item Arbitrary trilinear singlet couplings in the scalar potential,
$Cs\phi\phi^{*}$
\item Arbitrary fermion masses
\end{itemize}
The two last items are in fact related (and fermion masses are
constrained only in models with singlets).
Hence, both items do not apply to the MSSM (but apply in the NMSSM).
We conclude that acceptable spectrum can be accommodated without
altering cancellation of divergences!
Such explicit breaking of supersymmetry is soft breaking
in the sense that no quadratic divergences are re-introduced above the
(explicit) supersymmetry breaking scale.
%

%552
Let us elaborate and show that sfermion mass parameters 
indeed are soft. For example, consider the effect of the soft operator
$m^{2}\phi\phi^{*}$ and its modification of our previous calculation
of the quadratic divergence due to the quartic scalar interaction
$y^{2}|\phi_{1}|^{2}|\phi_{2}|^{2}$:
\begin{eqnarray}
y^{2}\int\frac{d^{4}q}{(2\pi^{4})}\frac{1}{q^{2}}  
&
\longrightarrow  
&
y^{2}\int\frac{d^{4}q}{(2\pi^{4})}\frac{1}{q^{2} - m^{2}} 
\sim 
y^{2}\int\frac{d^{4}q}{(2\pi^{4})}\left\{\frac{1}{q^{2}}
+\frac{m^{2}}{q^{4}}\right\}  
\nonumber\\
&
\sim
&
y^{2}\int\frac{d^{4}q}{(2\pi^{4})}\frac{1}{q^{2}}
+ i\frac{y^{2}}{16\pi^{2}}m^{2}\ln\frac{\Lambda_{\rm UV}^{2}}{m^{2}}.
\label{soft1}
\end{eqnarray}
Since the quadratically divergent term is as in the $m = 0$ case, 
it is still canceled by supersymmetry. Hence, eq.~(\ref{soft1})
translates to a harmless logarithmically divergent mass correction 
\begin{equation}
\Delta m^{2}_{i} = -y^{2}\frac{m^{2}_{j}}{16\pi^{2}}\ln 
\frac{\Lambda_{\rm UV}^{2}}{m^{2}},
\label{soft2}
\end{equation}
and similarly for the $g^{2}$ quartic interaction. Note the negative
over-all sign of the correction! The logarithmic correction is nothing
but the one-loop renormalization of the mass squared parameters due to
its Yukawa interactions, and it is negative. The implications are
clear - a negative squared mass, for example in the weak-scale Higgs
potential, may be a a result of quantum corrections - a subject
which deserves a dedicated discussion, and which we will return to in
Chapter~\ref{c7}.
(Of course, integrating (\ref{soft1}) properly one finds also 
finite corrections to the mass parameters which we do not discuss here.)
%

%553
An important implication of the above discussion is that light particles 
are protected from corrections due to the decoupling of heavy particles
(as heavy as the ultraviolet cut-off scale), as long as the
decoupling is within a supersymmetric regime, \ie 
$(m_{\rm heavy-boson}^{2} -m_{\rm heavy-fermion}^{2})/
(m_{\rm heavy-boson}^{2}  +m_{\rm heavy-fermion}^{2}) \rightarrow 0$.
This ensures the sensibility of the discussion of grand-unified
theories or any other theories with heavy matter and in which there
is tree-level coupling between light and heavy matter (leading to
one-loop corrections). Supersymmetry-breaking  corrections due to
the decoupling of heavy particles are still proportional only to the
soft parameters and do not destabilize  the theory.
(The only caveat being mixing among heavy and light singlets \cite{SINGLET}.)
This persists also at all loop orders. 
%

%554
Our potential $V = V_{F} + V_{D} + V_{\rm SSB}$ now contains 
superpotential contributions $V_{F} = |F|^{2}$, gauge contributions
$V_{D} = |D|^{2} + $ gaugino-Yukawa interactions, and contributions
that explicitly but softly break supersymmetry, the soft supersymmetry
breaking (SSB) terms. The SSB potential most generally consists of
\begin{eqnarray}
V_{\rm SSB}& = & 
m_{j^{*}}^{i\, 2}\phi_{i}\phi^{j*} + \left\{ B^{ij}\phi_{i}\phi_{j} + 
A^{ijk}\phi_{i}\phi_{j}\phi_{k} + 
C^{jk}_{i*}\phi^{i*}\phi_{j}\phi_{k} + \right.
\nonumber\\
&+& 
\left.
\frac{1}{2}M_{\alpha}\lambda_{\alpha}\lambda_{\alpha} + {h.c.}\right\}.
\label{VSSB}
\end{eqnarray}
The soft parameters were originally classified by Inoue {\it et al.}~\cite{INOUE}
and by Girardello and Grisaru \cite{GRGR},
while more recent and general discussions include Refs.~\cite{HARA,BFPT,JJ}.
Note that the trilinear couplings $A$ and  $C$ and the gaugino mass $M$
carry one mass dimension, and also carry phases. The parameters $C$
(often denoted $A^{\prime}$)
are not soft if the model contains singlets (e.g. in the NMSSM) but
are soft otherwise, in particular in the MSSM. In fact it can be shown
that they are equivalent to explicit supersymmetry breaking matter
fermion masses. Nevertheless, they appear naturally only in special
classes of models and will be omitted unless otherwise noted.
The scale of the soft supersymmetry breaking parameters 
$m \sim \sqrt{|B|} \sim |A| \sim |M| \sim m_{\rm SSB}$ is dictated by
the quadratic divergence which is cut off by the mass scale they set
(see eq.~(\ref{soft2})) and hence is given by
\begin{equation}
m_{SSB}^{2} \lesssim \frac{16\pi^{2}}{y_{t}^{2}} \mweak^{2}
\simeq (1\,{\rm TeV})^{2}.
\label{soft3}
\end{equation}
The mass scale $m_{\rm SSB}$  plays the role and
provides an understanding of the mass scale $M_{\rm SUSY}$
discussed in Sec.~\ref{sec:moreonsusy}.
Ultimately, the explicit breaking is to be understood as the imprints
of spontaneous supersymmetry breaking at higher energies
rather than be put by hand. (See Chapter~\ref{c10}.)
For the purpose of defining the MSSM,
however, a general parameterization is sufficient.
%

%555
Once substituting all flavor indices in the potential (\ref{VSSB}),
e.g.
$B^{ij}\phi_{i}\phi_{j} \rightarrow m_{3}^{2}H_{1}H_{2}$ (see Chapter~\ref{c8})
and (suppressing family indices)
$A^{ijk}\phi_{i}\phi_{j}\phi_{k}  \rightarrow
A_{U}H_{2}\widetilde{Q}\widetilde{U} +
A_{D}H_{1}\widetilde{Q}\widetilde{D} +
A_{E}H_{1}\widetilde{L}\widetilde{E}$,
one finds that  the MSSM contains many more parameters
in addition to the 17 free parameters of the SM. The Higgs sector
can be shown to still be described by only two free parameters.
However, the gauge sector contains three new gaugino mass parameters,
which could carry three independent phases; the scalar spectrum is
described by five  $3 \times 3$ hermitian matrices with six independent
real parameters and three independent phases each (where we constructed
the most general matrix in family space for the $Q$, $U$, $D$, $L$,
and $E$ ``flavored'' sfermions); and trilinear interactions are described (in
family space) by three $3 \times 3$ arbitrary matrices $A_{U},\,A_{D},
A_{E}$ with nine real parameters and nine phases each
(and equivalently for the $C$ matrices). A more careful examination
reveals that four phases can be eliminated by field redefinitions and
hence are not physical. Hence, the model, setting all $C = 0$ ($A^{\prime} = 0$)
(allowing arbitrary $C$ or $A^{\prime}$ coefficients) is described by 77 (104)
parameters if all phases are zero, and by 122 (176) parameters if
phases (and therefore CP violation aside from that encoded in the CKM matrix,
often referred to as soft CP violation)
are admitted.

\section{R-Parity and Its Implications}
\label{sec:s55}

%561
While supersymmetrizing the SM we followed a simple guideline of writing
the minimal superpotential that consistently reproduces the SM Lagrangian.
Once we realized that supersymmetry must be broken explicitly at the
weak scale, we introduced SSB parameters which conserved all of the
SM local and global symmetries. While it is  the most
straightforward procedure, it does not lead to the most
general result. In the SM lepton $L$ and baryon $B$ number are accidental
symmetries and are preserved to all orders in perturbation theory.
(For example, models for baryogenesis at the electroweak phase
transition rely on non-perturbative baryon number violation.)
Once all fields are elevated to superfields, one can interchange a
lepton and Higgs doublets $L \leftrightarrow H_{1}$ in the
superpotential and in the scalar potential, leading to violation of
lepton number by one unit by renormalizable operators. The  $\Delta L
= 1$ superpotential operators, for example, are
\begin{equation}
W_{\Delta L = 1} =  
\pm \mu_{L_{i}}L_{i}H_{2} +
\frac{1}{2}\lambda_{ijk}\left[L_{i},L_{j}\right]E_{k}
+ \lambda_{ijk}^{\prime}L_{i}Q_{j}D_{k},
\label{WdeltaL}
\end{equation}
where we noted explicitly the antisymmetric nature of the $\lambda$
operators. Similarly, one can write SSB $\Delta L = 1$ operators.
%

%562
Lepton number violation is indeed constrained by experiment, but
is allowed at a reasonable level, and in particular, electroweak-scale
$\Delta L = 1$ operators could lead to a $\Delta L = 2$
Majorana neutrino spectrum, hence, extending the SM and the MSSM in a
desired direction. However, baryon number is also not an accidental
symmetry in the MSSM and the
\begin{equation}
W_{\Delta B = 1} = 
\frac{1}{2}\lambda_{ijk}^{\prime\prime}U_{i}\left[D_{j},D_{k}\right]
\label{WdeltaB}
\end{equation}
operators are also allowed by gauge invariance.
The combination of lepton and baryon number violation
allows for tree-level decay of the proton from $\tilde{s}$ or $\tilde{b}$
exchange $(U)UD \rightarrow (U)\widetilde{D} \rightarrow (U)QL$, \ie
$p \rightarrow \pi l,\, Kl$. (The squark plays the role of a
lepto-quark field in this case and the model is a special case of a
scalar lepto-quark theory. See the scalar exchange diagram in
Fig.~\ref{fig:proton} below.)
%

%563
The proton life time is given by 
\begin{equation}
\tau_{p \rightarrow e\pi} \sim 10^{-16 \pm 1}{\rm
yr}(m_{\tilde{f}}/ 1\,{\rm
TeV})^{4}(\lambda^{\prime}\lambda^{\prime\prime})^{-2}. 
\end{equation}
It is constrained
by the observed proton stability $\tau_{p \rightarrow e\pi} \gtrsim
10^{33}$ yr, leading to the constraint 
$\lambda^{\prime}\lambda^{\prime\prime} \lesssim 10^{-25}$. The
constraint is automatically satisfied if either 
$\lambda^{\prime} = 0$ (only $B$ violation) or $\lambda^{\prime\prime}
= 0$ (only $L$ violation), leaving room for many possible and interesting
extensions of the MSSM. (For a review, see Ref.~\cite{RPV}, as well as
Chapter~\ref{c11}.)
%

%564
The $L$ and $B$ conserving MSSM, however, is again the most general
allowed extension (with minimal matter) if one imposes $L$ and $B$ by
hand. It is sufficient for that purpose to postulate a discrete
$Z_{2}$ (mirror) symmetry, {\it matter parity},
\begin{equation}
P_{M}(\Phi_{I}) = (-)^{3(B_{I} - L_{I})}.
\label{Pm}
\end{equation}
Matter parity is a discrete subgroup of the anomaly free $U(1)_{B-L}$
Abelian symmetry discussed in Chapter~\ref{c1}, 
and hence can be an exact symmetry if 
$U(1)_{B-L}$ is gauged at some high energy. Equivalently,
one often impose a discrete $Z_{2}$ $R$-parity \cite{FF},
\begin{equation}
R_{P}(O_{I}) = P_{M}(O_{I})\times(-)^{2S_{I}} = (-)^{2S_{I}+ 3B_{I} + L_{I}},
\label{Rp}
\end{equation}
where $O_{I}$ is an object with spin $S_{I}$. $R$-parity is a discrete subgroup
of a $U(1)_{R}$ Abelian symmetry under which the supersymmetric
coordinate transforms with a charge $R(\theta) = -1$ (which is a
conventional normalization), \ie 
$\theta \rightarrow e^{-i\alpha}\theta$. $R$-symmetry does not commute with
supersymmetry (and it extends its algebra \cite{LocalU1R}).
In particular, it distinguishes the superfield components.
Hence, $R$-parity efficiently separates
all SM (or more correctly, its 2HDM type II extension) particles with 
charge $R_{P}(O_{SM}) = (+)$ from their superpartners, the sparticles,
with charge $R_{P}({\rm sparticle}) = (-)$, and hence, is more often used.
Matter or $R$-parity correspond to a maximal choice that guarantees
stability of the proton. Other choices that do not conserve $B-L$, but
only $B$ or $L$, exist and correspond to $Z_{3}$ or higher symmetries
(for example, baryon and lepton parities \cite{RI,HPNK} and the
$\theta$-parity \cite{HASU}). 
%

%567
Though the minimal supersymmetric extension is defined by its minimal
matter content, it is often defined by also its minimal interaction
content, as is the case for $R_{P}$ invariant models.
We will assume for now, as is customary, that $R_{P}$ is an exact symmetry
of the low-energy theory (unless otherwise is stated).
%

%568
Imposing $R$-parity on the model dictates to large extent the
phenomenology of the model. In particular, the lightest superpartner (LSP)
(or equivalently, the lightest $R_{P} = (-)$ particle) must be stable:
It cannot decay to only $R_{P} = (+)$ ordinary particles since such an
interaction would not be invariant under the symmetry.
(This is not the case in other examples of symmetries given above 
which can stabilize the proton while admitting additional
superpotential or potential terms.) This is a most important
observation with strong implications:
\begin{enumerate}
\item
The LSP is stable and hence has to be neutral (e.g. a bino, wino,
Higgsino, sneutrino, or gravitino) or at least bound to a stable
neutral state (as would be the case if the gluino is the LSP and it
then hadronizes)
\item
Sparticles are produced in the collider in pairs
\item
Once a sparticle is produced in the collider, 
it decays to an odd number of sparticles.
In particular, its decay chain must
conclude with the LSP. The stable LSP escapes the detector,
leading in many cases to a distinctive large missing energy signature.
(If a light gravitino is the LSP, for instance, other distinctive signatures 
such as hard photons exist.) 
\item
The neutral LSP (in most cases) is a weakly interacting massive
particle (WIMP) and it
could constitute cold dark matter (CDM) with a sufficient density
\end{enumerate}
%

%569
The latter point requires some elaboration. 
Roughly speaking, the relic density of a given dark matter component
is proportional to the inverse of its annihilation rate, which in turn
is given by a cross section. The cross section
is typically proportional to the squared inverse of
the exchanged sparticle mass squared. Thus, the relic density is
proportional to sparticle mass scale to the fourth power,  
a relation which leads to model-dependent upper bounds of the order of
${\cal{O}}$(TeV) on the sparticle mass scale. 
In practice, the density is a complicated function
of various sparticle mass parameters,
leading to the relation for
the relic density $\Omega h^{2}$ \cite{DMeq1,DMeq2}
\begin{equation}
\Omega_{\rm LSP}h^{2} = \sigma_{0}
\frac{\left(m_{\tilde{l}}^{2} + m_{\rm LSP}^{2}\right)^{4}}
{m_{\rm LSP}^{2}\left(m_{\tilde{l}}^{4} + m_{\rm LSP}^{4}\right)},
\label{OmegaLSP}
\end{equation}
where $\sigma_{0} \approx (460\,{\rm GeV})^{-2}/\sqrt{N}$, $N$ here is
the number of degrees of freedom at the (cosmological)
decoupling of the LSP,
and this relation holds in a scenario in which the LSP
is the bino which annihilated primarily to leptons $l$ via a $t$-channel 
slepton $\tilde{l}$ exchange. (Recall the $\widetilde{B}l\tilde{l}^{*}$ vertex.)
Indeed, for $m_{\tilde{l}} \gg m_{\rm LSP}$ one has $\Omega h^{2} \simeq 
\sigma_{0} m_{\tilde{l}}^{4}/ m_{\rm LSP}^{2}$.
A cautionary note is in place: In certain cases the annihilation
process may depend on $s$-channel resonance annihilation and
the resonance enhancement of the cross section may allow
for a sufficient annihilation of a  WIMP which is a few fold
heavier than in the usual case \cite{DMs}. The upper bound is therefore
model dependent.
%

%5610
It should be stressed that the proximity of this
upper bound and the one derived from the fine-tuning of the Higgs
potential is tantalizing and suggestive. More generally, it relates
the CDM to the sparticle mass scale. Particles in
various sectors of the theory (which may be linked to the SM sector only
gravitationally or by other very weak interactions) 
could also have masses which are controlled by
this SSB scale (for example, by the gravitino mass $m_{3/2}$ discussed
in Sec.~\ref{sec:sugra}): 
Supergravity interactions can
generate such masses in many different sectors. Therefore,
CDM candidates may be provided by various sectors of the theory,
not necessarily the observable (MS)SM sector. 
An interesting proposal raised recently is that
of an axino dark matter \cite{axino}, which would weaken any CDM 
imposed upper-bound on the mass of the ordinary superpartners.
(The axino is the fermion partner of the axion
of an anomalous Peccei-Quinn or $R$-symmetry, on which we do no
elaborate in these notes.)
If, however, one more traditionally chooses to 
assume that the WIMP is the LSP, collider phenomenology
and cosmology need to be confronted \cite{DMlhc},
providing a future avenue to test such a hypothesis,
and furthermore, the nature of the WIMP.
(LSP assumptions may also be confronted with a variety
of low-energy phenomena, for example, see Ref.~\cite{DMbsgamma}.)
%

%5611
The collider phenomenology of the models was addressed 
in Zeppenfeld's \cite{ZEP} Tata's \cite{TATA2,TATA1} lecture notes. 
Models of CDM are reviewed, for
example, in Refs.~\cite{CDM1,CDM2}, and the search for energetic neutrinos from LSP
annihilation in the sun was discussed by Halzen \cite{HAZ}.
These issues will not be studied here.

\section{Mass Eigenstates and Experimental Status}
\label{sec:s56}

%571
We conclude this chapter with a transformation from current to mass
eigenstates, which is not a trivial transformation given electroweak
symmetry breaking (EWSB): 
As in our ``warm-up'' case of a supersymmetric limit
(Sec.~\ref{sec:s53}), interaction eigenstates with different
electroweak charges (and hence, chirality) mix once electroweak
symmetry is broken.
%

%572
The Higgs mass matrix is now 
\vspace*{0.4cm}
\begin{eqnarray}
M^{2}_{{H}}
& = &
\left(\begin{tabular}{cc}
$m_{H_{1}}^{2} + \mu^{2}$&$ m_{12}^{2}$\\
$m_{12}^{*\, 2}$& $m_{H_{2}}^{2} + \mu^{2}$
\end{tabular}\right), 
\label{mH2matrix}
\end{eqnarray}
and it has a negative eigenvalue
for $m_{1}^{2}m_{2}^{2} < |m_{12}^{2}|^{2}$, where $m_{i}^{2} =
m_{H_{i}}^{2} + \mu^{2}$, so that electroweak symmetry can be broken.
This condition is automatically satisfied for $m_{H_{2}}^{2} \lesssim
-\mu^{2}$ which often occurs due to negative quantum corrections proportional
to the $t$-quark Yukawa couplings eq.~(\ref{soft2}). (This is the radiative symmetry
breaking mechanism, which we will return to in Chapter~\ref{c7}.)
One neutral CP odd and two charged $d.o.f.$ 
are absorbed in the $Z$ and $W^{\pm}$ gauge bosons, respectively,
and two CP even (the lighter $h^{0}$ and the heavier $H^{0}$), 
one CP odd ($A^{0}$) and one complex charged ($H^{+}$) Higgs bosons remain
in the physical spectrum.
Folding in EWSB constraints, the spectrum is described by only two
parameters which are often taken to be $\tan\beta$ and $m_{A^{0}}^{2}
=m_{1}^{2} + m_{2}^{2}$. 
%

%573
One of the CP even states is the (model-independent)
light Higgs boson of supersymmetry which parmeterizes
the $\tan\beta = 1$ flat direction mentioned above.
This is readily seen in the limit in which all other Higgs $d.o.f.$
form (approximately) a degenerate  $SU(2)$ doublet 
which is heavy with a mass $\sim |\mu| \gg M_{W}$.
In this case, EWSB is SM-like, with the remaining physical CP-even
state receiving mass which is proportional to its quartic coupling $\lambda$,
now given by the gauge couplings $\lambda = (g_{2}^{2} +
g^{\prime\,2})/8 \times \cos^{2}2\beta$. 
In general, its mass $m_{h^{0}} \leq M_{Z}|\cos 2
\beta| \leq M_{Z}$ at tree-level, where $M_{Z}$ reflects 
the $D$-term nature of the quartic  coupling,
and the angular dependence reflects the flat direction.
Large ${\cal{O}}(100\%)$ loop corrections, again
proportional to the large $t$-quark Yukawa coupling, lift the (flat
direction and the) bound to  $m_{h^{0}} \lesssim \sqrt{2} M_{Z} \sim
130$ GeV. (Note that perturbation theory does not break
down. It is the tree-level term which is small rather than
the loop corrections being exceptionally large.)
Though the quartic coupling, and hence the mass, can be
somewhat larger in extended models, e.g. in some versions of the the
NMSSM, as long as perturbativity is maintained one has
$m_{h^{0}} \lesssim 160 - 200$ GeV where the upper range is achieved
only in a small class of (somewhat ad hoc) models.
The only exception is models with low-energy supersymmetry breaking
in which the breaking is not necessarily soft \cite{Su2}.
We discuss the Higgs sector in more detail in Chapter~\ref{c8}
where also references are given.
%

%574
Sparticles rather than Higgs bosons, however, will provide the
evidence for supersymmetry (though the absence of a light Higgs boson
can rule most models of perturbative low-energy supersymmetry out). The Dirac-like
neutral Higgsinos mix
with the other two neutral fermions, the Majorana bino and
neutral wino. (The latter can be rewritten as a photino and zino, \ie
as linear combinations aligned with the photon and the $Z$).
The physical eigenstates are the neutralinos $\widetilde{\chi}^{0}$. 
Their mass and mixing is given
by the diagonalization of the neutralino (tree-level) mass matrix
\vspace*{0.4cm}
\begin{eqnarray}
M_{\widetilde{\chi}^{0}}
& = &
\left(\begin{tabular}{cccc}
$M_{1}$ & 0 & $-M_{Z}c_{\beta}s_{W}$ &$M_{Z}s_{\beta}s_{W}$ \\
0 & $M_{2}$ & $M_{Z}c_{\beta}c_{W}$ &$M_{Z}s_{\beta}c_{W}$ \\
$-M_{Z}c_{\beta}s_{W}$ &$M_{Z}c_{\beta}c_{W}$& 0 & $\mu$ \\
$-M_{Z}s_{\beta}s_{W}$ &$-M_{Z}s_{\beta}c_{W}$& $\mu$ & 0\\
\end{tabular}\right), 
\end{eqnarray}
where $M_{1}$ and $M_{2}$ are the bino and wino SSB mass parameters,
respectively, and $s_{\beta} = \sin\beta,\,c_{\beta} = \cos\beta$, and
similarly for the weak angle denoted by a subscript $W$. The
neutralino mass matrix is written in the basis $(-i\widetilde{B},\,
-i\widetilde{W}^{0},\, \widetilde{H}_{1}^{0},\, \widetilde{H}_{2}^{0})$.
The EWSB off-diagonal terms correspond the the gauge-Yukawa interaction
terms with the Higgs replaced by its {\it vev}.
Note that in the limit $\mu \rightarrow 0$ the wino and bino do not
mix at tree-level. 
Similarly, the charged Higgs and charged gaugino states mix to form
the physical mass eigenstates, the charginos $\widetilde{\chi}^{\pm}$.
The chargino mass and mixing is determined by the chargino mass matrix
\vspace*{0.4cm}
\begin{eqnarray}
M_{\widetilde{\chi}^{\pm}}
& = &
\left(\begin{tabular}{cc}
$M_{2}$ & $\sqrt{2}M_{W}s_{\beta}$\\
$\sqrt{2}M_{W}c_{\beta}$& $-\mu$ \\
\end{tabular}\right), 
\end{eqnarray}
The gluino, of course, cannot mix and has a SSB
Majorana mass $M_{3}$.
%

%575
Finally, we can rewrite the sfermion mass matrix
for the sfermions 
$(\tilde{f}_{L},\,\tilde{f}_{R})$.
(Recall that $\tilde{f}_{R}$ is a shorthand notation for $\tilde{f}_{R}^{*}$.)
The mass-squared matrix  
\vspace*{0.4cm}
\begin{eqnarray}
M^{2}_{\tilde{f}}
& = &
\left(\begin{tabular}{cc}
$m^{2}_{LL}$&$m^{2}_{LR}$\\
$m^{2\,\,*}_{LR}$& $m_{RR}^{2}$
\end{tabular}\right) 
\end{eqnarray}
was previously given in the supersymmetric limit. Including the SSB
interactions one has
\begin{eqnarray}
m_{LL}^{2} &=& m_{\tilde{f}_{L}}^{2}+ m^{2}_{f} + 
M_{Z}^{2}\cos 2\beta\left[T_{3}^{f} -
Q_{f_{L}}\sin^{2}\theta_{W}\right], 
\label{m2LL} \\
m_{RR}^{2} &=& m_{\tilde{f}_{R}}^{2} + 
m^{2}_{f} + M_{Z}^{2}\cos 2\beta \times Q_{f_{R}}\sin^{2}\theta_{W}, 
\label{m2RR} \\
m_{LR}^{2} &=& m_{f}\left( A_{f} - \mu^{*}\tan\beta\right)\,\,\,[{\rm or}\,\, 
m_{f}\left(A_{f} - \mu^{*}/\tan\beta \right) ],
\label{m2LR}
\end{eqnarray}
where the first term in (\ref{m2LL}) and in (\ref{m2RR}) is the SSB
mass-squared parameter, the triliner SSB parameter is implicitly
assumed to be proportional to the Yukawa coupling $\hat{A}_{f} =
y_{f}A_{f}$, which is then factored out (the assumption is trivial
in the case of one generation but it constitutes a strong constraint in the case
of inter-generational mixing), and $\mu\tan\beta$ ($\mu/\tan\beta$)
terms appear in the down-squark and slepton mass matrices (up-squark
mass matrix), as before.
%

%576
Neglecting fermion masses, one has
the sum rule $m_{\widetilde{E}_{L}}^{2} - m_{\widetilde{N}_{L}}^{2}=
m_{\widetilde{D}_{L}}^{2} - m_{\widetilde{U}_{L}}^{2} =
-\cos^{2}\theta_{W}M_{Z}^{2}\cos 2 \beta = -M_{W}^{2}\cos 2 \beta >0$. 
This sum rule is modified if there is an extended gauge structure
with more than just the electroweak  $D$-terms. 
The $A$-terms are not invariant under $SU(2) \times U(1)$ 
and the sfermion doublet masses further split. 
The mass matrices can be combined to 
three $6 \times 6$ matrices and one $3 \times 3$ matrix (for the sneutrinos)
just as in the supersymmetric limit discussed in Sec.~\ref{sec:s53}.
Note that the SSB $A$-terms $AH\tilde{f}_{L}\tilde{f}_{R}$
are holomorphic (and do not involve 
complex conjugate fields) unlike the supersymmetric trilinear interactions 
$y\mu H^{*}\tilde{f}_{L}\tilde{f}_{R}$
(and SSB $C$-terms), a property which is particularly relevant for 
the stability of the vacuum discussed in Chapter~\ref{c12}.
(Note that each holomorphic $A$-term correpsonds to a flat direction
in the scalar potential.)
It is also relevant for the dependence of the
couplings of the physical eigenstates on $\tan\beta$. 
%

%577
The sfermion left-right mixing angle is conveniently given by
\begin{equation}
\tan 2 \theta_{\tilde{f}} = \frac{2m_{LR}^{2}}{m_{LL}^{2} -
m_{RR}^{2}}.
\end{equation}
Obviously, significant left-right mixing is possible if either the corresponding
fermion is heavy (as in the case of the stop squarks) or if
$\tan\beta$ is large and the corresponding fermion is not very
light (as could be the case for the sbottom squarks).
Observe, however, that in the limit $m^{2}_{\tilde{f}} \gg \langle
H_{i}^{0} \rangle^{2}$ the sfermion mixing is suppressed and 
mass eigenstates align with the current eigenstates. 
(This is true assuming $A < m_{RR}^{2}/\langle H_{i}^{0}\rangle,\,\,
m_{LL}^{2}/\langle H_{i}^{0}\rangle$, a constraint which is typically
impose by the stability of the vacuum. See Chapter~\ref{c12}.)
%

%578
Model-independent limits (\ie independent of the decay mode)
on the sparticles were given by the total
width measurement at the $Z$ pole at LEP. These limits constrain the
sparticles to be heavier than $40 - 45$ GeV, with the exception of the 
lightest neutralino, whose couplings to the $Z$ could be substantially
suppressed\footnote{One could also fine-tune the $\sin\theta_{\tilde{f}}/\sin\theta_{W}$ 
ratio so that a particular sfermion $\tilde{f}_{i}$ decouples from the $Z$.
Such tuning, however, is not scale invariant.} 
and could still be as light as $\sim 20$ GeV.
Also, no significant limit on the (heavy) gluino, which does not couple to the
$Z$ at tree level, is derived. 
LEP runs at higher energy further constrain many of the sparticles
to be heavier than $\sqrt{s}/2 \sim {\cal{O}}(100)$ GeV. 
($\sqrt{s}$ is the experiment center of mass energy, and it is
divided by the number of colliding leptons or partons.)
However, now the constraints are
model-dependent since off-resonance production involves also a
$t$-cahnnel exchange (which introduce strong model dependence) and,
in the absence of an universal tool such as the $Z$ width, searches
must assume in advance the decay chain and its final products.
The FNAL Tevatron can constrain efficently the strongly interacting squarks
and gluino with lower bounds of $\sqrt{s}/6 \sim
{\cal{O}}$(200-300) GeV, but again,
these bounds contain model-dependent assumptions.
Many specialized searches assuming unconventional decay chains
were conducted in recent years both at the Tevatron and LEP.
They are summarized and corresponding limits are updated periodically by the 
Particle Data Group \cite{PDG}. Obviously, a significant gap remains between
current limits (and particularly so, the model-independent limits)
and the theoretically suggested range of ${\cal{O}}(1\,{\mbox{TeV}})$. 
Though some sparticles may still be discovered at
future Tevatron runs, it is the Large Hadron Collider,
currently under consttruction at CERN, which will explore the 
${\cal{O}}(1\,{\mbox{TeV}})$ regime and the ``TeV World''.
%

%E5
\section*{Exercises}
\addcontentsline{toc}{section}{Exercises}
\markright{Exercises}

\subsubsection*{5.1}
Draw all the Feynamn diagrams that stem from that of the vector-fermion-fermion 
interaction and which describe the gauge-matter
interactions. What are the Feynman rules for the case of a $U(1)$? 

\subsubsection*{5.2}
Confirm that Tr$Y^{3} = 0$  for hypercharge 
in the MSSM, as well as the mixed anomaly traces
Tr$SU(N)^{2}U(1)$ where $N=2,\,3$. 

\subsubsection*{5.3}
Derive the complete MSSM component-field Yukawa Lagrangian.

\subsubsection*{5.4}
Derive the complete MSSM quartic potential, including gauge ($D$) and
Yukawa  ($F$) contributions. Use the relation
$\sigma^{a}_{ij}\sigma^{a}_{kl} = 2\delta_{il}\delta_{jk} -
\delta_{ij}\delta_{kl}$ among the $SU(2)$ generators
to write the Higgs quartic potential explicitly.
Map it onto the general form of a 2HDM potential
eq.~(\ref{s2V4}) and show that (at tree level)
$\lambda_{1} = \lambda_{2} = (1/8)(g^{\prime 2} + g^{2}_{2})$;
$\lambda_{3} = -(1/4)(g^{\prime 2} - g^{2}_{2})$; 
$\lambda_{4} = -(1/2) g_{2}^{2}$; and $\lambda_{5} = \lambda_{6}
= \lambda_{7} = 0$ (where $g^{\prime}$ and $g_{2}$ are the
hypercharge and $SU(2)$ couplings, respectively).

\subsubsection*{5.5}
Show that at the quantum level there could be non-diagonal
corrections to the gauge kinetic function 
$f_{\alpha\beta}$, for example, in a theory
involving $U(1) \times U(1)^{\prime}$ \cite{ZZprime}. 
(Consider a loop correction that mixes the gauge boson propagators.)

\subsubsection*{5.6}
Derive the $D$-term expectation value eq.~(\ref{Dvev}).
Rewrite the hypercharg in terms of the electric charge, weak isospin,
and the weak angle to derive the $D$ term contribution
to the sfermion mass squared. Confirm its flavor-dependent numerical
coefficient.

\subsubsection*{5.7}
Derive the Higgs mass-squared matrix in the supersymmetric limit.

\subsubsection*{5.8}
Derive the Higgs mass-squared matrix in the model with a singlet $S$
which is described in Sec.~\ref{sec:s53}.
Show that it has an off-diagonal element and that it has a negative eigenvalue.
Show, by considering $F_{H_{i}}$ contributions to the scalar
potential, that indeed $\langle S \rangle = 0$ and no left-right
sfermion mixing arises.

\subsubsection*{5.9}
Show that all the soft parameters are indeed soft by naive
counting, where possible (e.g. in the case of trilinear scalar
interactions) or otherwise by integration.

\subsubsection*{5.10}
Show that a supersymmetry breaking fermion mass, if allowed by the
gauge symmetries (for example, consider a toy model with two singlets
$W = S_{1}S_{2}^{2} + mS_{1}^{2}$, and a supersymmetry breaking
mass to the fermion component of $S_{2}$ 
$V_{\rm SSB} = \widetilde{\mu}\psi_{2}\psi_{2}$) can be recast,
after appropriate redefinitions, as a $Cs_{1}s_{2}s_{2}^{*}$ term in the
scalar potential (with $C = -\widetilde{\mu}$). 

\subsubsection*{5.11}
Show that in a model with a singlet (consider, for example, our toy
model above) an interaction $Cs\phi_{1}\phi_{2}^{*}$ is not soft
but leads to a quadratically divergent linear term in the scalar potential.
Return to Ex.~5.10 and write the tadpole diagram
which leads to a quadratically-divergent linear term for the singlet
$s_{1}$ in either langauge, and confirm their equivalence.

\subsubsection*{5.12}
Substitute the SM flavor indices in the soft potential (\ref{VSSB})
and confirm our counting of free parameters.

\subsubsection*{5.13}
Write the most general gauge invariant ($R$-parity violating)
SSB potential. How many parameters describe the MSSM if $R$-parity is
not imposed?

\subsubsection*{5.14}
Using the (continuous) $R$-charge assignment of the coordinate $\theta$,
what are the $R$-charges of the various component fields of the chiral
and vector superfields? and of the superpotential? Show that requiring
that the potential is $R$-invariant forbids gaugino masses and
trilinear scalar terms. Therefore, these parameters must carry
$R$-charge and their presence breaks the $U(1)_{R}$ symmetry.
Note that all of the above terms are invariant under the 
discrete $R_{P}$ subgroup!

\subsubsection*{5.15}
Calculate the upper bound on the slepton mass implied by the approximation
eq.~\ref{OmegaLSP} for $\Omega_{\rm LSP}h^{2} \leq 1$ and $m_{\rm LSP} = 100$ GeV.

\subsubsection*{5.16}
Diagonalize the neutralino and chargino mass matrices in the limits
$\mu \rightarrow \infty$ (gaugino region), $M_{1} \sim M_{2}
\rightarrow \infty$ (Higgsino region), and 
$\mu \sim M_{1} \sim M_{2} \rightarrow 0$.

\subsubsection*{5.17}
Extend the neutralino and chargino mass matrices to include
mixing with one generation of neutrinos and charged leptons, respectively, for
$\mu_{L_{3}} \neq 0$. The neutrino mass is given by the ratio
of the determinant of this matrix and that of the usual neutralino
mass matrix. Can you identify the limit in which one mass
eigenvalue (the neutrino mass) is zero (at tree-level)?
What is the effective LSP in these models?
Allow neutralino mixing with three generations of neutrinos
$\mu_{L_{a}} \neq 0$ for $a = 1,\,2,\,3$. Show that still only one
neutrino could be massive at tree level.

\subsubsection*{5.18}
Show that each holomorphic $A$-term corresponds to a flat $D$-term direction.
Estimate the upper bound on $A$ 
by considering tachion states in the sfermion mass matrix.

\subsubsection*{5.19}
{\bf (a)}
In the MSSM, like in the SM,  the decay $h^{0} \rightarrow b\bar{b}$ 
is often the dominant decay mode of the light CP-even Higgs boson.
Nevertheless, other decay channels may also be open, e.g. $h^{0} \rightarrow \gamma\gamma$
plays an important role is search strategy at the LHC.
Write the Feynman diagram for a supersymmetric invisible decay mode
of $h^{0}$, both assuming $R$-parity conservation and violation.
The stable or meta-stable invisible states escape the detector. \newline
\hspace*{0.5cm}{\bf (b)}
In some models the Goldstino is the LSP,
and the next to lightest supersymmetric particle (NLSP)  is charged
and stable on collider scales, for example the stau. 
What would be a potential signature of such a scenario?

%SP2
\chapter*{Summary}
\addcontentsline{toc}{section}{Summary}
\markright{Summary}

%S21
In this part of the notes
supersymmetry was motivated, constructed, and applied
to the Standard model of particle physics.
The minimal model was defined according to its particle content and 
superpotential, but it was shown to contain many arbitrary
parameters that explicitly break supersymmetry near the Fermi scale.
Indeed, many questions still remain unanswered:
\begin{itemize}
\item
Does the model remain perturbative and if so, up to what scale? 
\item
What is the high-energy scale which
we keep referring to as the ultraviolet cut-off scale? 
\item
What is the origin of the soft supersymmetry breaking parameters 
and can the number of free parameters be reduced? 
\item
Are there signatures of supersymmetry (short of sparticle discovery)
that could have been tested at past and current low-energy experiments?
\item
Could one distinguish a MSSM Higgs boson from a SM one?
\item
Can the model be extended to incorporate neutrino mass and mixing?
\end{itemize}
%

%S22
In addition, one would like to understand how all the different
aspects of supersymmetry -- perturbativity, renormalization, ultraviolet origins,
etc. come together to explain
the weak scale and its structure.
This and related issues
have been extensively studied in recent years.
While some possible answers and proposals were put forward,
no standard {\it high-energy} supersymmetric model exists. 
On the contrary, the challenge ahead is 
the deciphering of the high-energy theory from the low-energy data once
supersymmetry is discovered and established. 
%

%S23
We will address these and similar questions in the remaining parts of these notes.
In particular, we will try to link the infrared and the ultraviolet.
%

%
%
%
%P3
\part{Supersymmetry Top-Down: 
\protect\newline Understanding the Weak Scale}
\label{p3}

%C6
\chapter{Unification}
\label{c6}

%601
Shortly after the SM was established and asymptotic freedom realized,
it was suggested that the SM semi-simple product group of $SU(3)
\times SU(2) \times U(1)$ is embedded at some high energies in a unique
simple group, for example $SU(5)$ (which like the SM group has rank 5  
and hence is the smallest simple group that can contain the SM
gauge structure).  Aside from implications for quark-lepton unification,
Higgs fields, and the proton stability (to which we will return)
it predicts first and foremost that the seemingly independent SM gauge
couplings originate from a single coupling of the unified simple group \cite{GG}
and that their infrared splitting is therefore due to only
the scaling (or renormalization) of the corresponding quantum field
theory from high to low energies \cite{WEIN}.
Independently, it was also realized in the context
of string theory that the theory just below the string scale
often (but not always)
has a unique (or unified) value for all gauge couplings \cite{DIEN,HPN97,DIENPP,IB},
regardless of whether the SM group itself truly unifies 
(in the sense of $(i)$ embedding all SM fields in representations of some
simple group, and $(ii)$ spontaneous breaking of that group). 
%

\begin{figure}[t]
\postscript{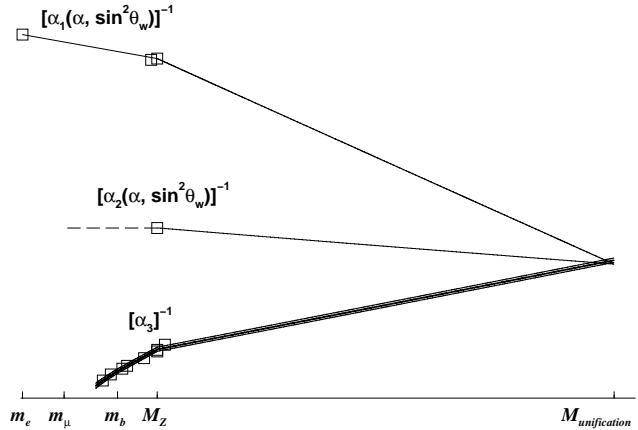}{0.75}
\caption{
The SM gauge couplings evolve from their
measured low-energy value (indicated by the square symbols)
to high energies using SM three-loop (for scales $\Lambda < M_{Z}$)
and MSSM two-loop (for scales $\Lambda > M_{Z}$) $\beta$-functions.
Their renormalization within the SM is confirmed by the data, while
extrapolation to higher energies, assuming the MSSM,
leads to their unification at nearly a point.
Note that (only) $\alpha_{3}$ exhibits asymptotically-free behavior.}
\label{fig:unification}
\end{figure}

%602
The question is then obvious: Do the measured low-energy couplings
unify (after appropriate scaling, \ie renormalization)
at some ultraviolet energies? Clearly, if this is the case
then their unification automatically defines an ultraviolet scale
(which is a reasonable choice to many of the exercises that we
will undertake in the following chapters),
and supergravity may further facilitate gauge-gravity unification. 
This question is readily addressed using the renormalization group
formalism (which is beautifully confirmed by the data 
in the case of QCD at energies 
up to the LEP center-of-mass energy of $\sim 200$ GeV).
Once the particle content of the model is specified, 
the $\beta$ function coefficients $b$ can be calculated and the couplings can
be extrapolated by integrating the renormaization group equation
$dg/d\ln\Lambda = (b/16\pi^{2})g^{3}$ (given here at one loop).
Such an integration is nothing but the scaling of the theory,
e.g. between the weak scale, where the 
couplings are precisely known, and some high-energy scale, assuming
a specific particle content. 
Such a scaling can be performed
in any perturbative theory as long as 
the particle content is known or fixed.
(For example, the SM scaling of QCD below the weak scale
agrees with the measurement of the QCD coupling at various energies.
See Fig.~\ref{fig:unification}.)
Its conclusions are meaningful if and
only if the low-energy couplings are measured to such precision
so that their experimental errors allow only a sufficiently small
range of values for each of the extrapolated high-energy couplings.
In particular, one needs to be able
to determine whether all three coupling intersects to a satisfactory
precision at a point, and if perturbativity (which is used in deriving
the equations) is maintained. 
%

%603
The 1990's brought about the most
precise determination of the gauge couplings at the $Z$ pole (see
summary by Langacker \cite{LEPFEST}), therefore enabling one to address the question.
Assuming only the SM matter content at all energies,
the three couplings fail to unify -- their integration curves
$g(\ln\Lambda)$ intersects in pairs only. 
(Of course, one may question the validity of such a framework to begin
with in the absence of any understanding as to how is the hierarchy problem 
resolved in this case.)
Extending the SM to the MSSM
and using the MSSM particle content begining at some point  between the weak
scale and a TeV scale, the three SM gauge coupling unify at a point at a
scale $M_{U} \simeq 3 \times 10^{16}$ GeV and with a value
$g_{U} \simeq 0.7$ or $\alpha_{U} \simeq 0.04$, defining the
unification scale (which is then a potential candidate for providing the ultraviolet cut off)
as well as confirming perturbativity, and hence consistency, of the framework.
This is illustrated in Fig.~\ref{fig:unification}. See also Ref.~\cite{PGLNP}.
%

%FIG
%%
\section{Gauge Coupling Unification}
%

%611
Let us repeat this exercise in some detail.
The renormalization of the couplings is given at two-loop order by
\begin{equation}
\frac{dg_{i}}{\ln\Lambda} = \frac{b_{i}}{16\pi^{2}}g_{i}^{3}
+\frac{b_{ij}}{(16\pi^{2})^{2}}g_{i}^{3}g_{j}^{2} -
\frac{a_{i\alpha}}{(16\pi^{2})^{2}}g_{i}^{3}y_{\alpha}^{2}.
\label{rge}
\end{equation}
The first term is the one-loop term given above, while the second and
third terms sum gauge and Yukawa two-loop corrections, respectively,
and we did not include any higher loop terms which are negligible in the MSSM. 
(The two-loop equation is free of any scheme dependences, though
negligible (finite) corrections appear when matching the scheme used to
describe the data to a scheme which can be used for extrapolation
in supersymmetry.)
The Yukawa terms, which are negative ($a_{i\alpha} > 0$),
correspond to only small ${\cal{O}}(1\%)$ but model ($\tan\beta$-) dependent corrections,
which we again neglect here. The gauge two-loop terms correspond
to ${\cal{O}}(10\%)$ corrections, which we will include in the results.
The one loop coefficients are the most important. In supersymmetry
they are conveniently written as a function of the chiral superfields quantum
numbers,
\begin{equation}
b_{i} = \sum_{a}T_{i}(\Phi_{a}) - 3C_{i},
\label{bi}
\end{equation}
where the Dynkin index
$T_{i}(\Phi_{a}) = 1/2\, (Q^{2})$ for a superfield $\Phi_{a}$ in the fundamental
representation of $SU(N)$ (with a $U(1)$ charge $Q$), and  
the Casimir coefficient for the adjoint representation $C_{i} = N\, (0)$
for $SU(N)$ ($U(1)$).
It is now straightforward to find in the MSSM
\begin{equation}
\left(\begin{tabular}{c}
$b_{1}$ \\ $b_{2}$ \\ $b_{3}$ \end{tabular}\right)
=
N_{\rm family}
\left(\begin{tabular}{c}
$2$ \\ $2$ \\ $2$ \end{tabular}\right)
+
N_{\rm Higgs}
\left(\begin{tabular}{c}
$\frac{6}{10}$ \\ $1$ \\ $0$ \end{tabular}\right)
-
\left(\begin{tabular}{c}
$0$ \\ $6$ \\ $9$ \end{tabular}\right)
=
\left(\begin{tabular}{c}
$+\frac{66}{10}$ \\ $+1$ \\ $-3$ \end{tabular}\right),
\label{biMSSM}
\end{equation}
where $N_{\rm family}$ and $N_{\rm Higgs}$ correspond to the number of 
chiral families and Higgs doublet pairs, respectively, and the index $i
= 1,2,3$ correspond to the $U(1)$, $SU(2)$, and $SU(3)$, coefficients.
%

%612
Here, we chose the ``unification'' normalization of the hypercharge
$U(1)$ factor, $Y \rightarrow \sqrt{3/5}Y$ and $g^{\prime} \rightarrow
\sqrt{5/3} \equiv g_{1}$. This is the correct normalization if each
chiral family is to be embedded in a representation(s) of the unified
group. In this case, the trace over a family over the
generators $T^{\alpha}$ for each of the SM groups $i$,
Tr$(T^{\alpha}_{i}(\Phi_{a})T^{\beta}_{i}(\Phi_{a}))
 =n\delta^{\alpha\beta}$, has to be equal to the same trace but when
taken over the unified group generators.
Reversing the argument, the trace has therefore
to be equal to the same $n$ for all subgroups. In the MSSM, both non-Abelian
factors have $n = 2$ and can indeed unify. The $U(1)$ normalization is chosen
accordingly such that Tr$(Q_{a}^{2}) = 2$ for each family. 
On the same footing, 
this condition and eq.~(\ref{bi}) imply that
adding any complete multiplet(s) (e.g. a SM family) of a unified
group modifies $b_{1}$, $b_{2}$, and $b_{3}$ exactly by the same amount.
(Note, however, that asymptotic freedom of QCD may be lost
if adding such multiplets. $SU(2)$ is already not asymptotically free
in the MSSM.) This relation also explains, in the context of unification, the
quantization of hypercharge which is dictated by the embedding of the
$U(1)$ in a non-Abelian group, and by the breaking 
pattern of the higher rank group.
(Such an embedding of the SM group in a
single non-Abelian group 
implies an anomaly free theory, and hence, this relation is somewhat equivalent
to the low-energy quantization based on the anomaly constraints.)
%

%613
Neglecting the Yukawa term, the coupled two-loop equations 
can be solved in iterations,
\begin{equation}
\frac{1}{\alpha_{i}(M_{Z})} = \frac{1}{\alpha_{U}} + b_{i}t +
\frac{1}{4\pi}\sum_{j = 1}^{3}\frac{b_{ij}}{b_{j}}(1 +
b_{j}\alpha_{U}t) -\Delta_{i}, 
\label{alphai}
\end{equation}
where the integration time is conveniently defined $t \equiv (1/2\pi)\ln(M_{U}/M_{Z})$,
and we assume, for simplicity, that the MSSM $\beta$-functions can be
used from the $Z$ scale and on. The threshold function $\Delta_{i}$
compensates for this naive assumption and takes into account the
actual sparticle spectrum. It can also contain contributions from
additional super-heavy particles and from operators (e.g., non-universal
corrections to $f_{\alpha\beta}^{i}$) that may appear
near the unification scale. 
%

%614
These equations can be recast in terms of
the fine-structure constant $\alpha(M_{Z})$ and the weak angle
$\sin^{2}\theta_{W} \equiv s^{2}_{W}(M_{Z})$. 
Using their precise experimental values one can then calculate
(up to threshold corrections) the unification scale, the value of the
unified coupling, and the value of the low-energy strong coupling.
(Since there are only two high-energy parameters, one equation can be
used to predict one weak-scale coupling.)
The first two are predictions that only test the perturbativity and
consistency of the extrapolation. (Note that the unification scale is
sufficiently below the Planck scale so that gravitational correction
may only constitute a small perturbation which can be summed, in
principle, in $\Delta_{i}$.)
%

%614
One finds, given current values,
\begin{eqnarray}
\alpha_{3}(M_{Z}) &=& \frac{5(b_{1} - b_{2})\alpha(M_{Z})}{
(5b_{1} + 3b_{2} - 8b_{3})s_{W}^{2}(M_{Z}) - 3(b_{2} - b_{3})} +
\cdots \\
& = & \frac{7\alpha(M_{Z})}{15s_{W}^{2}(M_{Z}) - 3} +
{\mbox {two-loop and threshold corrections}} \nonumber \\
&=& 0.116 + 0.014 - (0.000 - 0.003)  \pm \delta \nonumber\\
& = & (0.127 - 0.130) \pm \delta, 
\label{alpha3}
\end{eqnarray}
where in the third line one-loop, two-loop gauge, two-loop Yukawa,
and threshold corrections (denoted by $\delta$) are listed.
Note that any additional  complete multiplets of the unified group
in the low-energy spectrum would shift all the $b_{i}$'s by the same amount and
therefore factor out from the (one-loop) prediction.
This is true only at one loop and only for the $\alpha_{3}(M_{Z})$ and
$t$ predictions.
%

%615
Comparing the predicted value to the experimental one
(whose precision increased dramatically in recent years)
$\alpha_{3}(M_{Z}) = 0.118 \pm 0.003$, one finds a $\sim 8\%$
discrepancy, which determines the role that any structure near the weak
scale, the unification/gravity scale, or intermediate scales, can play.
In fact, only $\sim 3\%$ corrections are allowed at the unification
scale, since the QCD renormalization amplifies the corrections to
the required $8\%$ at the electroweak scale. 
GUT-scale corrections are the most likely 
conclusion from the discrepancy,
but we will not discuss here in detail the many possible sources of
such a small perturbation which could appear in the form of
a super-heavy non-degenerate spectrum, non-universal corrections to the
gauge kinetic function once the grand-unified theory is integrated
out, string corrections (in the case of a string interpretation), etc.
Let us instead comment on the possible threshold structure at the weak
scale (ignoring the possibility of additional 
complete multiplets at intermediate energies 
which will modify the two-loop correction \cite{complete}).
%

%616
Unless some particles are within tens of GeV from the weak scale, only
the (leading-)logarithm  corrections need to be considered (either by
direct calculations or using the renormalization group formalism). 
This is only the statement that the sparticle spectrum is typically expected
to preserve the custodial $SU(2)$ symmetry of the SM and hence not to contribute
to the (universal) oblique (e.g. $\rho$) and other (non-universal) parameters 
that measure its breaking and which would contribute 
non-logarithmic corrections to $\delta$.
%

%617
The leading logarithm threshold corrections can be summed in a straightforward but
far from intuitive way in the threshold parameter \cite{PGLNP} $\hat{M}_{SUSY}$, 
\begin{equation}
\delta = -\frac{19}{28}\frac{\alpha_{3}^{2}(M_{Z})}{\pi}\ln\frac{ 
\hat{M}_{SUSY}}{M_{Z}} 
\simeq -0.003\ln\frac{\hat{M}_{SUSY}}{M_{Z}}. 
\label{msusy}
\end{equation} 
If all sparticles are degenerate and heavy with a mass
$\hat{M}_{SUSY} \simeq 3$ TeV then the predicted and experimental
values of the strong coupling are the same. However, more careful
examination shows that $\hat{M}_{SUSY}$ is more closely related to the
gaugino and Higgsino spectrum (since the sfermion families correspond to
complete multiplets of $SU(5)$) and furthermore, typical models for the
spectrum give $10 \lesssim \hat{M}_{SUSY} \lesssim 300$ GeV even though
the sparticles themselves are in the hundreds of GeV range.
The threshold correction may then be even positive! (This is
particularly true when non-logarithmic corrections are included.)
While high-energy contributions to $\delta$ are likely to resolve
the discrepancy, they also render the unification result insensitive
to the exact value of $\hat{M}_{SUSY}$, since, e.g. 
an additional  deviation of $\sim 3\%$
from $\hat{M}_{SUSY} \ll M_{Z}$ corresponds to only $\sim 1\%$
corrections at the unification scale. Hence, even though
the sparticle threshold corrections may not resolve the discrepancy, they are very
unlikely destroy the successful unification. 
%

%618
This low level of ultraviolet
sensitivity (or rather high insensitivity) more then allows one to trust the result. 
(If the factor of $2\pi t \simeq 30$ corresponds to a
power-law (rather than logarithmic) renormalizations, as in some model
in which the gauge theory is embedded in a theory with
intermediate-energy extra dimensions \cite{DDG}, the sensitivity is amplified
by more than an order of magnitude, undermining any predictive power in
that case.) The predictive power in the minimal framework is unique
and one of the pillars in its support.

One could then take the point of view that the SM is to be embedded at
a grand-unification scale $M_{U} \simeq 3 \times 10^{16}$ GeV in a simple grand-unified
(GUT) group, $SU(3)_{c} \times SU(2)_{L} \times U(1)_{Y} \subseteq
SU(5);\,SO(10);\,SU(6);\,E_{6} \cdots$ where the rank five and six
options were specified. The GUT group may or may not be further embedded
in a string theory, for example. Alternatively, the unification scale may
be interpreted as a direct measurement of the string scale. 
The latter interpretation is one of the forces that brought about
a revolution in string model building, which generically had the string
scale an order of magnitude higher, and motivated $M$-theory model building
and other non-traditional approaches \cite{HPN97}, which were reviewed recently
by Dienes \cite{DIEN}. In particular, the 11-dimensional M-theory, with a fundamental
scale $M_{11} \gtrsim M_{U}$, allows 
a finite and small (six-dimensional) compactification volume $V \sim M_{U}^{-6}$
and a finite and small separation $R_{11}$ between the six- and four- 
dimensional ``walls''
(see Sec.~\ref{sec:extra}) such that 
$\alpha_{U} \sim M_{11}^{-6}/V \sim (M_{U}/M_{11})^{6}$
can be adjusted to fit its ``measured'' value of $\sim 0.04$ \cite{HPN97}.
%

%619
A GUT group with three chiral generations;
and the appropriate Higgs representations (that can break it to the SM
group as well as provide the SM Higgs doublets)
can be further embedded, in principle,  in a string theory
a decade or so above the unification scale.
However, this idea encounters many difficulties and was not yet
demonstrated to a satisfactory level \cite{DIEN}. 
In the reminder of our
discussion of unification we will outline 
the embedding in a GUT group, leaving open the question of string theory embedding.
\section{Grand Unification}
%

%621
Grand-unification, in contrast to only coupling unification,
requires one to embed all matter and gauge fields in representations of
the large GUT group. For example, all the SM gauge (super)fields are
embedded in the case of $SU(5)$ ($SO(10)$)
in the ${\bf 24}$ (${\bf 45}$) dimensional adjoint representation, 
while each family can be  embedded in $\{ D,L \subseteq {\bf \bar{5}}\} + 
\{Q,U,E \subseteq {\bf 10}\}$ ($ \{Q,U,D,L,E,N \subseteq {\bf
16}\}$) anti-fundamental $+$ antisymmetric (spinor) representations.
The Higgs fields of the SM can be embedded
in ${\bf \bar{5} + 5}$ (one or more fundamental ${\bf 10}$'s and anti-fundamentals).
The Higgs sector must also contain a higher dimensional representation(s)
(which may be taken to be the adjoint) that can break the GUT symmetry
spontaneously to its SM subgroup, and whose interactions must be
described by the appropriate superpotential.
%

%622
Note that the GUT theory resides in the globally supersymmetric SM
sector with all minima corresponding
to $\langle V \rangle = 0$. Therefore, the GUT $\rightarrow$ SM minimum is 
at best degenerate with the GUT conserving or any other minimum. It is the 
supergravity effects, which  typically can be parameterized by the (small) soft 
parameters, that must lift the degeneracy and pick the correct SM minimum. 
%

%623
Various issues stem from the embedding, and we will touch upon the most
important ones next. 
Recommended readings include Ref.~\cite{PGLNP}, which lists many of the  earlier
works. A sample of other recent research papers is given in
Ref.~\cite{S61}. Grand-unified theories were reviewed by Mohapatra \cite{MOH}.
\subsection{Yukawa Unification}
\label{sec:btau}
%

%%FIG
\begin{figure}[t]
\postscript{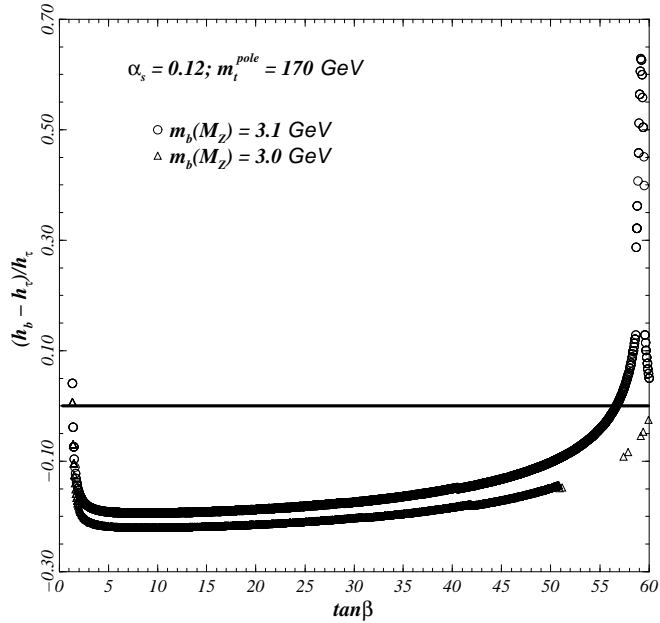}{0.75}
\caption{
The unification-scale difference
$y_{b} - y_{\tau}$ (denoted here $h_{b} - h_{\tau}$)
is shown in $y_{\tau}$ (denoted here $h_{\tau}$) units
for $m_{b}(M_{Z}) = 3 $ GeV, $\alpha_{3}(M_{Z}) = \alpha_{s}(M_{Z}) = 0.12$, 
$m_{t}^{pole} = 170$ GeV and as a function of $\tan\beta$.
The zero line corresponds to $b-\tau$ unification.
For comparison, we also show the difference for
$m_{b}(M_{Z}) = 3.1 $ GeV (which for $\alpha_{s}(M_{Z}) = 0.12$
is inconsistent with $m_{b}(m_{b}) < 4.45 $ GeV).
Note the rapid change near the (naive) small and large 
$\tan\beta$ solutions, which is a measure
of the required tuning in the absence of threshold corrections.
Also note that in most of the parameter space the unification-scale
leptonic coupling is the larger coupling. 
Taken from Ref.~\protect\cite{btaufig}.
}
\label{fig:btau}
\end{figure}
%%

%624
First and foremost, quarks and leptons unify in the sense that they
are embedded in the same GUT representations (e.g. the down singlet
and the lepton doublet are embedded in the ${\bf \bar{5}}$ of $SU(5)$).
This translates to simple boundary conditions for ratios of their Yukawa
couplings, e.g. $y_{d}/y_{l} = 1$ at the unification scale \cite{btau}. 
This relation (applied to a given generation) can 
then be renormalized down to the weak scale,
\begin{equation}
\frac{d}{d\ln{\Lambda}}\left(\frac{y_{d}}{y_{l}}\right) =
\frac{1}{16\pi^{2}}\left(\frac{y_{d}}{y_{l}}\right) 
\left\{
y_{u}^{2} + 3(y_{d}^{2}-y_{l}^{2}) - \frac{16}{3}g^{2}_{3} - \frac{4}{3}g^{2}_{1}
\right\},
\label{btauRGE}
\end{equation}
and tested. This is shown in Fig.~\ref{fig:btau} taken from Ref.~\cite{btaufig}.
(Note that the fermion masses are the current masses evaluated at $M_{Z}$
where $m_{b} \simeq 3$ GeV and $m_{\tau} \simeq 1.7$ GeV.)
Indeed, it is found to be correct for the third family couplings
($b - \tau$ unification) for either $\tan\beta \simeq 1 - 2$
(large top Yukawa coupling $y_{t}(M_{Z}) \simeq 0.95/\sin\beta$) 
or $\tan\beta \gtrsim 50$ (large bottom Yukawa coupling 
$y_{b}(M_{Z}) \simeq 0.017\tan\beta$),
and for a much larger parameter space once finite corrections
to the quark masses from sparticle loops (not included here)
are considered, e.g. $|\Delta(m_{b})/m_{b}| \lesssim 2\%\tan\beta$.
%

%625
Clearly, the successful renormalization of the unification relation
$y_{d}/y_{l} = 1$ requires large Yukawa couplings (which renormalize $y_{d}$).
The large Yukawa couplings are needed to counterbalance the QCD corrections.
Henceforth, it is not surprising that
these relations fail for the lighter families: 
Yukawa unification (in its straightforward form) applies to and distinguishes
the third family.
The first and
second generation fermion
masses may be assumed to vanish at the leading order, 
\begin{center}
$Y_{f_{ab}} \simeq y_{f_{3}}\left(\begin{tabular}{ccc}
0&0&0\\
0&0&0\\
0&0&1
\end{tabular}\right)$,
\end{center} 
and
could be further understood as setting the magnitude of the perturbations
for any such relations (either from higher dimensional Higgs
representations or from Planck-mass suppressed operators, or both).
\subsection{Proton Decay}
%
 
%%FIG
\begin{figure}[t]
\begin{center}
\leavevmode
\epsfxsize= 10 truecm
\epsfbox[105 590 520 720]{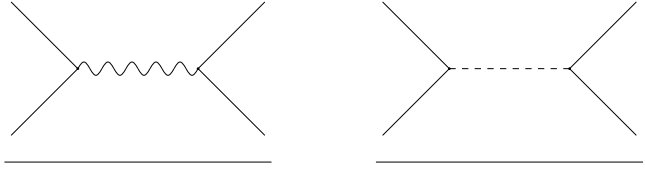}
\end{center}
\caption{Tree-level proton decay via ``lepto-quark'' gauge boson and scalar
exchange $qq^{\prime} \rightarrow q^{\prime\prime}l$. 
The straight line represents the spectator quark which is 
contained in proton and after its decay hadronizes 
with $q^{\prime\prime}$ to form a meson. 
$q$ ($l$) denotes a generic quark (lepton).}
\label{fig:proton}
\end{figure}
%%

%626
Secondly, additional non-SM matter and gauge (super)fields
appear, and in most cases must be rendered heavy. For example, in $SO(10)$
the {\bf 16} contains also a singlet right-handed neutrino $N$,
which is useful for understanding neutrino masses.
(We return to neutrino masses in Chapter~\ref{c11}.)
%

%627
Other heavy particles, however, are more troublesome.
The Higgs doublets are embedded together with color triplets
which interact with SM matter with the same Yukawa couplings as the
Higgs doublets. E.g. the matter ${\bf 10}_{a}+ {\bf \bar 5}_{a}$
superpotential in minimal $SU(5)$ 
(recall $y_{d} = y_{l}$ in the minimal model)
reads $W \sim y_{u_{ab}}5_{H}10_{a}10_{b} + y_{d_{ab}}\bar{5}_{H}10_{a}\bar{5}_{b}$,
and the 
${\bf 5}_{H}$ (${\bf \bar 5}_{H}$) 
contains the $H_{2}$ ($H_{1}$) Higgs doublet and a 
down-type color triplet $H^{C}_{2}$  ($H^{C}_{1}$) 
with a hypercharge $-1/3$ ($+1/3$). 
Also, the {\bf 24} contains not only the SM gauge
content but also the  $(3,\,2)_{-5/6}$ $(\bar X,\,\bar Y)$ vector superfields, 
as well as the complex conjugate representation $(X,\,Y)$, 
all of which must become massive after the spontaneous (GUT) symmetry breaking. 
%

%628
The $X$ and $Y$  are lepto-quark
(super)fields which connect lepton and quark fields and therefore lead to
proton decay, e.g. $p \rightarrow \pi^{0}e^{+}$, at
tree level. So do the color triplet Higgs fields, which therefore must
also become massive with a mass near the unification scale.
(Relevant diagrams are illustrated in Fig.~\ref{fig:proton}.)
The constraint from the proton life time measurement,
\begin{equation}
\tau_{P} \simeq 3 \times 10^{31\pm 1}{\mbox{ yr
}}\left(\frac{M_{U}}{4.6 \times 10^{14} {\mbox{ GeV}}}\right)^{4} \gtrsim 10^{33}
{\mbox{ yr}},
\label{proton}
\end{equation}
is, however, easily satisfied for $M_{U} \simeq 3 \times 10^{16}$ GeV.
%

%629
Nevertheless, supersymmetry implies also heavy color triplet Higgsinos
$\widetilde{H}^{C}_{i}$ with lepto-quark intergenerational
Yukawa couplings, for example, $y_{ab}\widetilde{H}^{C}_{1}\widetilde{Q}_{a}L_{b}$
(derived from $W \sim y_{d_{ab}}\bar{5}_{H}10_{a}\bar{5}_{b}$).
These terms induce proton decay $p \rightarrow K^{+}\bar\nu$ 
radiatively from loop (box) diagrams\footnote{These diagrams are similar to the 
meson mixing diagram Fig.~\ref{fig:box} only with fermion Majorana (gaugino)
and Dirac (colored Higgsino) mass insertions instead of the sfermion 
mass insertions.}
with the colored Higgsino and
the MSSM gauginos and sfermions circulating in the loop. 
The colored Higgsino interaction in this diagrams can be described 
by the  effective dimension five operators in the superpotential
$W \sim (QQQL + UDUE)/M_{U}$ whose $F$ component leads to dimension-five
vertices of the form $\tilde{f}\tilde{f}ff/M_{U}$.
In this alternative description the diagram can be cast as a triangle
sparticle loop: The sparticle dressing loop.
In order for the first operator not to vanish one $Q$ field
must carry a different generation label, hence, $p\rightarrow Kl$. (See exercise.)
%

%6210
Such diagrams $\propto (g^{2}y^{2}_{ab}/16\pi^{2})(1/\widetilde{m}M_{U})$,
where $\widetilde{m} \sim \mweak$ is a typical superpartner mass scale.
In comparison the tree-level diagrams Fig.~\ref{fig:proton}
$\propto (g^{2}/M_{U}^{2}), \, (y^{2}_{ab}/M_{U}^{2})$.
Indeed, radiative decay
proved to generically dominate proton decay in these models since 
the corresponding amplitude is suppressed by only two (rather than four) 
powers of the superheavy
mass scale. (The additional suppression by small intergenerational
Yukawa couplings is therefore crucial.)
It is interesting to note that these operators are forbidden if
instead of $R$-parity a discrete $Z_{3}$ baryon  parity is imposed,
which truly conserve baryon number but allows $R$-parity lepton
number violating operators \cite{RI}.
%

%6211
The radiative proton decay places one of the most severe constraints on the models
and eventually will determine their fortune.
This is particularly true after significant improvements in experimental constraints
from the Super-Kamiokande collaboration \cite{superK2} which yield
comparable life-time bounds $\sim{\cal{O}}(10^{33})\,$yr 
on the $K\nu$ and $\pi e$ decay modes.
On the other hand, the large predicted (in relative terms) amplitudes for
radiative proton decay provide an opportunity to test the unification
framework and supersymmetry simultanoeusly.
\subsection{Doublet - Triplet Splitting} 
%

%6212
Our discussion above leads to the other most difficult problem facing the
grand-unification framework. Supersymmetry guarantees that once the
Higgs doublets and color triplets are split so that the former are light
and the latter are heavy, this hierarchy is preserved to all orders
in perturbation theory.
Nevertheless, it does not specify how such a split may occur.
This is the doublet-triplet splitting problem which is
conceptually, though 
(because of supersymmetry non-renormalization theorems)
not technically, a manifestation of the hierarchy problem.
%

%6213
More generally, it is an aspect of the problem of fixing the 
$\mu$-parameter (\ie the doublet mass)
$\mu = {\cal{O}}(\mweak)$,
which was  mentioned briefly in the previous chapter. 
Like all other issues raised, extensive 
model-building efforts and many innovative solutions exist \cite{MOH},
but will not be reviewed here.
They typically involve extending the model representations,
symmetries and/or (unification-scale) space-time dimensions.
%

%6214
It is intriguing, however, that the fundamental problems of this framework,
the light fermion spectrum (e.g. $y_{ab}$) and the doublet-triplet splitting
(e.g. the colored Higgsino mass) combine to determine the 
proton decay amplitude $p \rightarrow K\nu$, which in turn
provides a crucial test of the framework, and consequently its potential downfall.
%

%E6
\section*{Exercises}
\addcontentsline{toc}{section}{Exercises}
\markright{Exercises}

\subsubsection*{6.1}
Calculate the $\beta$-function coefficients in the MSSM (eq.~(\ref{biMSSM})).
At what order the assumption of $R_{P}$ conservation affects the calculation?

\subsubsection*{6.2}
Confirm the hypercharge $U(1)$ GUT normalization.

\subsubsection*{6.3}
Solve in iterations the two-loop renormalization group equation for the gauge
coupling, neglecting Yukawa couplings. Rewrite the solutions as
predictions for $t$, $M_{U}$ and $\alpha_{3}(M_{Z})$.

\subsubsection*{6.4}
Use the strong coupling to predict the weak angle. Show that
at the unification scale one has the boundary condition
$s^{2}_{W}(M_{U}) = g^{\prime\,2}/(g^{\prime\, 2} +
g_{2}^{2})|_{M_{U}} = 3/8$. Compare your prediction to the data.

\subsubsection*{6.5}
Count degrees of freedom and show that the {\bf 16} of $SO(10)$
contains a singlet, the right-handed neutrino.

\subsubsection*{6.6}
Calculate the $\beta$-function coefficients in the MSSM (eq.~(\ref{biMSSM}))
as a function of the number of Higgs doublets.
Rewrite the one-loop solutions and the one-loop
predictions for $t$, $M_{U}$ and $\alpha_{3}(M_{Z})$ as 
a function of the number of Higgs doublets as well.
Examine the variation of the predictions as you decrease/increase
that number. The sharp change in the predictions is because
the Higgs doublets do not form a complete GUT representations.
Add the down-type color triplet contribution (so that the extra Higgs doublet
are embedded in ${\bf {5}} + {\bf \bar{5}}$ of $SU(5)$, e.g. the messenger
fields in Sec.~\ref{sec:GM})
to the $\beta$-functions
and repeat the exercise. The triplet completes the doublet GUT
representation.

\subsubsection*{6.7}
Solve the gauge part of the Yukawa unification equation (\ref{btauRGE})
and show that in the limit of small Yukawa couplings $y \ll g$
\begin{equation}
\left.\left(\frac{y_{d}}{y_{l}}\right)\right|_{M_{Z}} =
\left(\frac{\alpha_{3}(M_{Z})}{\alpha_{G}}\right)^{\frac{8}{9}}
\left(\frac{\alpha_{1}(M_{Z})}{\alpha_{G}}\right)^{\frac{10}{90}}.
\end{equation}
Compare the predictions for $m_{d}/m_{e}$ and $m_{s}/m_{\mu}$ 
(are they distinguished?) to the data. This is essentially the fermion mass
problem in GUT's.

\subsubsection*{6.8}
Introduce group theory (QCD and $SU(2)$) as well as
 generation indices to the proton decay
operator $QQQL$ and show that one $Q$ must be of the second
generation or otherwise the group theory forces the operators  to vanish
once antisymmetric indices are properly summed. (Why not third generation?)
Write the corresponding
one-loop (box) proton decay diagram with a gaugino, a colored Higgsino,
and sfermions circulating in the loop and derive the proportionality
relation given in the text. (Note that the two colored Higgsinos
form a heavy Dirac fermion.)

%
%C7
%
\chapter{The Heavy Top 
\protect\newline and Radiative Symmetry Breaking}
\label{c7}
%

%701
Now that we are familiar with the notion
and practice of renormalization group evolution, we
will continue and renormalize the MSSM in order
to understand the possible ultraviolet origins of its infrared structure.
(For example, see Ref.~\cite{UV}.)
We begin with an intriguing feature of the equations,
the quasi-fixed point. We then discuss the renormalization
of the spectrum parameters. 
Though we will use the quasi-fixed point regime for demonstration,
this is done for simplicity only and the radiative-symmetry-breaking
results derived below apply in general.
\section{The Quasi-Fixed Point}

\begin{figure}[t]
\postscript{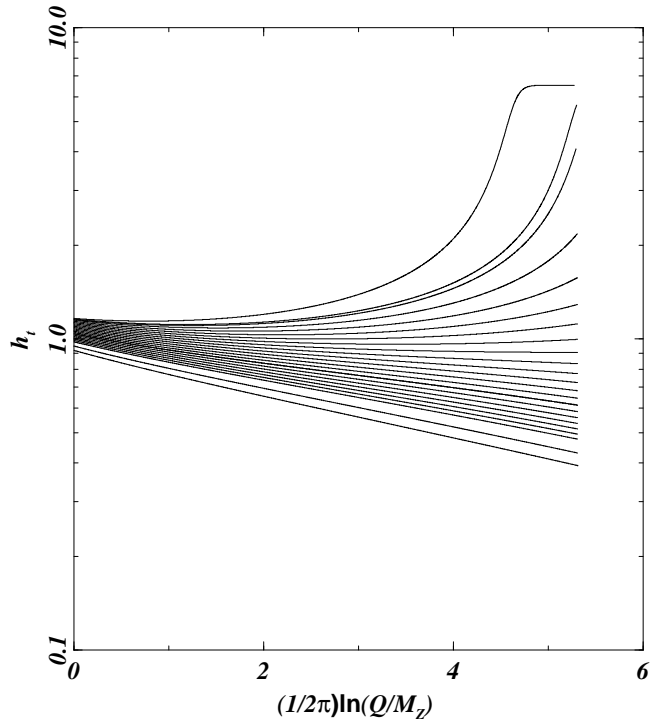}{0.75}
\caption{Convergence of the 
the top Yukawa coupling (denoted here $h_{t}$)
to its quasi-fixed point value, 
given a large range of initial values at the unification scale,
is illustrated (at two-loop order) as a function of the evolution time $t$,
$t=0$ at $M_{Z}$. For very large initial values perturbative validity
does not hold and the curves are shown for illustration only.
(The flat behavior is a two-loop effect.) Note the logarithmic scale.
}
\label{fig:fp}
\end{figure}

%711
Aside from gauge couplings, the top mass, and hence the top Yukawa
(baring in mind the possibility of $\pm{\cal{O}}( 5 - 10\%)$ 
finite corrections from sparticle loops),
is also measured quite precisely, 
$y_{t}(m_{t}) \simeq 0.95/\sin\beta$. Again, one can ask whether such a large
weak-scale coupling remains perturbative (when properly renormalized)
up to the unification scale. The answer is positive for the following reason:
Upon examining the one-loop renormalization group
equation for $y_{t}$,
\begin{equation}
\frac{dy_{t}}{d\ln\Lambda} \simeq \frac{y_{t}}{16\pi^{2}}
\left\{ -\frac{16}{3}g_{3}^{2} + 6y_{t}^{2}\right\},
\label{yt}
\end{equation}
where $g_{3} \equiv g_{s}$ is the QCD gauge coupling and
we neglected all other couplings, one finds a semi fixed-point
behavior \cite{QFP}.
For $y_{t_{\rm fixed}}^{2} = (16/18)g_{3}^{2} \simeq (1.15)^{2}$ the
right hand side equals zero (where we used, as an approximation, the weak-scale
value of $g_{3}$) and $y_{t}$ freezes at this value and is not 
renormalized any further. 
%

%712
For large enough values of $y_{t}(M_{U})$, for example, $y_{t}$ decreases with energy
according to eq.~(\ref{yt}) until it reaches $y_{t_{\rm fixed}}$.
This is the
top-Yukawa quasi-fixed-point value, \ie convergence from above.
The quasi-fixed-point behavior is illustrated in Fig.~\ref{fig:fp}.
(The value of $y_{t_{\rm fixed}}$ diminishes if there are any
other large Yukawa couplings, 
for example a large $y_{b}$, right-handed
neutrino couplings, $R_{P}$-violating couplings, singlet couplings in
the NMSSM, etc., which modify eq.~(\ref{yt}):
$\sum c_{i}y_{i_{\rm fixed}}^{2} = (16/18)g_{3}^{2}$ for some coefficients $c_{i}$.)
On the other hand, weak-scale values $y_{t} >
y_{t_{\rm fixed}}$ imply $y_{t} \gg 1$ at intermediate
energies below the unification scale. This then gives a lower bound on
$\tan\beta \gtrsim 1.8$, referred to following  eq.~(\ref{tanbeta}). 
(An upper bound $\tan\beta \lesssim 60$  
is derived by applying similar consideration to the bottom-Yukawa coupling.)
Renormalization group study, however, shows that the fixed point
is not reached for any (arbitrarily small) initial value of
$y_{t}(M_{U})$, and in fact, even when it is reached it is not an exact value. 
Rather, a small ${\cal{O}}(1\%)$ sensitivity for the boundary conditions remains. 
Hence, it is a quasi-fixed point\footnote{
Note that the equations also contain a true fixed point \cite{FP} which, however,
is not reached given the values of the physical parameters.} 
\cite{QFP} (which serves more
as an upper bound to insure consistency of perturbation theory up
to the unification scale). Nevertheless, its existence  can beautifully explain
$y_{t}(m_{t}) \sim 1$ for a large range of initial values at the
unification scale. It is this behavior that plays a crucial role in
the successful prediction of $b - \tau$ unification discussed in Sec.~\ref{sec:btau}.
%

%713
Current bounds on the Higgs mass, however, limit $\tan\beta$ from below
$\tan\beta \gtrsim 2$ (see the next chapter)
so that the quasi-fixed point scenario may not
be fully realized, at least not in its simplest version. 
Nevertheless, whether or not the top-Yukawa saturates its upper bound,
given the heavy top mass it is a large coupling.
Its renormalization curve is attracted\footnote{
It can be shown that more generally
$y_{t}(m_{t}) \simeq {\cal{O}}(1)$ for $y_t(M_{U}) \gtrsim 0.5$.
See Fig.~\ref{fig:fp}.}  
towards the quasi-fixed point and $y_{t}(\Lambda) \simeq {\cal{O}}(1)$
even though it may not actually reach the fixed point. (See Fig.~\ref{fig:fp}.)
This is an essential and necessary ingredient
in the renormalization of the SSB parameters from some (at this point, arbitrary) 
boundary conditions, for example, at the unification scale.
(It is implicitly assumed here that these parameters appear in the
effective theory and are ``hard'' already at high-energies, but we will
return to this point in Chapter~\ref{c10}.)
%

%714
Next, we focus (only) on the 
renormalization of the terms relevant for electroweak symmetry
breaking; the radiative symmetry breaking (RSB) mechanism.
A large $y_{t}$ coupling was postulated more than a decade ago as a mean to
achieve RSB \cite{RSB,ccb2} and in that sense is a prediction of the MSSM framework
which is successfully confirmed by the experimentally measured heavy
top mass $m_{t} \simeq \nu$.
\section{Radiative Symmetry Breaking}
%

%721
In order to reproduce the
SM Lagrangian properly, a negative mass squared 
in the Higgs potential is required. (See also Chapter~\ref{c12}.)
Indeed, the $m_{H_{2}}^{2}$ parameter is
differentiated from all other squared mass parameters once we include
the Yukawa interactions. 
Consider the coupled renormalization group equations, including, for
simplicity,  only gauge and top-Yukawa effects.
(More generally, the $b$-quark, $\tau$-lepton, right-handed neutrino,
singlet and $R_{P}$-violating couplings may not be negligible.) 
Then, the one-loop
evolution of $m_{H_{2}}^{2}$ (and of the coupled parameters
$m_{U_{3}}^{2}$ and $m_{Q_{3}}^{2}$) with respect to the logarithm of
the momentum-scale is given by
\begin{equation}
\frac{d m_{H_{2}}^{2}}{d \ln \Lambda} = \frac{1}{8\pi^{2}}
(3y_{t}^{2}\Sigma_{m^{2}} - 3g_{2}^{2}M_{2}^{2} - g_{1}^{2}M_{1}^{2}),
\label{rge1a}
\end{equation}
and
\begin{equation}
\frac{d m_{U_{3}}^{2}}{d \ln \Lambda} = \frac{1}{8\pi^{2}}
(2y_{t}^{2}\Sigma_{m^{2}} - \frac{16}{3}g_{3}^{2}M_{3}^{2} -
\frac{16}{9}g_{1}^{2}M_{1}^{2}),
\label{rge1b}
\end{equation}
\begin{eqnarray}
\frac{d m_{Q_{3}}^{2}}{d \ln \Lambda} = \frac{1}{8\pi^{2}}
(y_{t}^{2}\Sigma_{m^{2}} - \frac{16}{3}g_{3}^{2}M_{3}^{2} -
3g_{2}^{2}M_{2}^{2} 
- \frac{1}{9}g_{1}^{2}M_{1}^{2}),
\label{rge1c} &&
\end{eqnarray}
where $\Sigma_{m^{2}} = [m_{H_{2}}^{2} + m_{Q_{3}}^{2} +
m_{U_{3}}^{2}+ A_{t}^{2}]$, and we denote the SM $SU(3)$, $SU(2)$ and 
(the GUT normalized) $U(1)$
gaugino masses  by $M_{3,2,1}$, as before.
%

%722
The one-loop gaugino mass renormalization obeys 
\begin{equation}
\frac{d}{d\ln\Lambda}\left(\frac{M_{i}^{2}}{g_{i}^{2}}\right) = 0,
\label{gauginorge}
\end{equation}
and its solution simply reads $M_{i}(M_{i}) =
(\alpha_{i}(M_{i})/\alpha_{i}(\Lambda_{\rm UV}))
M_{i}(\Lambda_{\rm UV})$ where a typical choice is $\Lambda_{\rm UV}  =
M_{U}$. Note that in  unified theories the gaugino mass boundary
conditions are given
universally by the mass of the single gaugino of the GUT group
so that $M_{3}:M_{2}:M_{1} \simeq 3:1:1/2$ at the weak scale (where
the numerical ratios are the ratios $\alpha_{i}(M_{Z})/\alpha_{U}$).
This is gaugino mass unification. (It also holds, but for different reasons,
in many string models.)
%
 
%723
Given the heavy $t$-quark, one has $y_{t} \sim 1 \sim
g_{3}$.  (In fact, for near quasi-fixed point
values typically $y_{t} > g_{3}$ at high energies.)  
While QCD loops still dominate the evolution of the stop masses squared 
$m_{Q_{3}}^{2}$ and $m_{U_{3}}^{2}$, Yukawa loops dominate the
evolution of $m_{H_{2}}^{2}.$ On the one hand, the stop squared masses
and $\Sigma_{m^{2}}$ increase with the decreasing scale. On the other
hand, the more they increase the more the Higgs squared mass decreases
with scale and, given the integration or evolution time
$2\pi t \sim 30$, it is rendered negative at or before the weak scale.  
(This mechanism hold more generally for different values of $t$. As $t$
decreases the ratio $|m^{2}_{Q,\,U}/m_{H_{2}}^{2}|$ must increase.)
The $m_{3}^{2}H_{1}H_{2}$
Higgs doublet mixing term ensures that both Higgs doublets have
non-vanishing expectation values.
%

\begin{figure}[t]
\postscript{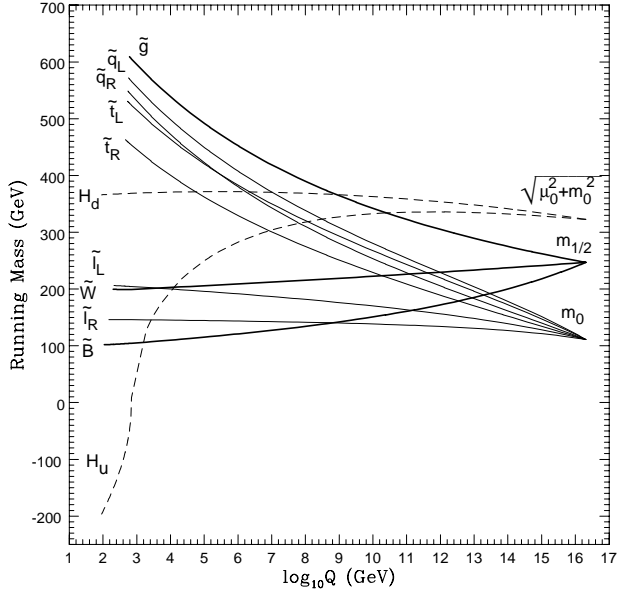}{0.75}
\caption{The renormalization group evolution of SSB masses and $\mu$
for a representative case with universal
boundary conditions at the unification scale $M_{U}$.
$\tilde{t}_{L} = \widetilde{Q}_{3}$ and $\tilde{t}_{R} = \widetilde{U}_{3}$,
and $\tilde{q}_{L\, (R)}$ is a left- (right-) handed squark of
the first two generations.
}
\label{fig:rsb}
\end{figure}

%724
This is a simplistic description of the
mechanism of radiative electroweak symmetry breaking.  In
fact, the sizeable $y_{t}$ typically renders the Higgs squared mass
too negative and some (fine?) tuning (typically of $\mu$) is required in
order to extract  correctly the precisely known electroweak scale.
The degree of acceptable tuning can be argued  to bound the 
mass parameters from above, e.g. $m_{\tilde{t}}\lesssim (4\pi/y_{t})M_{Z}$,
which is a rephrasing of the generic bound eq.~(\ref{soft3}).
(The Higgs potential and its minimization are discussed in the next chapter.) 
An example of the renormalization group evolution of the SSB (and
$\mu$) parameter is illustrated in Fig.~\ref{fig:rsb}, taken from
C.~Kolda. For illustration, universal (see Chapter~\ref{c10}) boundary conditions
are assumed at the unification scale $M_{U}$.
%

%%FIG
%%

%725
In the quasi-fixed point scenario, 
it is possible to solve analytically for the low
energy values of the soft scalar masses in terms of the high scale
boundary conditions. 
For illustration,
we conclude our discussion with those solutions.
We include, for completeness, also the solutions for sfermions and Higgs
SSB masses which are not affected by the large $y_{t}$ (for example,
see Carena {\it et al.} in Ref.~\cite{S61}):
\begin{eqnarray}
m_{H_2}^2 &\simeq& m_{H_2}^{2}(M_{U})+ 0.52 M_{1/2}^2 - 3 \Delta m^2
  \nonumber \\
m_{H_1}^2 &\simeq& m_{H_1}^{2}(M_{U})+ 0.52 M_{1/2}^2 \nonumber \\
m_{Q_i}^2 &\simeq& m_{Q_i}^{2}(M_{U})+ 7.2 M_{1/2}^2 - \delta_i
  \Delta m^2 \nonumber \\
m_{U_i}^2 &\simeq& m_{U_i}^{2}(M_{U})+ 6.7 M_{1/2}^2 - \delta_i 2
  \Delta m^2 \label{carena} \\
m_{D_i}^2 &\simeq& m_{D_i}^{2}(M_{U})+ 6.7 M_{1/2}^2 \nonumber \\
m_{L_i}^2 &\simeq& m_{L_i}^{2}(M_{U})+ 0.52 M_{1/2}^2 \nonumber \\
m_{E_i}^2 &\simeq& m_{E_i}^{2}(M_{U})+ 0.15 M_{1/2}^2 \nonumber \ ,
\end{eqnarray}
where
\begin{eqnarray}
\Delta m^2 &\simeq& \frac{1}{6} \left[ m_{H_2}^{2}(M_{U})+
m_{Q_3}^{2}(M_{U})+ m_{U_3}^{2}(M_{U})\right] r \nonumber \\
 && + M_{1/2}^2 \left( \frac{7}{3} r - r^2
\right) + \frac{1}{3} A_0 \left(\frac{1}{2} A_0 - 2.3 M_{1/2}
\right) r \left( 1- r \right) \ ,
\label{deltam}
\end{eqnarray}
and, for simplicity, and as is customary, we have assumed 
a common gaugino mass $M_{1/2}$
and trilinear scalar coupling $A_0$ at the high scale, which is
conveniently  identified here with the unification scale
$M_{U}$. (Numerical coefficients will change otherwise) 
We will not discuss in detail the renormalization of the $A$-parameters,
which are assumed here to be proportional to the Yukawa
couplings. However, we note that their renormalization also exhibits fixed points.
The subscript $i$ is a generational index; $\delta_1 = \delta_2 = 0$ and
$\delta_3 = 1$.  Finally, the parameter $r = \left[ y_t/y_{t_{\rm fixed}}
\right] ^2 \leq 1$ is a measure of the proximity of the top Yukawa
coupling to its quasi-fixed point  value at the weak scale.
%

%726
It is interesting to observe that for vanishing gaugino
masses the system (\ref{rge1a})--(\ref{rge1c}) has a zero fixed point
(\ie it is insensitive to the value of $m_{0}^{2}$) as long as
$\left[ m_{H_2}^2 (M_{U}), m_{U_3}^2 (M_{U}), m_{Q_3}^{2}(M_{U})\right]
= m_0^2 \, \left[ 3,2,1 \right]$. (See exercise.) Such fixed points, in the presence of
Yukawa quasi-fixed points, are a more general phenomenon with important
consequences for the upper bound on $m_{0}$ \cite{FKP} and hence, ``(fine-)tuning''. 
This also relates to the concept of focus points \cite{focus} which may be 
similarly realized
$(i)$ for particular boundary-condition patterns,
$(ii)$ and for particular values of $y_{t}$
(which experimentally coincide with intermediate values of $\tan\beta \gtrsim 10$),
$(iii)$ and if the ultraviolet scale is (or near) the unification scale.
If the ``focusing'' is realized then
$m_{H_{2}}^{2}(M_{Z}) \simeq  - M_{Z}^{2}$ regardless of the exact
magnitude of its boundary condition at the unification scale.
The focus point behavior also
undermines  (fine-)tuning considerations to some degree.
Both, the fixed and the focus points, hold at one loop
and their application is constrained by the two-loop corrections.
%

%727
The general (two-loop) renormalization group equations are given in
Ref.~\cite{TLOOP}. Other related papers include Ref.~\cite{S62}.
%

%E7
%
\section*{Exercises}
\addcontentsline{toc}{section}{Exercises}
\markright{Exercises}

\subsubsection*{7.1}
Confirm the one-loop gaugino-mass unification relation
$M_{3}:M_{2}:M_{1} \simeq 3:1:1/2$ at the weak scale.

\subsubsection*{7.2}
Omitting all terms aside from QCD in the stop renormalization group
equation,
and using the renormalization group
equations for the gauge couplings and for the gaugino mass, 
show that if $m_{Q}(\Lambda_{\rm UV}) = 0$ then 
$m_{Q}^{2}(m_{Q}) \simeq 7M_{3}^{2}(\Lambda_{\rm UV}) \simeq
M_{3}^{2}(M_{3})$, where $m_{Q}$ correspond to the stop doublet mass
and $M_{3}$ to the gluino mass. Compare to the analytic solution with
$\delta_{i} = 0$.
Include all gauge terms and assume gaugino mass unification,
can you derive the difference between the renormalized $m_{Q}^{2}$ and
$m_{U}^{2}$ ?

\subsubsection*{7.3}
Write the one-loop renormalization group 
equations (\ref{rge1a})--(\ref{rge1c}) in a matrix form.
This defines a three-dimensional space. Show that 
the vector $\left[ m_{H_2}^2 (\Lambda_{\rm UV}), 
m_{U_3}^2 (\Lambda_{\rm UV}), m_{Q_3}^{2}(\Lambda_{\rm UV}) \right]
= m_0^2 \, \left[ 3,2,1 \right]$ is an eigenvector of the coefficient
matrix in the limit of vanishing gaugino masses (and $A_{0} = 0$). 
Find its eigenvalue and solve the matrix equation along this direction
for a fixed $y_{t}$. Compare to eq.~(\ref{carena}).
Study the behavior of the solution and show that it has a zero 
stop-Higgs fixed point in this case. In practice, it implies that the
gaugino mass, rather than the the sfermion boundary conditions, dictate the
low-energy value of the vector. Find all eigenvalues and eigenvectors
and write the general solution in this limit. 

\subsubsection*{7.4}
Generalize eqs.~(\ref{rge1a})--(\ref{rge1c}) to the case
$y_{t} \simeq y_{b} \simeq {\cal{O}}(1)$, as is appropriate in the large
$\tan\beta$ regime. (Note: $t \to b,\, Q \to Q,\,
U \to D,\, H_{2} \to H_{1}$.) Next, generalize eq.~(\ref{carena})
accordingly. (For simplicity, denote $\Delta m^{2} \to
\Delta m^{2}_{t},\, \Delta m^{2}_{b}$.) Finally, repeat Ex.~7.3 
for this case (assume $y_{t} = y_{b}$).
This exercise can be further generalized for the
``$SO(10)$ unification''  case, $y_{\nu_{\tau}} = y_{t} = y_{b} = y_{\tau}$.

\subsubsection*{7.5}
Consider the scale-dependent behavior of $m_{H_2}^2(\Lambda)$
in the case $m_{H_2}^2(M_{U}) < 0$. ($M_{U}$ serves as the ultraviolet boundary.)
Show that, in general, $m_{H_2}^2(\Lambda)$ is bounded from below.

%
%
%

%C8
\chapter{The Higgs Potential 
\protect\newline and the Light Higgs Boson}
\label{c8}

%801
In the previous chapter it was demonstrated that a negative mass squared 
in the Higgs potential is generated radiatively for a large range of boundary conditions.
We are now in position to write  and minimize the Higgs potential and examine 
the mass eigenvalues and eigenstates and their characteristics.
We will do so in this chapter.
\section{Minimization of the Higgs Potential}
%

%811
In principle, it is far from clear that the Higgs bosons rather than
some sfermion receive {\it vev's}. Aside from the sneutrino (whose
{\it vev} only breaks lepton number, leading to the generation of
neutrino masses) all other sfermions cannot have non-vanishing
expectation values or otherwise QED and/or QCD would be spontaneously broken.
Furthermore, there could be some direction in this many field space
in which the complete scalar potential (which involves all Higgs and
sfermion fields) is not bounded from below. We reserve the  discussion of
these issues to Chapter~\ref{c12} and only state 
here that they lead to constraints on the
parameter space, for example, on the ratios of the SSB parameters
$|A_{f}/m_{\tilde{f}}|^{2} \lesssim 3 - 6$.
For the purpose of this chapter, let us simply assume that any such
constraints are satisfied and let us focus on the Higgs potential.
%

%812 
The Higgs part of the MSSM (weak-scale) scalar potential
reads, assuming for simplicity that all the  parameters are real, 
\begin{eqnarray}
V(H_{1},\,H_{2}) = (m_{H_{1}}^{2} + \mu^{2})|H_{1}|^{2} +
(m_{H_{2}}^{2} + \mu^{2})|H_{2}|^{2}  & & \nonumber \\
- m_{3}^{2}(H_{1}H_{2} + h.c.) 
+\frac{{\lambda}^{\mbox{\tiny MSSM}}}{2}(|H_{2}|^{2} - |H_{1}|^{2})^{2}
+\Delta V, & &
\label{pot}
\end{eqnarray}
where $m_{H_{1}}^{2}$, $m_{H_{2}}^{2}$, 
and $m_{3}^{2}$ ($\mu$) 
are the soft (supersymmetric) mass parameters
renormalized down to the weak scale (\ie eq.~(\ref{pot})
is the one-loop improved tree-level potential), 
$m_{3}^{2} > 0$,
$\lambda^{\mbox{\tiny MSSM}} = (g_{2}^{2} + g^{\prime\, 2})/4$
is given by the hypercharge and $SU(2)$ $D$-terms,
and we suppress $SU(2)$ indices. (Note that $F$-terms
do not contribute to the pure Higgs quartic potential in the MSSM.
They do contribute in the NMSSM.)
The one-loop correction 
$\Delta V = \Delta V^{\rm one-loop}$
(which, in fact, is a threshold correction
to the one-loop improved tree-level potential)
can be absorbed to a good approximation
in redefinitions of the tree-level parameters.
%

%813
A broken $SU(2)\times U(1)$ 
(along with the constraint 
$m_{H_{1}}^{2} + m_{H_{2}}^{2} + 2\mu^{2} \geq 2|m_{3}^{2}|$
from vacuum stability, \ie the requirement of a bounded potential along the flat
$\tan\beta = 1$ direction)
requires 
\begin{equation}
(m_{H_{1}}^{2} + \mu^{2})(m_{H_{2}}^{2} + \mu^{2}) \leq |m_{3}^{2}|^{2}.
\label{cond}
\end{equation}
Eq.~(\ref{cond}) is automatically satisfied
for $m_{H_{2}}^{2} + \mu^{2} <0$ (given $m_{H_{1}}^{2}
> 0$), which was the situation discussed in Chapter~\ref{c7}.
The minimization conditions then give 
\begin{equation}
\mu^{2} = \frac{m_{H_{1}}^{2} - m_{H_{2}}^{2}\tan^{2}\beta}{\tan^{2}\beta - 1}
-\frac{1}{2}M_{Z}^{2},
\label{min1}
\end{equation}
\begin{equation}
{m_{3}^{2}} = -\frac{1}{2}\sin2\beta\left[
{m_{H_{1}}^{2} + m_{H_{2}}^{2} + 2\mu^{2}}\right].
\label{min2}
\end{equation}
%

%814
By writing eq.~(\ref{min1}) we subscribed to the convenient notion that $\mu$
is determined by the precisely known value $M_{Z} = 91.19$ GeV.
This is a mere convenience. Renormalization
cannot mix the supersymmetric $\mu$ parameter (which is protected by
non-renormalization theorems which apply to the superpotential,
$d\mu/d\ln\Lambda \propto \mu$) 
and the SSB parameters, and hence the independent $\mu$ can be treated as
a purely low-energy parameter. Nevertheless it highlights the
$\mu$-problem, why is a supersymmetric mass
parameter exactly of the order of magnitude of 
the SSB parameters (rather than $\Lambda_{\rm UV}$, for example) \cite{KIM}.
We touched upon this point in the context of GUT's and doublet-triple
splitting, but it is a much more general puzzle 
whose  solution must encode some information on the ultraviolet
theory which explains this relation. 
(Several answers were proposed in the literature, 
including Ref.~\cite{KIM,GM,KPP,muTalk} and various variants of the NMSSM.) 
%

%815
The above form of eq.~(\ref{min1}) also highlights the  fine-tuning issue
whose rough measure is the ratio $|\mu/M_{Z}|$.
Typically $|m_{H_{2}}^{2}|$ is a relatively large parameter controlled by the stop
renormalization, which itself is controlled by QCD and gluino loops. One often finds
that a phenomenologically acceptable value of $\mu$ is $|\mu(M_{Z})| \simeq 
|M_{3}(M_{Z})|$
and that $M_{Z}$ is then determined by a cancellation between two
${\cal{O}}$(TeV) parameters, e.g. 
$(1/2)M_{Z}^{2} \simeq - (m_{H_{2}}^{2} + \mu^{2})$
in the large $\tan\beta$ limit.
Clearly, this is a product of our practical decision to fix
$M_{Z}$ rather than extract it. All it tells us is that $M_{Z}$ (or
$\nu$) is a special rather than arbitrary value.
The true tuning problem is instead in the relation
$|\mu| \simeq |M_{3}|$ which is difficult to understand.
Fine-tuning is difficult to quantify, and each of its definitions in
the literature
has its own merits and conceptual difficulties. Caution is in place
when applying such esthetic notions to actual calculations, an
application which we will avoid.
\section{The Higgs Spectrum and Its Symmetries}
%

%821
Using the minimization equations, the pseudo-scalar mass-squared 
matrix (\ref{mH2matrix}) (the corresponding CP-even and charged Higgs
matrices receive also contribution from the $D$-terms, or equivalently,
from the quartic terms) is now 
\begin{equation}
M^{2}_{PS} =  
m^{2}_{3}\left( \begin{array}{cc}  \tan\beta & -1 \\ -1 & 1/\tan\beta\end{array} \right).
\end{equation}Its determinant vanishes due to the massless Goldstone
boson.
It has a positive mass-squared eigenvalue
$m_{A^{0}}^{2} = {\rm Tr}M^{2}_{PS} = m_{3}^{2}/((1/2)\sin 2\beta) =
m_{1}^{2} + m_{2}^{2}$, where as before
$m_{i}^{2} \equiv m_{H_{i}}^{2} + \mu^{2}$ for $i=1,2$.
Electroweak symmetry breaking is then confirmed.
The angle $\beta$ is now seen to be the rotation angle
between the current and mass eigenstates.
%

%822
The CP-even Higgs tree-level mass matrix reads
\begin{equation}
M^{2}_{H^{0}} =  
m^{2}_{A^{0}}\left( \begin{array}{cc}
s_{\beta}^{2}&-s_{\beta}c_{\beta} \\ -s_{\beta}c_{\beta} &
c^{2}_{\beta}\end{array} \right) +
M^{2}_{Z}\left( \begin{array}{cc}c_{\beta}^{2}&-s_{\beta}c_{\beta} \\ -s_{\beta}c_{\beta} &
s^{2}_{\beta}\end{array} \right), 
\end{equation}
with (tree-level) eigenvalues
\begin{equation}
m_{h^{0},\,H^{0}}^{2\,T} = \frac{1}{2}\left[
m_{A^{0}}^{2} + M_{Z}^{2} \mp
\sqrt{(m_{A^{0}}^{2} + M_{Z}^{2})^{2} -
4m_{A^{0}}^{2} M_{Z}^{2}\cos^{2}2\beta \,} \right].
\label{tree}
\end{equation}
Note that at this level there is a sum rule
for the neutral Higgs eigenvalues: $m_{H^{0}}^{2} + m_{h^{0}}^{2} =
m_{A^{0}}^{2} + M_{Z}^{2}$.
%

%823
There are two particularly interesting limits to eq.~(\ref{tree}).
In the limit $\tan\beta \rightarrow 1 $ one has $|\mu| \rightarrow \infty$ 
and the $SU(2)\times U(1)$
breaking is driven by the $m_{3}^{2}$ term.
In practice, one avoids
the divergent limit by taking
$\tan\beta \gtrsim 1.1$, as is also required from perturbativity of
the top-Yukawa coupling and by the experimental lower bound
on the Higgs boson mass (discussed in the next section).
For $\tan\beta \rightarrow \infty$ 
one has $m_{3} \rightarrow 0$ 
so that the symmetry breaking is driven by $m_{H_{2}}^{2} < 0$.
%

%824
The $\tan\beta \rightarrow 1$ case corresponds to an approximate
$SU(2)_{L + R}$ custodial symmetry of the vacuum:
Turning off hypercharge
and flavor mixing, 
and if $y_{t} = y_{b} = y$, then one can rewrite
the $t$ and $b$ Yukawa terms in a 
$SU(2)_{L}\times SU(2)_{R}$ invariant form \cite{C1}, 
\begin{equation}
y\left(\begin{array}{c}t_{L} \\ b_{L} \end{array} \right)_{a}
\epsilon_{ab}
\left(\begin{array}{cc}H_{1}^{0} & H_{2}^{+} \\ 
H_{1}^{-} & H_{2}^{0} \end{array} \right)_{bc}
\left(\begin{array}{c}-b^{c}_{L} \\ t^{c}_{L} \end{array} \right)_{c}
\label{matrixeq}
\end{equation}
where in the SM $H_{2} = i\sigma_{2}H_{1}^{\ast}$. 
For $\nu_{1}  = \nu_{2}$ 
(as in the SM or in the $\tan\beta \rightarrow 1$ limit) 
the symmetry is spontaneously broken
$SU(2)_{L}\times SU(2)_{R} \rightarrow SU(2)_{L+R}$. 
However, $y_{t} \neq y_{b}$ and the different hypercharges of
$U_{3} = t^{c}_{L}$ and $D_{3} = b^{c}_{L}$ explicitly break
the left-right symmetry, and therefore the residual custodial
symmetry. 
%

%825
In the MSSM, on the other hand,  $H_{1}$ is distinct from $H_{2}$
and if $\nu_{1} \neq \nu_{2}$  (where $\nu_{i} = \langle H_{i}^{0} \rangle$)
$SU(2)_{L}\times SU(2)_{R} \rightarrow U(1)_{T_{3L} + T_{3R}}$.
Therefore, the $SU(2)_{L + R}$ symmetry 
is preserved  if $\beta = \frac{\pi}{4}$ ($\nu_{1}  = \nu_{2}$) 
and  is maximally broken if $\beta = \frac{\pi}{2}$ ($\nu_{1} \ll \nu_{2}$).
(This is the same approximate custodial symmetry which was mentioned above
in the context of the smallness of quantum corrections to electroweak
observables and couplings, but as manifested in the Higgs sector.)
The symmetry is broken at the loop level
so that one expects in any case $\tan\beta$ above unity.
As a result of the symmetry, 
\begin{equation}
M^{2}_{H^{0}} \approx  \mu^{2} \times
\left( \begin{array}{rr}1 & -1 \\ -1 & 1\end{array} \right),
\end{equation}
and it has a massless tree-level eigenvalue, $m_{h^{0}}^{T} \approx 0$.
This is, of course, a well known result of the tree-level formula
when taking $\beta \simeq \frac{\pi}{4}$. 
The mass is then determined by the loop corrections which are well
known (to two-loop) $m_{h^{0}}^{2} \approx \Delta^{2}_{h^{0}} \propto h_{t}^{2}m_{t}^{2}$
(see the next section).
The heavier CP-even Higgs boson mass eigenvalue 
equals approximately $\sqrt{2}|\mu|$.
(The loop corrections are less relevant here as typically
$m_{H^{0}}^{2} \gg \Delta_{H^{0}}^{2}$). 
The custodial symmetry (or the large $\mu$ parameter)
dictates in this case a degeneracy 
$m_{A^{0}} \approx m_{H^{0}} \approx m_{H^{+}} \approx  \sqrt{2}|\mu|$. 
(The tree-level corrections to that relation are 
${\cal{O}}(M_{W,\,Z}/m_{A^{0}})^{2}$.)
That is, at a scale $\Lambda \approx \sqrt{2}|\mu|$ 
the heavy Higgs doublet $H$ is decoupled, and the 
effective field theory below that scale
has only one 
SM-like ($\nu_{h^{0}} = \nu$) Higgs doublet, $h$ ($  = H$ of Chapter~\ref{c1})
which contains a light physical state. 
This is a special case of the MSSM in which all other Higgs bosons
(and possibly sparticles) decouple. (The decoupling limit typically holds
for $m_{A^{0}} \gtrsim 300$ GeV, and is realized more generally. See exercise.)
%

%826
The Higgs sector in the large $\tan\beta$ case  exhibits an
approximate $O_{4} \times O_{4}$ symmetry \cite{C2}. 
For $m_{3} \rightarrow 0$ (which is the situation in case $(2)$) 
there is no mixing 
between $H_{1}$ and $H_{2}$ and
the Higgs sector respects the $O_{4} \times O_{4}$ symmetry
(up to gauge-coupling corrections), \ie
invariance under independent rotations of each doublet.
The symmetry is broken to $O_{3} \times O_{3}$ 
for $\nu_{1} \neq \nu_{2} \neq 0$ and the six Goldstone bosons
are the three SM Goldstone bosons, $A^{0}$, and $H^{\pm}$.
The symmetry is explicitly broken
for $g_{2} \neq 0$ (so that $m_{H^{+}} = M_{W}$)
and is not exact
even when neglecting gauge couplings 
(\ie $m_{3} \neq 0$). 
Thus, $A^{0}$ and $H^{\pm}$ are massive pseudo-Goldstone bosons,
$m_{H^{+}}^{2} - M_{W}^{2} \approx m_{A^{0}}^{2} = C\times m_{3}^{2}$. 
However, $C = -2/\sin2\beta$ and it can be large,
which is a manifestation of the fact that 
$O_{4} \times O_{4} \rightarrow O_{4} \times O_{3}$  for $\nu_{1} = 0$.
(The limit $m_{3} \rightarrow 0$ corresponds also to a $U(1)$
Peccei-Quinn symmetry under which the combination $H_{1}H_{2}$ is charged.)
In the case $\beta \rightarrow  \frac{\pi}{2}$
one has $m_{h^{0}}^{T} \approx M_{Z}$ (assuming $m_{A^{0}} \geq M_{Z}$). 
When adding the loop  corrections
$m_{h^{0}} \lesssim \sqrt{2}M_{Z} \approx 130$ GeV. (See the next section.) 
\section{The Light Higgs Boson}
%

%%FIG
\begin{center}
\begin{figure}[thb]
\begin{center}
    \parbox{95mm}{\epsfxsize=\hsize\epsffile[82 207 550 568]
           {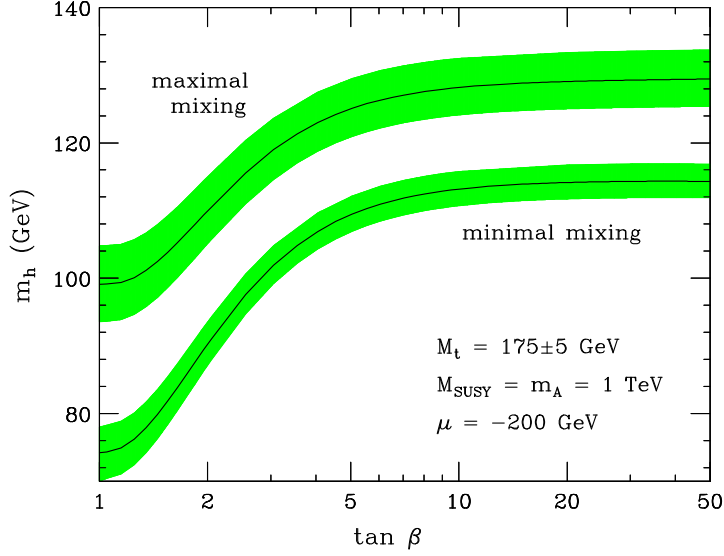}}
\caption[0]{The radiatively corrected
light CP-even Higgs mass is plotted
as a function of $\tan\beta$, 
for the maximal squark left-right mixing (upper band) and minimal squark
mixing cases. The impact of the top-quark
mass is exhibited by the shaded bands; the central value corresponds
to $m_t=175$~GeV, while the upper (lower) edge of the bands
correspond to increasing (decreasing) $m_t$ by 5~GeV. $M_{\rm SUSY}$
is a typical superpartner mass and $\mu$ is the Higgsino mass parameter.
Taken from Ref.~\protect\cite{HIGGSFIG}.}
\label{fig:higgs1}
\end{center}
\end{figure}
\end{center}
%%

%831
Before concluding the discussion of the Higgs sector,
let us examine the lightness of the SM-like Higgs boson from a
different perspective, as well as the one-loop corrections to its mass.
Including one-loop corrections, the general upper bound is derived                  
\begin{equation}
m_{h^{0}}^{2} \leq M_{Z}^{2}\cos^{2}2\beta 
 + \frac{3\alpha m_{t}^{4}}{4\pi s^{2} (1 -s^{2})M_{Z}^{2}}
\left\{\ln\left(\frac{m_{\tilde{t}_{1}}^{2}m_{\tilde{t}_{2}}^{2}}
{m_{t}^{4}}\right) + \Delta_{\theta_{t}} \right\}
\label{mh}
\end{equation}
where
\begin{eqnarray}
&\Delta_{\theta_{t}} =  \left(m_{\tilde{t}_{1}}^{2}-m_{\tilde{t}_{2}}^{2}\right)
\frac{\sin^{2}2\theta_{t}}{2m_{t}^{2}}\ln\left(
\frac{m_{\tilde{t}_{1}}^{2}}{m_{\tilde{t}_{2}}^{2}}\right) 
& \nonumber \\
&  + \left(m_{\tilde{t}_{1}}^{2}-m_{\tilde{t}_{2}}^{2}\right)^{2}
\left(\frac{\sin^{2}2\theta_{t}}{4m_{t}^{2}}\right)^{2}
\left[2 - \frac{m_{\tilde{t}_{1}}^{2}+m_{\tilde{t}_{2}}^{2}}
{m_{\tilde{t}_{1}}^{2}-m_{\tilde{t}_{2}}^{2}}
\ln\left(\frac{m_{\tilde{t}_{1}}^{2}}{m_{\tilde{t}_{2}}^{2}}\right)
\right], &
\label{mixing}
\end{eqnarray}
and where $m_{\tilde{t}_{i}}^{2}$ are the eigenvalues of the
stop $\tilde{t}$
mass-squared matrix, $\theta_{t}$ is the left-right stop mixing angle, 
and we have neglected other loop contributions. 
The tree-level mass squared, ${m_{h^{0}}^{T\, 2}}$, and the loop correction,
$\Delta_{h^{0}}^{2}$, are bounded by the first
and second terms on the right-hand side of eq.~(\ref{mh}), respectively.
In the absence of mixing, $\Delta_{\theta_{t}} = 0$.
For $\tan\beta \rightarrow 1$ one obtains for the tree-level mass
$m_{h^{0}}^{2\,T} \rightarrow 0$, and thus 
$m_{h^{0}}^{2} \approx \Delta_{h^{0}}^{2}$.
%

%832
Clearly, and as we observed before, the tree-level mass vanishes 
as $\tan\beta \rightarrow 1$ ($\cos 2 \beta \rightarrow 0$).
In this limit, the $D$-term (expectation value) vanishes as well as
the value of the tree-level potential
which is now quadratic in the fields. It corresponds to a flat
direction of the potential and the massless real-scalar $h^{0}$ is its ground
state. Now that  we have identified the flat direction it is clear that the
upper bound must be proportional to $\cos 2 \beta$ so that 
$h^{0}$, which parameterizes this direction, is massless
once the flat direction is realized. The proportionality to
$M_{Z}$ is only the manifestation that the quartic couplings are the
gauge couplings. Hence, the lightness of the Higgs boson is a model-independent statement. 
%

%833
The flat direction is always lifted by quantum
corrections, the most important of which is given in eq.~(\ref{mh}).
These corrections may be viewed as effective quartic couplings that have
to be introduced to the effective theory once the stops $\tilde{t}_{i}$, for
example, are integrated out of the theory at a few hundred GeV or
higher scale. These couplings are proportional to the large Yukawa couplings
(for example from integrating out loops induced by
$(y_{t}\tilde{t}H_{i})^{2}$ quartic $F$-terms in the scalar
potential, as in Fig.~\ref{fig:Hdiv2}). Note that even though one finds in many cases
${\cal{O}}(100\%)$ corrections to the light Higgs mass
(and hence $m_{h^{0}} \leq M_{Z}  \rightarrow m_{h^{0}} \lesssim
\sqrt{2}M_{Z}$) this does not signal the breakdown of perturbation
theory. It is only that the tree-level mass (approximately) vanishes.
Indeed, two-loop corrections are much smaller (and shift $m_{h^{0}}$
by typically only a few GeV) and are often negative.
%

%834
The above upper bound is modified if and only if
the Higgs potential contains terms 
(aside from the loop corrections)
that lift the flat direction, for example,
this is the case in the NMSSM 
(see, for example, Ref.~\cite{DreesS,NMSSM}, and references therein)
or if the gauge structure is extended by
an Abelian factor \cite{CVE} ${\rm SM} \rightarrow {\rm SM} \times U(1)^{\prime}$ 
(an extension that could still be consistent with gauge coupling
unification \cite{PGLW} though this is not straightforward).
However, as long as one requires all coupling to stay perturbative
in the ultraviolet then the additional contributions to the Higgs mass
are still modest leading to $m_{h^{0}} \lesssim 150 - 200$ GeV
(including loop corrections), where
the upper bound of $190 - 200$ GeV  was shown \cite{UpperBound} to be saturated 
only in somewhat contrived constructions (which still preserve
unification).
In certain supersymmetric ${\rm SM} \rightarrow {\rm SM} \times U(1)^{\prime}$ frameworks 
(which contain also SM singlets) SM-like Higgs boson as heavy as 180 GeV
(including loop corrections) was found \cite{LPW}.
This corresponds to roughly ${\cal{O}}(100\%) \simeq M_{Z}$ corrections
due to a perturbative singlet and ${\cal{O}}(100\%)$ corrections from
a weakly coupled Abelian factor, so that 
(adding in quadrature) $m_{h^{0}}^{2} \lesssim 4M_{Z}^{2}$.
The only known caveat in the argument for model-dependent
lightness of the Higgs boson is the case of low-energy
supersymmetry breaking where new tree-level terms may appear
in the Higgs potential \cite{Su2}. (See Sec.~\ref{sec:hardhiggs}.) 
%

%835
The existence of a model-independent light Higgs boson is therefore
a prediction of the framework (with the above caveat). 
It is encouraging to note that it seems
to be consistent with current data. The $W$ mass measurement and other
electroweak observable strongly indicate that the SM-like Higgs is
light $m_{h^{0}} \lesssim$ 200-300 GeV where the best fitted
values are near $100$ GeV \cite{PGLER,LEPFEST}. 
(Some caveats, however, remain \cite{Pes,N23}, as discussed in Chapter~\ref{c1}.)
Searches at the LEP experiments bound
the SM-like Higgs mass from below $m_{h^{0}} \gtrsim 115$ GeV \cite{HiggsLimit}.
As noted earlier (see Sec.~\ref{sec:moreonsusy}),
both theory and experiment indicate a (SM-like) Higgs mass range
\begin{equation}
g_{\rm Weak}\nu \lesssim m_{h^{0}} \lesssim \nu
\label{gnunu} 
\end{equation}
(where $g_{\rm Weak}\nu \equiv M_{Z}$). This is an encouraging hint
and an important consistency check.
Depending on final luminosity,
the model-independent light Higgs may be probed already in 
the current Tevatron run, but it may be that its discovery (or exclusion)
must wait until the LHC is operative\footnote{
Detection of a Higgs boson in
this mass range may be difficult in the LHC environment as it relies
on the suppressed radiative diphoton decay channel $h^{0} \rightarrow \gamma\gamma$.
}.
%

%836
The predicted mass range for the light Higgs boson in the MSSM is
illustrated in 
Fig.~\ref{fig:higgs1} taken from Ref.~\cite{HIGGSFIG}.
Further Discussion of the Higgs sector and references can also be found,
for example, in Refs.~\cite{HIGGSREV,S63}.

\vspace*{1cm}

%E8
\section*{Exercises}
\addcontentsline{toc}{section}{Exercises}
\markright{Exercises}

\subsubsection*{8.1}
Derive the minimization conditions of the Higgs potential given above
by first organizing  the minimization equations as
\begin{equation}
m_{1}^{2} = m_{3}^{2}\frac{\nu_{2}}{\nu_{1}} + \frac{1}{4}\left(
g_{2}^{2} + g^{\prime 2}\right)(\nu_{2}^{2} - \nu_{1}^{2})
\end{equation}
\begin{equation}
m_{2}^{2} = m_{3}^{2}\frac{\nu_{1}}{\nu_{2}} + \frac{1}{4}\left(
g_{2}^{2} + g^{\prime 2}\right)(\nu_{2}^{2} - \nu_{1}^{2}),
\end{equation}
where $m_{i}^{2} \equiv m_{H_{i}}^{2} + \mu^{2}$ for $i=1,2$.

\subsubsection*{8.2}
Derive the tree-level charged Higgs mass-squared matrix 
$M_{H^{\pm}}^{2} = M_{PS}^{2}(1 + (1/2)M_{W}^{2}\sin 2 \beta)$,
show that it has a massless eigenvalue (the Goldstone boson)
and derive the sum rule $m_{H^{\pm}}^{2} = m_{A^{0}}^{2} + M_{W}^{2}$.

\subsubsection*{8.3}
Show that the rotation angle of the CP-even Higgs states is given by
\begin{equation}
\sin 2 \alpha = - \frac{m_{A^{0}}^{2} + M_{Z}^{2}}{m_{H^{0}}^{2} -
m_{A^{0}}^{2}}\sin 2 \beta.
\end{equation}
One can define a decoupling limit $\alpha \rightarrow \beta -
(\pi/2)$, which is reached, in practice,  for $m_{A^{0}} \gtrsim 300$ GeV. 
Show that in this limit the heavy Higgs eigenstates form a doublet $H$
that does not participate in electroweak symmetry breaking.
The other (SM-like) Higgs doublet is 
roughly given in this limit 
(up to ${\cal{O}}(M_{Z}^{2}/M_{H}^{2})$ corrections and a phase) 
by $h \simeq H_{1}\cos\beta + H_{2}\sin\beta$.

\subsubsection*{8.4}
Derive the tree-level bound 
$m_{h^{0}} \leq M_{Z}|\cos2\beta|$ 
from eq.~(\ref{tree}).

\subsubsection*{8.5}
Use the experimental
lower bound on the mass of a SM-like Higgs boson and the Higgs mass formulae
to derive a lower bound on $\tan\beta$. Repeat for various values of the parameters
entering the loop correction. Show that the 2000 experimental lower bound 
\cite{HiggsLimit} quoted above
implies $\tan\beta \gtrsim 2$. For caveats, see Ref.~\cite{Su2}.

\subsubsection*{8.6}
In a common version of the NMSSM, $W = \mu H_{1}H_{2}$ is replaced
with $W = \lambda SH_{1}H_{2} + (\kappa/3)S^{3}$, where $S$ is a singlet Higgs field.
Derive the extended ($F$, $D$, and SSB) Higgs potential (for the doublets and singlet)
in this case, and its minimization conditions.
Show that the potential exhibits a discrete $Z_{3}$ symmetry
broken only spontaneously by the Higgs {\it vev's},
which is the (cosmological) downfall of the model due to the
associated domain wall problems at a post inflationary epoch.
Write down the CP-even Higgs and neutralino extended mass matrices in
this model. Find the tree-level light Higgs mass in the limit
$\tan\beta = 1$ and show that it does not vanish (and that indeed the
potential does not have a flat direction in this case).
This is because a new term $\propto \sin^{2}2\beta$ appears alongside
the $\cos^{2}2\beta$ term \cite{DreesS}.

%
%
%

%SP3
\chapter*{Summary}
\addcontentsline{toc}{section}{Summary}
\markright{Summary}
%

%S31
Though no evidence for supersymmetry has been detected as of the
writing of this manuscript, indications in support
of the framework have been accumulating
and include the unification of gauge couplings; the heavy $t$-quark;
and electroweak data preference of $(i)$ a light Higgs boson
and $(ii)$ no ``new-physics'' quantum modifications to various observables.
While individually each argument carries little weight, the combination
of all indications is intriguing. 
The first three topics were explored in this part of the notes.
We did so in a way that provided the reader with a critical
view, as well as with a road map to many of the issues
not discussed here in detail. The current status of the different 
models and proposals was  emphasized.
%

%S32 
The first issue, unification, also served as a warm up
exercise in renormalization. The second issue 
provided the foundation to understanding the weak scale in the context
of supersymmetry: 
Electroweak symmetry breaking, and hence
the induction the weak scale, occurs dynamically and
can be argued to be unique and to be based
on perturbativity and renormalization rather than on ultraviolet details.
The third issue addressed the perhaps
most important aspect of any electroweak-scale model, the
Higgs boson and its characteristics.
Electroweak quantum corrections were investigated in detail in Ref.~\cite{EP}
and will not be discussed here.
Useful reviews not mentioned so far include Refs.~\cite{BAG,DAW}.
%

%S33
A different set of relations between the infrared and ultraviolet,
namely, the (very) low-energy constraints on the sparticle spectrum and couplings,
will be discussed next in Part~\ref{p4}. In particular, we will
demonstrate how low-energy constraints translate to ultraviolet principles.
%

%P4
\part{An Organizing Principle}
\label{p4}
%

%C9
\chapter{The Flavor Problem}
\label{c9}
%

%901
The most general MSSM Lagrangian is still described by many 
parameters. In particular, an arbitrary choice of parameters can lead to unacceptably 
large flavor changing neutral currents (``The Flavor Problem''), e.g.
meson mixing and rare flavor-changing decays.  Also,
if large phases are present, unacceptable  (either flavor conserving or violating)  
contributions to CP-violating amplitudes  could arise (``The CP Problem''), 
e.g. the flavor conserving electron and neutron dipole moments.
%

%902
Here, we focus on the flavor problem. The flavor-conserving CP problem \cite{CP}
is, in practice, less constraining: It may be resolved
independently if no new large physical phases are present in the
weak-scale Lagrangian, \ie they are absent in the high-energy theory 
due to a symmetry or coincidence (see the next chapter),
or alternatively large relative phases could be diminished by
renormalization effects and could be ``renormalized away'' \cite{CP1}. 
It may even be resolved in some (very) special cases by 
cancellations \cite{CP2}, though tuning
cancellations is at best equivalent to assuming small phases to begin with.
(Nevertheless, the physical implications may be different.)
%

%903
The flavor problem stems from the very basic supersymmetrization of
the SM. Recall that even though each fermion flavor is coupled in the
SM with an arbitrary $3\times 3$ Yukawa (or equivalently,
mass) matrix, the unitary diagonalization matrix guarantees that the fermion-fermion
coupling to a neutral gauge boson is basis independent and flavor blind
at tree-level. 
This is also true to a
sfermion-sfermion coupling to a neutral gauge boson which is now
multiplied by the sfermion unitary rotation matrices.
However, by supersymmetry, there exists
also a fermion-sfermion-gaugino vertex. If both fermion and sfermion
mass matrices are diagonalized, then the gaugino vertex is rotated
by two independent unitary matrices and, in general, their product
$U_{f}U^{\dagger}_{\tilde{f}} \neq I$ is not trivial. 
The rotations maintain in this case some (non-trivial) flavor structure and the
gaugino-fermion-sfermion vertex is not flavor diagonal. That is,
generically the theory  contains,
for example,  a $g_{s}\tilde{g}d\tilde{s}^{*}$ vertex! 
At the loop level, such new flavor changing 
couplings lead to new contributions
to SM FCNC observables, just as flavor changing charged currents 
do within the SM.
%

%904
The flavor problem, however, is a blessing in disguise: It provides one with 
an efficient organizing principle which eliminates some
of the arbitrariness in the parameter space. We will apply
this insight in the next chapter. 
Let us first proceed with some concrete examples
which will reveal these principles. 
\section{Meson Mixing}
%

%%FIG
\begin{figure}[ht]
\begin{center}
\leavevmode
\epsfxsize= 10 truecm
\epsfbox[195 515 435 620]{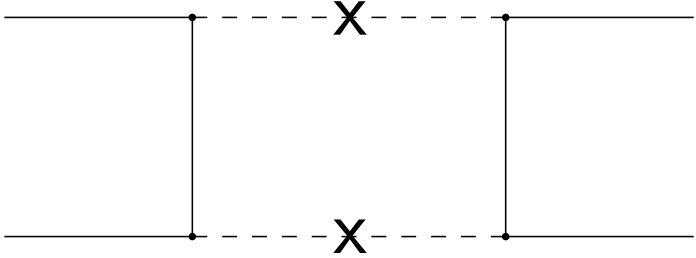}
\end{center}
\caption{A generic ``box'' diagram contribution to meson mixing with gauginos
and squarks circulating in the loop. The Flavor changing is
approximated here as flavor changing squark-mass
insertions, which  are denoted explicitly. In the case of  kaon mixing,
non-trivial phases would lead
to a contribution to the parameter $\epsilon_{K}$.} 
\label{fig:box}
\end{figure}

%911
In the SM, contributions
to meson mixing from gauge-boson - quark couplings arise only at one loop.
Even then, given the small mass differences among the light fermions
and the smallness of light-heavy fermion mixing,
such contributions, particularly to meson mixing, 
are highly suppressed (the GIM mechanism). 
However, given the flavor violating gaugino vertex in supersymmetry,
it is straightforward to construct quantum corrections
for example, to $K- \bar{K}$ mixing, which violate flavor by two
units: See Fig.~\ref{fig:box} for illustration. Such corrections could be
{\it a priori} arbitrarily large.
(Adding phases, new contributions to the SM parameter $\epsilon_{K}$
arise and provide a strong constraint.)
%

%912
Such effects are more conveniently evaluated in the fermion mass basis
in which flavor violation is encoded in the sfermion mass-squared matrix, 
which retains inter-generational off-diagonal entries,
\begin{equation}
M_{\tilde{f}}^{2} = \left(
\begin{tabular}{ccc}
$m^{2}_{\tilde{f}_{1}}$&
$\Delta_{\tilde{f}_{1}\tilde{f}_{2}}$&
$\Delta_{\tilde{f}_{1}\tilde{f}_{3}}$\\
$\Delta_{\tilde{f}_{2}\tilde{f}_{1}}$&
$m^{2}_{\tilde{f}_{2}}$&
$\Delta_{\tilde{f}_{2}\tilde{f}_{3}}$\\
$\Delta_{\tilde{f}_{3}\tilde{f}_{1}}$&
$\Delta_{\tilde{f}_{3}\tilde{f}_{2}}$&
$m^{2}_{\tilde{f}_{3}}$
\end{tabular}\right).
\label{sfermionmatrix}
\end{equation}
In this basis the sfermions of a given sector $f=Q,\,U,\,D,\,L,\,E$ mix
at tree level.
The matrix (\ref{sfermionmatrix}) corresponds to the CKM rotated
original sfermion mass-squared matrix
($\Delta_{\tilde{f}_{i}\tilde{f}{j}} \propto
m^{2}_{\tilde{f}_{i}} - m^{2}_{\tilde{f}_{j}}$). 
Additional flavor-violation arises, in principle, from the
$A$-matrices. However, we ignore chirality labels,
and left-right mixing
(which would further complicate matters) is not included here. 
%

%913
When the above (insertion) approximation is applied to $K$ meson mixing
one has, for example, for the $K_{L} - K_{S}$ mass difference
\begin{equation}
\frac{\Delta m_{K}}{m_{K}f_{K}^{2}} \simeq
\frac{2}{3}\frac{\alpha_{3}}{216}
\frac{1}{\widetilde{m}^{2}}
\left(\frac{\Delta_{\tilde{d}\tilde{s}}}{\widetilde{m}^{2}}\right)^{2},
\end{equation}
where we used standard notation for the kaon mass $m_{K}$ and for the
relevant form factor $f_{K}$, and for  simplicity dimensionless 
functions are omitted. 
This leads to the constraint on the ratio of a typical sqaurk mass $\widetilde{m}$
and the inter-generational mixing
\begin{equation}
\left(\frac{500 {\mbox{ GeV}}}{\widetilde{m}^{2}}\right)
\left(\frac{\Delta_{\tilde{d}\tilde{s}}}{\widetilde{m}^{2}}\right)
\lesssim 10^{-(2-3)},
\label{KKmixing}
\end{equation}
where the complete amplitude, including higher order corrections, was
evaluated \cite{KKmix} in deriving (\ref{KKmixing}). 
Weaker constraints are derived from $B$- and from $D$-meson mixing.
\section{Magnetic and Electric Moments}
\label{sec:SMM}
%

%%FIG
\begin{figure}[t]
\begin{center}
\epsfxsize= 6.5 cm
\leavevmode
\epsfbox[200 500 425 640]{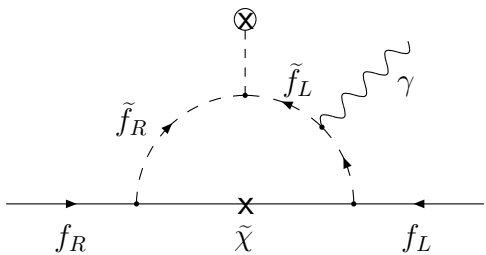}
\end{center}
\caption[f1]{One-loop radiative 
 magnetic moment operator (with the chiral violation provided by the 
sfermion left-right mixing). $\widetilde{\chi}$ is a neutralino or chargino,
and in the quark case there exists also a gluino loop.}
\label{fig:mmoment}
\end{figure}
%

%921
Another set of observables, which is of particular importance, is given by
magnetic moments with virtual sparticles circulating in the loop,
shown in Fig.~\ref{fig:mmoment}.
SM contribution to any
anomalous and transition magnetic moments
is a one-loop effect.
This is also the case for the supersymmetry contribution which is
therefore of the same order of magnitude. 
Henceforth, magnetic moments could shed
light on any new physics in general, and supersymmetry
and its parameter space, in particular.
Here, we review a few (experimentally) promising examples.
Note that contributions to the various magnetic
operators are strongly correlated within a given framework. (See the next chapter
for frameworks.) Hence, a combined analysis provides a stronger probe
than examining each observable individually. However, this is
difficult to do in a  model-independent fashion.
\subsection{Flavor Conserving Moments}
%

%922
To begin with consider flavor conserving magnetic moments\footnote{
Note that the ``pure fermionic'' form of the operator, which dependes on
the Pauli matrix combination $\sigma_{\mu\nu}$, 
suggests that it vanishes in the supersymmetric limit.}, 
\begin{center}
$(eQ_{f}/2m_{f})a_{f}\bar{f}\sigma_{\mu\nu}f F^{\mu\nu} $,
\end{center}
for example,  the anomalous muon magnetic
moment $a_{\mu}$ \cite{gmu}. 
The flavor-conserving muon magnetic moment  is
currently probed  far beyond current bounds at the Brookhaven E821 experiment,
offering a rare opportunity for evidence of
new physics \cite{E821}. Sparticle loops  generically contribute
\begin{equation}
a_{\mu}^{\mbox{\tiny SUSY}} \sim \pm
\frac{\alpha_{2}}{4\pi}\frac{m_{\mu}^{2}M_{Z}\tan\beta}{\widetilde{m}^{3}},
\label{amu}
\end{equation}
where $\widetilde{m}$ is a typical superpartner (in this case, chargino)
mass scale, and the sign is determined by the sign of the $\mu$ parameter. 
(In the special case of models in which the muon mass comes about radiatively
rather from a tree-level Yukawa coupling one has \cite{BFPT}
$a_{\mu}^{\mbox{\tiny SUSY}} \sim + m_{\mu}^{2}/{3\widetilde{m}^{2}}$, which is
significantly enhanced.)
This is a most interesting ``flavor-conserving'' $\tan\beta$-dependent
constraint/prediciton in supersymmetry.
In particular, as $\tan\beta$ increases, a measurement of 
$|a_{\mu}^{\mbox{\tiny SUSY}}| > 0$ can place a decreasing upper bound
on the sparticle scale $\widetilde{m}$.  
%

%923
Similarly to the flavor conserving
magnetic moment, the  same loop diagram, but with an added phase\footnote{
If phases are present then the flavor violating
parameter $\epsilon^{\prime}_{K}/\epsilon_{K}$ may also receive 
``soft'' contributions. $\epsilon^{\prime}_{K}/\epsilon_{K}$ 
has been recently measured with
some accuracy \cite{PDG} and could actually accommodate new-physics 
contributions.}, 
corresponds to a
(flavor-conserving) dipole moment. For example, the electron dipole
moment constraints slepton masses to be in the multi-TeV range if the
corresponding ``soft'' phase $\phi$ (for example a gaugino or $A$-parameter
phase, or their relative phase) is large, 
\begin{equation}
\frac{
\widetilde{m}}{\sqrt{(M_{Z}/\widetilde{m})\tan\beta \sin\phi}}
\gtrsim 2 {\mbox{ TeV}}.
\end{equation} 
\subsection{Transition Moments: $b \rightarrow s \gamma$}
%

%924
Returning to flavor violation,
the same magnetic operator could connect two different fermion flavors
leading to magnetic transition operators 
which violate flavor by one unit and
where again the SM and supersymmetry
contributions arise at the same order. The most publicized case is that
of the $b\rightarrow s\gamma$ operator \cite{bsgamma};
the first such operators whose amplitude was measured.
The projected precision of the next generation of experiments is expected
to allow one to disentangle SM from new-physics contributions.
The generic supersymmetry contribution to the $b\rightarrow s\gamma$
branching ratio arises (just as in the SM and 2HDM) due to CKM
rotations $V$. It reads
\begin{eqnarray}
BR(b\rightarrow s\gamma)
& \simeq & 
{BR(b\rightarrow ce\bar\nu)}
\frac{|V_{ts}^{*}V_{tb}|^{2}}{|V_{cb}|^{2}}\frac{6\alpha}{\pi}
\left[ {\mbox{SM}} + {\mbox{CH}} 
\pm
\frac{m_{t}^{2}\tan\beta}{\widetilde{m}^{2}} \right]^{2}.
\label{bsgamma}
\end{eqnarray}
The first contribution is the SM $W$-loop (written for reference).
Currently, the SM term is consistent  by itself with the experimental measurement,
but the theoretical and experimental
uncertainties accommodate possible new physics contributions.
The second  term is from a charged-Higgs loop.
The third term is the one arises in supersymmetry
from  the SM CKM rotations which, by
supersymmetry, apply also, e.g.  to chargino vertices.
%

%925
The charged Higgs
contribution, also not written  explicitly,
is positive, and comparable to the SM contribution for a relatively
light charged Higgs $m_{H^{\pm}} \lesssim$ 200-300 GeV
(\ie below the higgs decoupling limit of the previous chapter).
Therefore, in the heavy sparticle limit, in which sparticle (in particular, squark)
contributions decouple,
one can constrain (already using current data \cite{PDG}) 
the charged Higgs mass from below.
This is the case, for example,
in the gauge mediation framework which is discussed in the next section.
%

%926
The third term is present in any supersymmetric theory.
Its sign is given by the relative sign of $A_{t}$ and $\mu$,
and $\widetilde{m}$ is of the order of the chargino and stop masses. 
There could also be a contribution in supersymmetry which is 
intrinsically flavor violating 
(see Ref.~\cite{bsgamma} and Gabbiani {\it et al.} in Ref.~\cite{S64}),
\begin{equation}
BR(b\rightarrow s\gamma) \simeq
\frac{2\alpha_{3}^{2}\alpha}{81\pi^{2}}
{\cal{F}}\frac{m_{b}^{3}\tau_{B}}{\widetilde{m}^{2}}
\left(\frac{\Delta_{\tilde{s}\tilde{b}}}{\widetilde{m}^{2}}\right)^{2}
\label{bsgamma2}
\end{equation}
with ${\cal{F}} \simeq {\cal{O}}(1 - 10)$, $\tau_{B} \sim 10^{-12}$
sec the $B$-meson
mean life time,  and $\widetilde{m}$ of the
order of the gluino and sbottom masses. 
We observe two quite different contributions: 
Eq.~(\ref{bsgamma}) with flavor violation 
induced by SM CKM (fermion) rotations of an otherwise flavor conserving 
chargino-quark-squark vertex,
for example; and eq.~(\ref{bsgamma2}) with explicit flavor violation
(described by the sfermion mass-squared matrix  (\ref{sfermionmatrix})). 
(It is possible that the two contributions cancel out to some degree.)
%

%927
While $b \rightarrow s \gamma$ has been observed and the branching
ratio  measured, it does not yet provide a signal of new physics, 
nor it provides strong constraints,
in particular, if both contributions
(\ref{bsgamma}) (including the charged-Higgs term) and (\ref{bsgamma2}) are considered.
Nevertheless, some constraints on relevant parameters
can be obtained (as a function of other parameters), which is useful
when analyzing specific models.
This situation may improve in the near future.
\subsection{Transition Moments: $\mu \rightarrow e \gamma$}
%

%928
Moderate to strong constraints arise 
in the  most interesting case of individual-lepton number (lepton-flavor)
violating  (but total lepton-number conserving) magnetic
transition operators such as $\mu \rightarrow e \gamma$. 
These operators vanish in the SM
in which lepton number is conserved. Thus,
an observation of such a process will provide a
clear indication for new physics!
%

%929
In supersymmetry slepton
mass-squared matrices and trilinear couplings may not respect these accidental
flavor symmetries of the SM, allowing for such processes.
Lepton-flavor violations could further be related
to specific models of total lepton number violation and neutrino
masses and/or grand-unified models (see Chapter~\ref{c11})
where quarks and leptons are predicted to
have similar flavor structure and rotations at high energies.
%

%9210
LFV experiments are therefore of extreme importance, in
particular, given that lepton-flavor violation in (atmospheric) neutrino oscillations
was established experimentally \cite{superK}. 
However, the experiments are also extremely
difficult given the small amplitudes expected for such processes
(which are not enhanced by large QCD or Yukawa couplings). 
In the case of $\mu \rightarrow e \gamma$, one currently has
\begin{equation}
\widetilde{m} \gtrsim 600 {\mbox{ GeV }}
\sqrt{\left(\frac{M_{Z}}{\widetilde{m}}\right) 
\left(\frac{\Delta_{\tilde{e}\tilde{\mu}}}{\widetilde{m}^{2}}\right) 
\tan\beta},
\end{equation}
where $\widetilde{m}$ here is of the order of the slepton mass.
\section{From Constraints to Principles}
%

%931 
Flavor violations (and magnetic moments) in supersymmetry 
and the corresponding signals and
constraints have been studied by many authors, and a sample of
papers is listed in Ref.~\cite{S64}.
Supersymmetry is not the only extension of the SM which
is constrained by flavor (and CP) conservation. In fact, such
constraints are generic to any extension since 
by its definition new-physics extends the SM unique structure
that eliminates tree-level FCNC's -- a property which is difficult to
maintain. Nevertheless, unlike strongly-coupled theories or theories
with extra dimensions (whether supersymmetric or not), in the case of perturbative
supersymmetry the theory is well-defined and calculable, 
and hence, the problem is well-defined and calculable 
(as illustrated above), leaving no place to hide.
%

%932
The satisfaction of the above and similar constraints derived from
low-energy observables leads to the consideration of a small number of
families of special models, and hence, provides an organizing principle
as well as restores predictive power. 
Clearly, realistic models require nearly flavor-conserving
gaugino couplings. Models with the right balance of the universality limit
$\Delta_{\tilde{f}_{i}\tilde{f}{j}} \rightarrow 0$ (in the physical
fermion mass basis) and the  decoupling limit $\widetilde{m} \gg M_{Z}$
(in which the sfermions and their couplings can be integrated out)
can achieve that. The former limit is difficult to justify without 
a governing principle (though such a principle may be a derivative of
the ultraviolet theory and only seem as arbitrary from the low-energy
point of view);  the latter limit may  seem naively as contradictory to the
notion of low-energy supersymmetry and its solution to the hierarchy
problem. Nevertheless, these are the keys for the resolution of the
flavor problem. 
%

%934
We continue in the next chapter with a brief review of the 
realization of these limits in various frameworks for the origins of the SSB
parameters.
%

%E9
\section*{Exercises}
\addcontentsline{toc}{section}{Exercises}
\markright{Exercises}

\subsubsection*{9.1}
Find $\Delta_{\tilde{f}_{i}\tilde{f}{j}}$ (eq.~(\ref{sfermionmatrix}))
by rotating the diagonalized sfermion mass-squared matrix to the fermion 
mass basis. Generalize eq.~(\ref{sfermionmatrix}) to include left-right mixing.

\subsubsection*{9.2}
Write tree-level meson mixing diagrams in theories with 
$R$-parity violation. Find a simple condition on the
$R$-parity violating Yukawa couplings
under which such tree-level contributions vanish.

\subsubsection*{9.3}
The magnetic moment loop is nearly identical
to the flavor conserving 
one-loop correction to a fermion mass
(with sparticles circulating in the loops).
Write and estimate these loops. Show that the leading QCD 
correction to the $b$-mass $\sim y_{b}(2\alpha_{3}/3\pi)\langle H_{2}^{0} \rangle
\sim m_{b}\tan\beta(2\alpha_{3}/3\pi)$ if all SSB (and $\mu$) parameters are equal.
(At what scale are the different parameters evaluated?)
In realistic models the correction is $\sim \pm 2\% m_{b}\tan\beta$, as alluded to
in Sec.~\ref{sec:btau}. What is the correction to the top mass in this limit?
Why are the corrections to lepton masses highly suppressed?

\subsubsection*{9.4}
Calculate the numerical coefficients of the CKM and intrinsic supersymetry
contributions to $BR(b\rightarrow s\gamma)$. (Use Ref.~\cite{PDG} for data inputs.)
Add appropriately in quadrature and (assuming a heavy charged Higgs boson) constrain
$\Delta_{\tilde{s}\tilde{b}}/\widetilde{m}^{2}$ as a function of
$\tan\beta/\widetilde{m}^{2}$, and vice versa. (Assume similar values for
the different $\widetilde{m}^{2}$'s.)

\subsubsection*{9.5}
Assume that at the GUT scale leptons and quarks are rotated
by the same CKM rotations, and that the only other source of
LFV is $m_{E_{3}}^{2} < m_{E_{1}}^{2} = m_{E_{2}}^{2}$.
Estimate the induced contributions to the $\mu \rightarrow e\gamma$
branching ratio. What parameters does it depend on?
%
%
%

%C10
\chapter{Models}
\label{c10}
%

%1001
We are now in position to use the flavor problem to organize
the SSB parameter space. First, however, the parameter space
itself needs to be efficiently parameterized.
We begin by postulating that supersymmetry is broken
spontaneously in some ``hidden'' sector of the theory which contains
the Goldstino and which interacts
with the SM ``observable'' sector only via a specific agent. The agent
is then the messenger of supersymmetry breaking. 
These messenger interactions cannot be renormalizable tree-level
interactions or otherwise the SM could couple directly to the Goldstino multiplet
and one would encounter many of the problems
that plagued our attempt  in Sec.~\ref{sec:s53}
to break supersymmetry spontaneously in the SM sector.
%

%1002
The messengers and their interactions then decouple at a scale
$\Lambda_{\rm mediation}$ which effectively serves the scale of
the mediation of the SSB parameters in the low-energy theory, \ie
this is the scale below which these parameters can be considered as
``hard'' and treated with the renormalization-group formalism (as we
did in Chapter~\ref{c7} in our discussion of radiative symmetry breaking).
This general ``hidden-messenger-observable'' phenomenological framework describes
most of the models and will suffice for our purposes. 
It is  the details of the messenger interactions and the mediation scale
that impact the spectrum parameters and the flavor problem and its solution.
(It is important to recall that gauge, Yukawa, and quartic interactions
are dictated by supersymmetry at all energies above the
actual sparticle mass scale, and that only the dimensionful parameters
interest us in this section.)
\section{Operator Expansion}
%

%1011
In order to review the different classes of models, let
us re-write the soft parameters (and for completeness, also the
$\mu$-parameter) in a generalized operator form, using the tools of
(global) supersymmetry described in Chapter~\ref{c4}.
We can express the most general set of operators as
an expansion in  powers of $\sqrt{F_{X,\,Z}}$, where $X$ and $Z$ are
singlet and non-singlet superfields which
spontaneously break (or parameterize such breaking) 
supersymmetry. They contribute (if not provide) to the Goldstino
component of the physical gravitino; 
their $F$-components are order parameters of supersymmetry
breaking as seen in the SM sector.
The operators are suppressed by powers of
$M \simeq {\cal{O}}(\Lambda_{\rm mediation})$.
The fields $X$ and $Z$ which couple to the SM sector
are the messengers of supersymmetry breaking,
and the interaction which couple them to the SM sector is its agent.
%

%1012
The gravitino mass can be shown (by the local-supersymmetry  condition of a 
vanishing cosmological constant)  to be $m_{3/2} = \sum_{i}F_{i}/\sqrt{3}M_{P}$
where $M_{P}$ is as before the local-supersymmetry expansion parameter: The reduced
Planck mass $M_{P} = \mplanck/\sqrt{8\pi} \simeq 2.4 \times 10^{18}$
GeV. Otherwise no  direct application of 
local supersymmetry is needed. 
%

%1013
The leading terms in this
expansion (see Chapter~\ref{c13} for a generalization)
which generate the $\mu$ parameter and soft terms have the
following form:
\begin{eqnarray}
{\rm Scalar \ masses:} && a\int d^2 \theta d^{2}\bar\theta \, 
\Phi_i^\dagger \Phi_i
\left[ \frac{X^\dagger X}{M^2} + \frac{Z^\dagger Z}{M^2} +
\cdots \right] \label{scalar}\\
\mu \ {\rm parameter:} && b\int d^2 \theta d^{2}\bar\theta \, H_1 H_2
\left[ \frac{X^\dagger}{M} + \hc + \cdots \right] \nonumber \\ 
{\rm and} &&
c\int d^2 \theta\, H_1 H_2
\left[ \frac{X^{n}}{M^{n-1}} \right. \nonumber \\
&& \left. + c^{\prime}\frac{\hat{W}^{\alpha}\hat{W}_{\alpha}}
{M^{2}} + \hc + \cdots \right] 
\label{mu}\\
{\rm Higgs \ mixing} \ (m_{3}^{2}): && d\int d^2 \theta d^{2}\bar\theta \, H_1 H_2
\left[ \frac{X^{\dagger}X}{M^{2}} + \frac{Z^\dagger Z}{M^2} + \cdots \right] 
\label{m3} \\ 
{\rm Gaugino \ masses:} && e\int d^2\theta\, W^{\alpha}
W_{\alpha} \left[ \frac{X}{M} + \hc + \cdots \right] \label{gaugino}\\
A{\rm -terms:} && f\int d^2 \theta\, \Phi_i \Phi_j \Phi_k
\left[ \frac{X}{M} + \hc + \cdots \right] \nonumber \\ 
{\rm and}&&
g\int d^2 \theta d^{2}\bar\theta \, \Phi_i^{\dagger} \Phi_i \left[ \frac{X}{M} +
\hc + \cdots \right] \label{Aterm}\ ,
\end{eqnarray}
where $a - g$ are dimensionless coefficients, 
the $W_{\alpha}$ ($\hat{W}_{\alpha}$) are spinorial gauge supermultiplets containing the
standard model gauginos (a hidden-sector gauge field which condenses), 
the $\Phi_i$ are SM (observable)  chiral
superfields, and $X$ and $Z$ represent supersymmetry-breaking gauge singlet and
non-singlet superfields, respectively. These terms give the SSB parameters
and the $\mu$ parameter when the $X$ and $Z$ fields acquire $F$-term
{\it vev's}: $X \to \theta^2 F_{X}$, $Z \to 
\theta^2 F_{Z}$ (and, in the second source for $A$-terms, $\Phi^{\dagger}_i
\to \bar{\theta}^2 F^*_{\Phi_i} \sim \bar{\theta}^2\Phi_j \Phi_k $).
%

%1014
The $\mu$-parameters could also arise from a  supersymmetry
conserving {\it vev} of $x$, $X \rightarrow x + \theta^{2} F_{X}$
The case $n = 1$ in eq.~(\ref{mu}) corresponds to the NMSSM
and one has to arrange $\langle x \rangle \simeq \mweak \simeq m_{\rm SSB}$.
In this case $m_{3}^{2} \sim A\langle x \rangle \sim A\mu$.
The case $n = 2$ corresponds to $\langle x \rangle \simeq
\sqrt{M\mweak}$, which coincides with the spontaneous supersymmetry
breaking scale in supergravity (see the next section).
The case $n = 3$ has  $X^{3} \equiv \langle W_{\rm
hidden} \rangle$, where $ \langle W_{\rm hidden} \rangle$
is a hidden superpotential {\it vev} 
which (in local supersymmetry) also  breaks supersymmetry.
It can be related to first operator in (\ref{mu})
via redefinitions (a K\"ahler transformation in local supersymmetry). 
%

%1015
Observe that if only one singlet field participates in supersymmetry
breaking then the SSB parameters have only one common phase,
and at most one additional phase arises in the $\mu$ parameter.
Since it can be shown that two phases can always be rotated away and
are not physical, there is no ``CP problem'' in this case.
\section{Supergravity}
\label{sec:supergravity}
%

%1021
It is widely thought that if supersymmetry is realized in nature then
it is a local (super)symmetry, \ie a supergravity (SUGRA) theory.
(In particular, in this case the cosmological constant is not an order
parameter and could be fine-tuned to zero.)
Supergravity interactions provide in this case the theory of (quantum) gravity
at some ultraviolet scale (though it is a non-renoemalizable theory
and hence is probably not the ultimate theory of gravity). As a theory of
gravity, supergravity interacts with all sectors of the theory and
hence provides an ideal agent of supersymmetry breaking. In fact,
even if other agents/messengers exist, supergravity mediation is
always operative once supersymmetry is gauged.
Its mediation scale is, however, always the reduced Planck mass.
From the operator equations (\ref{scalar}) -- (\ref{Aterm}) 
and from $m_{3/2} \sim F/M_{P}$ one observes that SUGRA
contribution to the SSB parameters is always $\sim m_{3/2}$, 
so it could be the leading contribution 
for a sufficiently heavy gravitino $m_{3/2}
\gtrsim M_{Z}$, but otherwise its contributions are suppressed.
It may be argued that the hierarchy problem is then the problem
of fixing the gravitino mass (which may be solved by dynamically
breaking supersymmetry in the hidden sector).
%

%1022
Let us then consider the case of a ``heavy'' gravitino,
$m_{3/2} \simeq {\cal{O}}(100 - 1000)$ GeV (and ${\cal{O}}(1)$ coefficients).
Supersymmetry is spontaneously broken in this case at a scale
$\sqrt{F} \sim \sqrt{m_{3/2}M_{P}} \sim 10^{10-11}$ GeV. The scale of
the mediation is always fixed in supergravity to $M_{P}$.
The hidden sector is truly hidden in this case
in the sense that it interacts only gravitationally with the observable
sector (though supergravity interactions could take various forms).
Supergravity mediation leads to some obvious observations.
First, if $\mu = 0$ in the ultraviolet theory, it is induced 
(the first term in eq.~(\ref{mu})) once
supersymmetry is broken and is of the order of the gravitino mass,
resolving the general $\mu$-problem \cite{KIM,GM,KPP}.
(More solutions arise if the model is such that $\langle x \rangle
\simeq \sqrt{F}$ or supersymmetry is broken by gaugino condensation
in the hidden sector \cite{FPT,CHO} 
${\langle \hat{W}^{\alpha}\hat{W}_{\alpha}\rangle} \simeq \langle F
\rangle^{3/2}$.) 
Secondly, since the gravitino mass determines the weak scale 
$m_{\tilde{f}} \simeq m_{3/2} \simeq M_{Z}$ , 
the decoupling limit  cannot be realized. 
Hence, in order for the SSB parameters to conserve flavor,
supergravity has to conserve flavor so that the universality limit can
be realized. 
%

%1023
Naively, gravity is flavor blind. In general, however, supergravity
(and string theory) is not!
Re-writing, for example, the quadratic operators with flavor indices one has
$a_{\alpha\beta ab}Z_{\alpha}Z^{\dagger}_{\beta}\Phi_{a}\Phi_{b}^{\dagger}$.
The universality limit corresponds to the special case of
$a_{\alpha\beta ab} = \hat{a}_{\alpha\beta}\delta_{ab}$,
where at least each flavor sector
is described by a unique $\hat{a}_{\alpha\beta}\delta_{ab}$ coefficient.
(Different flavor sectors could be distinguished by an overall factor.)
In order to maintain universality one has to forbid any other
hidden-observable mixing such as 
$(ZZ^{\dagger})(\Phi\Phi^{\dagger})^{2}/M^{4}$ which lead to
quadratically divergent corrections (proportional to the gravitino mass).
The latter, in turn,  spoil the universality at the 
percentile level \cite{HPNNP,CHO1,CHO} (which is the experimental sensitivity level),
$a_{\alpha\beta ab} = \hat{a}_{\alpha\beta}(\delta_{ab} + (N_{ab}/16\pi^{2}))$
for some flavor-dependent counting factor $N_{ab}$.
It was argued that the universality may be a result of a flavor
(contentious or discrete) symmetry which is respected by the hidden
sector and supergravity or a result of string theory.
If the symmetry is exact (which most likely requires it to be
gauged) then universality can hold to all orders. If universality
is a result of a ``string miracle'' (such as supersymmetry breaking
by a stabilized dilaton \cite{LOUIS}) then it is not expected to be exact
beyond the leading order. 
%

%1025
A minimal assumption that is often made as a ``best first guess''
is that of total universality (of scalar masses squared), $A$-term proportionality (to Yukawa
couplings), and gaugino mass unification, which lead to only
four ultraviolet parameters (in addition to possible phases and the
sign of the $\mu$-parameter): $m^{2}_{\phi}(\Lambda_{\rm UV}) 
= m_{0}^{2}$ ({\it universality}), $A_{ijk}(\Lambda_{\rm UV})  = A_{0}y_{ijk}$
({\it universality and proportionality}),
$M_{i}(\Lambda_{\rm UV}) = M_{1/2}$ ({\it gaugino unification}), 
$m_{3}^{2}(\Lambda_{\rm UV})  =
B_{0}\mu(\Lambda_{\rm UV})$,  which
are all of the order of the gravitino mass \cite{SUGRA}.
($\mu$ is not a free parameter in this case but is fixed
by the $M_{Z}$ constraint, eq.~(\ref{min1}),
only its sign is a free parameter.)
This framework is sometimes called minimal supergravity.
It may be that only the operator ({\ref{gaugino}) is present at
the mediation scale ($m_{0} = A_{0} = 0$)
and that all other SSB parameters are induced
radiatively by gaugino loops. (For example, 
see Kelley {\it et al.} in Ref.~\cite{S62}.)
In this case, the gauge interactions guarantee
universality in $3\times 3$ subspaces: GIM universality.
Such models are highly predictive.
Universality (whether in each $3\times 3$
subspace or for all fields) is the  mystery of supergravity models,
but since the mediation itself and the $\mu$-parameter are trivially
given in this framework, it cannot be discounted and the price may be
worth paying. 
%

%1026
It was also proposed that supergravity mediation is
carried out only at the quantum level \cite{SUGRA2}  (anomaly mediation) 
via supergravity quantum corrections which
generically appear in the theory. The relevant coefficients are then 
given by loop factors with specific pre-factors
which are determined by the low-energy theory, e.g. $e \sim 
b_{i}g_{i}^{2}/16\pi^{2}$ is given by the low-energy (\ie weak-scale)
one-loop $\beta$-function (and there is no gaugino mass unifcation). 
These proposals, though economic and elegant
and with unique signatures (the winos are lighter than the bino, for example)
face difficulties in deriving a consistent scalar spectrum and
the $\mu$-parameter. 
In particular, the coefficients for the sfermions squared masses
are given by the respective two-loop anomalous dimensions,
which though universal
(up to Yukawa-coupling corrections)
are negative in the case of the sleptons (aside from the $\widetilde\tau$
if $y_{\tau}$ is sufficiently large). This scenario therefore predicts
negative squared masses for sleptons.
Known cures are contrived and/or reintroduce ultraviolet dependencies.
A lesson from these proposals, however, is that gaugino unification cannot be exact.
(For a general study of corrections to gaugino unification, see Ref.~\cite{KRIB}.)
More generally, quantum mediation could extend beyond anomaly mediation \cite{SUGRA3,KK}
and could lead to viable models.
\section{Gauge Mediation}
\label{sec:GM}
%

%1031
It may be that supergravity mediation is sufficiently
suppressed by  a light gravitino mass (and hence no assumption on its
flavor structure is needed).
Then, a new mediation mechanism and messenger sector are
required. An attractive option is that the messenger interactions are
gauge interactions so that universality is an automatic consequence
(e.g. gaugino mediation and anomaly mediation in SUGRA).
This is the gauge-mediation (GM) framework. 
The hidden sector (which is not truly hidden now) 
communicates via, e.g.  new (messenger) gauge interactions
with a messenger sector, which, in turn, communicates
via the ordinary gauge interactions with the observable sector.
%

%1032
It is sufficient to postulate that the new messenger gauge and  
Yukawa interactions mediate the supersymmety breaking 
to a (SM) singlet messenger $X = x + \theta^{2}F_{X}$,  which
parameterizes the supersymmetry breaking in the messenger sector.
(Some other hidden fields, however, with $F > F_{X}$ may dominate the
massive gravitino.)
The singlet $X$ interacts also with SM non-singlet messenger fields $V$ 
and $\bar{V}$.
The Yukawa interaction $y X V \bar{V}$ in turn communicates the supersymmetry
breaking to the messengers $V$ and $\bar{V}$ via the mass matrix,
\begin{equation}
M^{2}_{v\bar{v}} \sim 
\left(\begin{tabular}{c c}
$y^{2}x^{2}$& $y F_{X}$\\
$y F_{X}^{*}$ & $y^{2}x^{2}$
\end{tabular}\right).
\label{Mmessenger}
\end{equation} 
In turn, the vector-like pair $V$ and $\bar{V}$, 
which transforms under the SM gauge group
(for example, they transform as ${\bf 5}$ and ${\bf \bar 5}$ of
$SU(5)$, \ie as down singlets and lepton doublets and their complex conjugates), 
communicates the supersymmetry breaking
to the ordinary MSSM fields via gauge loops.
The gauge loops commute with flavor and, thus, 
the spectrum is charge dependent  but flavor diagonal,
if one ensures that all other possible 
contributions to the soft spectrum are absent or are strongly
suppressed. This leads to (GIM-)universality.
Such models  are often referred to as 
``gauge mediation of supersymmetry breaking'' or messenger models \cite{MESSE}.
%

%1033
This scenario is conveniently described by the above operator equations
with $M \simeq \langle x  \rangle$, $e \sim \alpha_{i}/4\pi$ a generic
one-loop factor and $a \sim (\alpha_{i}/4\pi)^{2}$ a two-loop factor.
(If the messenger dynamics itself is non-perturbative, by dimensionless analysis,
$e \sim \alpha_{i}$  and $a \sim (\alpha_{i})^{2}$ \cite{4pi}.)
All other coefficients generically equal zero at the messenger scale.
The sparticle spectrum, and hence, the weak scale, are given
in this framework
by $M_{Z}  \sim m_{\rm SSB} 
\sim (\alpha_{i}/4\pi)(F_{X}/x)$,
where $\alpha_{i}$ is the relevant SM gauge coupling at the scale
$\Lambda_{\rm mediation} \sim x$.
One then has  $F_{x}/x \sim (4\pi/\alpha_{i})M_{Z}
\sim 10^{5}$ GeV. In the minimal version one  assumes 
$\Lambda_{\rm mediation} \sim  F_{x}/x
\sim 10^{5}$ GeV; a similar
scale for the spontaneous supersymmetry breaking in the hidden sector;
as well as $F_{X} \sim x^{2}$; resulting in an one-scale model. 
The latter assumption could be relaxed as long as the
phenomenologically determined ratio $F_{x}/x \sim (4\pi/\alpha)M_{Z}
\sim 10^{5}$ GeV remains fixed. In general, 
$\Lambda_{\rm mediation} \sim  x$ could be at a much higher
scale \cite{RAB}. Also, the scale of spontaneous 
supersymmetry breaking in the hidden sector
could be one or two orders of magnitude higher than $\sqrt{F_{X}}$,
for example, if $F_{X}$ is induced radiatively
by a much larger $F$-term of a hidden field
(in which case $X$ contributes negligibly to the massive gravitino).
%

%1034
The messengers induce gaugino masses at one-loop 
($e \sim \alpha_{i}/4\pi$)
and scalar masses at two-loops ($a \sim (\alpha_{i}/4\pi)^{2}$). 
(The messengers obviously cannot
couple  via gauge interactions to the other chiral fields at one-loop.)
This leads to the desired relation $m_{\tilde{f}}^{2} \sim M_{\lambda}^{2}$
between scalar and gaugino masses. In addition, there is a mass hierarchy
$\sim  \alpha_{3}/\alpha_{2}/\alpha_{1}$ between the heavier strongly
interacting sparticles and the only weakly interacting sparticles which are lighter.
(In detail, it depends on the charges of the messengers).
In particular, gaugino mass relations reproduce those of gaugino mass unification.
The $A$-parameters
arise only via mixed Yukawa-gauge renormalization and are of no concern
(since renormalization-induced $A$-parameters are proportional to Yukawa coupling).
However, the Higgs mixing parameters also do not arise from gauge interactions,
the Achilles heal of the framework.
%

%1035
One may introduce new hidden-observable Yukawa interactions for the purpose
of generating $\mu$ and $m_{3}^{2}$,  but then generically
$b \simeq d \simeq (y^{2}/16\pi^{2})$ for some generic Yukawa coupling
$y$, leading to a hierarchy problem 
$|m_{3}^{2}| \sim |\mu|\Lambda_{\rm mediation}$. 
This overshadows the otherwise success of gauge mediation, and the
possible resolutions \cite{muMESS} are somewhat technically involved
and will not be presented here. It was also shown that the resulting uncertainty
in the nature of the Higgs-messenger interactions introduces
important corrections to  the Higgs potential and mass \cite{Su2}.
%

%1036
Note that this framework has a small
number of ``ultraviolet'' parameters and it is highly predictive
(less so in extended versions).
The gravitino is very light and is the LSP (with signatures such as 
neutralino decays to an energetic photon and Goldstino missing energy
or a charged slepton escaping the detector, decaying only outside the
detector to a lepton and a Goldstino).
Also, the SSB parameters are not ``hard'' at higher scales such as the
unification scale. Radiative symmetry breaking relies in this case on 
mass hierarchy mentioned above,
$m_{\tilde{q}}^{2}(\Lambda_{\rm mediation}) \sim
(\alpha_{3}/4\pi)^{2}\Lambda_{\rm mediation}^{2}\gg
m_{H_{i}}^{2}(\Lambda_{\rm mediation})
\sim (\alpha_{2}/4\pi)^{2}\Lambda_{\rm mediation}^{2}$, rather
than on a large logarithm, and its solutions take a slightly different
form than eqs.~(\ref{carena}). 
\section{Hybrid Models I} 
\label{sec:hybrid1}
%

%1041
Can the gauge-mediation order parameter $F_{X}/x$ be induced
using only supergravity interactions? (Supergravity serves in such a
case only as a trigger for the generation of the SSB parameters.)
The answer is positive.
%

%1042
An operator of the form
\begin{equation}
\int d^2 \theta d^{2}\bar\theta \, 
 X\left[\frac{Z^\dagger Z\Phi^{\dagger}\Phi}{M^3_{P}} + h.c. +
\cdots \right] \label{tadpole}\\
\end{equation}
leads to a quadratically divergent tadpole (see Fig.~\ref{fig:tadpole}) loop 
(with $\Phi$ circulating in the loop)
which is cut-off at $M_{P}$, and hence to a scalar potential of the form
$V({x})  = (|F_{Z}|^{2}/M_{P})x + |\partial W/\partial X|^{2}
\sim m_{3/2}^{2}M_{P}x + x^{4}$ (where we omit loop and counting factors and
in the last step we assumed $W(X) = X^{3}/3$). 
%

%1043
This is only a sketch of 
this hybrid frameork \cite{HPNNP} in which a supergravity linear term in a singlet
$X$ can trigger a gauge mediation framework for
$\sqrt{F_{Z}} \sim 10^{8 \pm 1}$ GeV (leading to $x \sim 10^{5}$ GeV
and $F_{X} \sim x^{2}$, as in minimal gauge mediation).
The clear benefit is that the hidden sector ($Z$ in this case) remains hidden 
while $X$, which couples as usual $XV\bar{V}$, is a true observable sector singlet field.
This leads to a simpler radiative structure and to a more stable model.
The triggering gravity mediation (leading to the linear term) is carried out, as in
anomaly mediation, only at the quantum level (but with a very
different source than in anomaly mediation). 
\section{Hybrid Models II} 
%

%1051
A different hybrid approach is that of superheavy supersymmetry (or
the $2-1$) framework, which as implied by its name, relies on
decoupling in order to weaken the universality constraint.
The conflict between naturalness and experimental constraints is
resolved in this case by observing that, roughly speaking, naturalness restricts
the masses of scalars with large Yukawa couplings, while experiment
constrains the masses of scalars with small Yukawa couplings \cite{TWOONE}.
Naturalness affects particles which are strongly coupled to the Higgs
sector, while experimental constraints are strongest in sectors with
light fermions which are produced in abundance.  This suggests that
naturalness and experimental constraints may be simultaneously
satisfied by an ``inverted hierarchy'' approach, in which light
fermions have heavy superpartners, and heavy (third family) fermions
have ``light'' ${\cal{O}}(M_{Z})$ superpartners (hence, $2-1$ framework).  Therefore,
the third generation scalars (and Higgs) with masses $m_{\rm light}
\lesssim 1$ TeV satisfy
naturalness constraints, while first and second generation scalars at
some much higher scale $m_{\rm heavy}$ avoid many experimental difficulties.
%

%1052
A number of possibilities have been proposed to dynamically generate
scalar masses at two hierarchically separated
scales. Usually one assumes that $m_{\rm light} \simeq m_{3/2}$ 
is generated by the usual supergravity mediation (and $\mu$
generation may follow SUGRA as well), while 
$m_{\rm heavy}$ is generated by a different ``more important'' (in terms of
the relative contribution) mechanism which, however, discriminates
among the generations. 
(Naturalness allows the stau, in some cases, to be heavy.)
%

%1053
Such a ``more important'' mechanism may arise from the $D$-terms of
a ultraviolet anomalous flavor $U(1)$ 
(we did not discuss in these notes the case of an anomalous $U(1)$ \cite{FI}) 
with a non-vanishing $D$-term,
$F_{Z}F_{Z}^{\dagger} \rightarrow \langle D \rangle^{2}$ in eq.~(\ref{scalar}),
and $\langle D \rangle/M \gg m_{3/2}$ has to be arranged \cite{DPOM}.
Alternatively, there could be a flavor (or horizontal) messenger
mechanism at some intermediate energies \cite{HPNNP,NNN} (and in this case the 
gauged flavor symmetry is anomaly free and it is broken only at
intermediate energies). The messenger model in this case follows
our discussion above only that the messengers are SM singlets
charged only under the flavor (horizontal) symmetry.
In both of these examples a gauge horizontal (flavor) symmetry discriminates among the
generations and does not affect the MSSM gaugino masses.
For example, the horizontally neutral third generation
sfermions and Higgs fields do not couple to the horizontal $D$-term (the first case)
or to the horizontal messengers (the second case). 
The coefficients $a$ in eq.~(\ref{scalar}) are in these cases
flavor dependent but do not mix the light and heavy sfermions,
a mixing which is protected by the horizontal gauge symmetry.
%

%1054
A fundamentally different approach
\cite{FKP} is that $F_{Z} \gg F_{X}$ (a limit which realizes an effective $U(1)_{R}$
symmetry in the low-energy theory) and all the boundary conditions for
all the scalars are in the multi-TeV range (while gauginos
are much lighter). The light stop squarks and Higgs fields, for
example, are then driven radiatively and asymptotically to $m_{\rm light}$.
Indeed, and not surprisingly, one find that in the presence of large Yukawa couplings there
is such a zero fixed point which requires, however, that the respective
sfermion and Higgs boundary conditions have specific ratios (\ie it
is realized along a specific direction in field space as pointed out
in Chapter~\ref{c7}).
Here, the hierarchy is indeed inverted as the light scale
is reached via renormalization by large Yukawa couplings. 
%

%1055
In practice, all of these solutions are constrained
by  higher-order terms which couple
the light and heavy fields and which are proportional to small Yukawa couplings
or which arise only at two-loops.
Such terms are generically suppressed, but are now enhanced by the heavy
sfermions. The importance of such effects is highly model dependent
and they constrain each of the possible realization in a different
fashion, leaving more than sufficient room for model building. 
%

%1056
Even though such a realization of supersymmetry would leave some
sfermions beyond the kinematic reach of the next generation of
collider experiments, some sfermion and gauginos should be discovered.
It was noted that by measuring the ratio of the gaugino-fermion-(light)sfermion
coupling and the gauge coupling, $g_{\widetilde{\chi}f\tilde{f}^{*}}/g$ 
(the superoblique parameters \cite{SUPOBL}),
one can confirm in many cases the presence of the heavy states
by measuring  logarithmically-divergent
quantum corrections to these ratios $\sim\ln m_{\rm heavy}$, providing an handle on these
and other models with multi-TeV fields (e.g., the squarks in GM).
The phenomenology of the $2-1$ approach was examined recently in some detail
in Ref.~\cite{BAERTATA}.
\section{Alignment}
%

%1061
Finally, it was proposed that the Yukawa and
sfermion squared mass (and trilinear $A$) matrices are aligned in
field space, and hence, are diagonalized simultaneously \cite{HORIZ}.
Thus, $\Delta_{ij} = 0$ without universality.
Such an alignment may arise dynamically once the
(Coleman-Weinberg) effective potential is minimized with respect to some
low-energy moduli $X$ and $Z$. However, such a mechanism tends to be unstable 
with regard to higher-order corrections. Alternatively, it could be that 
such alignment is a result of some high-energy symmetry principle.
Realization of the latter idea tend be cumbersome, in contrast to the
simplicity of the assumption, and it is difficult to 
envision conclusive tests of specific symmetries (though the
alignment idea itself may be ruled out, in principle, in sfermion
oscillations \cite{FFF} which cannot arise for $\Delta_{ij} \equiv 0$.) 
Nevertheless, this is another possibility in which case $F_{X}/M$ and
its square are aligned with the Yukawa matrices and are diagonal
(only) in the physical mass basis of fermions.
%

%E10
\section*{Exercises}
\addcontentsline{toc}{section}{Exercises}
\markright{Exercises}

\subsubsection*{10.1}
Integrate the operator equations (\ref{scalar})--(\ref{Aterm})
to find the SSB and $\mu$-parameters.

\subsubsection*{10.2}
Solve for the sfermion and Higgs mass-squared  parameters
in the case that supergravity mediation generates only the gaugino
mass. (See eq.~(\ref{carena}).)
What is the LSP (assume $|\mu| \simeq M_{3}$),
and is it charged or neutral?
Assume instead complete universality at the supergravity scale
$m_{\tilde{f}}^{2} = m^{2}_{H} = m_{0}^{2}$ for all sfermion and Higgs fields.
In what fashion gaugino masses and Yukawa couplings break
universality? Why is the breakdown of univesality by corrections
proportional to Yukawa couplings ``relatively safe'' (as far as FCNC are concerned)?

\subsubsection*{10.3}
In the spirit of the previous exercise, show that
squark mass squared matrices are never truly universal.
What are the implications, for example, for gluino couplings?
Flavor-changing gluino couplings are important for
meson mixing, $b\rightarrow s \gamma$, and proton decay amplitudes.

\subsubsection*{10.4}
Supergravity mediation in the presence of a grand unified theory
induces the SSB parameters for that theory, not for the MSSM.
How many independent soft parameters are in this case
at the unification scale 
(with and without universality) for minimal $SU(5)$; $SO(10)$?
Renormalization within a grand-unified epoch introduces
radiative corrections 
to the stau squared mass ($m^{2}_{\widetilde{\tau}_{L}}$ or
$m^{2}_{\widetilde{\tau}_{R}}$ ?)
which are proportional to the large top-Yukawa coupling. 
The corrections break universality in the
slepton sector. The breakdown of slepton universality induces, in
turn, contributions to low-energy lepton flavor violation such as 
the $\mu \rightarrow e\gamma$ process (Ex.~9.5). For sufficiently light
sleptons ($m_{\tilde{l}} \lesssim$ 300-400 GeV) it may be observable
in the next generation of the relevant experiments, if built.

\subsubsection*{10.5}
Show that the winos are heavier than the bino in anomaly mediation
(as defined in the text). 
Consider the chargino-LSP mass difference, 
assuming both are  gaugino-like.

\subsubsection*{10.6}
Derive the messenger mass matrix (\ref{Mmessenger}).

\subsubsection*{10.7}
Show that in the minimal gauge mediation model described above
$e = (\alpha_{i}/4\pi)T_{i}$ for gaugino $i$, and
normalizing Dynkin index $T_{i}$ to unity  
(e.g. for ${\bf 5}$ and ${\bf \bar 5}$  of $SU(5)$ messengers) 
that for each sfermion (and Higgs) $a = 2\sum_{i}(\alpha_{i}/4\pi)^{2}C_{i}$ where
$C_{1} = (3/5)Y^{2}$, $C_{2} = 3/4$ for an $SU(2)$ doublet and
$C_{3} = 4/3$ for an $SU(3)$ triplet are the Casimir operators.

\subsubsection*{10.8}
Derive the relation
$|m_{3}^{2}| \sim |\mu|\Lambda_{\rm mediation}$
in gauge mediation.

\subsubsection*{10.9}
Calculate $F_{X}$ in the hybrid supergravity - gauge-mediation
model. Calculate the gravitino mass in this and in the ``traditional''
gauge-mediation models and show that supergravity effects $\sim
m_{3/2}^{2}$ are indeed negligible in both cases.

\subsubsection*{10.10}
Compare typical sparticle mass patterns in minimal SUGRA
and in minimal GM.

%SP4 
\chapter*{Summary}
\addcontentsline{toc}{section}{Summary}
\markright{Summary}
%

%S41
The electroweak/Higgs  scale was already (technically) explained, 
and in a fairly model-independent way, 
by the discussion in Part~\ref{p3}.
Nevertheless, its magnitude 
depends on at least two free parameters; the scale of 
spontaneous supersymmetry breaking $\sqrt{F}$ and
the scale of its mediation to the SM fields $M$.
The two scales combine to give the SSB scale $\sim F/M$. 
Henceforth, their  ratio is fixed by the electroweak scale constraint.
On the other hand, the scales of supersymmetry breaking and of its mediation
relate to the ultraviolet origins of the soft terms in the fundamental theory,
and are fixed by the dynamics of a given framework.
This  provides a useful linkage between the infrared and the ultraviolet.
%

%S42
In this part of the notes we parameterized and expanded the ultraviolet theory 
in terms of these scales and discussed different realizations,
including various variants of supergravity and gauge-mediation models.
In doing so we  related some of the finer details
of the infrared  and ultraviolet physics.  
In order to perform  this exercise, however, we first had to search for
an organizing principle applicable to the vast SSB ``model space''. 
%

%S43
Indeed, the most general MSSM Lagrangian is described by many arbitrary
parameters, in particular, once flavor and CP conservation
are not imposed on the soft parameters. Aside from the obvious
loss of predictive power, many generic models are  actually in an
apparent conflict with the low-energy data: An arbitrary choice
of parameters can result in unacceptably large 
FCNC (``The Flavor Problem'') and, if large phases are present, 
also large (either flavor conserving or violating)  contributions to 
CP-violating amplitudes (``The CP Problem'').
Clearly, one has to identify those special cases
which evade these constraints, and by doing so the whole framework
regains predictive power. 
We therefore turned in our search of an organizing principle
to the question of lack of low-energy evidence for supersymmetry.
%

%S44
The absence of observable contributions to FCNC
was shown to indeed provide 
the desired organizing principle for high-energy frameworks, 
which themselves attempt to organize the model many parameters.
Understanding the solution of the flavor problem
(which is straightforward in the SM,
given is special structure, but not in any of its extensions) goes to the heart of 
the question of origins of the theory.
Though the number of options was narrowed down,
the origin of the SSB parameters remains elusive. Discovery of
supersymmetry will open the door to probing the mediation scale
and mechanism, and hence, to a whole new (ultraviolet) arena. 
%

%S45
In the remaining part of these notes we will return to 
and expand on a small number of topics. (The selected topics
include a generalization of the operator classification given in this
part of the notes.) 
%

%P5
\part{Selected Topics}
\label{p5}
%

%C11
\chapter{Neutrinos}
\label{c11}

%1101
The confirmation of neutrino oscillations \cite{superK}
provides a concrete and first
proof of physics beyond the SM. If nature realizes
supersymmetry, an option which we explore in this manuscript, then neutrino
mass and mixing (and LFV) must be realized at some energy scale 
supersymmetrically . We review in this chapter
avenues for neutrino mass generation
in the framework of supersymmetry. 
%

%1102
Unless neutrinos have ``boring'' Dirac masses 
$y_{\nu_{ab}}\langle H_{2}^{0} \rangle L_{a}N_{b}$, 
which only imply super-light sterile right-handed neutrinos $N_{a}$
and an extension of the usual
fermion mass hierarchy problem, then 
the  neutrinos have a $\Delta L = 2$ Majorana mass and
there must be some source of 
total lepton number violation (LNV), $\Delta L \neq 0$. We will review both 
scenarios,
$R_{P}$-conserving  $\Delta L = 2$ terms and
$R_{P}$-violating  $\Delta L = 1$ terms.
%

%1103
We will also illustrate 
the ``rewards'' of the different scenarios: Supersymmetry provides
somewhat delicate relations between neutrino physics and other arenas
(such as LFV and $\tan\beta$-dependent observables in general)
in the $\Delta L = 2$ case of a heavy right-handed neutrino, and 
many new channels for sparticle production and decay
arise in the $\Delta L = 1$ $R_{P}$-violating case. Therefore,
neutrino physics in supersymmetry
is not independent but rather linked to other ``new physics''
observables, and it may be probed via various avenues.
\section{$\Delta L = 2$ Theories}
%

%1111
It is straightforward to ``supersymmetrize''
old ideas of a heavy right-handed (SM-singlet) neutrino 
with a large $\Delta L = 2$ Majorana mass $M_{R}$. The right-handed
neutrino mixes with the SM left-handed
neutrinos via the usual  ${\cal{O}}(\mweak)$ Dirac mass term,
\begin{eqnarray}
M_{(\nu_L\,\nu_R)}
& = & 
\left(\begin{tabular}{cc}
&$\mweak$\\
$\mweak$& $M_{R}$
\end{tabular}\right) 
\nonumber 
\label{mseesaw}
\end{eqnarray}
leading to $m_{\nu} \sim \mweak^{2}/M_{R}$: The {\it see-saw}  mechanism 
\cite{seesaw}. (Note that the mass and interaction eigenstates are nearly identical.)
This mechanism does not require supersymmetry,
but is trivially embedded in a supersymmetric framework by extending
the superpotential (above the heavy neutrino scale $M_{R}$)
$W \rightarrow W + y_{\nu_{ab}}H_{2}L_{a}N_{b} + M_{R_{ab}}N_{a}N_{b}$ (and $N_{a}$ here
is the heavy right-handed neutrino superfield of generation $a=1,\,2,\,3$).
% 

%1112
Unified models (Chapter~\ref{c6}) typically imply further that the Dirac
$\tau$-neutrino mass is of the order of the top mass (top - neutrino
unification), $m_{\nu_{\tau}} \sim m_{t}^{2}/M_{R}$.
A natural choice for the
scale $M_{R}$ is the unification scale, and indeed
simple fits to the data favor \cite{LAN,VB} $M_{R} \sim 10^{13-15}$ GeV. 
Model building along these lines was a subject of intense activity 
following the observation of neutrino oscillations.
For an example of a (unified) see-saw model, see
Ref.~\cite{BPW}, where a double see-saw mechanism
$M_{R} \sim M_{U}^{2}/M_{P}$ and $m_{\nu} \sim M_{W}^{2}/M_{R} \sim
M_{W}^{2}M_{P}/M_{U}^{2}$ was proposed, with $M_{P}$ is replaced by
a lower ``string scale''.
Another possibility \cite{KPP} is that the scale $M_{R}$ is related to
the scale of spontaneous supersymmetry breaking in the hidden sector,
in which case the Dirac mass may not be identical to $m_{t}$.
``See-saw'' models of neutrino masses were reviewed recently
in Ref.~\cite{SeeSawR}.
%

%1113
Top -- neutrino unification $y_{t}(M_{U}) = y_{\nu_{\tau}}(M_{U})$
is natural in many unified models (but not $SU(5)$ where $N$ is a singlet) 
in which the top and the right-handed neutrino are embedded in the same representation.
We note that it excludes 
(independent of all other considerations such as the Higgs mass)
the $\tan\beta \sim 1$ solutions to $b - \tau$ unification (Fig.~\ref{fig:btau}).
This double Yukawa unification constrains $\tan\beta$ from below
due to cancellation of large top-Yukawa effects
by the large neutrino-Yukawa effects \cite{hnuR}: In these models
wave-function renormalization of the $b$ by a top loop
is  balanced (above the heavy-neutrino decoupling scale)
by the wave-function renormalization of the $\tau$ by a neutrino loop
with $y_{\nu} \simeq y_{t}$. Eq.~(\ref{btauRGE}) is modified accordingly
(in the decades between the unification $M_{U}$ and neutrino $M_{R}$ scales) to read
\begin{eqnarray}
\frac{d}{d\ln{\Lambda}}\left(\frac{y_{d}}{y_{l}}\right) &=&
\frac{1}{16\pi^{2}}\left(\frac{y_{d}}{y_{l}}\right) \times
\nonumber \\
&& \left\{
(y_{u}^{2} - y_{\nu}^{2})+ 3(y_{d}^{2}-y_{l}^{2}) 
- \frac{16}{3}g^{2}_{3} - \frac{4}{3}g^{2}_{1}
\right\}. 
\label{btauRGE2}
\end{eqnarray}
In particular, it predicts $\tan\beta \gg 1$ and therefore
that the light-Higgs mass saturates its upper 
bound of $130$ GeV or so, consistent with experimental bounds.
It also constrains other $\tan\beta$ dependent observables such as the 
anomalous muon magnetic moment (Sec.~\ref{sec:SMM}) where large values
of $\tan\beta$ may lead to upper bounds on sparticle masses.
%

%1114
If indeed the right-handed neutrino is present, in supersymmetry there
is also a right-handed sneutrino. Since the heavy right-handed
neutrino decouples in a global supersymmetric regime, the sneutrino is
also heavy and decouples. Nevertheless, 
it still has a ``small'' SSB mass. (This is not as straightforward
in the case of low-energy gauge mediation.) 
It does not affect the decoupling yet it renormalizes other SSB terms
of the light fields:
We have  already seen above that
if the right-handed neutrino
superfield  couples with a large Yukawa coupling it can affect the renormalization
of the superpotential (e.g. $b - \tau$ unification).
Its renormalization of the SSB parameters is of similar origins and importance.
It occurs in the decades in between the SSB mediation scale and
the right-handed neutrino decoupling scale.
For example, consider  $(i)$ RSB and the $m_{H_{2}}^{2}$ parameter which is now driven
negative by both top and neutrino terms,  or $(ii)$ the slepton spectrum $m_{\tilde{l}}^{2}$.
%

%1115
A right-handed neutrino is therefore expected
to leave its imprints in  the weak-scale parameters such as $m_{\tilde{l}}^{2}$.
In particular, since $y_{\nu_{a}}  \neq y_{\nu_{b}}$ for $a \neq b$ 
then the affect are not flavor blind and one expects that it would lead
to LFV effects.
Particularly interesting is the 
disturbance of slepton (mass) universality (when assumed) and mass relations. 
(Relevant references were included
in Ref.~\cite{S64}.) This in turn offers an interesting
complementarity between neutrino physics, the slepton spectrum, and
lepton flavor violation experiments (e.g. $\mu \rightarrow e \gamma$
discussed in Sec.~\ref{sec:SMM} on the one hand and slepton oscillations \cite{FFF}
in future collider experiments on the other hand).
\section{$\Delta L = 1$ Theories}
%

%1121
The relation between neutrino physics and supersymmetry
is even more fundamental and extensive if the neutrino 
mass originates from $R$-partiy
violation \cite{HASU,RPV,HPNNP2,RPV2}. 
In this case the origin of the mass is at the weak scale
and no right-handed neutrinos are necessary.
%

%1122
As we argued earlier, supersymmetry essentially
encodes lepton - Higgs duality, which is typically removed by
hand when imposing $R$-parity. Generically, the theory contains
explicit $\Delta L = 1 $ breaking via superpotential operators,
SSB operators, and as a consequence, $L$ could also be broken
spontaneously by sneutrino {\it vev's}. 
One needs to apply $\Delta L =1$
operation twice in order to induce a Majorana neutrino mass. In
general, such models can admit \cite{HASU}
a one-loop radiative mass ($\propto \lambda^{\prime\,2}$)
and  a tree-level mass induced by
neutrino-neutralino tree-level mixing 
$\propto \langle \tilde \nu \rangle \mu_{L}$ (Ex.~5.17).
(Our notation follows eq.~(\ref{WdeltaL}).)
%

%1123
The tree-level mass is intriguing: Only a supersymmetric neutrino mass 
$\mu_{L}LH_{2}$ is allowed 
in the superpotential (aside from the usual Higgs mass),
while the neutrino is the only SM fermion which is left massless
by the superpotential Yukawa terms. Hence, it offers an interesting
complementarity between Yukawa and mass terms in the superpotential,
and an elegant and economic realization of $R$-parity violation.
It can also be shown to be related to the $\mu$-problem.
(For a discussion see Ref.~\cite{HPNNP2,HempflingRp,Valle} and references therein.) 
In the basis in which the electroweak breaking {\it vev} is only
in the Higgs fields $\langle \widetilde{\nu}\rangle = 0$ 
one has for the tree-level Majorana mass
$m_{\nu} \simeq (\mu_{L}^{2}M_{W}^{2}\cos^{2}\beta)/(2\mu^{2}M_{2})$.
Note that only a single neutrino mass can be generated from tree-level
neutralino-neutrino mixing. (This is because a chiral-like $SU(4)$ symmetry of the mass
matrix is broken at tree-level at most to $SU(2)$, guaranteeing two massless states.)
%

%F
%
\begin{figure}[t]
\begin{center}
\epsfxsize= 6.0cm
\leavevmode
\epsfbox[200 565 420 685]{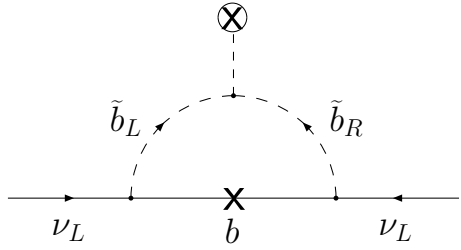}
\end{center}
\caption[f1]{A one-loop contribution to the Majorana neutrino mass 
 arising from the $\Delta L = 1$ $\lambda^{\prime}_{i33}$ Yukawa operator.}
\label{neutrinomass}
\end{figure}
%

%1124
In the remaining of this chapter,
however, we focus on
the radiative  Majorana neutrino mass illustrated in Fig.~\ref{neutrinomass}.
One obtains
\begin{equation}
\frac{m_{\nu}}{{\rm MeV}} 
\sim \lambda^{\prime\,2}
 \left(\frac{300\, {\rm GeV}}{m_{\tilde{b}}}\right)
 \left(\frac{m_{LR}^{2}}{m_{b}\,m_{\tilde{b}}}\right) \,,
\label{neumass}
\end{equation} 
where a $b$-quark and $\tilde{b}$-squark are assumed to circulate in
the loop (so that $\lambda^{\prime} =\lambda^{\prime}_{i33}$), 
and $m_{LR}^{2}$ is the $\tilde{b}$ left-right mixing
squared mass.  (The neutrino mass may vanish in the limit of a continuous $U(1)_{R}$
symmetry, which suppresses $A$ and $\mu$ terms and corresponds in our case to 
$m_{LR}^{2} \ll m_{b}\,m_{\tilde{b}}$.)  
Imposing laboratory limits on the $\nu_{e}$ mass one can, for example, derive
severe constraints on $\lambda^\prime_{133}$~\cite{MNU},
but $\lambda_{333}^{\prime}$ (\ie $m_{\nu} \rightarrow m_{\nu_{\tau}}$)
could still be ${\cal{O}}(1)$ \cite{EFP}. 
(Note that a heavy ${\cal{O}}$(MeV) $\tau$-neutrino has not been ruled out \cite{PDG}
but it cannot be stable on a cosmological time scales.)
One has to consider the full $\lambda^{\prime}$ matrices in order 
to obtain neutrino mixing, but clearly, the complexity of the
parameter space admits many possible scenarios for mixing\footnote{
Note that in the case that $R$-parity violation originating
from a $\mu$-term $W_{\Delta L = 1} = \mu_{L}LH_{2}$
then the $\lambda$ and $\lambda^{\prime}$
couplings appear by rotations of the usual Yukawa couplings
once leptons and Higgs bosons are appropriately defined at low energies,
e.g. $H_{1}$ is customarily defined along the relevant EWSB {\it vev}.
Flavor mixing arises in this case as well at the loop level only.}.
%

%1125
$R$-partiy violation can then explain the neutrino spectrum
as an electroweak (rather than GUT) effect.
This comes at a  price and with a reward. 
The price is that the sparticles are not stable 
and the LSP is not a dark matter candidate
(but some other sector may still provide
a dark matter candidate whose mass is of the order of the gravitino mass).
The obvious reward is the many new channels are available at 
colliders to produce squarks and sleptons \cite{RPV}. 
Consequently, the corresponding cross-section, branching-ratio,
and asymmetry measurements
may provide information on neutrino physics. We will conclude
our discussion of LNV with an explicit example.

\section{Neutrinos vs. Collider Physics}

%1131
As mentioned above, rich collider phenomena arise in the $\Delta L = 2$ case 
(e.g. slepton oscillations) and even more so in the case of $\Delta L = 1$ 
operators. In this section we focus on the latter.
Perhaps the most relevant example,
which can be probed in the current Tevatron run, 
is given by new exotic top decays.
Application of $SU(2)$ and supersymmetry rotations to the vertex $\lambda^{\prime}\nu 
b\tilde{b}$ assumed in (\ref{neumass}), gives (the relevant vertex for) the decay
channel $t_{L} \rightarrow b_R \widetilde{\tau}_{L}$, assuming that it is
kinematically allowed. Such decays were studied in Ref.~\cite{Top1,EFP},
and more recently in Ref.~\cite{Top2}.
%

%1132
For $m_t = 175$ GeV one finds \cite{EFP}
\begin{equation}
\frac
{\Gamma_{t \rightarrow b\widetilde{\tau}}}
{\Gamma_{t \rightarrow bW}}=
1.12 \, \lambda'^2_{333}\left[
1 - \left(\frac{m_{\widetilde{\tau}_L}}{175\rm{ GeV}}\right)^2
\right]^{2},
\label{rt}
\end{equation}
where explicit generation indices were introduced for clarity.
The $\tilde\tau$-decay modes are highly model-dependent in the 
case of LNV. 
In particular, all superpartners typically decay in the collider 
and the typical large missing energy signature is replaced with 
multi-$b$ and lepton signatures, which may be used for 
identification. (See, for example, Ref.~\cite{RPTEV}.)
If the sneutrino is the LSP 
then the three-body decays $\widetilde{\tau} \to
\widetilde{\nu}_{\tau} f\bar{f}'$ and $\widetilde{\tau} \to Wb\bar{b}$ are
sufficiently phase space-suppressed 
(recall that $SU(2)$ invariance requires
$m_{\widetilde{\tau}_{L}} \simeq m_{\widetilde{\nu}_{\tau}}$)
so that the dominant decay mode is
either $\widetilde{\tau} \to \bar{c}b$ or $\widetilde{\tau} \to e \bar{\nu}_e$.
%

%1133
In order to study the  constraints on
$\lambda'_{333}$, which are otherwise weak,
let us assume that the $\bar{c}b$ mode is
dominant.  This new decay mode alters the number of $t\bar{t}$ events
expected in each of the channels
which characterize top decays (\ie $tt \rightarrow$
jet jet, lepton lepton, lepton jet, where jet refers to hadronic activity) 
both through an enhancement of the percentage
of hadronic decays and through the increased probability of $b$-tagging
events given the assumption of  $b$-rich $\widetilde{\tau}$ decays.  
For each (final-product) channel, 
one can constraint $\lambda'_{333}$ (as a function of the stau mass) from 
the number of events expected in the presence of $\widetilde{\tau}$ decays
relative to the number expected in the SM. (Currently, 
limited data on top decays is available and
the constraints are still quite weak.)
A more promising approach is to examine kinematic parameters in
$t\bar{t}$ events, {\em e.g.}, the reconstructed $W$ mass in lepton +
jets events with a second loosely tagged $b$ \cite{Top3}. (The two
untagged jets define $M_W$.)   At the time of the writing of 
this manuscript, with just
10 events, this gives $\lambda'_{333}\lesssim 0.4 \, (1.0)$ for
$m_{\widetilde{\tau}_{L}} = 100 \, (150)$ GeV \cite{EFP}.  
Such kinematic analysis may
therefore provide strong constraints on LNV
couplings in the future.
%

%1134
Similarly to top decays, a $\widetilde{\tau}$ may be produced in a larger 
mass range if it is radiated of a $t$ or $b$ quark, e.g. 
$gg, qq \rightarrow t(t \rightarrow b\widetilde{\tau})$.
(These channels compete with resonant ($s$-channel) production
$qq \rightarrow \widetilde{\tau}$ \cite{Schannel}.)
This was studied recently in Ref.~\cite{STRA}.
The inclusive production cross sections 
$p\bar{p} \to t \widetilde{\tau}_{L} X$, $p {p} \to t \widetilde{\tau}_{L} X$ 
are obtained by combining the production cross sections arising from
the $2 \to 2$ elementary process $ g b \to t \widetilde{\tau}_L$ to those
induced by $2\to 3$ partonic processes $ gg,\, qq \to t b\widetilde{\tau}_L$. 
(They may be independently measured if relatively complicated final
states could be distinguished.)  The corresponding inclusive cross sections
are obtained by convolution of the
hard-scattering cross section of quark- and gluon-initiated processes
with the quark and gluon distribution functions in $p$ and
$\bar{p}$, and can be found in Ref.~\cite{STRA} where they are evaluated
for the Tevatron and LHC parameters:
See Figs.~\ref{TeVrviol} and \ref{LHCrviol}, respectively.
% 

%F
\begin{figure}[ht]
\begin{center}
\epsfxsize= 11.7cm
\leavevmode
\epsfbox[25 165 585 565]{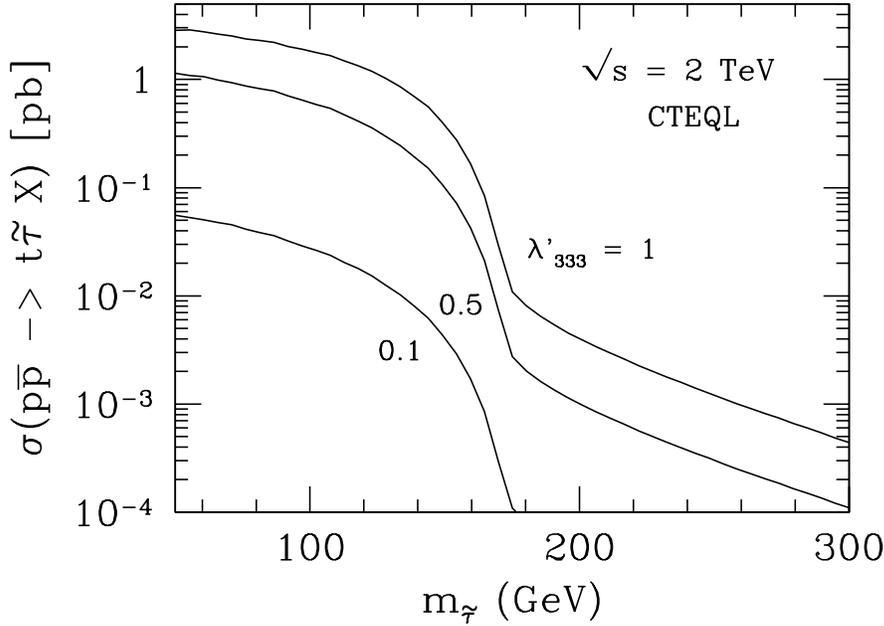}
\end{center}
\caption[f1]{The leading-order production cross section
 $\sigma(p\bar{p}\to t \widetilde{\tau} X)$, for
 $\sqrt{s}=2\,$TeV, as a function of the $\widetilde{\tau}$ mass, is
 shown  for different values of 
 $\lambda^\prime_{333}$. Renormalization and factorization scales
 are fixed as $\mu_R\! =\! \mu_f\! =\! m_t+m_{\widetilde{\tau}}$.
 Taken from Ref.~\cite{STRA}.}
\label{TeVrviol}
\end{figure}
%

%F
\begin{figure}[htb]
\begin{center}
\epsfxsize= 11.7cm
\leavevmode
\epsfbox[25 165 585 585]{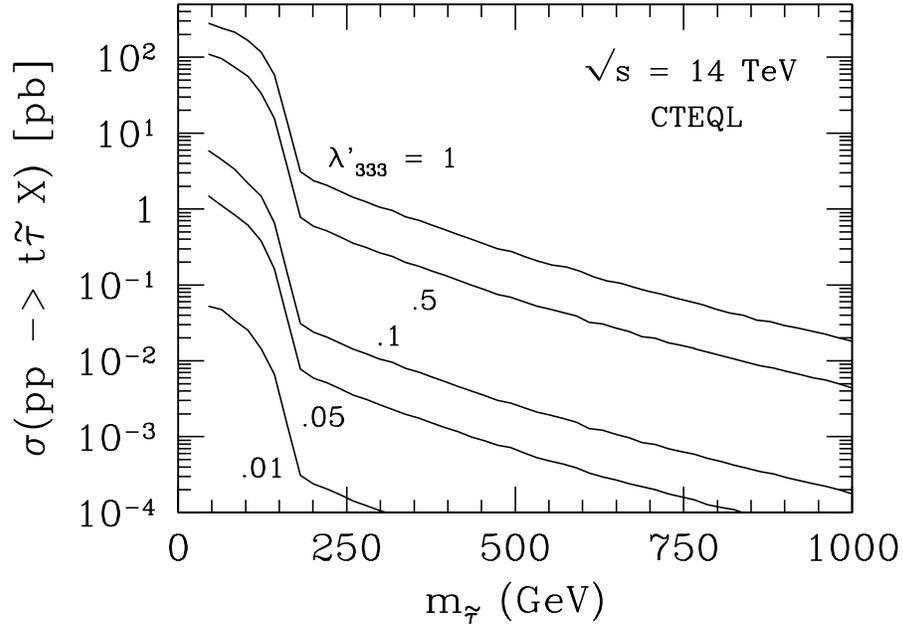}
\end{center}
\caption[f1]{The leading-order production cross section
 $\sigma(p p\to t \widetilde{\tau} X)$, for
 $\sqrt{s}=14\,$TeV, as a function of the $\widetilde{\tau}$ mass is
 shown for different values of 
 $\lambda^\prime_{333}$. Renormalization and factorization scales
are fixed as $\mu_R\! =\! \mu_f\! =\! m_t+m_{\widetilde{\tau}}$.
Taken from Ref.~\cite{STRA}.}
\label{LHCrviol}
\end{figure}
%

%1135
The large cross section obtained in the case of $\sqrt{s} = 14\,$TeV
implies that at the LHC, with a luminosity of 100 fb$^{-1}$ per year,
light $\widetilde{\tau}$'s may be produced in abundance even for couplings
as small as 0.01, whereas for large couplings, they may be produced up
to masses of ${\cal{O}}(1)\,$TeV. (Of course, background studies
are needed before any final conclusions are drawn.)
The Tevatron amplitude, on the other hand,
is still dominated by the on-shell top decays discussed above, and no
significant improvement in sensitivity is achieved.
%

%C12
\chapter{Vacuum Stability}
\label{c12}
%

%1211
In general, the global minimum of the scalar potential need not be
the EWSB minimum, and it is far from clear that it is
justified to consider 
only the Higgs potential when determining the vacuum\footnote{
Sneutrino {\it vev's} discussed in the previous chapter are
a (trivial) example.}:  
The vacuum could break, in principle, color and/or charge. 
%

%1212
Consider, as an example \cite{ccb1}, the superpotential
\begin{equation}
W = y_{t}U_{L_{3}}U_{3}H_{2}^{0} + \mu H_{1}^{0}H_{2}^{0},
\label{wccb}
\end{equation}
and we have performed an $SU(2)$ rotation 
so that $H_{2}^{+}$ has no vacuum expectation value. 
(The inclusion of $H_{1}^{-}$ is therefore immaterial:
We assume below that $m_{1}^{2} > 0$.)
In practice, eq.~(\ref{wccb}) includes the relevant terms
in the superpotential when searching for the global minimum
of the scalar potential in the large 
$y_{t}$ limit, which will suffice here as an example.
Using standard techniques we arrive at the corresponding
scalar potential:
\begin{equation}
y_{t}^{2}V = M^{2} - \Gamma + \lambda^{2},
\label{pot1}
\end{equation}
where the bilinear, trilinear, and quartic terms are
\begin{equation}
M^{2} = m_{1}^{2}{H_{1}^{0}}^{2} +
m_{2}^{2}{H_{2}^{0}}^{2} -
2m_{3}^{2}H_{1}^{0}H_{2}^{0} +
m_{{\tilde{t}}_{L}}^{2}{\tilde{t}}_{L}^{2} +
m_{{\tilde{t}}_{R}}^{2}{\tilde{t}}_{R}^{2},
\label{bigm}
\end{equation}
\begin{equation}
\Gamma = \left| 2\left(|A_{t}|H_{2}^{0} + s|\mu|H_{1}^{0}\right)
{\tilde{t}}_{L}{\tilde{t}}_{R}\right|,
\label{gamma}
\end{equation}
and
\begin{equation}
\lambda^{2} = ({\tilde{t}}_{L}^{2} + {\tilde{t}}_{R}^{2}){H_{2}^{0}}^{2}
+ {\tilde{t}}_{L}^{2}{\tilde{t}}_{R}^{2}
+ \frac{1}{y_{t}^{2}}(\frac{\pi}{2}\times``D-{\mbox{terms''}}),
\label{lambda2}
\end{equation}
respectively.
Here, ${\tilde{t}}_{L} = \widetilde{U}_{L_{3}},\,\, 
{\tilde{t}}_{R} = \widetilde{U}_{3} = \widetilde{U}_{R_{3}}$.
All fields were scaled $\phi \rightarrow \phi/y_{t}$ and are 
taken to be real and positive (our phase  
choice for the fields, which fixed $m_{3}^{2} > 0$
and $\Gamma > 0$, \ie maximized the negative 
contributions to $V$)
and all parameters are real.
$m_{1,\,2}^{2} = m_{H_{1,\,2}}^{2} + \mu^{2}$,
$s = \mu A_{t}/|\mu A_{t}|$, and the expression for the $``D-$terms''
is 
\begin{eqnarray}
4\alpha^{\prime}
\left[-\frac{{H_{1}^{0}}^{2}}
{2} + \frac{{H_{2}^{0}}^{2}}{2} + \frac{{\tilde{t}}_{L}^{2}}{6}
- \frac{2{\tilde{t}}_{R}^{2}}{3}\right]^{2} && \nonumber \\
+ \alpha_{2}\left[{H_{1}^{0}}^{2}
- {H_{2}^{0}}^{2} + {\tilde{t}}_{L}^{2}\right]^{2} && \nonumber \\ 
+ \frac{4}{3}\alpha_{3}\left[{\tilde{t}}_{L}^{2} 
- {\tilde{t}}_{R}^{2}\right]^{2}.  & & 
\label{dterms}
\end{eqnarray}
We can parameterize the four fields in terms
of an overall scale $\phi$ and three angles
$0 \leq \alpha,\,\beta,\, \gamma \leq \frac{\pi}{2}$:
$H_{1}^{0} = \phi\sin{\alpha}\cos{\beta}$,
$H_{2}^{0} = \phi\sin{\alpha}\sin{\beta}$,
${\tilde{t}}_{R} = \phi\cos{\alpha}\cos{\gamma}$,
${\tilde{t}}_{L} = \phi\cos{\alpha}\sin{\gamma}$,
and redefine 
\begin{equation}
y_{t}^{2}V(\phi) = M^{2}(\alpha , \,\beta , \,\gamma)\phi^{2} - 
\Gamma(\alpha , \,\beta , \,\gamma)\phi^{3} + 
\lambda^{2}(\alpha , \,\beta , \,\gamma)\phi^{4}.
\label{pot2}
\end{equation}
Then, for fixed angles, $V(\phi)$ will have a minimum
for $\phi \neq 0$ provided the condition
$32M^{2}\lambda^{2} < 9\Gamma^{2}$ is satisfied.
In that case,
\begin{equation}
\phi_{\rm min} = \frac{3}{8}\frac{\Gamma}{\lambda^{2}}\left[1 +
\left(1 - \frac{32M^{2}\lambda^{2}}{9\Gamma^{2}}\right)^{\frac{1}{2}}
\right] \geq 0,
\label{xmin}
\end{equation}
and
\begin{equation}
y_{t}^{2}V_{\rm min} = -\frac{1}{2}\phi_{\rm min}^{2}\left(
\frac{\Gamma}{2}\phi_{\rm min} - M^{2} \right).
\label{pot3}
\end{equation}
%

%1213
The SM minimum corresponds to $\alpha = \frac{\pi}{2}$ and
$\beta = \beta^{0}$ ($\gamma$ is irrelevant 
and $\tan\beta^{0} = \nu_{1}/\nu_{2}$ is the angle (\ref{tanbeta})
used to fix $\mu$, $m_{3}^{2}$, as well as the Yukawa couplings). 
It is easy to convince
oneself that in that limit the $4 \times 4$ second-derivative
matrix is $2 \times 2$ block diagonal 
(otherwise baryon number is violated).
Thus, it is sufficient to confirm that the four physical
eigenvalues are positive to ensure that 
it is a minimum.
If these conditions are satisfied then the SM is at least
a local (negative-energy) minimum, and one has
$\Gamma_{\mbox{\tiny SM}} = 0$, 
$M^{2}_{\mbox{\tiny SM}} < 0$, and eqs.~(\ref{xmin})
and (\ref{pot3}) reduce to the usual results
$\phi_{\rm min}^{\mbox{\tiny SM}} = \sqrt{- M^{2}_{\mbox{\tiny SM}}/2
\lambda^{2}_{\mbox{\tiny SM}}}$, 
$y_{t}^{2}V_{\rm min}^{\mbox{\tiny SM}}=
- M^{4}_{\mbox{\tiny SM}}/4\lambda^{2}_{\mbox{\tiny SM}}$. 
%

%1214
Let us now consider the possibility
of additional color and/or charge breaking (CCB) 
minima defined by $\cos\alpha \neq 0$.
Assuming $m_{\tilde{t}_{L,\,R}}^{2} > 0$, this
requires $\Gamma \neq 0$, which we assume hereafter. 
From (\ref{pot2})--(\ref{pot3})
it is easy to classify the possible CCB minima for definite
$\alpha$, $\beta$, $\gamma$.
One finds
that for $\Gamma^{2} \leq 32\lambda^{2} M^{2}/9$ 
there is no CCB minimum, while for
$32\lambda^{2} M^{2}/9 < \Gamma^{2} \leq 4\lambda^{2} M^{2}$
the CCB minimum exists but has 
$V_{\rm min}^{\mbox{\tiny CCB}} > 0 > V_{\rm min}^{\mbox{\tiny SM}} $,
which is presumably safe.
For $4\lambda^{2} M^{2} < \Gamma^{2}$
(including the more rare case 
$M^{2} < 0$, that must fall in this category)
there is a negative-value CCB minimum, which may however be
either local (presumably safe), \ie
$V_{\rm min}^{\mbox{\tiny CCB}} > V_{\rm min}^{\mbox{\tiny SM}}$,
or global (probably  unacceptable), \ie
$V_{\rm min}^{\mbox{\tiny CCB}} < V_{\rm min}^{\mbox{\tiny SM}}$.
Here, a sufficient (but not necessary)
condition for an acceptable model is
\begin{equation}
\Gamma^{2} \leq 4\lambda^{2} M^{2}.
\label{master}
\end{equation}
In principle, the above discussion holds for any number 
of fields (\ie any number of angles),
only the explicit expressions for $M^{2}$, $\Gamma$,
and $\lambda^{2}$ are more complicated.
If constraint (\ref{master})
holds for every choice of $\alpha$, $\beta$, and $\gamma$ 
($\cos\alpha \neq 0$)
then there is no negative-valued
color and/or charge breaking minimum, 
global (GCCB) or local.
In the special case $M^{2} < 0$ the constraint cannot be satisfied
and there will be a negative-energy local CCB minimum.
%

%F
%
\begin{figure}[ht]
\begin{center}
\epsfxsize= 11.8 cm
\leavevmode
\epsfbox[54 175 565 575]{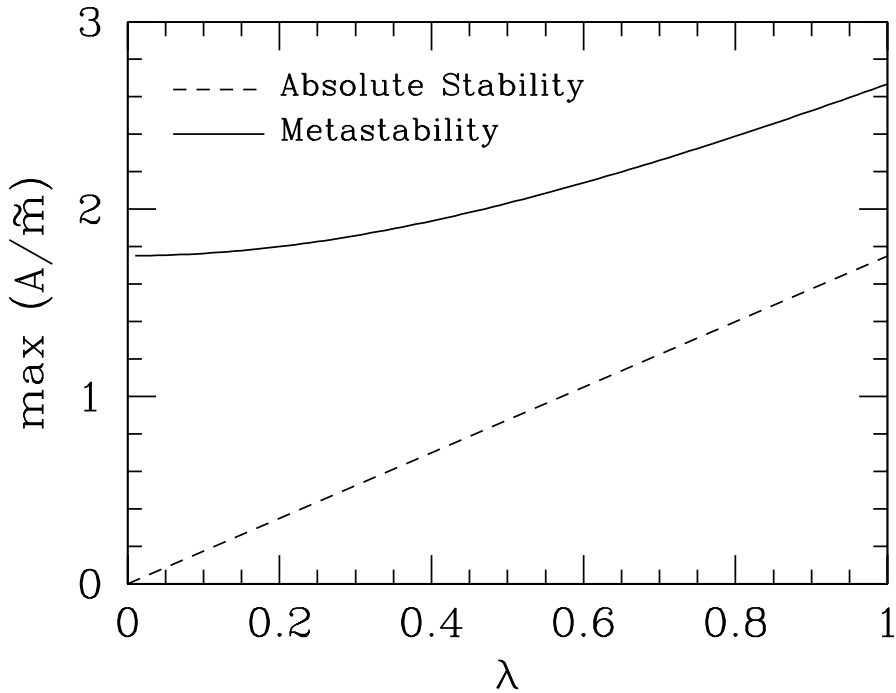}
\end{center}
\caption[f1]{Maximal value of $A/\widetilde{m}$ for absolute
 stability and metastability of tree-level scalar
 potential as a function of the square root of the quartic coupling, $\lambda$.
 Absolute stability refers to the absence of global
 charge-breaking minima, while metastability refers to a
 lifetime of the charge preserving vacuum greater than the
 age of the Universe, corresponding to a bounce action
 $S > 400$. Taken from Ref.~\cite{BFPT}.}
 \label{fig:bouncefig}
\end{figure}
%

%1215
Eq.~(\ref{master}) is illustrated 
(for $M^{2} = 3\widetilde{m}^{2}$ and $\Gamma = 2A$,
as is appropriate for $H_{2}^{0} = \tilde{t}_{L} =\tilde{t}_{R}$
and $H_{1}^{0} = 0$; see below)
by the lower curve 
in Fig.~\ref{fig:bouncefig} (taken from Ref.~\cite{BFPT}).
The Higgs mass term $ m_{i}^2$ receives a contribution from
the $\mu$-term.
Hence, $\widetilde{m}^{\,2} \gg   m_{\tilde{t}_{L},\,\tilde{t}_{R}}^2$
is possible in principle
with a large Dirac mass, $\mu^2 \gg m_{\tilde{t}_{L},\,\tilde{t}_{R}}^2$.
However, minimal
tuning of electroweak symmetry breaking (see Chapter~\ref{c8}) usually implies
$|m_{i}^2| \sim M_Z^2 \lesssim m_{\tilde{t}_{L},\,\tilde{t}_{R}}^2$.  
In what follows
$\widetilde{m} \sim m_{\tilde{t}_{L},\,\tilde{t}_{R}}$ is implicitly assumed.  
%

%1216
If eq.~(\ref{master}) does not hold, further investigation
is needed to determine whether the minimum is global
or local. As long as $m_{2}^{2}$ is the only possible 
negative mass-squared parameter then eq.~(\ref{master})
is satisfied for $\sin\beta = 0$, \ie there is no negative-valued
CCB minimum for $H_{2}^{0} = 0$.
We will therefore
restrict our attention to the case $H_{2}^{0} \neq 0$,
in which case it is convenient to reparameterize
$\phi = H_{2}^{0}$
and rescale all fields by an additional 
factor of $1/H_{2}^{0}$,
which simplifies the discussion.
%

%1217
One can derive analytic 
constraints from (\ref{master}), which are typically
relevant only for specific regions of the parameter space.
The negative contribution of the trilinear terms to the total
potential is maximized in a direction along which all scalar fields
{\bf that appear in the trilinear term} have equal expectation values
(and all other fields vanish), so that the 
$D$-terms $\propto 1/y_{t}^{2}$ may vanish\footnote{Note that 
unlike  the case of  $A$-terms discussed here,
the hypercharge $D$-term 
does not vanish along an equal-field direction 
$H_{1}^{0} = \tilde{t}_{L}=\tilde{t}_{R}$ ($H_{2}^{0} = 0$) 
corresponding to the operator $C H_{1}^{0\,*}\tilde{t}_{L}\tilde{t}_{R}$.
Therefore, models with $C$-terms are generically more stable.}.
For example, if we fix $H_{1}^{0} = 0$, $\tilde{t}_{L}=\tilde{t}_{R}=1$
(all in $H_{2}^{0}$ units),
then (\ref{master}) gives the well-known result \cite{ccb2}
\begin{equation}
A_{t}^{2} \leq 3(m_{\tilde{t}_{L}}^{2} +m_{\tilde{t}_{R}}^{2}
+ m_{H_{2}}^{2} + \mu^{2}).
\label{ccb3}
\end{equation}
The equal-field direction is not a relevant
requirement, however, for $y_{t} \approx 1$
(\ie there may exist a deeper minimum
with non-vanishing $D-$terms).
It is also obvious that eq.~(\ref{ccb3}) is trivial in the limit
$|\mu| \rightarrow \infty$.
A more useful constraint in this case \cite{ccb3}
is [taking  $H_{1}^{0} \approx 1$, $\tilde{t}_{L} \approx \tilde{t}_{R} 
 \equiv t \ll 1$
and using (\ref{master})]
\begin{equation}
\left( |A_{t}| + s|\mu| \right)^{2} \leq 
2(m_{\tilde{t}_{L}}^{2} +m_{\tilde{t}_{R}}^{2}).
\label{ccb4}
\end{equation}
Note that if the order $t^{2}$ corrections
to (\ref{ccb4}) are not negligible then $V$ is less negative,
which motivated our choice $t \ll 1$, \ie
the more dangerous direction. Other constraints are derived in different
parts of the parameter space, particularly when the whole
superpotential is considered.
%

%1218
Lastly, there are two caveats in our discussion above.
The first is that we used 
in this section the (one-loop 
improved) tree-level potential.
Calculations should therefore be performed
at the $\tilde{t}$-scale 
to eliminate large loop corrections.
Secondly, a GCCB minimum may be ``safe'' if separated from the local
standard-model minimum by a tunneling time greater
than the age of the universe.
%

%1219
Cosmological selection of the color and charge preserving
vacuum at the origin is natural, since this is a point of enhanced
symmetry and always a local minimum of the free energy at high enough
temperature.  It is sufficient therefore to ensure
a lifetime greater than the present age of the
universe. This  corresponds to a bounce action out of the metastable vacuum
of $S \gtrsim 400$~\cite{STABILITY}.  
The bounce action for the potential under discussion may be
calculated numerically~\cite{STABILITY,SARID} and it 
interpolates between the thin-wall limit, 
$\lambda^2 {\widetilde{m}^2}/A^2\rightarrow 1/3^-$, in which the two 
vacua are nearly degenerate
$$
S_{\rm thin} \simeq \frac{9 \pi^2}{2}
 \left(\frac{\widetilde{m}^{\,2}}{A^2}\right)
 \left( 1- 3 \lambda^2 \frac{\widetilde{m}^{\,2}}{A^2}\right)^{-3}\,,
$$
and the thick-wall limit,
$\left\vert\lambda^2 {\widetilde{m}^{2}}/{A^2}\right\vert \ll 1$,
in which the quartic term is not important
$$
S_{\rm thick} \simeq 1225
\left(\frac{\widetilde{m}^{\,2}}{A^2}\right) \,.
$$
(Here, we again assume for simplicity  $M^{2} = 3\widetilde{m}^{2}$, $\Gamma = 2A$,
and the equal field direction of eq.~(\ref{ccb3}).)
The maximum value of $A/\widetilde{m}$ for which $S > 400$ is shown in
fig.~\ref{fig:bouncefig} as a function of the square root of the quartic coupling, 
$\lambda$ (upper curve). (The figure is
taken from Ref.~\cite{BFPT} and the empirical fit to the numerically
calculated bounce action given in Ref.~\cite{SARID} was employed.)  The maximum
allowed $A/\widetilde{m}$ continues as a smooth monotonically
decreasing function for $\lambda^2 <0$ (as might be the case if
$\lambda^{2}$ arises along certain direction in field space
only at one loop), even though the renormalizable
potential is unbounded from below in this case.  For 
$|\lambda^2| \ll 1$ the weak requirement of metastability on a
cosmological time scale is met for $A/\widetilde{m} \lesssim 1.75$
($|\Gamma/\sqrt{M^{2}}| \lesssim 2$).
%

%12110
Recommended readings on the subject of vacuum stability
include Refs.~\cite{ccb1,STABILITY,CCB}.
%

%C13
\chapter{Supersymmetry Breaking Operators}
\label{c13}
%

%1301
While supersymmetric extensions of the Standard Model
can be fully described in terms of explicitly broken 
global supersymmetry, this description is only effective.
Once related to spontaneous breaking
in a more fundamental theory, the effective parameters
translate to functions of two distinct scales,
the scale of spontaneous supersymmetry breaking 
and the scale of its mediation to the standard-model fields.
Here, the scale dependence will be again written explicitly, 
and the full spectrum of supersymmetry breaking operators 
which emerges will be explored, expanding on the discussion in Chapter~\ref{c10}. 
%

%1302
Scale-dependent operators can play an important role
in determining the phenomenology.
For example, theories with low-energy supersymmetry breaking,
such as gauge mediation, may correspond to a scalar potential
which is quite different than in theories with  
high-energy supersymmetry breaking, such as gravity mediation.
As a concrete example, the Higgs mass prediction 
(Chapter~\ref{c8}) will be 
discussed in some detail and its upper bound will be shown to
be sensitive to the supersymmetry breaking scale.
%

%1303
Spontaneous supersymmetry breaking is most conveniently parameterized in terms
of a spurion field $X = \theta^{2}F_{X}$ with a non vanishing $F$-{\it vev},
which, as discussed previously, is an order parameter of (global) supersymmetry breaking. 
All SSB parameters can be written as non-renormalizable operators 
which couple the spurion $X$ to the SM superfields.
The operators are suppressed by the scale of
the mediation of supersymmetry breaking from the 
(hidden) sector, parameterized by the spurion, to the SM (observable) sector.
The coefficients of the various operators are dictated
by the nature of the interaction between the sectors (the agent) and
by the loop-order at which it occurs. In the following, we will
explicitly write all operators which couple the two sectors.
This will allow us to consider a general 
form of the supersymmetry breaking potential, which still
resolves the hierarchy problem. 
While we will again identify the operators 
which correspond to the SSB parameters,
our focus will be on those operators
which are not included in the minimal form of
the potential (\ref{VSSB}). 
%

%1304
Even though the SSB parameters appearing in (\ref{VSSB})
are functions of the relevant ultraviolet scales,
their magnitude is 
uniquely determined by the assumption that supersymmetry stabilizes
the weak scale against divergent quantum corrections,
$m_{\rm SSB} \sim {\cal{O}}(\mweak)\sim {\cal{O}}(100)$ GeV etc.   
This places a constraint on the ratio
of the supersymmetry breaking scale $\sqrt{F}$ and the scale of
its mediation $M$ such that $F/M \simeq \mweak$,
and it implies that no information
can be extracted
regarding either scale from the SSB parameters.
(Specific ultraviolet relations between
the various SSB parameters could
still be studied using the renormalization group formalism.)
The generalization of (\ref{VSSB}) introduces dimensionless
hard supersymmetry breaking (HSB) parameters in the potential.
Their magnitude depends on 
the various scales in a way which both
provides useful information (unlike the SSB parameters)
and does not destabilize
the solution to the hierarchy problem.

\section{Classification of Operators}
\label{sec:hard}
%

%1311
It is convenient to contain, without loss of generality,
all observable-hidden interactions in the non-holomorphic
K\"ahler potential
$K$, ${\cal{L}} 
= \int d^{2}\theta d^{2}\bar{\theta} K$ (which is not protected by
non-renormalization theorems).
We therefore turn to a general classification of $K$-operators,
originally presented in Ref.~\cite{N22}. (See also Refs.~\cite{SUSY30,Martin}.)
We do not impose any global symmetries, which can obviously eliminate
some of the operators, and we keep all operators which 
survive the superspace integration.

%1312
The derivation here differs from the one in Chapter.~\ref{c10} in that that
$(a)$ all operators are contained, for convenience, in the K\"ahler potential $K$
(and superpotential $W$ operators are rewritten as $K$ operators
$O_{W} \rightarrow X^{\dagger}O_{W}$, 
with the appropriate adjustment of power and size
of $M$); and $(b)$ the spurion is  assumed to have only a 
$F$-{\it vev}, $X = \theta^{2}F_{X}$.
In Chapter~\ref{c10} a non-singlet field $Z = \theta^{2}F_{Z}$
was considered as well. For simplicity, here we consider only
a singlet $X = \theta^{2}F_{X}$. In general, operators $\sim (XX^{\dagger})^{n}$
imply also $(ZZ^{\dagger})^{n}$-operators, with the possible
exception of $Z$ charged under non-linear symmetries.
Also, in order to not confuse gauginos $\lambda$ with quartic couplings,
the latter are denoted in this chapter by\footnote{The symbol $\kappa$ is usually reserved
for non-renormalizable couplings. Indeed, the quartic couplings here, 
though renormalizable themselves, arise
from integrating out non-renormalizable hidden-observable interactions.} 
$\kappa$.
%

%1313
The effective low-energy K\"ahler potential of a rigid 
$N = 1$ supersymmetry theory
is given by
\begin{eqnarray}
K &=&  K_{0}(X,X^{\dagger}) + K_{0}(\Phi,\Phi^{\dagger}) 
\nonumber \\
&+& 
\frac{1}{M}K_{1}(X,X^{\dagger},\Phi,\Phi^{\dagger})
\nonumber \\
&+& \frac{1}{M^{2}}K_{2}(X,X^{\dagger},\Phi,\Phi^{\dagger})
\nonumber \\
&+& 
\frac{1}{M^{3}}K_{3}(X,X^{\dagger},\Phi,\Phi^{\dagger},D_{\alpha},W_{\alpha})
\nonumber \\
&+& 
\frac{1}{M^{4}}K_{4}(X,X^{\dagger},\Phi,\Phi^{\dagger},D_{\alpha},W_{\alpha})
+ \cdots 
\label{Kgeneral}
\end{eqnarray}
where, as before, $X$ is the spurion and 
$\Phi$ are the chiral 
superfields of the low-energy theory. 
$D_{\alpha}$ is the covariant derivative with respect 
to the superspace chiral coordinate $\theta_{\alpha}$, 
and $W_{\alpha}$ is the gauge supermultiplet in its chiral 
representation (Ex.~4.7). 
Once a separation between supersymmetry breaking fields 
$X$ and low-energy
$\Phi$ fields is imposed
(\ie the assumption that supersymmetry is broken in a hidden sector),
there is no tree-level renormalizable
interaction between the two sets of fields, and their mixing
can arise only at the non-renormalizable level $K_{l \geq 1}$.
%

%1314
The superspace integration  ${\cal{L}}_{D} = 
\int d^{2}\theta d^{2}\bar{\theta}K$
reduces $K_{1}$ and $K_{2}$ to the usual SSB terms, as well as the 
superpotential $\mu$-parameter $W \sim \mu\Phi^{2}$, which were 
discussed in Chapter~\ref{c10}. It also contains Yukawa operators
$W \sim  y\Phi^{3}$ which can appear in the effective low-energy 
superpotential. These are summarized
in Tables~ \ref{table:o1} and~\ref{table:o1b}.
(We did not include linear terms that may appear (see Sec.~\ref{sec:hybrid1}
for an example) in the case of a low-energy singlet superfield $\Phi_{\rm singlet}$.) 
Finally, the last term in Table~\ref{table:o1} 
contains correlated quartic and Yukawa couplings. 
They are soft as they involve at most logarithmic divergences.
%

%1315
Integration over $K_{3}$ produces non-standard soft terms, 
for example, the $C$ (or $A^{\prime}$) terms.
These terms are soft unless the theory contains a pure singlet field,
in which case they can induce a quadratically divergent linear term.
They are summarized in Table~\ref{table:o2}.
The integration over $K_{3}$ 
also generates contributions to the (``standard'') 
$A$ and gaugino-mass terms.
These terms could arise at lower orders in $\sqrt{F}/M$ from integration 
over holomorphic functions 
(and in the case of $A$, also from $K_{1}$), e.g., eqs.~(\ref{gaugino}) and 
(\ref{Aterm}).
Note that in the presence of superpotential Yukawa couplings, a 
supersymmetry breaking
Higgsino mass term $\tilde{\mu}\widetilde{H}_{1}\widetilde{H}_{2}$
can be rotated to a combination of $\mu$ and $C-$terms,
and vice versa (Ex.~5.10).  
%

%1316
Lastly, superspace integration over $K_{4}$ leads to dimensionless
hard operators. These are summarized in Table~\ref{table:o3},
and will occupy the remaining of this chapter.
Table~\ref{table:o3} also contains supersymmetry breaking gauge-Yukawa
interactions $\sim\lambda\psi\phi^{*}$. This is equivalent to 
the HSB kinetic term for the gauginos which were discussed recently
in Ref.~\cite{KK}. (Note that HSB gaugino couplings \cite{SUPOBL,HIK} as well as
quartic \cite{HH} and other HSB couplings
are also generated radiatively in the presence of SSB.) 
%

%1317
Higher orders in $(1/M)$ can be safely neglected
as supersymmetry and the superspace integration allow only
a finite expansion in $\sqrt{F_X}/M$, that is ${\cal{L}}
= f[F^{n}_X/M^{l}]$ with $n \leq 2$ and
$l$ is the index $K_{l}$ in expansion eq.~(\ref{Kgeneral}). 
Hence, terms with $l > 4$ are suppressed by at least 
$(x /M)^{l-4}$.
We assume the limit $x \ll M$
for the supersymmetry preserving {\it vev}  $x$
so that all such operators can indeed be neglected
and the expansion is rendered finite.
%

%1318
It is interesting to identify two phenomenologically
interesting groups of terms in $K$,
$(i)$ those terms which can break the chiral symmetries
and can generate Yukawa terms in the low-energy effective theory,
and $(ii)$ new sources for quartic interactions.
%

%1319
The relevant chiral symmetry breaking 
terms in tables~\ref{table:o1} and~\ref{table:o2} 
can be identified with $A$ and $C$ terms
which couple the matter sfermions
to the Higgs fields of electroweak symmetry breaking.
The chiral symmetry breaking originates in this case in the
scalar potential and propagates to the fermions at one loop \cite{BFPT}.
More interestingly,
a generic K\"ahler potential is also found to contain 
tree-level chiral Yukawa couplings.
These include ${\cal{O}}(F_X/M^{2})$ supersymmetry conserving and SSB
couplings and ${\cal{O}}(F^{2}_X/M^{4})$ HSB chiral symmetry breaking 
couplings, leading to new avenues for fermion mass generation \cite{N22}.
%

%13110
Quartic couplings arise at  ${\cal{O}}(F_X/M^{2})$,
from supersymmetry conserving operators in  Table~\ref{table:o1}
(depending on $F_{\Phi}$), and at
${\cal{O}}(F^{2}_X/M^{4})$ from HSB
couplings in  Table~\ref{table:o3}. 
They can potentially alter the supersymmetry
conserving nature of the quartic potential, in general
(e.g., in (\ref{pot}),(\ref{lambda2}),(\ref{dterms})).
%

%13111
The relative importance of the HSB operators
relates to a more fundamental question:
What are the scales $\sqrt{F_X}$ and $M$? 
This will be addressed in an example below.
However, before doing so we need to address a different
question regarding the potentially destabilizing properties
of the different HSB operators,
which relates to the nature
of the cut-off scale $\Lambda_{\rm UV}$. 
Indeed, one has to confirm that a given theory  is not destabilized
when the hard operators  are included,
an issue which is interestingly model independent.
In order to do so, consider the implications of the hardness of the operators
contained in $K_{4}$. Yukawa and quartic couplings
can destabilize the scalar potential by corrections $\Delta m^{2}$ 
to the mass terms of the order of
\begin{equation}
\Delta m^{2} \sim \left\{\begin{array}{c}
\frac{\kappa}{16\pi^{2}}\Lambda^{2}_{\rm UV} \sim 
\frac{1}{16\pi^{2}}\frac{F^{2}_{X}}{M^{4}}\Lambda^{2}_{\rm UV} \sim 
%\frac{1}{16\pi^{2}}\frac{F^{2}_{X}}{M^{2}} \sim 
\frac{1}{16\pi^{2}c_{m}}m^{2}
\\
\\
\frac{y^{2}}{16\pi^{2}}\Lambda^{2}_{\rm UV}  \sim 
\frac{1}{16\pi^{2}}\frac{F^{4}_{X}}{M^{8}}\Lambda^{2}_{\rm UV}  \sim 
\frac{1}{16\pi^{2}c_{m}}m^{2}\frac{m^{2}}{M^{2}},
\end{array} \right.
\label{QD}
\end{equation}
where we identified $\Lambda_{\rm UV} \simeq M$ and $c_{m}$ is a dimensionless
coefficient omitted in Table~\ref{table:o1}, $m^{2}/2 = c_{m}F_{X}^{2}/M^{2}$.
The hard operators were substituted by the appropriate
powers of $F_{X}/M^{2}$.
Once $M$ is identified as the cut-off scale above which
the full supersymmetry is restored,
then these terms are harmless as the contributions are bound
from above by the tree-level  scalar  mass-squared parameters.
In particular, the softness condition imposed 
on the supersymmetry breaking terms in Chapter~\ref{c5} was
sufficient but not necessary. 
(This observation extends to the case of non-standard soft operators
such as $C \sim F_{X}^{2}/M^{3}$ in the presence of a singlet). 
%

%13112
In fact, such hard divergent corrections are well known in supergravity
with $\Lambda_{\rm UV} = M = M_{P}$, where they perturb
any given set of tree-level boundary conditions for the SSB 
parameters (see Sec.~\ref{sec:supergravity} and references therein).
Given the supersymmetry breaking scale in this case, $F \simeq \mweak M_{P}$,
the Yukawa (and quartic) operators listed below are 
proportional in these theories 
to $(\mweak/M_{P})^{n}$, $n=1,\,2$, 
and are often omitted.
Nevertheless, they can shift any boundary conditions
for the SSB by ${\cal{O}}(1 - 100\%)$ 
due to quadratically divergent one-loop corrections. 
%

%13113
We conclude that, in general, quartic couplings and chiral Yukawa couplings
appear once supersymmetry is broken, and if supersymmetry
is broken at low energy
then these couplings could be sizable yet harmless.
We will explore possible implications of the HSB quartic couplings in the next section.
In the next chapter we will touch upon some implications of
HSB chiral Yukawa couplings.
%
%
%

%T 
\begin{center}
\begin{table}
\caption{The soft supersymmetry breaking terms as operators
contained in $K_{1}$ and $K_{2}$. $\Phi = \phi + \theta\psi + \theta^{2}F$
is a low-energy superfield while $X$, $\langle F_{X} \rangle  \neq 0$, 
parameterizes supersymmetry breaking. 
$F^{\dagger} = \partial W/ \partial \Phi$.
The infrared operators are obtained by superspace integration
over the ultraviolet operators.
}
\label{table:o1}
\vspace*{0.3cm}
\begin{center}
\begin{tabular}{cc}\hline
Ultraviolet $K$ operator & Infrared ${\cal{L}}_{D}$ operator  \\ \hline
&\\
$\frac{X}{M}\Phi\Phi^{\dagger} + \hc$ & $A \phi F^{\dagger}+ \hc$ \\
&\\
$\frac{XX^{\dagger}}{M^{2}}\Phi\Phi^{\dagger}$ + \hc& 
$\frac{m^{2}}{2}\phi\phi^{\dagger} + \hc $ \\
&\\
$\frac{XX^{\dagger}}{M^{2}}\Phi\Phi + \hc $ & $B\phi\phi + \hc$ \\
&\\
$\frac{X^{\dagger}}{M^{2}}\Phi^{2}\Phi^{\dagger}
+ \hc $ & $\kappa\phi^{\dagger}\phi F + \hc$ \\
 & $y\phi^{\dagger}\psi\psi + \hc$ \\

\end{tabular}
\end{center}
\end{table}
\end{center}
%

%T
\begin{center}
\begin{table}
\caption{
The effective renormalizable
$N = 1$ superpotential $W$ operators
contained in $K_{1}$ and $K_{2}$, ${\cal{L}} = \int d^{2}\theta W$. 
Symbols are defined in Table \ref{table:o1}.
The infrared operators are obtained by superspace integration
over the ultraviolet operators.
}
\label{table:o1b}
\vspace*{0.3cm}
\begin{center}
\begin{tabular}{cc}\hline
Ultraviolet $K$ operator & Infrared $W$ operator  \\ \hline
&\\
$\frac{X^{\dagger}}{M}\Phi^{2}+ \hc$ & $\mu\Phi^{2}$ \\
&\\
$\frac{X^{\dagger}}{M^{2}}\Phi^{3}+ \hc$ & $y\Phi^{3}$ \\

\end{tabular}
\end{center}
\end{table}
\end{center}
%

%T
\begin{center}
\begin{table}
\caption{The non-standard or semi-hard  
supersymmetry breaking terms as operators
contained in $K_{3}$. 
$W_{\alpha}$ is the $N=1$ chiral
representation of the gauge supermultiplet and $\lambda$ is the respective
gaugino. $D_{\alpha}$ is the covariant derivative with respect to the 
superspace coordinate $\theta_{\alpha}$.
All other symbols are as in Table \ref{table:o1}.
The infrared operators are obtained by superspace integration
over the ultraviolet operators.
}
\label{table:o2}
\vspace*{0.3cm}
\begin{center}
\begin{tabular}{cc}\hline
Ultraviolet $K$ operator & Infrared ${\cal{L}}_{D}$ operator  \\ \hline
&\\
$\frac{XX^{\dagger}}{M^{3}}\Phi^{3} + \hc$ 
& $A\phi^{3} +\hc$ \\
&\\
$\frac{XX^{\dagger}}{M^{3}}\Phi^{2}\Phi^{\dagger} + \hc$ 
& $C\phi^{2}\phi^{\dagger} +\hc$ \\
&\\
$\frac{XX^{\dagger}}{M^{3}}D^{\alpha}\Phi D_{\alpha}\Phi + \hc $ & 
$\tilde{\mu}\psi\psi + \hc$\\ 
&\\
$\frac{XX^{\dagger}}{M^{3}}D^{\alpha}\Phi W_{\alpha} + \hc $ & 
$M^{\prime}_{\lambda}\psi\lambda + \hc$ \\ 
&\\
$\frac{XX^{\dagger}}{M^{3}}W^{\alpha} W_{\alpha} + \hc $ & 
$\frac{M_{\lambda}}{2}\lambda\lambda + \hc$ \\ 

\end{tabular}
\end{center}
\end{table}
\end{center}
%

%T
\begin{center}
\begin{table}
\caption{The dimensionless hard  
supersymmetry breaking terms as operators
contained in $K_{4}$. 
Symbols are defined as in Table~\ref{table:o1} and Table~\ref{table:o2}.
The infrared operators are obtained by superspace integration
over the ultraviolet operators.
}
\label{table:o3}
\vspace*{0.3cm}
\begin{center}
\begin{tabular}{cc}\hline
Ultraviolet $K$ operator & Infrared ${\cal{L}}_{D}$ operator  \\ \hline
&\\
$\frac{XX^{\dagger}}{M^{4}}\Phi D^{\alpha}\Phi D_{\alpha} \Phi + \hc$ 
& $y\phi\psi\psi  +\hc$ \\
&\\
$\frac{XX^{\dagger}}{M^{4}}\Phi^{\dagger}D^{\alpha}\Phi D_{\alpha} \Phi + \hc$ 
& $y\phi^{\dagger}\psi\psi  +\hc$ \\
&\\
$\frac{XX^{\dagger}}{M^{4}}\Phi D^{\alpha}\Phi W_{\alpha}  + \hc$ 
& $\bar{y}\phi\psi\lambda  +\hc$ \\
&\\
$\frac{XX^{\dagger}}{M^{4}}\Phi^{\dagger}D^{\alpha}\Phi W_{\alpha} + \hc$ 
& $\bar{y}\phi^{\dagger}\psi\lambda  +\hc$ \\
&\\
$\frac{XX^{\dagger}}{M^{4}}\Phi W^{\alpha} W_{\alpha} + \hc$ 
& $\bar{y}\phi\lambda\lambda  +\hc$ \\
&\\
$\frac{XX^{\dagger}}{M^{4}}\Phi^{\dagger}W^{\alpha} W_{\alpha} + \hc$ 
& $\bar{y}\phi^{\dagger}\lambda\lambda  +\hc$ \\
&\\
$\frac{XX^{\dagger}}{M^{4}}\Phi^{2}\Phi^{\dagger\, 2} + \hc$ 
& $\kappa(\phi\phi^{\dagger})^{2} +\hc$ \\
&\\
$\frac{XX^{\dagger}}{M^{4}}\Phi^{3}\Phi^{\dagger} + \hc$ 
& $\kappa\phi^{3}\phi^{\dagger} +\hc$ \\

\end{tabular}
\end{center}
\end{table}
\end{center}
\section{The Higgs Mass vs. The Scale of Supersymmetry Breaking}
\label{sec:hardhiggs}
%

%T
\begin{center}
%\begin{table*}[htb]
\begin{table}[t]
\caption{Frameworks for estimating $\kappa_{\rm hard}$.
(Saturation of the lower bound on $M$ is assumed.)}
\label{table:frameworks}
\begin{center}
\begin{tabular}{llll}
\hline
Framework &$\hat{\kappa}$& $M$ & ${\kappa}_{\rm hard}$  \\ \hline
&&& \\
TLM ($n= 0$)&$\sim 1$& $\gtrsim \widetilde{m}$  & $(\widetilde{m}/M)^{2} \sim 1$ \\
&&& \\
NPGM ($n=1/2$) & $\sim 1$& $\gtrsim 4\pi \widetilde{m}$  & $(4\pi \widetilde{m}/M)^{2} \sim 1$ \\
&&& \\
MGM  ($n = 1$) & $\lesssim \frac{1}{16\pi^{2}}$ & $\gtrsim 16\pi^{2}\widetilde{m}$ &
$(4\pi \widetilde{m}/M)^{2} \sim \frac{1}{16\pi^{2}}$ \\ 
&&& \\ \hline
\end{tabular}\\[2pt]
\end{center}
%\end{table*}
\end{table}
\end{center}
%

%1321
As shown above,
in general, HSB quartic couplings $\kappa_{\rm hard}$ arise
in the scalar potential
(from non-renormalizable operators in the K\"ahler potential, for example).
Assuming that the SSB
parameters are characterized by a parameter $\widetilde{m} \sim 1\, {\rm TeV}$
then
\begin{equation}
\kappa_{\rm hard} = 
\hat{\kappa}\frac{F^{2}}{M^{4}} \simeq
\hat{\kappa}(16\pi^{2})^{2n}\left(\frac{\widetilde{m}}{M}\right)^{2},
\label{lambdahard}
\end{equation}
where $M$ is a dynamically determined scale parameterizing the communication
of supersymmetry breaking to the SM sector, which is distinct
from the supersymmetry breaking scale $\sqrt{F} \simeq
(4\pi)^{n}\sqrt{\widetilde{m}M}$. 
The exponent $2n$ 
is the loop order at which the mediation of supersymmetry breaking 
to the (quadratic) scalar potential occurs. 
(Non-perturbative dynamics
may lead to different relations that can be described instead 
by an effective value of $n$.)
The coupling $\hat{\kappa}$ is an unknown dimensionless coupling
(for example, in the K\"ahler potential). As long as such quartic couplings
are not arbitrary but are related to the source of the SSB parameters
and are therefore described by (\ref{lambdahard}), then 
they do not destabilize the scalar potential
and do not introduce quadratic dependence on the ultraviolet cut-off
scale, which is identified with  $M$. 
This was demonstrated in the previous section.
Stability of the scalar potential only constrains
$\hat{\kappa} \lesssim \min{\left((1/16\pi^{2})^{2n-1}, 1\right)}$
(though calculability and predictability are diminished).
%

%1322
The $F$- and $D$-term-induced 
quartic potential (e.g. in eq.~(\ref{pot})) gives
for the (pure $D$-induced) 
tree-level Higgs coupling, $V = \kappa h^{4}$,
\begin{equation}
\kappa = \frac{g\prime^{2} + g^{2}}{4}\cos^{2}2\beta,
\label{lambda}
\end{equation}
where we work in the decoupling limit in which
one physical Higgs doublet $H$ is sufficiently heavy and
decouples from electroweak symmetry breaking 
while a second SM-like Higgs doublet is 
roughly given by $h \simeq H_{1}\cos\beta + H_{2}\sin\beta$.
(This was explained in Chapter~\ref{c8}.)
The HSB  coupling corrects this relation.
Given the strict tree-level upper bound that follows from (\ref{lambda}),
$m_{h^{0}}^{2} \leq M_{Z}^{2}\cos^{2}2\beta$, it is suggestive that
HSB may not be only encoded in, but also measured via, the Higgs mass.
We will explore this possibility, originally
pointed out and studied in Ref.~\cite{Su2}, in this section.
%

%1323
In the case that supergravity interactions mediate
supersymmetry breaking from some hidden sector (where supersymmetry
is broken spontaneously) to the SM sector, one has $M = M_{P}$.
The corrections are therefore negligible whether the mediation
occurs at tree level ($n=0$) or loop level ($n \geq 1$) and can be 
ignored for most purposes. (For exceptions, see discussion and references
of quantum effects in supergravity in Chapter~\ref{c10} as well as Ref.~\cite{Martin}.)
In general, however, the scale of supersymmetry breaking is an 
arbitrary parameter and depends on the dynamics that mediate the SSB
parameters.
For example, in the case of $N=2$ supersymmetry
(see the next chapter)
one expects $M \sim 1\,{\rm TeV}$ \cite{N22}. 
Also, in models with extra large dimensions
the fundamental $\mplanck$ scale can be as low as  a few TeV,
leading again to $M \sim 1\,{\rm TeV}$.
(For example, see Ref.~\cite{POM}.) A ``TeV-type'' mediation scale
implies a similar supersymmetry breaking scale and provides an
unconventional possibility.
(For a discussion, see Ref.~\cite{N22}.) 
If indeed $M \sim 1\,{\rm TeV}$ then
$\kappa_{\rm hard}$ given in (\ref{lambdahard}) is ${\cal{O}}(1)$
(assuming tree-level mediation (TLM) and
${\cal{O}}(1)$ couplings $\hat{\kappa}$ in the K\"ahler potential).
The effects on the Higgs mass must be considered in this case.
%

%1324
Though one may argue that TLM models represent a theoretical extreme,
this is definitely a viable possibility.
A more familiar and surprising example is given by
the (low-energy) gauge mediation (GM) framework. (See Sec.~\ref{sec:GM}). 
In GM, SM gauge loops communicate between
the SM fields and some messenger sector(s), mediating the SSB potential.
The Higgs sector and the related operators, however, are poorly
understood in this framework \cite{muMESS} and therefore 
all allowed operators should be considered.
In its minimal incarnation (MGM) $2n = 2$, and $M \sim
16\pi^{2}\widetilde{m} \sim 100\, {\rm TeV}$ parameterizes both the mediation 
and supersymmetry breaking scales. 
The constraint (\ref{QD}) corresponds
to $\kappa_{\rm hard} \sim \hat{\kappa} \lesssim 1/16\pi^{2}$
and the respective contribution
to the Higgs mass could be comparable to the contribution of
the supersymmetric coupling (\ref{lambda}).
A particularly interesting case is that of non-perturbative messenger dynamics
(NPGM) in which case $n_{eff} = 1/2$,
$M \sim 4\pi \widetilde{m} \sim 10\, {\rm TeV}$ \cite{4pi}, 
and the constraint 
on $\hat{\kappa}$ is relaxed to 1. Now
$\kappa_{\rm hard}\lesssim 1$ terms could
dominate the Higgs mass. The various frameworks are summarized in
Table~\ref{table:frameworks}.
%

%1325
In order to address the $\beta$-dependence
of the HSB contributions (which is different from that of all other terms)
we recall the general two-Higgs-doublet model (2HDM).
The Higgs quartic potential was already given in a general form in eq.~\ref{s2V4}.
It can be written down as (e.g. see Ref.~\cite{HH})
\begin{eqnarray}
V_{\phi^{4}}&=&
\frac{1}{2}\kappa_1(H_1^{\dagger}H_1)^2
+\frac{1}{2}\kappa_2(H_2^{\dagger}H_2)^2 \nonumber \\
&+&\kappa_3(H_1^{\dagger}H_1)(H_2^{\dagger}H_2)
+\kappa_4(H_1 H_2)(H_2^{\dagger}H_1^{\dagger}) \nonumber \\
&+&\left\{\frac{1}{2}\kappa_5(H_1 H_2)^2+[
\kappa_6(H_1^{\dagger}H_1) \right. \nonumber \\
&+& 
\left. \kappa_7(H_2^{\dagger}H_2)]H_1 H_2+{ h.c.}\right\}.
\label{Vphi4}
\end{eqnarray}
In the decoupling limit it simply reduces to the
SM with one ``light'' physical Higgs boson $h^{0}$,
$m_{h^0}^2=\kappa\nu^2$,
$\kappa=c_{\beta}^4\kappa_1+s_{\beta}^4\kappa_2+
2s_{\beta}^2c_{\beta}^2(\kappa_3+\kappa_4+\kappa_5)+
4c_{\beta}^3s_{\beta}\kappa_6+4c_{\beta}s_{\beta}^3\kappa_7$,
where $s_{\beta} \equiv \sin\beta$ and $c_{\beta} \equiv \cos\beta$,
and $\nu = \langle h^{0} \rangle = 174$ GeV is the SM Higgs {\it vev}
(normalized consistently with Chapter~\ref{c8}).
%

%1326
Allowing additional HSB quartic terms besides the 
usual gauge ($D$-)terms and loop contributions, 
$\kappa_{1\ldots{7}}$ can be written out 
explicitly as 
\begin{eqnarray}
\kappa_{1,2}&=&\frac{1}{2}(g\prime^2+g^2)+\kappa_{\rm soft\,1,2}
+\kappa_{\rm hard\,1,2},\\
\kappa_3&=&-\frac{1}{4}(g\prime^2 - g^2)+\kappa_{\rm soft\, 3}
+\kappa_{\rm hard\, 3}, \\
\kappa_4&=&-\frac{1}{2}g^2+\kappa_{\rm soft\, 4}
+\kappa_{\rm hard\, 4}, \\
\kappa_{5,6,7}&=& \kappa_{\rm soft\, 5,6,7}
+\kappa_{\rm hard\, 5,6,7},
\end{eqnarray}
where $g\prime$ and $g$ are again the SM hypercharge and $SU(2)$ gauge couplings, and 
$\kappa_{{\rm soft}\,i}$ sums the loop effects
due to soft supersymmetry breaking effects $\sim \ln{\widetilde{m}}$, eqs.~(\ref{mh}),
(\ref{mixing}).
The effect of the
HSB contributions $\kappa_{{\rm hard}\,i}$ 
is estimated next.
%

%1327
While Ref.~\cite{Su2} explores the individual contribution
of each of the $\kappa_{\rm hard}$ couplings,
here we will assume, for simplicity, that 
$\kappa_{hard\, i} = \kappa_{\rm hard}$ are all equal
and positive.
The squared Higgs mass $m_{h^0}^{2}$ reads in this case
\begin{equation}
m_{h^0}^2 =
M_Z^2\cos^{2}{2\beta}+\delta{m}^{2}_{\rm loop}+(c_{\beta}+s_{\beta})^4\nu^2
\kappa_{\rm hard},
\label{mhall}
\end{equation}
where $\delta{m}^{2}_{\rm loop} \lesssim M_{Z}^{2}$. 
(Note that no new particles or gauge interactions were introduced.)
%

%1328
Given the relation (\ref{mhall}), one can
evaluate the HSB contributions to the Higgs mass for
an arbitrary $M$ (and $n$). 
We define an effective scale
$M_{*} \equiv (M/(4\pi)^{2n}\sqrt{\hat{\kappa}})
({\rm TeV} /\widetilde{m})$.
The HSB contributions decouple  
for $M_{*} \gg  \widetilde{m}$, and the results reduce to the MSSM limit
with only SSB (e.g. supergravity mediation).
However, for smaller values of $M_{*}$ the Higgs mass is 
dramatically enhanced.  For $M$=1 TeV and TLM or 
$M$=$4\pi$ TeV and NPGM, both of which correspond to $M_{*} \simeq 1$ TeV, 
the Higgs mass  could be as heavy as  475 GeV for 
$\tan\beta=1.6$ and 290 GeV for $\tan\beta=30$.
This is to be compared with 104 GeV and 132 GeV \cite{carena}, 
respectively, if HSB are either ignored or negligible.
(A SM-like Higgs boson $h^{0}$ may be as heavy as 180 GeV
in certain $U(1)^{\prime}$ models \cite{LPW} with only SSB.)
%

%1329
In the MGM case $\hat{\kappa} \lesssim 1/16\pi^{2}$ so that
$M_{*}\sim 4\pi\, {\rm TeV}$
(unlike the NPGM where $M_{*} \sim 1$ TeV).
HSB effects are now more moderate but can increase the Higgs mass by 40 (10) 
GeV for $\tan\beta=1.6\ (30)$ (in comparison to the case with only SSB.)
Although the increase in the Higgs mass in this case
is not as large as in the TLM and NPGM
cases, it is of the same order of magnitude as, 
or larger than, the two-loop 
corrections due to SSB \cite{carena} 
(which are typically of the order of a few GeV), 
setting the uncertainty range on any such calculation.
Also, it is more difficult now to set a
model-independent lower bound on $\tan\beta$ based on the Higgs mass.
%

%Figure
%
\begin{figure}[t]
\begin{center}
\epsfxsize= 7.5 cm
\leavevmode
\epsfbox[40 180 520 600]{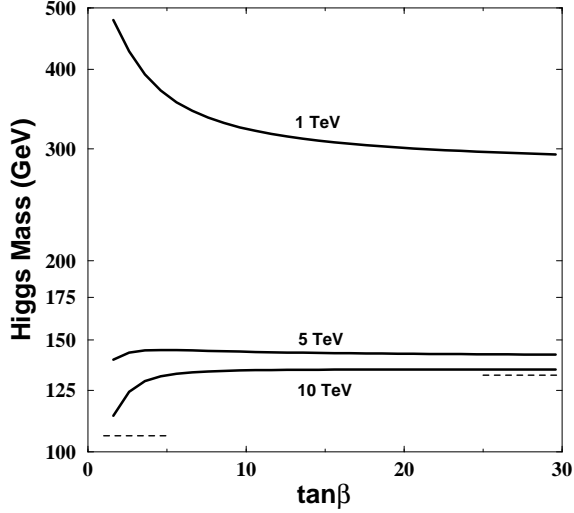
}
\end{center}
\caption[f1]{The light Higgs boson mass (note the logarithmic scale)
is shown as a function of $\tan\beta$ for $M_{*}= 1,\,5,\,10$ TeV
(assuming equal HSB couplings).
The upper bound when considering only SSB ($M_{*} \rightarrow \infty$)
is indicated for comparison (dashed lines)
for $\tan\beta = 1.6$ (left) and $30$ (right).
Taken from Ref.~\protect\cite{Su2}.}
\label{fig:m1510}
\end{figure}
%

%13210
In Fig.~\ref{fig:m1510}, $m_{h^{0}}$ dependence on $\tan\beta$ for fixed 
values of $M_{*}$
is shown.  The $\tan\beta$ dependence is from
the tree-level mass and from the HSB corrections, 
while the loop corrections to $m_{h^{0}}^2$ are fixed, for simplicity,
at $9200\, {\rm GeV}^2$ \cite{carena}.
The upper curve effectively corresponds to $\kappa_{\rm hard} \simeq 1$.
The HSB contribution dominates the Higgs mass and 
$m_{h^{0}}$ decreases with increasing $\tan\beta$.  As indicated above,
$m_{h^{0}}$ could be in the range of 300-500 GeV,
dramatically departing from  calculations
which ignore HSB terms.
The lower two curves illustrate the range\footnote{
Given the many uncertainties, e.g. the messenger quantum numbers
and multiplicity and $\sqrt{F}/M$ \cite{MESSE}, we identify the MGM with
a $M_{*}$-range which corresponds to a factor of two
uncertainty in the hard coupling.
} 
of the corrections in the MGM, 
where the tree-level 
and the HSB contributions compete.  
The $\cos2\beta$ dependence of the tree-level term
dominates the $\beta$-dependence of these two curves. 
Clearly, 
the Higgs mass could discriminate 
between the MGM and NPGM and help to better
understand the origin of the supersymmetry breaking. 
%

%13211
{}Following the Higgs boson discovery, 
it should be possible to extract information on the mediation scale $M$.
In fact, some limits can already be extracted.
Consider the upper bound on the Higgs mass derived from a fit to electroweak 
precision data: $m_h^{0}\ <\ 215$ GeV at 95$\%$ confidence level \cite{PDG}.
(Such fits are valid in the decoupling limit discussed here.) 
A lower bound on the scale $M$ in MGM could be obtained 
by rewriting eq.~(\ref{mhall}) as
\begin{equation}
m_Z^2\cos^22\beta  + \delta{m}^2_{\rm loop} +
(c_{\beta} + s_{\beta})^4\nu^2\left(\frac{4\pi{m}_0}{M}\right)^2 
\leq  (215\ {\rm GeV})^2,
\label{limit}
\end{equation}
assuming equal $\kappa_{\rm hard}$'s.  
For $\tan\beta=1.6$, it gives 
$M\geq$ 31 TeV while for $\tan\beta=30$ the lower bound is $M\geq$ 19 TeV.
Once $m_{h^{0}}$ is measured, 
more stringent bounds on $M$ could be set. 
%

%13212
Our discussion illustrates that the scale of the mediation of
supersymmetry breaking explicitly appears in the prediction
of the Higgs mass (and with a distinct $\beta$-dependence).
In turn, it could lead in certain cases
to a much heavier Higgs boson than usually anticipated
in supersymmetric theories.
It could also distinguish models, e.g. supergravity mediation from 
other low-energy mediation and 
weakly from strongly interacting messenger sectors.
Given our ignorance of the (K\"ahler potential and) HSB terms,
such effects can serve for setting the uncertainty on any
Higgs mass calculations and can be used to
qualitatively constrain the scale of mediation of supersymmetry breaking
from the hidden to the SM sector. 
%
%
%
%

%C14
\chapter{$N=2$ Supersymmetry}
\label{c14}

%
%

%1411
We conclude these notes by considering an extended supersymmetry framework,
specifically, $N=2$ supersymmetry, and by further entertaining the possibility
of embedding the SM in such a framework.
%

%1412
The  N=2 supersymmetry algebra has two spinorial generators 
$Q^i_{\alpha}$, $i=1,2$, 
satisfying 
\begin{equation}
\{{Q}_{\alpha}^{i}, {{\bar{Q}}}_{\dot{\beta}}^{j}\}=
2\sigma^{\mu}_{\alpha\dot{\beta}}P_{\mu}\delta^{ij},
\end{equation}
where $\sigma^{\mu}$ are, as usual, the Pauli matrices 
and $P_{\mu}$ is the momentum. 
The $N=2$ theory can be described in the $N=1$ formulation, as we will do here,
as long as one imposes a non-Abelian $R$-symmetry, the exchange $SU(2)_{R}$ symmetry, 
on the $N=1$ description. The supercharges ${Q}_{\alpha}^{i}$
form a doublet of the (exchange) $SU(2)_{R}$ $R$-symmetry, and the $N=1$
description can be viewed as a rotation acting on it (while the Lagrangian preserves the 
full symmetry).
%

%1413
The lowest $N =2$ spin representations, which are the relevant 
ones for embedding the SM, 
are the hypermultiplet and vector multiplet. 
Written in the familiar $N=1$ language, the  hypermultiplet is 
composed of two $N=1$ chiral multiplets $X=(\phi_{x},\ \psi_x)$ and $Y=(\phi_{y},\ \psi_y)$,
with $Y$ occupying  representation ${\cal{R}}$ of the gauge groups
which are conjugate to that of $X$, ${\cal{R}}(X) = {\cal{R}}(Y^{\dagger})$.
Schematically, the hypermultiplet is described by a ``diamond'' plot
\vspace*{0.25cm}
\begin{center}
\begin{tabular}{cccccccc}
&&$\psi_{x}$ &&&&& $+\frac{1}{2}$\\
&$\nearrow$&&&&&&\\
$\phi_{x}$ &&&& $\phi_{y}^{\dagger}$&&&$0$ \\
&&&$\swarrow$&&&&\\
&&$\bar{\psi}_{y}$&&&&&$-\frac{1}{2}$ 
\end{tabular}
\end{center}
\vspace*{0.25cm}
where the first, second and third rows
correspond to helicity $-1/2$, $0$, and $+1/2$ states, respectively.
The vector multiplet contains a  $N=1$ vector multiplet 
$V=(V^{\mu}, \lambda)$, where  $\lambda$ is a gaugino,
and a $N=1$ chiral multiplet $\adjoint=(\sadjoint,\ 
\fadjoint)$ in the adjoint representation of the gauge group (or a singlet
in the Abelian case).  Schematically, it is described by
\vspace*{0.25cm}
\begin{center}
\begin{tabular}{cccccccc}
&&$V^{\mu}$ &&&&& $1$ \\
&$\nearrow$&&&&&&\\
$\lambda$ &&&& $\fadjoint$&&&$\frac{1}{2}$ \\
&&&$\nearrow$&&&\\
&&$\sadjoint$ &&&&& $0$
\end{tabular}
\end{center}
\vspace*{0.25cm}
where the first, second and third rows
correspond to helicity $0$, $1/2$, and $1$ states, respectively.
The $N=1$ superfields are given by the two $45^{\circ}$ 
sides of each diamond (indicated by arrows),
with the gauge field arranging itself in its chiral representation $W_{\alpha}$.
The particle content is doubled in comparison to the $N=1$ supersymmetry case
and it is four times that of the SM: 
For each of the usual chiral fermions $\psi_{x}$ and
its complex-scalar partner $\phi_{x}$, there are a conjugate
mirror fermion $\psi_{y}$  and complex scalar  $\phi_{y}$
(so that the theory is vectorial).
For each gauge boson and 
gaugino, there is a {\it mirror} gauge boson  $\sadjoint$ and 
a {\it mirror} gaugino $\fadjoint$.  
%

%1414
The $N=0$ boson and fermion components of the 
hyper and vector-multiplet form $SU(2)_{R}$ representations.
States with equal helicity form a $SU(2)_{R}$ doublet $(\phi_{x},\phi_{y}^{\dagger})$
and an anti-doublet $(\fadjoint,\lambda)$, 
while all other states are $SU(2)_{R}$ singlets.
In fact, the full $R$-symmetry is $U(2)_{R}$
of which the exchange $SU(2)_{R}$ is a subgroup. 
There are additional $U(1)_{R}^{N=2}$, 
$U(1)_{J}^{N=2}$ subgroups such that the $R$-symmetry is either
$SU(2)_{R}\times U(1)_{R}^{N=2}$, or in some cases only
a reduced $U(1)_{J}^{N=2}\times U(1)_{R}^{N=2}$.
The different superfields $X \sim \phi_{x} +\theta\psi_{x}$, etc.
transform under the $U(1)$ 
symmetries with charges $R$ and $J$ given by
\begin{equation}
R({X}) = r = -R({Y}), \ \ \ R({\adjoint}) = -2, \ \ \ 
\end{equation}
\begin{equation}
J({X}) = -1 = J({Y}), \ \ \ J({\adjoint}) = 0, \ \ \ 
\end{equation}
and $R(W_{\alpha}) = J(W_{\alpha}) = -1$.
The (manifest) supercoordinate $\theta$ 
has, as usual, charge $R({\theta}) = J({\theta}) = -1$. 
%

%1415
The $SU(2)_{R}\times U(1)_{R}^{N=2}$ invariant $N=2$ Lagrangian can be written 
in the $N=1$ language as  
\begin{eqnarray}
{\cal L} =\int d^{2}\theta \left\{
\frac{1}{2g^2}W^{\alpha}W_{\alpha}+\sqrt{2}igY\adjoint{X} + h.c. \right\} 
&& 
\nonumber \\
%&+ &
+\int d^{2}\theta d^{2}\bar{\theta} \left\{
{\rm{Tr}}(
\adjoint^{\dagger}{\rm{e}}^{2gV}\adjoint{\rm{e}}^{-2gV}+X^{\dagger}
{\rm{e}}^{2gV}X+Y^{\dagger}{\rm{e}}^{-2gV^T}Y)\right\},
&&
\label{L}
\end{eqnarray}
where here we wrote the gauge coupling $g$ explicitly, the first (second)
integral is the $F$- ($D$-) term, and
we allow for non-Abelian gauge fields
(not included in Chapter~\ref{c4}):
 $\adjoint=\Phi^a_VT^a$ and $V=V^aT^a$, 
$T^a$ being the respective generators.
The second $F$-term is the superpotential. 
The only free coupling is the gauge coupling $g$: The coupling 
constant of the Yukawa term in the superpotential is 
fixed by the gauge coupling due to a global $SU(2)_{R}$. 
In particular, the $SU(2)_{R}$ symmetry forbids any chiral Yukawa terms
so that fermion mass generation is linked to supersymmetry breaking
(we return to this point below).
Note that the $U(1)_{R}^{N=2}$ forbids any mass terms $W\sim \mu^{\prime} XY$
(and the full $R$-symmetry forbids the usual $N=1$ $\mu$-term 
$W\sim \mu \hd\hu$.)
Unlike the $SU(2)_{R}$, $U(1)_{R}^{N=2}$ can survive supersymmetry breaking.
%

%1416
The $N=2$ Lagrangian (\ref{L}) also exhibits
several discrete symmetries, 
which  may or may not be broken in the broken supersymmetry regime.
There is a trivial 
extension of the usual $N=1$ $R$-parity
$Z_{2}$ symmetry which does not distinguish the 
ordinary fields from their mirror partners: 
\begin{equation}
\theta\rightarrow{-}\theta, \ \ \ X_M\rightarrow{-}X_M, \ \ \ 
Y_M\rightarrow{-}Y_M,
\end{equation}
where all other supermultiplets are $R_{P}$-even and
where the hypermultiplets have been divided into the odd matter multiplets
$(X_M,\ Y_M)$ and the even Higgs multiplets $(X_H,\ Y_H)$.
(Note that $V$ is even but $W_{\alpha}$ is odd.)
As in the $N=1$ case, all the ordinary and mirror quarks, leptons and Higgs 
bosons are $R_{P}$-even, while the ordinary and mirror
gauginos are $R_{P}$-odd. $R_{P}$ is conveniently used to define
the superpartners (or sparticles) as the $R_{P}$-odd particles. 
The LSP is stable if $R_{P}$ remains unbroken.
This was all discussed in Sec.~\ref{sec:s55}.
%

%1417
A second parity, called mirror parity ($M_{P}$), 
distinguishes the mirror particles
from their partners:
\begin{equation}\theta\rightarrow \theta, \ \ \
Y_M\rightarrow{-}Y_M, \ \ \ Y_H\rightarrow{-}Y_H, \ \ \ 
\adjoint\rightarrow{-}\adjoint,
\end{equation}
and all other superfields (including $W_{\alpha}$) are $M_{P}$-even.
It is convenient to use mirror parity to define the mirror particles
as the $M_{P}$-odd particles. (This definition should not be confused
with other definitions of mirror particles 
used in the literature and which are
based on a left-right group
$SU(2)_{L} \times SU(2)_{R}$ or a mirror world which interacts
only gravitationally with the SM world.)
The lightest mirror parity odd particle (LMP) 
is also stable in a theory with 
unbroken mirror parity.  However, if supersymmetry breaking
does not preserve mirror parity,  
mixing between the ordinary matter and the mirror fields is allowed. 
%

%1418
There is also a reflection (exchange) symmetry 
(which must be broken at low energies), the 
mirror exchange symmetry:
\begin{equation}
X\leftrightarrow{Y},\ \ \ \adjoint\leftrightarrow\adjoint^{\rm{T}}, \ \ \ 
V\leftrightarrow{-}V^{\rm{T}}.
\end{equation}
Like in the case of this continuous $SU(2)_{R}$,
if the reflection symmetry remains exact after supersymmetry breaking
then for each left-handed fermion  there would be a degenerate 
right handed mirror fermion in the conjugate gauge 
representation, which is phenomenologically not acceptable.
%

%1419
For easy reference, we list 
in Table~\ref{table:content} the minimal particle content of 
the MN2SSM: The minimal $N=2$ supersymmetric\footnote{Note that here
we discuss the case of global $N=2$ supersymmetry. In the local
case there are also two gravitini and the relation
between their mass and the SSB parameters is not obvious.} 
extension of the SM.  A mirror partner $Y$ ($\Phi_{V}$)
exists for every ordinary superfield $X$ ($V$) of the  $N = 1$  MSSM. 
One could eliminate one Higgs hypermultiplet and treat $\hd$ and $\hu$ as 
mirror partners.  However, this could lead to the spontaneous breaking of  
mirror parity when the Higgs bosons acquire {\it vev's}, 
and as a result, to more complicated mixing and radiative structures than 
in a theory with two Higgs hypermultiplets. 
%

%
%T
\begin{table}
\caption{Hypermultiplets and vector multiplets in the MN2SSM.
Our notation follows that of Table~\ref{table:mattersm}.}
\label{table:content}
\begin{tabular}{c|c|c}
&{$X\, / \,V$}&{$Y\, / \Phi_{V}$}\\ 
\hline &&\\
&$Q=(\sQ, Q)=({ 3},{ 2})_{ \frac{1}{6}}$&
$\mrQ=(\mrsQ, \mrQ)=({ \bar{3}},{ \bar{2}})_{-\frac{1}{6}}$
\\ &&\\ \cline{2-3} &&\\
Matter&$U=(\sU, U)=({ \bar{3}},{ 1})_{-\frac{2}{3}}$&
$\mrU=(\mrsU,\mrU)=({ {3}}, { {1}})_{\frac{2}{3}}$
\\ &&\\ \cline{2-3} &&\\
(hyper-)&$D=(\sD, D)=({ \bar{3}}, { 1})_{\frac{1}{3}}$&
$\mrD=(\mrsD,\mrD)=({ {3}}, { {1}})_{-\frac{1}{3}}$
\\&&\\ \cline{2-3} &&\\
multiplets&$L=(\sL, L)=({ 1}, { 2})_{-\frac{1}{2}}$&
$\mrL=(\mrsL, \mrL)=({ 1},{ \bar{2}})_{\frac{1}{2}}$
\\ &&\\ \cline{2-3} &&\\
&$E=(\sE, E)=({ 1}, { 1})_{1}$&
$\mrE=(\mrsE, \mrE)=({ {1}}, { {1}})_{-1}$
\\ &&\\ \hline &&\\
Higgs (hyper -)&$H_1 =(H_1, \hinod)=({ 1}, { {2}})_{-\frac{1}{2}}$& 
$\mrhd =(\mrhd, \mrhinod)=({ 1},{ \bar{2}})_{\frac{1}{2}}$
\\ &&\\ \cline{2-3} &&\\
multiplets&$H_2 = (H_2, \hinou)=({ 1},{ \bar{2}})_{\frac{1}{2}}$& 
$\mrhu = (\mrhu, \mrhinou)=({ 1},{ {2}})_{-\frac{1}{2}}$
\\ &&\\ \hline &&\\
Vector& $g = (g,\gluino)=({ 8},{ 1})_{0}$&
$\gadjoint = (\gsadjoint,\gfadjoint)=({ 8},{ 1})_{0}$ 
\\ &&\\ \cline{2-3} &&\\
multiplets&$W = (W,\wino)=({ 1},{ 3})_{0}$&
$\wadjoint = (\wsadjoint,\wfadjoint)=({ 1},{ 3})_{0}$ 
\\ &&\\ \cline{2-3} &&\\
&$B=(B,\bino)=({ 1},{ 1})_{0}$&
$\badjoint = (\bsadjoint,\bfadjoint)=({ 1},{ 1})_{0}$ 
\\ &&\\ \hline
\end{tabular}
\end{table}
%
%

%14110
For the above particle content, and 
imposing the full $U(2)_{R}$ on the superpotential, 
the theory is scale invariant and is given by
the superpotential (after phase redefinitions) 
\begin{eqnarray}
W/\sqrt{2}
&=&
g^{\prime}(\frac{1}{6}\mrQ\badjoint{Q}-\frac{2}{3}\mrU\badjoint{U}
+\frac{1}{3}\mrD\badjoint{D}-\frac{1}{2}\mrL\badjoint{L}
+\mrE\badjoint{E} \nonumber \\
&-&\frac{1}{2}\mrhd\badjoint{H}_1+\frac{1}{2}\mrhu\badjoint{H}_2)
\nonumber \\
&+&
g_2(\mrQ\wadjoint{Q}+\mrL\wadjoint{L}+\mrhd\wadjoint{H}_1+
\mrhu\wadjoint{H}_2) \nonumber \\
&+&
g_3(\mrQ\gadjoint{Q}+\mrU\gadjoint{U}+\mrD\gadjoint{D}). 
\label{WN2}
\end{eqnarray} 
After substitution in the Lagrangian (\ref{L}),
the superpotential (\ref{WN2}) gives rise in the usual manner 
to gauge-quartic and gauge-Yukawa interactions.
All interactions are gauge interactions!
Table~\ref{table:content} and the superpotential (\ref{WN2}) define
the MN2SSM (in the supersymmetric limit).
As expected, one missing ingredient in the vector-like
superpotential (\ref{WN2}) is chiral Yukawa
terms. This is the fermion mass problem of extended supersymmetry frameworks.

%1411
The fermion mass problem in these models has many facets. First and
foremost, the generation of any chiral spectrum must be a result of
supersymmetry breaking.  Secondly, the two sectors have to be
distinguished with sufficiently heavy mirror fermions and (relatively)
light ordinary fermions, with any mixing between the two sectors
suppressed, at least in the case of the first two families.  In
addition there are the issues of the heavy third family and of the
very light neutrinos in the ordinary sector, and subsequently, of
flavor symmetries and their relation to supersymmetry breaking.  One
possible avenue to address these issues is to utilize the HSB
Yukawa couplings of the previous section (in the framework of
low-energy supersymmetry breaking) \cite{N22}.  We note that 
unlike in the ($N = 1$) case with high-energy supersymmetry breaking,
it is possible in this case that only one Higgs doublet acquires a {\it vev}.  
this is because HSB Yukawa terms are not necessarily constrained by holomorphicity.
%

%1412
These and other technical challenges in embedding the SM in a $N=2$
framework are intriguing, but will not be pursued here.
The $N=2$ framework outlined here was originally proposed
in Ref.~\cite{N21}, and more recently in Ref.~\cite{N22}.
Experimental status of models with three additional families
was addressed in Ref.~\cite{N23}. 

\clearemptydoublepage

%%%%%%%%%%%%%%%%%%%%%%%%%%%%%%%%%%%%%%%%%%%%%%%%%%%%%%%%%%%
%% sample LaTeX REFERENCES file for English Physics Books
%% to be compiled via the sample file input.tex
%% Vers. 07/99
%%%%%%%%%%%%%%%%%%%%%%%%%%%%%%%%%%%%%%%%%%%%%%%%%%%%%%%%%%%

\clearemptydoublepage

%%Back_Matter%%%%%%%%%%%%%%%%%%%%%%%%%%%%%%%%%%%%%%%%%%%%%%%%%%%%%%%%
\addcontentsline{toc}{chapter}{Index}

\flushbottom
\printindex

%%%%%%%%%%%%%%%%%%%%%%%%%%%%%%%%%%%%%%%%%%%%%%%%%%%%%%%%%%%%%%%%%%%%%%

\end{document}